\documentclass{aa}  

\usepackage{graphicx}
\usepackage{txfonts}
\usepackage{xcolor}
\usepackage{ulem}

\defcitealias{Polletta2007}{P07}
\defcitealias{Gruppioni2013}{G13}
\defcitealias{Vallini2015}{V15}
\defcitealias{Bisigello2021}{B21}
\usepackage[flushleft]{threeparttable}
\usepackage{scrextend}
\usepackage{multirow}
\usepackage{placeins}
\usepackage{longfigure}

\begin{document} 

\title{Delving deep: a population of extremely dusty dwarfs observed by JWST} 

\author{L. Bisigello
          \inst{1,2}\thanks{laura.bisigello@unipd.it} \and
          G. Gandolfi \inst{3,4,5} \and
          A. Grazian\inst{2} \and
          G. Rodighiero\inst{1,2} \and
          L. Costantin\inst{6} \and
          A. R. Cooray\inst{7} \and
          A. Feltre\inst{8} \and
          C. Gruppioni\inst{8} \and
          N. P. Hathi\inst{9} \and
          B. W. Holwerda\inst{10} \and
          A. M. Koekemoer\inst{9} \and
          R. A. Lucas\inst{9} \and
          J. A. Newman\inst{11} \and
          P. G. P\'erez-Gonz\'alez\inst{6} \and
          L. Y. A. Yung\inst{12} \and
          A. de la Vega\inst{13} \and
          P. Arrabal Haro\inst{14} \and
          M. B. Bagley\inst{15} \and
          M. Dickinson\inst{14} \and
          S. L. Finkelstein\inst{15} \and
          J. S. Kartaltepe\inst{16} \and
          C. Papovich\inst{17,18} \and
          N. Pirzkal\inst{19} \and
          S. Wilkins\inst{20,21}
          }

\institute{Dipartimento di Fisica e Astronomia, Università di Padova, Vicolo dell'Osservatorio, 3, I-35122, Padova, Italy
\and
INAF--Osservatorio Astronomico di Padova, Vicolo dell'Osservatorio 5, I-35122, Padova, Italy
\and
Scuola Internazionale Superiore Studi Avanzati (SISSA), Physics Area, Via Bonomea 265, 34136 Trieste, Italy
\and
Institute for Fundamental Physics of the Universe (IFPU), Via Beirut 2, 34014 Trieste, Italy
\and
Istituto Nazionale Fisica Nucleare (INFN), Sezione di Trieste, Via Valerio 2, 34127 Trieste, Italy
\and
Centro de Astrobiolog\'ia (CAB), CSIC-INTA, Ctra de Ajalvir km 4, Torrej\'on de Ardoz, 28850, Madrid, Spain 
\and
Department of Physics \& Astronomy, University of California, Irvine, 4129 Reines Hall, Irvine, CA 92697, USA
\and
INAF-Osservatorio di Astrofisica e Scienza dello Spazio, via Gobetti 93/3, I-40129, Bologna, Italy
\and
Space Telescope Science Institute, 3700 San Martin Dr., Baltimore, MD 21218, USA
\and
Physics \& Astronomy Department, University of Louisville, 40292 KY, Louisville, USA
\and
Department of Physics and Astronomy and PITT PACC, University of Pittsburgh, Pittsburgh, PA 15260, USA
\and
Astrophysics Science Division, NASA Goddard Space Flight Center, 8800 Greenbelt Rd, Greenbelt, MD 20771, USA
\and
Department of Physics and Astronomy, University of California, 900 University Ave, Riverside, CA 92521, USA
\and
NSF's National Optical-Infrared Astronomy Research Laboratory, 950 N. Cherry Ave., Tucson, AZ 85719, USA
\and
Department of Astronomy, The University of Texas at Austin, Austin, TX, USA
\and
Laboratory for Multiwavelength Astrophysics, School of Physics and Astronomy, Rochester Institute of Technology, 84 Lomb Memorial Drive, Rochester, NY 14623, USA
\and
Department of Physics and Astronomy, Texas A\&M University, College Station, TX, 77843-4242 USA
\and
George P.\ and Cynthia Woods Mitchell Institute for Fundamental Physics and Astronomy, Texas A\&M University, College Station, TX, 77843-4242 USA
\and
ESA/AURA Space Telescope Science Institute
\and
Astronomy Centre, University of Sussex, Falmer, Brighton BN1 9QH, UK
\and
Institute of Space Sciences and Astronomy, University of Malta, Msida MSD 2080, Malta
}

   \date{Received ; accepted }

 
  \abstract
   {}
   {We take advantage of the NIRCam photometric observations available as part of the Cosmic Evolution Early Release Science survey (CEERS) to identify and analyse very red sources in an effort to discover very dusty star forming galaxies. }  
   {We select red galaxies as objects with a $S/N>3$ at 4.4 $\mu$m and a $S/N<2$ in all JWST and \textit{HST} filters at $\lambda\leq2\mu$m, which corresponds to $\rm[F200W]-[F444W]>1.2$ considering CEERS depths. This selection is ideal to identify very dusty ($A_V>1$ mag) galaxies with stellar masses between $10^6$ to $10^{10}\, \rm M_{\odot}$ at $z<5$, more massive dusty galaxies at $z=5-18$ and galaxies at $z>18$ due to the Lyman absorption, independently of their dust extinction.}
   {Our sample of F200W-dropouts contains no strong candidates at $z>6.5$, instead it consists almost completely ($\sim81\%$) of $z<2$ low-mass galaxies, with a median stellar mass of $10^{7.3} \rm M_{\odot}$. These galaxies show an exceptional dust extinction with median value of $A_{V}=$4.9 mag, completely unexpected given their low stellar mass. The remaining galaxies, which are at z<6.5, show similar large dust extinction ($A_V>1$), but they are generally more massive $>10^{7.5}\rm M_{\odot}$.}
   {}


   \maketitle
%
\section{Introduction}
Dust is an ubiquitous component of galaxies. It has a significant impact on the observed spectra as well as on the derived physical properties and, in turn, on our understanding of galaxy evolution. Indeed, around half of the cumulative energy emitted by stars at all epochs is absorbed by the surrounding dust and re-emitted at infrared (IR) wavelengths \citep{Hauser2001}. Therefore, the dust content and its evolution with redshift remains one of the main topic in extragalactic astrophysics. \par
For example, dust attenuation is one of the major uncertainties when deriving the cosmic star-formation-rate density (SFRD), particularly, but not only, at $z>3$ \citep{Madau2014}. On the one hand, studies based on rest-frame ultra-violet (UV) data \citep[e.g.][]{Bouwens2015,Oesch2018}, which are largely affected by dust but have been possible up to $z=10$, have reported a sharp decline of the SFRD at $z>2$. On the other hand, far-IR or radio observations \citep[][]{Rowanrobinson2016,Novak2017,Gruppioni2020,Enia2022}, which have been generally limited to $z<7$ due to observational challenges, have shown that the SFRD may show a less drastic decline, or even a plateau, at $z>2$. At the same time, the comparison between IR and rest-frame UV and optical observations have highlighted the presence of sources that were completely missed by even the deepest \textit{HST} observations \citep[e.g.][]{Simpson2014,Franco2018,Wang2019,Talia2021,Gruppioni2020}. The presence of such sources show how the absence of deep IR observations may have biased our past view of the stellar mass assembly of the Universe. Moreover, dusty galaxies are interlopers when searching for $z>10$ galaxies, as their red continuum could be misinterpreted as a Lyman-break at ultra-high redshift \citep{Wilkins2014,Holwerda2015,Holwerda2020,Zavala2022}. \par
The dust attenuation present in a star-forming galaxy has a positive correlation with the stellar mass \citep[e.g.][]{Pannella2015,McLure2018}, with more massive galaxies being in general more dust attenuated and galaxies below 10$^{8.5}\,\rm M_{\odot}$ having a little dust attenuation. This is not surprising given that a large stellar mass generally correspond to intense star-formation activity \citep{Brinchmann2004,Noeske2007} and, therefore, to large dust content, as the dust is produced by stellar activity such as pulsating moderate mass stars and supernovae \citep[e.g.][]{Scalo1980}. However, the search for dusty low-mass galaxies requires very deep IR observations, which are not possible with \textit{HST} or \textit{Spitzer}, as they are extremely faint in the rest-frame optical given the combination of low stellar mass and high dust extinction. In addition, the mentioned relation between stellar mass and dust extinction seems not to evolve with redshift up to $z\sim3$ \citep[e.g.,][]{Whitaker2017,Shapley2022}, with a possible decrease at higher redshifts \citep[e.g.][]{Fudamoto2020}. 
\par
The launch of JWST has opened, as expected, a new observational window in the Universe. On the one hand it has allowed for identifying a large number of $z>10$ candidates just in the first few months of observations \citep[e.g.][]{Adams2022,Carnall2022,Castellano2022,Donnan2022,Finkelstein2022a,Naidu2022,PerezGonzales2023b}, some of which have shown a puzzling, high content of dust \citep{Rodighiero2023}. On the other hand, it has enabled statistical investigation of the nature of optically-faint and optically-dark galaxies \citep{Barrufet2022,PerezGonzales2023}, which was impossible to do with previous near-IR observatory such as \textit{Spitzer}. In particular, such analysis can be pushed now to intermediate and small stellar masses, without being limited to the most massive and IR-luminous galaxies. \par
In this work we take advantage of some of the first photometric observations obtained by JWST to study some of the reddest galaxies observable. Such red colours could be caused by a high redshift or large dust content. In particular, we consider the first observations of the Cosmic Evolution Early Release Science (CEERS\footnote{\href{https://ceers.github.io/}{https://ceers.github.io/}}), which cover an area of $\sim$36 arcmin$^{2}$ that is larger than the area of the early release observations on SMACS0723 ($\sim9\, \rm arcmin^{2}$) or the GLASS-JWST early release science \citep[18 arcmin$^{2}$;][]{Treu2022}, thus allowing the selection of rare sources and increasing the number statistics of previous surveys. The structure of the paper is the following. In Section \ref{sec:data} we describe the used data and the adopted selection criterion. In Section \ref{sec:Physprop} we describe the method used to derive the physical property associated to each object and we analyse them in Section \ref{sec:results}. We then summarise our work and report our conclusive remarks in Section \ref{sec:summary}.
Throughout the paper, we assume a Chabrier initial mass function \citep{Chabrier2003} and a $\Lambda$CDM cosmology with $H_0=70\,{\rm km}\,{\rm s}^{-1}{\rm Mpc}^{-1} $, $\Omega_{\rm m}=0.3$ and $\Omega_\Lambda=0.7$. All magnitudes are in the AB system \citep{Oke1983}.

\section{Data and sample selection}\label{sec:data}
\subsection{CEERS survey and catalogues}
CEERS is a JWST early release science (ERS) program (Proposal ID $\#$1345, Finkelstein et al. \textit{in prep.}). CEERS covers approximately 100 arcmin$^{2}$ in the Extended Groth Strip field (EGS) with both imaging and spectroscopic observations. In particular, it consists of 10 pointings with the Near Infrared Camera (NIRCam) combined with 6 pointings with the Near InfraRed Spectrograph instrument (NIRSpec) and 8 pointings with the Mid InfraRed Instrument (MIRI). \par
In the same field there are also archival \textit{HST} observations from the All-wavelength Extended Groth Strip International Survey \citep{Davis2007}, the Cosmic Assembly Deep Extragalactic Legacy Survey \citep[CANDELS;][]{Grogin2011,Koekemoer2011} and 3D-\textit{HST} \citep{Momcheva2016}. These catalogues have been updated (v1.9\footnote{\href{https://ceers.github.io/releases.html}{https://ceers.github.io/releases.html}}) aligning their astrometry with that of Gaia DR3. \par
For this work we consider the CEERS NIRCam catalogue presented in \citet{Finkelstein2023}. In particular, the catalogue is based on the NIRCam images of the first CEERS public data release \citep[Data Release 0.5;][]{Bagley2022}, taken as part of CEERS on the 21 June 2022. These images, for which no complete overlapping MIRI observations are available, consists of four pointings (1, 2, 3 and 6) in three short wavelength filters (i.e. F115W, F150W, and F200W)  and four long-wavelength bands (i.e. F277W, F356W, F410M, and F444W). Source detections were based on the inverse-variance-weighted sum of the F277W and F356W images, after matching them by their point-spread-function (PSF) and removing the background. PSFs were derived stacking observed stars present in the field.\par
Photometric measurements were performed after matching all images to the F444W PSF, which has the largest full-width at half-maximum (FWHM) among the considered JWST filters. In particular, we made use of the PyPHER routine \citep{Boucaud2016} to derive the kernel necessary to match each PSFs to the F444W one and then we convolved each image with its respective kernel. This process was applied also to \textit{HST}/ACS F606W and F814W imaging, as their PSF FWHMs are smaller than the F444W one. We instead did not convolve the \textit{HST}/WFC3 band as their PSF FWHMs are larger than F444W one, but we applied an additional aperture correction to take this into account. \par
More details on the catalogue creation and image reductions are presented in \citet{Finkelstein2023} and \citet{Bagley2022}, respectively, while we list in Table \ref{tab:depths} the filters included in the catalogue and the respective observational depths. The latter were derived, as explained in details in \citet{Finkelstein2022aa,Finkelstein2023}, by extracting from each image the fluxes inside non-overlapping circular apertures of different radius, from 0.1" to 3", all placed avoiding real sources. These fluxes were then used to derived how the noise varies as a function of the number of pixels in each aperture. The final 5$\sigma$ depths were estimated using the mentioned noise function in an aperture of diameter 0.2", then corrected to total flux using the PSF.\par

\begin{table}
    \centering
    \begin{tabular}{cc|cc}
    JWST & depth(5$\sigma$) & \textit{HST} & depth(5$\sigma$) \\
    \hline
       F115W & 29.2 & F606W & 28.6 \\
       F150W & 29.0 & F814W & 28.3 \\
       F200W & 29.2 & F105W & 27.1 \\
       F277W & 29.2 & F125W & 27.3 \\
       F356W & 29.2 & F140W & 26.7 \\
       F410M & 28.4 & F160W & 27.4 \\
       F444W & 28.6 & & \\
    \end{tabular}
    \caption{Summary of the filters and 5$\sigma$ depths of the catalogue considered in this work.}
    \label{tab:depths}
\end{table}
\subsection{Sample selection}\label{sec:selection}
We select galaxies that have a $\rm S/N>3$ in the F444W filter and a $\rm S/N<2$ in every filter equal or below 2 $\mu$m (i.e. \textit{HST}/F606W, \textit{HST}/F814W, \textit{HST}/F105W, \textit{HST}/F125W, \textit{HST}/F140W, \textit{HST}/F160W, JWST/F115W, JWST/F150W, and JWST/F200W)\footnote{Using a magnitude limit equivalent to a S/N$=2$ in all filters at $\lambda\leq2\mu$m, instead of a S/N cut, results in the selection of a different galaxy sample, because of local noise fluctuation. However, we verify with pointing 3 that the general properties of the samples selected using a S/N cut and a magnitude cut are consistent.}. Considering the CEERS observational depths this corresponds roughly to a colour cut [F200W]-[F444W]>1.2. We checked that results are consistent if we consider a $\rm S/N>5$ in the F444W filter, even if the resulting sample is reduced in number. We highlight here that this method does not select equally red galaxies at all magnitudes, as objects that are brighter in the F444W filter must be intrinsically redder in the [F200W]-[F444W] colour to meet our S/N limits. The selection method has been however chosen to include $z>10$ galaxies, without limiting the sample only to blue, relatively dust-free, objects. \par
We also limit our analysis to galaxies with a magnitude [F444W]$<29$ mag, which roughly corresponds to the 3$\sigma$ limit of the four pointings \citep{Bagley2022}. For this selection we consider the photometry derived with the Kron aperture, with a Kron factor and a minimum radius set to 1.1 and 1.6, respectively. These values were chosen via simulations to maximise the signal-to-noise value. We also correct all fluxes in order to retrieve the total flux, using larger apertures \citep[see][for further details]{Finkelstein2023}. We then visually checked each selected source to remove artefacts, for example, due to the image edges or to the presence of saturated stars checking also each co-added F277W, F356W, and F444W band images. Finally, we co-added the F115W, F150W and F200W band images and exclude every source with a S/N$\geq$2 in it. Since the original catalogue has been selected in the F277W+F356W co-added image, we checked for additional sources detected in the F444W filter and not present in catalogue by \citet{Finkelstein2023}, using the same extraction methodology and selection criteria. \par
The final sample of F200W-dropouts based on the catalogue by \citet{Finkelstein2023} consists of 135 objects, of which 15, 13, 48, and 59 are present in pointing 1, 2, 3 and 6, respectively. Among these sources, 128 have $S/N>3$ in at least two filters, indicating that they are not spurious sources. In addition, after removing artefacts (e.g., edges, cosmic rays, bad pixels) and sources with contaminated photometry, we found nine sources (three in pointing 2 and 3, two in pointing 1 and one in pointing 6) not present in the catalogue based on the F277W+F356W co-added image as they are faint in one or both of these bands. However, four of these sources are detected only in the F444W filter, so we do not attempt to derive any physical property for them. We call this sub-sample F200W-dropouts-extra hereafter. The total sample of F200W-dropout, including the F200W-dropouts-extra sub-sample, is therefore made of 144 objects. In the next sections we will analyse all 144 objects when looking at their colours, but we will consider only the 133 of them detected in at least two filters when performing the SED fitting. \par
\begin{figure*}
    \centering    \includegraphics[width=0.99\linewidth,keepaspectratio]{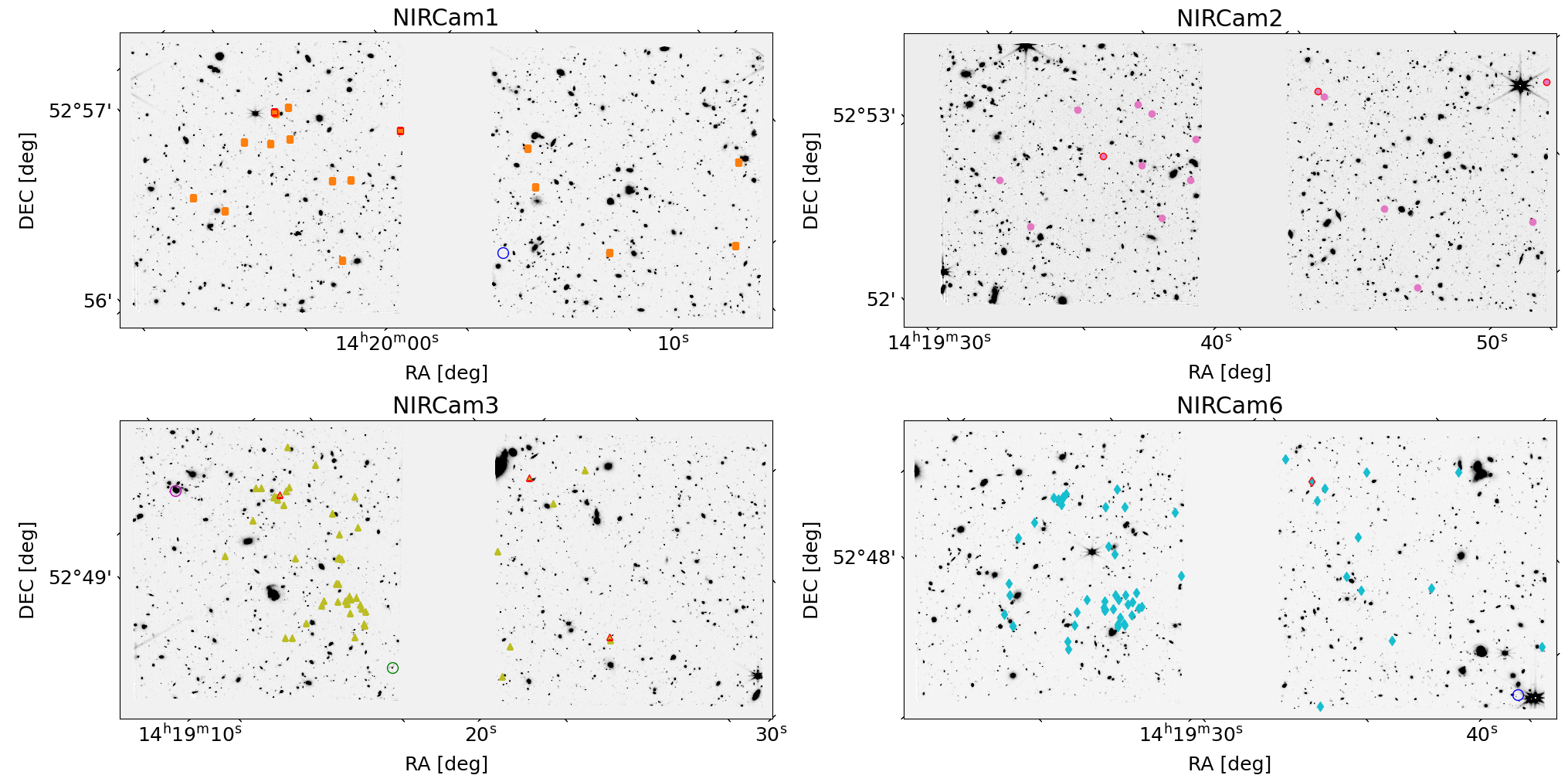}
    \caption{Sky position of the F200W-dropouts in the four NIRCam pointings over-plotted to the F444W image. Points with a red edge are objects in the F200W-dropouts-extra sub-sample. Open circles indicate the position of known groups by \citet[][blue]{Erfanianfar2013}, \citet[][green]{Jin2022} and \citet[][magenta]{Coogan2023}. }
    \label{fig:sky}
\end{figure*}

 
Differences in numbers in the four pointings may be due to the presence of over-densities, as visible in Figure \ref{fig:sky}. However, these possible over-densities do not correspond to known galaxy groups from X-ray observations \citet{Erfanianfar2013} or density analysis \citep{Coogan2023,Jin2022}, therefore additional spectroscopic observations are necessary to confirm them.\par
Some differences are visible in the number of sources detected in the two NIRCam detectors, i.e. 32 in one detector and 112 in the other. First, as verified in \citet{Bagley2022}, the photometric residuals from detector-to-detector are at the level of $\sim2-5\%$ in all filters and sources do not overlap with any feature visible in the flat fields of the long filters. Second, if we apply a higher magnitude cut, i.e. [F444W]<27.8 corresponding to a 10$\sigma$ level, the difference is still present, as four galaxies are present in one detector and ten in the other one. We can therefore exclude that the observed difference is due to different noise level and calibration.  \par

\subsection{What do we expect from this selection?}\label{sec:modelsel}
Before moving on analysing the observed sample, to investigate our selection criterion we derived the expected JWST fluxes for all the spectral energy distribution (SED) templates used later to derive physical properties, but considering a coarser grid in the different physical parameters. We then use these fluxes to understand which model is compatible with the F200W-dropout selection described in the previous paragraph. In particular, we included \citet{Bruzual2003} stellar population models with stellar metallicity from 0.005 solar up to solar (i.e., 0.005 Z$_{\odot}$, 0.02 Z$_{\odot}$, 0.2 Z$_{\odot}$, 0.5 Z$_{\odot}$, Z$_{\odot}$). We allow the redshift to vary up to $z=20$ ($z=$0.1, 0.3, 0.5 and from 1 to 20 with steps of 1) and the stellar mass to range from 10$^{6}\,\rm M_{\odot}$ up to 10$^{12.5}\,\rm M_{\odot}$, with steps of $\rm log(M_{*}/M_{\odot})=0.5$. Nebular continuum and emission lines are included based on a pre-computed model grid, already included in the SED fitting code (see Sec. \ref{sec:SEDfitting}), based on the photoionisation code CLOUDY \citep{Ferland2017} assuming a ionisation parameter (i.e. dimensionless ratio of densities of
ionising photons to hydrogen) of $\log_{10}U\in{-4,-3,-2}$, a Hydrogen density of 100 atoms cm$^{-3}$ \citep[see][for more details]{Carnall2018}. We considered the same reddening law \citep[i.e.,][]{Calzetti2000} for both the stellar continuum and the nebular emission lines with dust extinction up to $A(V)=6$, in steps of 0.5 mag. Finally, we considered a delayed exponential (i.e. SFR$\propto t\,e^{-(t/\tau)}$) star-formation history with ages ranging from 1 Myr to the age of the Universe (0.001, 0.01, 0.1, 1, 3, 5, 7, and 15 Gyr) and $\tau=$0.01 to 10 Gyr (i.e., 0.01, 0.1, 1, 3, 5, and 10 Gyr). To identify SED templates not detected at $\lambda\leq 2\,\mu$m, we consider 5$\sigma$ limits of 29.2, 29.0 and 29.2 mag for the F115W, F150W and F200W filters (Table \ref{tab:depths}). \par
\begin{figure}
    \centering
    \includegraphics[trim=55 0 90 45,clip,width=0.99\linewidth,keepaspectratio]{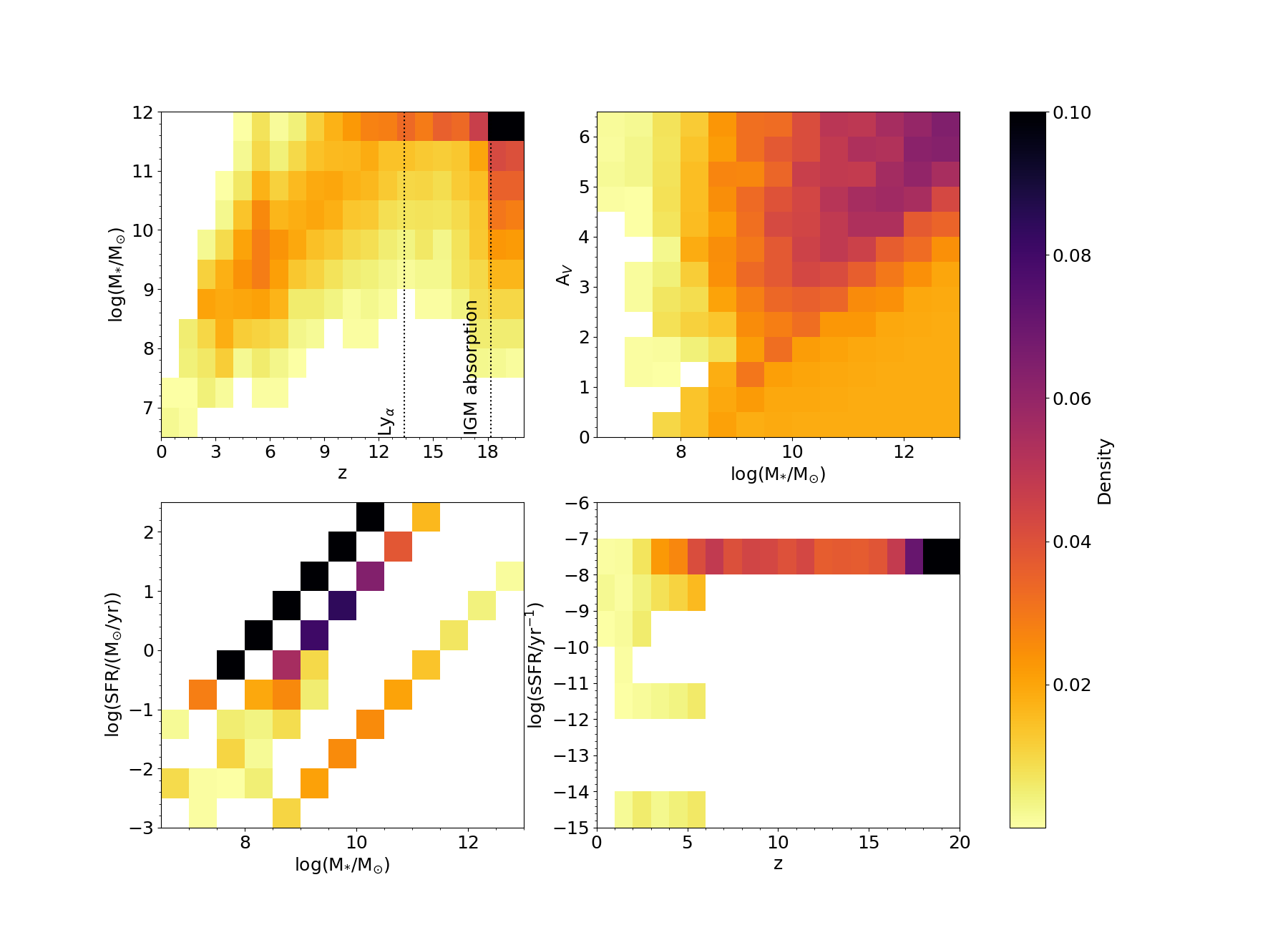}
    \caption{Physical properties associated to the SED templates that would be selected as part of our F200W-dropout sample. Top left: redshift and stellar mass. Top right: stellar mass and A$_{V}$. Bottom left: stellar mass vs. SFR. Bottom right: redshift and sSFR. For a visual purpose, we associated a $\rm log(sSFR/(M_{\odot}/yr))=-15$ to all galaxies with lower sSFR. Colour indicates the density of templates; i.e. number of templates in one bin and corresponding to our F200W-dropout selection respect to the total number of models corresponding to our F200W-dropout selection. SFR and sSFR are discrete as we select for this test a coarser distribution in star-formation histories and ages than the ones used for the SED fitting.}
    \label{fig:tracks}
\end{figure}

The physical properties of the templates corresponding to the F200W-dropout selection are reported in Figure \ref{fig:tracks}. First, the considered selection is expected to identify every galaxy candidate at $z\geq18.2$ (considering the wavelength range where the F200W throughput is at least 50$\%$ of the maximum value) massive enough to be observed at 4.4$\mu$m, as there is no flux at $\lambda\leq2\mu$m due to the inter-galactic medium absorption (IGM). Second, at $13.4\leq z<18.2$ the flux in the F200W is only partially absorbed from intergalactic hydrogen, so we expect to observe galaxies with stellar masses above 10$^9$ M$_{\odot}$ and various dust-extinction. At $z<13.4$ only some combinations of stellar mass and dust extinction can make a galaxy visible at 4.4$\mu$m, but not at $\lambda\leq2\mu$m. For example, at $z<2$, this happens only to galaxies with $\rm M_{*}<10^{8.5}\, M_{\odot}$ and $A_{V}>3$, while galaxies with with $\rm M_{*}>10^{11}\, M_{\odot}$ are possible at $z>3.7$, as otherwise they would be detected at $\lambda\leq2\mu$m. Third, we may select passive galaxies with $\rm log(sSFR/(M_{\odot}/yr))\leq-11$, only up to $z=6$. 
Overall, the chosen selection criteria is biased toward very high-$z$ galaxies and toward dusty sources. 

\section{Physical property derivation} \label{sec:Physprop}
\subsection{SED fitting} \label{sec:SEDfitting}
\begin{figure}
    \centering
    \includegraphics[trim={1.cm 0.5cm 2.8cm 1.5cm},clip,width=0.99\linewidth,keepaspectratio]{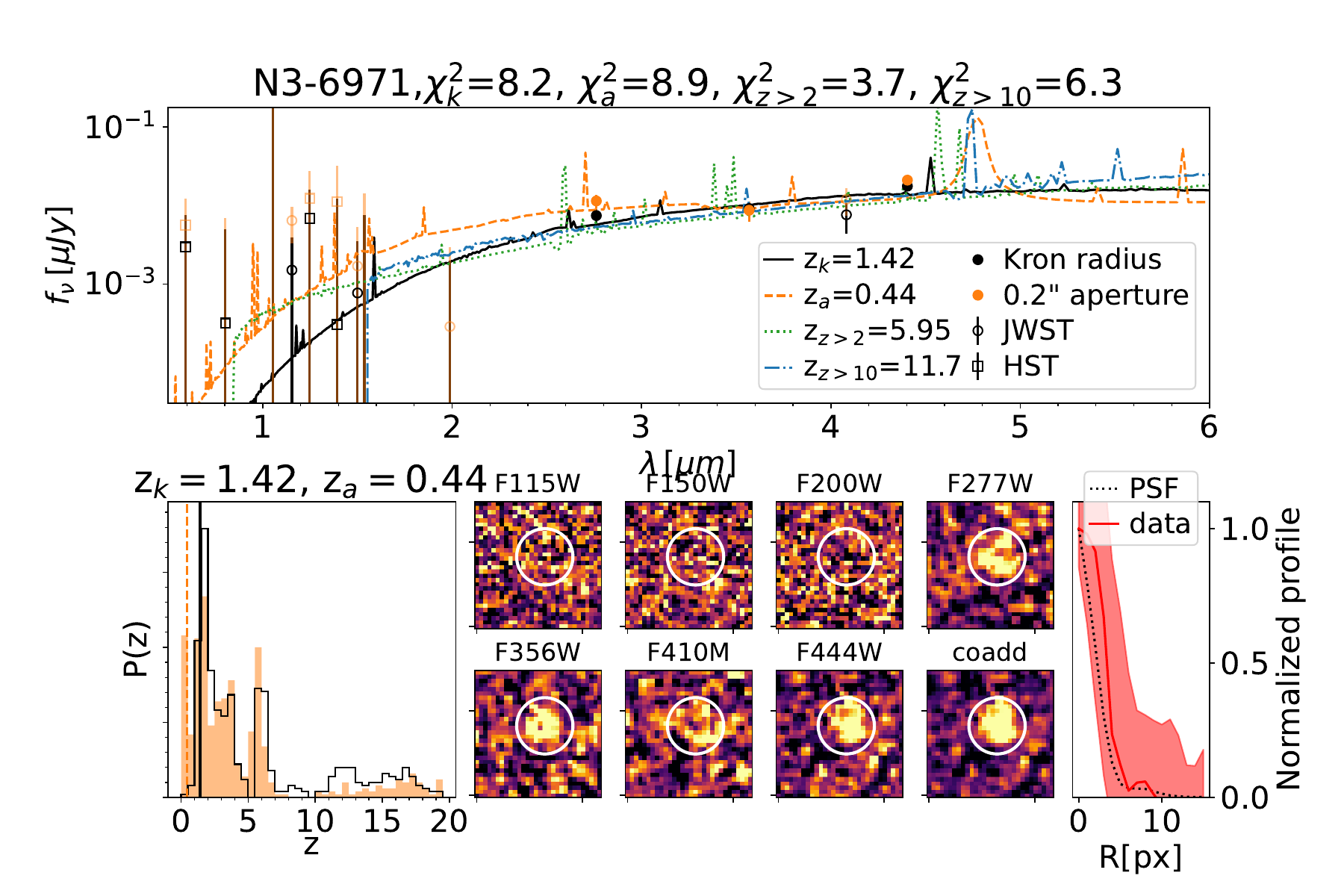}
    \caption{Example of the SED fitting applied to a galaxy (N3-6971) in our F200W-dropout sample. In the top panel we report the photometry and the best fit derived with the 0.2"-radius aperture (orange) and the Kron aperture leaving the redshift free (black), forcing it at $z>2$ (green dotted) and at $z>10$ (blue dash-dotted). Empty symbols indicate fluxes with $S/N<3$. Similarly, in the bottom left panel we report the redshift probability distribution function (PDF) derived with the two photometric sets, with the peak values highlighted by vertical lines. On the bottom, we include eighth cutout images in the JWST filters and in the co-added F277W, F356W, and F444W band images. A white circle indicates the 0.2"-radius aperture. On the bottom right we report the light profile of the target (red solid line and shaded area) compared with the observed PSF of the F444W filter (black dotted line).}
    \label{fig:SEDexample}
\end{figure}

We use the Bayesian Analysis of Galaxies for Physical Inference and Parameter EStimation \citep[BAGPIPES;][]{Carnall2018} to derive
both the redshift and galaxy physical properties (e.g. stellar mass, SFR) using the models anticipated in Section \ref{sec:modelsel}, but with a finer grid of values for each physical parameter. We include all filters listed in Table \ref{tab:depths} and for fluxes with low S/N values (i.e., $S/N<3$), we do not impose any hard upper limit in the SED fitting, but we include the measured photometry with the associated uncertainties. This SED fitting procedure was performed using the photometry derived with the Kron aperture, but in Appendix \ref{sec:fixedpaerture} we repeat the analysis using a 0.2"-radius aperture. This change in the input photometry sometimes results in different physical properties for the individual galaxies, but the general results for the overall F200W-dropout sample remain unchanged. In Figure \ref{fig:SEDexample} we report, as example, the SED fitting procedure applied to one galaxy in the F200W-dropout sample. In the entire sample, 132/133 objects results in a good SED fit with a $\chi^{2}<10$. For comparison, the same SED fitting procedure described for the F200W-dropout was applied also to the entire sample in the CEERS.\par 
To verify the robustness of our results we repeated the BAGPIPES fitting changing some key parameters. The details of these additional results are reported in Appendix \ref{sec:SMC}. First, we assumed the reddening law observed in the Small Magellanic Cloud, instead of the reddening law by \citet{Calzetti2000}, but this change has a minor impact on the results. Second, in our SED fitting we assumed a \citet{Draine2007} dust emission models, but \citet{Leja2017} pointed out that this model sometimes over-predicts the 3.3 $\mu$m polycyclic aromatic hydrocarbon (PAH) emission line strength. This spectral feature is inside the F444W filter indicatively between $z=0.1$ and $z=0.5$ and we therefore decided to repeat the BAGPIPES fit limiting the mass associated to PAH to only 10\% of the total dust mass. Even in this case the impact on our results is minor. Finally, we have changed the star-formation history to an exponentially declining one (i.e. SFR$\propto e^{-(t/\tau)}$), which result on differences in the physical properties of each galaxy, while leaving the general results of the paper unchanged.

\subsection{Are we delving too shallow?}
In this section we explore the possibility that some of the objects in our F200W-dropout sample are not extra-galactic sources but, instead, galactic brown dwarfs. To perform this test we fit our candidates F200W-dropouts with L and T dwarf models from \citet{Burrows2006}. In particular, these templates span metallicities between [Fe/H]=-0.5 and 0.5, effective temperatures between 700 K and 2300 K and gravities between $10^{4.5}$ and $10^{5.5}\,\rm cm\, s^{-2}$. Of the full F200W-dropout sample, only two objects have a $\chi^{2}$ derived with the brown dwarf templates that is better than the $\chi^{2}$ derived with the galaxy models. However, the normalisation of the best brown dwarf templates for these objects indicates that they should be all within 10 pc from the Sun. \par
In the solar neighbourhood, \citet{McKee2015} found a surface density of brown dwarfs of $\sim1.2\,\rm M_{\odot}/pc^2$, which corresponds to a number density of 16-40 $pc^{-2}$ if we associate to them a stellar mass between 0.03 and 0.075 $\,M_{\odot}$. This implies that we would expect between 3 and 7 brown dwarfs in the area covered by four NIRCam pointings, which is consistent with the found numbers.
However, additional near-IR spectroscopic data are necessary to verify the presence of the absorption features characteristic of brown-dwarf and confirm their nature. These two possible stellar objects are further discussed in Appendix \ref{sec:BD}.\par

\section{Results}\label{sec:results}
In this Section we analyse the observed colours and the physical properties of our F200W-dropouts sample, comparing them with other literature samples and with the full CEERS sample. 

\subsection{Colours}
\begin{figure}
    \centering
    \includegraphics[width=0.99\linewidth,keepaspectratio]{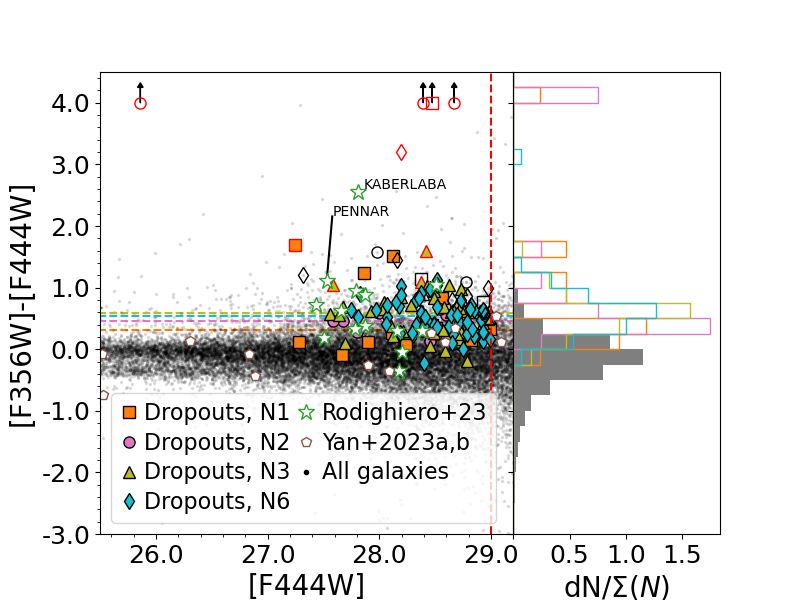}
    \caption{Left: [F444W] magnitude vs. [F356W]-[F444W] colour of the F200W-dropout sample in the four NIRCam pointings (coloured points and histograms, named N1, N2, N3, and N6) and for full CEERS sample (black dots and grey histogram). Points with a red edge are objects in the F200W-dropout-extra sub-sample. Empty symbols indicates objects in our sample with a S/N$<$3 in the F356W filter and we assigned, for visual purposes, a colour of [F356W]-[F444W]=4 to galaxies with redder colours. We also report the sample of F200W-dropouts by \citet[][green empty stars]{Rodighiero2023}, highlighting their two candidates at $z>10$, namely KABERLABA and PENNAR, and the sample by \citet[brown empty pentagons][]{Yan2023}. Right: Normalised [F356W]-[F444W] colour density distribution. Each histogram is normalised so that its integral is equal to one.}
    \label{fig:f444color}
\end{figure}

As a first test, in Figure \ref{fig:f444color} we report the [F444W] magnitude and the [F356W]-[F444W] colour of our F200W-dropout sample, divided into the four NIRCam pointings. We also show the full CEERS galaxy sample for comparison. We see no major differences between the F200W-dropouts in the four pointings, excluding any strong bias in the photometry. At the same time, F200W-dropouts have generally redder [F356W]-[F444W] colours with respect to the full CEERS sample, i.e. a median value of 0.55 (0.56, including objects in the F200W-dropout-extra sub-sample) against -0.15, respectively. Moreover, only 4/144 objects in the F200W-dropout sample have blue $\rm [F356W]-[F444W]<0.0$ colours. Analysing the SED models presented in Section \ref{sec:modelsel}, a blue [F356W]-[F444W] colour in our F200W-dropout sample is generally associated to $z>16$ templates with almost no dust extinction. However, once the photometric errors in both filters are taken into account all four objects are consistent with red colours within 1$\sigma$. All the nine objects in the F200W-dropouts-extra sub-sample have $\rm [F356W]-[F444W]>1$ showing that they are the red tail of our F200W-dropout sample.\par
\begin{figure}
    \centering
    \includegraphics[width=0.99\linewidth,keepaspectratio]{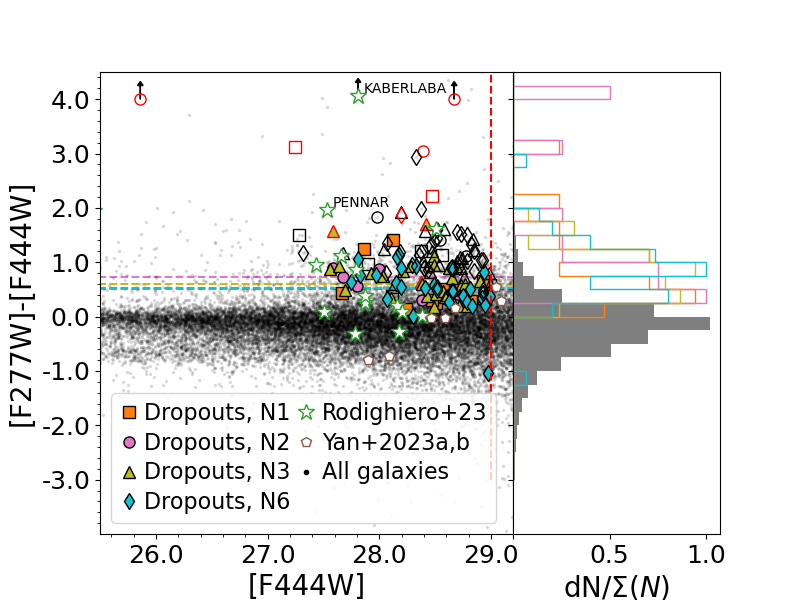}
    \caption{Same as Figure \ref{fig:f444color}, but showing the [F277W]-[F444W] colour. Empty symbols indicates objects with a S/N$<$3 in the F277W filter.}
    \label{fig:f277color}
\end{figure}

Looking at the [F277W]-[F444W] colours (Figure \ref{fig:f277color}), our F200W-dropout sample shows again generally redder colours than the full CEERS sample, with median values of 0.83 (0.85 including the F200W-dropout-extra sub-sample) against -0.19 mag, respectively. However, these values are affected by large uncertainties, given than 46$\%$ of our F200W-dropout sample has a S/N$<$3 in the F277W filter. Again, the nine objects in the F200W-dropouts-extra sub-sample are the red tail of our F200W-dropout sample, having $\rm [F277W]-[F444W]>0.6$. We do not show any additional colour, as the F410M observations are shallower than the ones in other filters and indeed only 28$\%$ of the F200W-dropout sample has a S/N$>$3 in this filter. At the same time, the remaining filters at $\lambda<2\,\mu$m are by construction associated to S/N$<$2.\par
Some galaxies in the general CEERS sample are very red in the [F277W]-[F444W] and [F356W]-[F444W] colours. These objects are not included in the F200W-dropout sample because their are bright at $\lambda\leq2$, i.e. $S/N>2$, or have $S/N<3$ in the F444W filter. \par
In both Figure \ref{fig:f444color} and \ref{fig:f277color} we can also see that our F200W-dropouts occupy the same color-magnitude space of the bulk of the F200W-dropouts presented in \citet{Rodighiero2023}. Only some of the sources in our F200W-dropout-extra sample has [F356W]-[F444W] and [F277W]-[F444W] colours similar or redder than one of their two candidates at $z>10$ (i.e. KABERLABA) and only a few of our F200W-dropout sources have colours and magnitudes similar to the other $z>10$ candidate (i.e. PENNAR). However, the only galaxies with [F356W]-[F444W] colour similar to KABERLABA is detected only in the F444W filter, making any further comparison impossible. Among the remaining four objects with [F356W]-[F444W] colours redder than KABERLABA, only the brightest one (N2-202826) in the F444W filter is detected in more than one filter (i.e. F444W and F410M), with a marginal detection in the F356W filter.  \par
We also compare our F200W-dropout sample with the F200W-dropout samples by \citet{Yan2023,Yan2023a}. Their sample selection is slightly different from the one adopted in our work, as they applied a colour cut [F200W]-[F356W]$\geq$0.8 mag, a S/N$<$2 only in the F150W band and a S/N$>$5 in the F277W filter. This selection results in galaxies with both [F277W]-[F444W] and [F356W]-[F444W] colours bluer than our sample. 

\subsection{Physical properties: a population of dusty dwarfs at $z<2$}
\begin{figure*}
    \centering
    \includegraphics[width=0.99\linewidth,keepaspectratio]{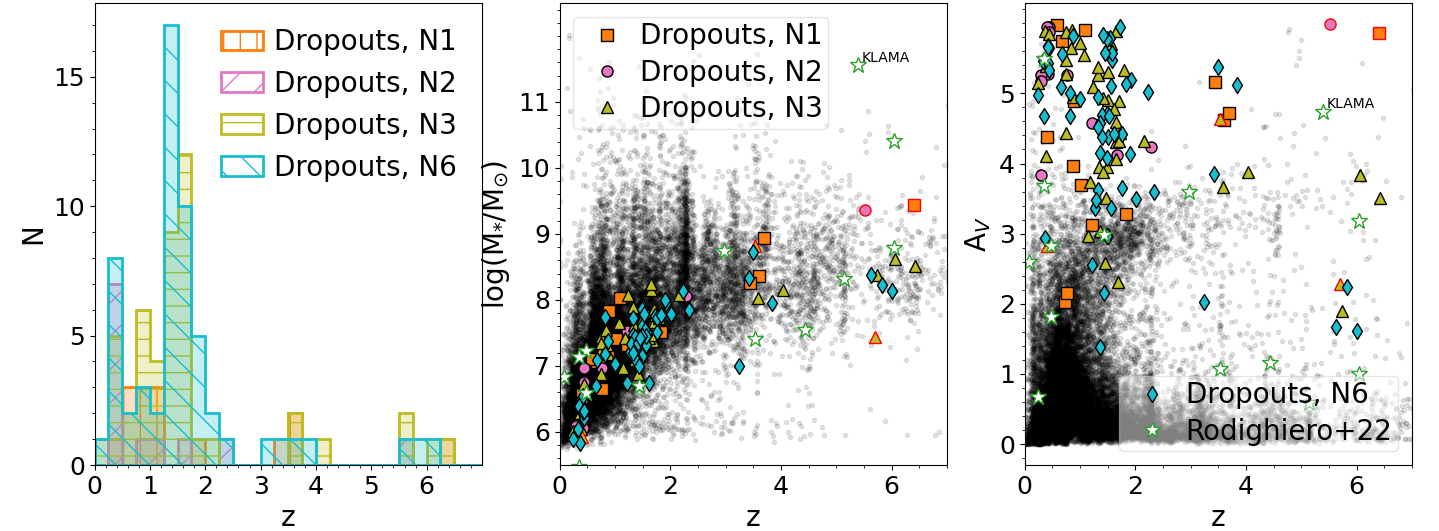}
    \caption{Redshift distribution, redshift vs. stellar mass and redshift vs. A(V) for F200W-dropouts in the four CEERS poitings (colour points) and for the entire sample in the NIRCam2 pointing (small black dots). Points with a red edge are objects in the F200W-dropout-extra sub-sample. Based on photometry derived using the Kron radius. We also report the sample of F200W-dropouts by \citet[][green empty stars]{Rodighiero2023}.}
    \label{fig:z_M_Av}
\end{figure*}

We now move on analysing the physical properties derived for our F200W-dropout sample, taking into account, however, that these are results based on few observational points, given the large number of non-detections, and additional data are necessary to give more robust constrains. In Figure \ref{fig:z_M_Av} we report the redshift, stellar mass and dust extinction for our F200W-dropout sample. In agreement with the test performed with the SED templates, our F200W-dropout sample is composed at 81$\%$ by galaxies at $z<2$, with a median stellar mass of 10$^{7.3}\,\rm M_{\odot}$ and a median dust extinction of 4.9 mag. The entire sample of F200W-dropouts has A$_V>$0.9 showing an extremely large dust content.\par 
The pointing N3 and N6 have a larger number of F200W-dropouts and they show an excess in the redshift distribution at $\rm z\sim1.5$. This secondary redshift peak is possibly due to over-densities present in these pointings, as previously shown in Figure \ref{fig:sky}, but additional spectroscopic data will be necessary to confirm it.\par
In our F200W-dropout sample we see a lack of galaxies between $z=4$ and $z=5.5$ (see left panel in Fig. \ref{fig:z_M_Av}), but this is neither motivated by the selection applied, as SED templates at these redshifts and with the colours of our F200W-dropout galaxies are indeed possible, neither by the presence of strong nebular emission lines, as the H$_{\alpha}$ nebular emission line is inside one of the red filters (i.e. F277W, F356W, and F444W) from $z\sim2.5$ to $z\sim6.5$ and should in case boost the flux of these filters. This is for example the case for two of the sources in the F200W-dropout-extra sub-sample, as the excess flux observed in these galaxies in the F444W filter could be explained by the H$_{\alpha}$ nebular line at $z=5.5-6.5$.\par
We note that the galaxy properties of our F200W-dropout sample is extremely different respect to the optical-dark populations presented in previous works using the same CEERS data \citep[i.e.][]{PerezGonzales2023b,Barrufet2022}. Those optical-dark galaxies are selected based on their [F150W]-[F444W] or [F150W]-[F356W] colours and correspond to a population of dust-obscured galaxies at $z>2$ with intermediate to high stellar masses ($M>10^{9} M_{\odot}$). While all galaxies in the F200W-dropout sample are faint at 1.5$\mu$m with [F150W]$>$28.5 mag, only $~40\%$ have [F150W]-[F444W]$>2$ mag or [F150W]-[F356W]$>1.5$ mag, as used by \citet{Barrufet2022} and \citet{PerezGonzales2023b}, respectively. These colours criteria are however merely indicative for our sample, given that galaxies in our F200W-dropout sample are completely undetected (i.e. S/N$<2$) in the F150W filter. At the same time, galaxies in the F200W-dropout sample are also quite faint in the F444W filter with magnitudes from 29.0 to 27.5, while the sample by \citet{Barrufet2022} has [F444W]$<$26.4 mag and the one by \citet{PerezGonzales2023b} has [F444W]4$<$27.5 mag.   \par

\begin{figure*}
    \centering
    \includegraphics[width=0.99\linewidth,keepaspectratio]{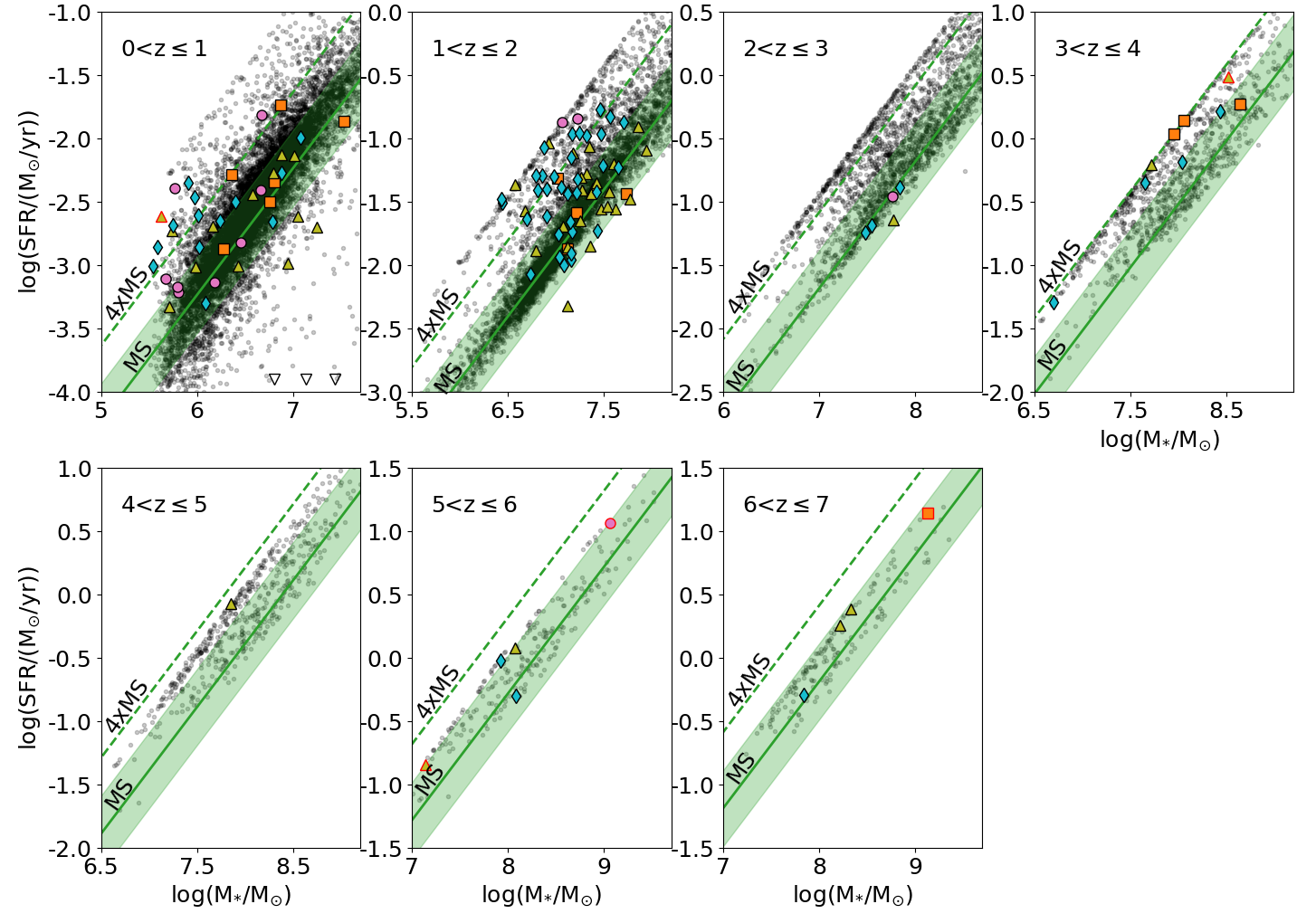}
    \caption{SFR vs. stellar mass from z=0 to 7 for F200W-dropout galaxies in the four CEERS poitings (coloured points) and the entire CEERS sample (black dots). Points with a red edge are objects in the F200W-dropout-extra sub-sample. Based on photometry derived using the Kron radius. Both SFR and stellar mass are derived fitting the observed optical and near-IR observations. We include the MS parametrization derived by \citet{Speagle2014} as reference. Symbols are the same reported in Figure \ref{fig:z_M_Av}.}
    \label{fig:MS}
\end{figure*}

In Figure \ref{fig:MS} we show the distribution of our F200W-dropout sample in the SFR vs. stellar mass plane. The majority of our galaxies are consistent with the main-sequence of star-forming galaxies \citep[MS,e.g.,][]{Brinchmann2004,Noeske2007,Whitaker2012,Steinhardt2014,Santini2017,Bisigello2018}, as extrapolated from the parametrisation by \citet{Speagle2014}. \par
If we concentrate on galaxies at $z<2$, we obtain that their median SFR is 0.01 M$_{\odot}/yr$. This value, if converted using the relation by \citet{Kennicutt1998}, results in a total IR luminosity (8-1000 $\mu$m) of L$_{IR}\sim9\times10^{7}\,\rm L_{\odot}$, generally below \textit{Herschel} capability beyond the local Universe. This shows why such galaxy population has been missed up to now: they are in general too red for \textit{HST} and to faint for \textit{Spitzer} and \textit{Herschel}. \par
Looking into more details in the SFR-M$_{*}$ plane (Fig. \ref{fig:MS}), the fraction of starburst, roughly defined as galaxies above 0.6 dex \citep{Rodighiero2011} from the MS, in our F200W-dropout sample is of 26$\%$ at $z\leq1$, 27$\%$ at $1<z\leq2$ with only few other cases around $z\sim3.5$. These results suggest that a burst of star-formation is probably not the cause of the large dust extinction. It is however necessary to take these results with caution given that the estimated SFR is based on the SED analysis of rest-frame UV-to-optical or optical-to-near-IR data, depending on the redshift. In the absence of far-IR observations these SFR estimates may be underestimated \citep[e.g.][]{Puglisi2017,Elbaz2018,Xiao2023}, particularly for the most dusty objects. At the same time, it is necessary to take into account that intrinsic correlations are present between the SFR and the stellar mass, given that both are derived from the SED fitting. Moreover, the assumed delayed SFH and the minimum allowed age of 1 Myr limit the sSFR to values below $10^{-7.9} yr^{-1}$, as observed also in other works \citep[e.g.][]{Ciesla2017}. Finally, the galaxy physical properties are based on a limited number of photometric points with large S/N values.

Of the full F200W-dropout, only five objects (4\% of the sample) have a SFR at least four times below the MS, showing therefore an exceptional low SFR. However, all these objects are well described by the SED templates of star-forming galaxies when considering the aperture photometry, indicating that a second not-passive solution is also plausible.



\begin{figure*}
    \centering
    \includegraphics[width=0.99\linewidth,keepaspectratio]{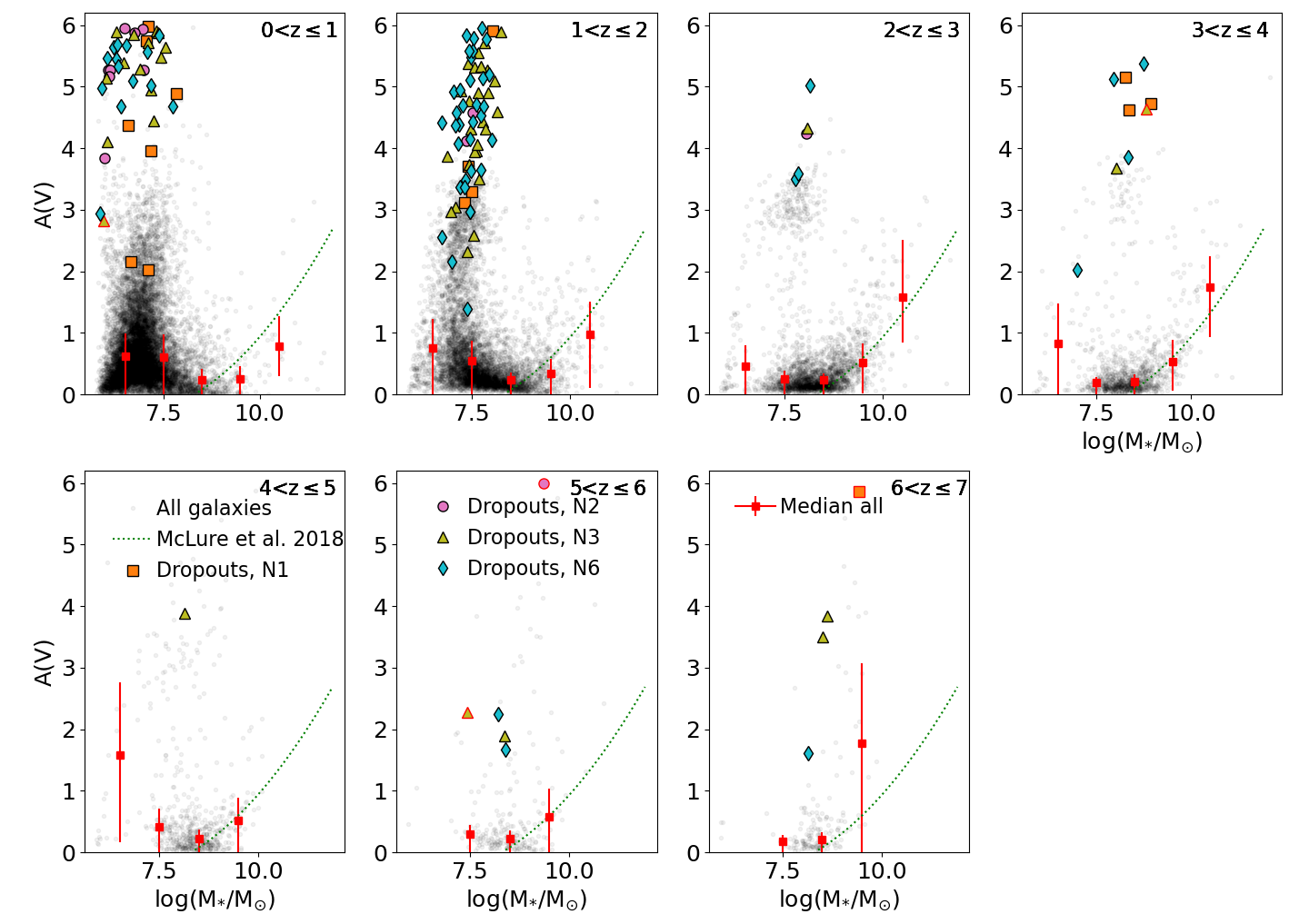}
    \caption{Mass and dust extinction A$_{V}$ from z=0 to 7 for the F200W-dropout sample (colored points) and the entire CEERS sample in NIRCam2 (black dots). We also report the median A$_{V}$ at different stellar masses for the general CEERS sample with good SED fit $\chi_{red}^{2}<10$ (red squares), when at least ten galaxies are present in the stellar mass bin. We also report the expected relation derived assuming a \citet{Calzetti2000}'s reddening law and an intrinsic UV slope of -2.3 \citep[green dotted lines;][]{McLure2018}.}
    \label{fig:M_Av}
\end{figure*}

As previously shown in Figure \ref{fig:z_M_Av} (right panel), our F200W-dropout sample show in general a very high dust extinction. This is even more striking if we compare it with galaxies of similar stellar mass (Fig. \ref{fig:M_Av}). Indeed when considering all redshifts, while our F200W-dropouts have median A$_V=4.7$, all the galaxies in CEERS have a median value of A$_V=0.48$. The striking difference is shown in Figure \ref{fig:M_Av}, where we also report the median A$_V$ for the entire CEERS sample in different stellar mass bins. This median values are in agreement with the average relation between dust extinction and stellar mass \citep[e.g.,][]{McLure2018}, showing that the large dust extinction of our F200W-dropout is not a direct artefact of the adopted SED fitting procedure, but is dictated by their red colours. Overall, the dusty dwarf population we identify represent 0.5$\%$ of the population of galaxies with similar stellar mass (i.e. $10^{5.8}-10^{7.2} M_{\odot}$) at $z<1$ and 0.8$\%$ of galaxies with $10^{6.8-8.2} M_{\odot}$ at $z=1-2$, assuming that the completeness of the entire CEERS sample and of the F200W-dropout at those stellar masses are the same. This show that these exceptional galaxies clearly are rare and not representative of the general dwarf population at $z<2$.\par 
In the same Figure it is possible to see that there are galaxies with $A_{V}>1$ that are not present in our F200W-dropout sample. On the one hand, the absence of galaxies with stellar masses above $\sim10^8\,\rm M_{\odot}$ at $z<2$ in our F200W-sample is purely due to a selection effect, as mentioned in Section \ref{sec:modelsel}. As an additional check, we search for dusty galaxies with $>10^8\,\rm M_{\odot}$ in the entire NIRCam pointing 2 and we found 167 objects with A$_V>$1 and 29 with A$_V>$3, the majority of which have $S/N>2$ in the F200W filter. \par
On the other hand, there are different reason beyond the non selection of some galaxies with stellar masses below $10^{8}\,\rm M_{\odot}$ and $A_{V}>1$. Around 44$\%$ of these objects are simply too faint in the F444W filter, as they have [F444W]$>29$ mag or a $S/N<3$. The remaining objects have a S/N$>$2 in at least one of the filter at $\lambda>2\mu$m. In general, all these not-selected low-mass dusty sources show in general bluer colour than our F200W-dropout sample, with median [F277W]-[F444W]=-0.21 and median [F356W]-[F444W]=-0.16. These results on lower stellar masses at a fixed dust extinction. 

\subsection{On the absence of $z>10$ galaxies}
As visible in Figure \ref{fig:z_M_Av}, no galaxies are present at $z>6.5$ ($z>9$ considering aperture photometry, see Appendix \ref{sec:fixedpaerture}) in our F200W-dropout sample. On the one hand, some blue $z>10$ galaxies, as the candidates presented in \citet{Finkelstein2023}, are discarded from our sample as they have $S/N>3$ in the F200W filter, given that the flux at these wavelength is not totally absorbed by the IGM. On the other hand, observations considered in this work cover an area four times larger than the observations in SMACS0723 analysed in \citet{Rodighiero2023} and would therefore expect to observe at least few dusty galaxies at $z>10$. However, it is necessary to take into account that one of their two $z>10$ candidates, i.e. KABERLABA, is extremely red and faint, e.g. [F356W]=30.5 mag, and it is therefore below the detection limit of CEERS in all bands except F444W. On the contrary, a galaxy similar to the other $z>10$ candidate, i.e. PENNAR, would be bright enough to be detected in at least three filters, even without the magnification of the SMACS0723 cluster. \par
\begin{figure}
    \centering
    \includegraphics[width=0.99\linewidth,keepaspectratio]{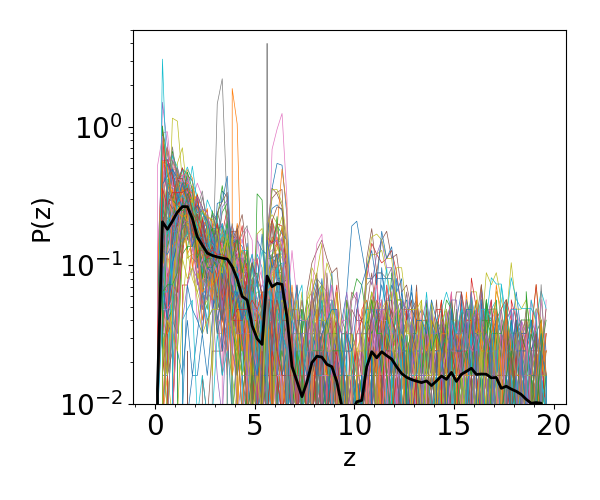}
    \caption{Redshift probability distributions P(z) of the entire F200W-dropout sample (coloured line) and average P(z) (black thick solid line).}
    \label{fig:PDF}
\end{figure}

Even if no galaxy are associated to $z>10$, 93/133 objects have a probability larger than 10$\%$ of being at this high-z, as visible from the average redshift probability distribution in Figure \ref{fig:PDF}. By integrating the probability distribution at $z>10$ of each of these objects and summing up all of them, we expect our sample to statistically include 17 galaxies at $z>10$, which however would require a spectroscopic follow-up to be identified and confirmed. Such secondary solutions correspond to a median stellar mass of 10$^{9.5}\,\rm M_{\odot}$, median $\rm log(SFR)=1.4$ and median dust extinction of $\rm A_{V}=2$ mag, which are in line with the physical properties of the $z>10$ candidate PENNAR from \citet{Rodighiero2023}.

\section{Summary and conclusions}\label{sec:summary}
In this work we have taken advantage of the first four NIRCam pointings available as part of the CEERS survey. Available observations include six NIRCam broad-band filters, from 1.15 $\mu$m to 4.4$\mu$m, the medium F410M NIRCam filter and five \textit{HST} filters, covering from 0.6$\mu$m to 1.6$\mu$m. Such photometric observations were considered to explore a population of extremely red galaxies, defined as objects with $S/N>3$ at 4$\mu$m and $S/N<2$ in all filters at $\lambda\leq2\mu$m and in the co-added image of the JWST blue filters (F115W, F150W and F200W). We verify, using available SED models, that such colour selection is particularly suited for identifying dusty (A$_V>$1) galaxies at $z<18.2$, but in principle also $z\geq18.2$ galaxies, independently by their dust extinction. \par
This search results in a sample of 144 F200-dropouts, with eight of them detected only in the F444w filter. These objects have in general both [F277W]-[F444W] and [F356W]-[F444W] colours redder than the general galaxy population and we found no galaxies with secure blue colours, excluding the presence of very high-z galaxies. Indeed, when performing an SED-fitting analysis we found no candidate galaxy at z$>$9, showing the rarity of dusty high-z sources, which have been instead identified in other fields \citep{Rodighiero2023}. We however speculated, using the redshift probability distribution associated to each galaxy, that our sample could statistically include 17 galaxies with $z>10$ misinterpreted as low-z one. \par
At the same time, $\sim81\%$ of the F200W-dropout sample corresponds to extremely dusty dwarf galaxies (median A$_{V}=$4.9 mag and median stellar mass of 10$^{7.3}\,\rm M_{\odot}$) at $z<2$. These are mainly star-forming galaxies, both inside and above the MS, with median SFR$=0.01\,\rm M_{\odot}/yr$. Such extreme dust content is striking, given that the dust content of galaxies, which correlates with the stellar mass, is expected to be negligible for galaxies with stellar masses below 10$^{8.5}\,\rm M_{\odot}$. However, these population of extremely dusty dwarfs have a minor contribution on the overall galaxy population, being less than 1$\%$ of the galaxy sample at similar stellar masses and redshifts, but they could contribute to the diffuse extragalactic background light \citep{Driver2016,Koushan2021,Kramer2022}. \par
We can speculate about different possible scenario causing this excess in dust extinction. First, these galaxies could be more compact than galaxies with similar size, causing a large dust extinction even if the overall dust content remain similar. However, galaxies with such stellar masses are expected to be too small to be resolved even with JWST, extrapolating the stellar-mass size relation \citep{Wel2014}, and we can therefore not test this scenario. Second, these galaxies could been experiencing a burst of star-formation, having therefore an increase in dust production. Our preliminary analysis shows that only a fraction of these galaxies are considerably above the main-sequence of star-formation, but such estimates are heavily limited by the low-number of photometric data available and the absence of far-IR observations. Third, they could be the remnants of more massive galaxies, but again this scenario can not be tested with the available data. Finally, these galaxies could host an active galactic nuclei, but we can not again verify this possibility with the few photometric point available. Therefore, we need to wait for future spectroscopic observations or data at longer wavelengths to investigate further these galaxy population and confirm their exceptional dust content.

\begin{acknowledgements}
LB  and GR acknowledge the support from grant PRIN MIUR $2017 - 20173ML3WW$\char`_$001$. GG acknowledge M. Giulietti for the helpful discussions. LC acknowledges financial support from Comunidad de Madrid under Atracci\'on de Talento grant 2018-T2/TIC-11612 and Spanish Ministerio de Ciencia e Innovaci\'on MCIN/AEI/10.13039/501100011033 through grant PGC2018-093499-B-I00
\end{acknowledgements}

%
\bibliographystyle{aa} 
\bibliography{main} 

\begin{thebibliography}{74}
\expandafter\ifx\csname natexlab\endcsname\relax\def\natexlab#1{#1}\fi

\bibitem[{{Adams} {et~al.}(2022){Adams}, {Conselice}, {Ferreira}, {Austin},
  {Trussler}, {Juod{\v{z}}balis}, {Wilkins}, {Caruana}, \& {Dayal}}]{Adams2022}
{Adams}, N.~J., {Conselice}, C.~J., {Ferreira}, L., {et~al.} 2022, arXiv
  e-prints, arXiv:2207.11217

\bibitem[{{Bagley} {et~al.}(2022){Bagley}, {Finkelstein}, {Koekemoer},
  {Ferguson}, {Arrabal Haro}, {Dickinson}, {Kartaltepe}, {Papovich},
  {P{\'e}rez-Gonz{\'a}lez}, {Pirzkal}, {Somerville}, {Willmer}, {Yang}, {Yung},
  {Fontana}, {Grazian}, {Grogin}, {Hirschmann}, {Kewley}, {Kirkpatrick},
  {Kocevski}, {Lotz}, {Medrano}, {Morales}, {Pentericci}, {Ravindranath},
  {Trump}, {Wilkins}, {Calabr{\`o}}, {Cooper}, {Costantin}, {de la Vega},
  {Hutchison}, {Lucas}, {McGrath}, {Wang}, \& {Wuyts}}]{Bagley2022}
{Bagley}, M.~B., {Finkelstein}, S.~L., {Koekemoer}, A.~M., {et~al.} 2022, arXiv
  e-prints, arXiv:2211.02495

\bibitem[{{Barrufet} {et~al.}(2022){Barrufet}, {Oesch}, {Weibel}, {Brammer},
  {Bezanson}, {Bouwens}, {Fudamoto}, {Gonzalez}, {Gottumukkala}, {Illingworth},
  {Heintz}, {Holden}, {Labbe}, {Magee}, {Naidu}, {Nelson}, {Stefanon}, {Smit},
  {van Dokkum}, {Weaver}, \& {Williams}}]{Barrufet2022}
{Barrufet}, L., {Oesch}, P.~A., {Weibel}, A., {et~al.} 2022, arXiv e-prints,
  arXiv:2207.14733

\bibitem[{{Bisigello} {et~al.}(2018){Bisigello}, {Caputi}, {Grogin}, \&
  {Koekemoer}}]{Bisigello2018}
{Bisigello}, L., {Caputi}, K.~I., {Grogin}, N., \& {Koekemoer}, A. 2018, \aap,
  609, A82

\bibitem[{{Boucaud} {et~al.}(2016){Boucaud}, {Bocchio}, {Abergel}, {Orieux},
  {Dole}, \& {Hadj-Youcef}}]{Boucaud2016}
{Boucaud}, A., {Bocchio}, M., {Abergel}, A., {et~al.} 2016, \aap, 596, A63

\bibitem[{{Bouwens} {et~al.}(2015){Bouwens}, {Illingworth}, {Oesch}, {Trenti},
  {Labb{\'e}}, {Bradley}, {Carollo}, {van Dokkum}, {Gonzalez}, {Holwerda},
  {Franx}, {Spitler}, {Smit}, \& {Magee}}]{Bouwens2015}
{Bouwens}, R.~J., {Illingworth}, G.~D., {Oesch}, P.~A., {et~al.} 2015, \apj,
  803, 34

\bibitem[{{Brinchmann} {et~al.}(2004){Brinchmann}, {Charlot}, {White},
  {Tremonti}, {Kauffmann}, {Heckman}, \& {Brinkmann}}]{Brinchmann2004}
{Brinchmann}, J., {Charlot}, S., {White}, S.~D.~M., {et~al.} 2004, \mnras, 351,
  1151

\bibitem[{{Bruzual} \& {Charlot}(2003)}]{Bruzual2003}
{Bruzual}, G. \& {Charlot}, S. 2003, \mnras, 344, 1000

\bibitem[{{Burrows} {et~al.}(2006){Burrows}, {Sudarsky}, \&
  {Hubeny}}]{Burrows2006}
{Burrows}, A., {Sudarsky}, D., \& {Hubeny}, I. 2006, \apj, 640, 1063

\bibitem[{{Calzetti} {et~al.}(2000){Calzetti}, {Armus}, {Bohlin}, {Kinney},
  {Koornneef}, \& {Storchi-Bergmann}}]{Calzetti2000}
{Calzetti}, D., {Armus}, L., {Bohlin}, R.~C., {et~al.} 2000, \apj, 533, 682

\bibitem[{{Carnall} {et~al.}(2022){Carnall}, {Begley}, {McLeod}, {Hamadouche},
  {Donnan}, {McLure}, {Dunlop}, {Bondestam}, {Cullen}, {Jewell}, \&
  {Pollock}}]{Carnall2022}
{Carnall}, A.~C., {Begley}, R., {McLeod}, D.~J., {et~al.} 2022, arXiv e-prints,
  arXiv:2207.08778

\bibitem[{Carnall {et~al.}(2018)Carnall, McLure, Dunlop, \&
  Davé}]{Carnall2018}
Carnall, A.~C., McLure, R.~J., Dunlop, J.~S., \& Davé, R. 2018, MNRAS, 480,
  4379

\bibitem[{{Castellano} {et~al.}(2022){Castellano}, {Fontana}, {Treu},
  {Santini}, {Merlin}, {Leethochawalit}, {Trenti}, {Mestric}, {Vanzella},
  {Bonchi}, {Belfiori}, {Nonino}, {Paris}, {Polenta}, {Roberts-Borsani},
  {Boyett}, {Calabro}, {Glazebrook}, {Grillo}, {Mascia}, {Mason}, {Mercurio},
  {Morishita}, {Nanayakkara}, {Pentericci}, {Rosati}, {Vulcani}, {Wang}, \&
  {Yang}}]{Castellano2022}
{Castellano}, M., {Fontana}, A., {Treu}, T., {et~al.} 2022, arXiv e-prints,
  arXiv:2207.09436

\bibitem[{{Chabrier}(2003)}]{Chabrier2003}
{Chabrier}, G. 2003, \pasp, 115, 763

\bibitem[{{Ciesla} {et~al.}(2017){Ciesla}, {Elbaz}, \& {Fensch}}]{Ciesla2017}
{Ciesla}, L., {Elbaz}, D., \& {Fensch}, J. 2017, \aap, 608, A41

\bibitem[{{Coogan} {et~al.}(2023){Coogan}, {Daddi}, {Le Bail}, {Elbaz},
  {Dickinson}, {Giavalisco}, {G{\'o}mez-Guijarro}, {de la Vega}, {Bagley},
  {Finkelstein}, {Franco}, {Cooray}, {Behroozi}, {Bisigello}, {Casey},
  {Ciesla}, {Dimauro}, {Finoguenov}, {Koekemoer}, {Lucas},
  {P{\'e}rez-Gonz{\'a}lez}, {Yung}, {Arrabal Haro}, {Kartaltepe}, {Jogee},
  {Papovich}, {Pirzkal}, \& {Wilkins}}]{Coogan2023}
{Coogan}, R.~T., {Daddi}, E., {Le Bail}, A., {et~al.} 2023, arXiv e-prints,
  arXiv:2302.08960

\bibitem[{{Davis} {et~al.}(2007){Davis}, {Guhathakurta}, {Konidaris}, {Newman},
  {Ashby}, {Biggs}, {Barmby}, {Bundy}, {Chapman}, {Coil}, {Conselice},
  {Cooper}, {Croton}, {Eisenhardt}, {Ellis}, {Faber}, {Fang}, {Fazio},
  {Georgakakis}, {Gerke}, {Goss}, {Gwyn}, {Harker}, {Hopkins}, {Huang},
  {Ivison}, {Kassin}, {Kirby}, {Koekemoer}, {Koo}, {Laird}, {Le Floc'h}, {Lin},
  {Lotz}, {Marshall}, {Martin}, {Metevier}, {Moustakas}, {Nandra}, {Noeske},
  {Papovich}, {Phillips}, {Rich}, {Rieke}, {Rigopoulou}, {Salim},
  {Schiminovich}, {Simard}, {Smail}, {Small}, {Weiner}, {Willmer}, {Willner},
  {Wilson}, {Wright}, \& {Yan}}]{Davis2007}
{Davis}, M., {Guhathakurta}, P., {Konidaris}, N.~P., {et~al.} 2007, \apjl, 660,
  L1

\bibitem[{{Donnan} {et~al.}(2022){Donnan}, {McLeod}, {Dunlop}, {McLure},
  {Carnall}, {Begley}, {Cullen}, {Hamadouche}, {Bowler}, {McCracken},
  {Milvang-Jensen}, {Moneti}, \& {Targett}}]{Donnan2022}
{Donnan}, C.~T., {McLeod}, D.~J., {Dunlop}, J.~S., {et~al.} 2022, arXiv
  e-prints, arXiv:2207.12356

\bibitem[{{Draine} \& {Li}(2007)}]{Draine2007}
{Draine}, B.~T. \& {Li}, A. 2007, \apj, 657, 810

\bibitem[{{Driver} {et~al.}(2016){Driver}, {Andrews}, {Davies}, {Robotham},
  {Wright}, {Windhorst}, {Cohen}, {Emig}, {Jansen}, \& {Dunne}}]{Driver2016}
{Driver}, S.~P., {Andrews}, S.~K., {Davies}, L.~J., {et~al.} 2016, \apj, 827,
  108

\bibitem[{{Elbaz} {et~al.}(2018){Elbaz}, {Leiton}, {Nagar}, {Okumura},
  {Franco}, {Schreiber}, {Pannella}, {Wang}, {Dickinson}, {D{\'\i}az-Santos},
  {Ciesla}, {Daddi}, {Bournaud}, {Magdis}, {Zhou}, \& {Rujopakarn}}]{Elbaz2018}
{Elbaz}, D., {Leiton}, R., {Nagar}, N., {et~al.} 2018, \aap, 616, A110

\bibitem[{{Enia} {et~al.}(2022){Enia}, {Talia}, {Pozzi}, {Cimatti},
  {Delvecchio}, {Zamorani}, {D'Amato}, {Bisigello}, {Gruppioni}, {Rodighiero},
  {Calura}, {Dallacasa}, {Giulietti}, {Barchiesi}, {Behiri}, \&
  {Romano}}]{Enia2022}
{Enia}, A., {Talia}, M., {Pozzi}, F., {et~al.} 2022, \apj, 927, 204

\bibitem[{{Erfanianfar} {et~al.}(2013){Erfanianfar}, {Finoguenov}, {Tanaka},
  {Lerchster}, {Nandra}, {Laird}, {Connelly}, {Bielby}, {Mirkazemi}, {Faber},
  {Kocevski}, {Cooper}, {Newman}, {Jeltema}, {Coil}, {Brimioulle}, {Davis},
  {McCracken}, {Willmer}, {Gerke}, {Cappelluti}, \& {Gwyn}}]{Erfanianfar2013}
{Erfanianfar}, G., {Finoguenov}, A., {Tanaka}, M., {et~al.} 2013, \apj, 765,
  117

\bibitem[{{Ferland} {et~al.}(2017){Ferland}, {Chatzikos}, {Guzm{\'a}n},
  {Lykins}, {van Hoof}, {Williams}, {Abel}, {Badnell}, {Keenan}, {Porter}, \&
  {Stancil}}]{Ferland2017}
{Ferland}, G.~J., {Chatzikos}, M., {Guzm{\'a}n}, F., {et~al.} 2017, \rmxaa, 53,
  385

\bibitem[{{Finkelstein} {et~al.}(2022{\natexlab{a}}){Finkelstein}, {Bagley},
  {Song}, {Larson}, {Papovich}, {Dickinson}, {Finkelstein}, {Koekemoer},
  {Pirzkal}, {Somerville}, {Yung}, {Behroozi}, {Ferguson}, {Giavalisco},
  {Grogin}, {Hathi}, {Hutchison}, {Jung}, {Kocevski}, {Kawinwanichakij},
  {Rojas-Ruiz}, {Ryan}, {Snyder}, \& {Tacchella}}]{Finkelstein2022aa}
{Finkelstein}, S.~L., {Bagley}, M., {Song}, M., {et~al.} 2022{\natexlab{a}},
  \apj, 928, 52

\bibitem[{Finkelstein {et~al.}(2023)Finkelstein, Bagley, Ferguson, Wilkins,
  Kartaltepe, Papovich, Yung, Haro, Behroozi, Dickinson, Kocevski, Koekemoer,
  Larson, Bail, Morales, Pérez-González, Burgarella, Davé, Hirschmann,
  Somerville, Wuyts, Bromm, Casey, Fontana, Fujimoto, Gardner, Giavalisco,
  Grazian, Grogin, Hathi, Hutchison, Jha, Jogee, Kewley, Kirkpatrick, Long,
  Lotz, Pentericci, Pierel, Pirzkal, Ravindranath, Ryan, Trump, Yang,
  Bhatawdekar, Bisigello, Buat, Calabrò, Castellano, Cleri, Cooper, Croton,
  Daddi, Dekel, Elbaz, Franco, Gawiser, Holwerda, Huertas-Company, Jaskot,
  Leung, Lucas, Mobasher, Pandya, Tacchella, Weiner, \&
  Zavala}]{Finkelstein2023}
Finkelstein, S.~L., Bagley, M.~B., Ferguson, H.~C., {et~al.} 2023, The
  Astrophysical Journal Letters, 946, L13

\bibitem[{{Finkelstein} {et~al.}(2022{\natexlab{b}}){Finkelstein}, {Bagley},
  {Haro}, {Dickinson}, {Ferguson}, {Kartaltepe}, {Papovich}, {Burgarella},
  {Kocevski}, {Huertas-Company}, {Iyer}, {Koekemoer}, {Larson},
  {P{\'e}rez-Gonz{\'a}lez}, {Rose}, {Tacchella}, {Wilkins}, {Chworowsky},
  {Medrano}, {Morales}, {Somerville}, {Yung}, {Fontana}, {Giavalisco},
  {Grazian}, {Grogin}, {Kewley}, {Kirkpatrick}, {Kurczynski}, {Lotz},
  {Pentericci}, {Pirzkal}, {Ravindranath}, {Ryan}, {Trump}, {Yang}, {Almaini},
  {Amor{\'\i}n}, {Annunziatella}, {Backhaus}, {Barro}, {Behroozi}, {Bell},
  {Bhatawdekar}, {Bisigello}, {Bromm}, {Buat}, {Buitrago}, {Calabr{\`o}},
  {Casey}, {Castellano}, {Ch{\'a}vez Ortiz}, {Ciesla}, {Cleri}, {Cohen},
  {Cole}, {Cooke}, {Cooper}, {Cooray}, {Costantin}, {Cox}, {Croton}, {Daddi},
  {Dav{\'e}}, {de La Vega}, {Dekel}, {Elbaz}, {Estrada-Carpenter}, {Faber},
  {Fern{\'a}ndez}, {Finkelstein}, {Freundlich}, {Fujimoto},
  {Garc{\'\i}a-Argum{\'a}nez}, {Gardner}, {Gawiser}, {G{\'o}mez-Guijarro},
  {Guo}, {Hamblin}, {Hamilton}, {Hathi}, {Holwerda}, {Hirschmann}, {Hutchison},
  {Jaskot}, {Jha}, {Jogee}, {Juneau}, {Jung}, {Kassin}, {Le Bail}, {Leung},
  {Lucas}, {Magnelli}, {Mantha}, {Matharu}, {McGrath}, {McIntosh}, {Merlin},
  {Mobasher}, {Newman}, {Nicholls}, {Pandya}, {Rafelski}, {Ronayne}, {Santini},
  {Seill{\'e}}, {Shah}, {Shen}, {Simons}, {Snyder}, {Stanway}, {Straughn},
  {Teplitz}, {Vanderhoof}, {Vega-Ferrero}, {Wang}, {Weiner}, {Willmer},
  {Wuyts}, {Zavala}, \& {CEERS Team}}]{Finkelstein2022a}
{Finkelstein}, S.~L., {Bagley}, M.~B., {Haro}, P.~A., {et~al.}
  2022{\natexlab{b}}, \apjl, 940, L55

\bibitem[{{Franco} {et~al.}(2018){Franco}, {Elbaz}, {B{\'e}thermin},
  {Magnelli}, {Schreiber}, {Ciesla}, {Dickinson}, {Nagar}, {Silverman},
  {Daddi}, {Alexander}, {Wang}, {Pannella}, {Le Floc'h}, {Pope}, {Giavalisco},
  {Maury}, {Bournaud}, {Chary}, {Demarco}, {Ferguson}, {Finkelstein}, {Inami},
  {Iono}, {Juneau}, {Lagache}, {Leiton}, {Lin}, {Magdis}, {Messias},
  {Motohara}, {Mullaney}, {Okumura}, {Papovich}, {Pforr}, {Rujopakarn},
  {Sargent}, {Shu}, \& {Zhou}}]{Franco2018}
{Franco}, M., {Elbaz}, D., {B{\'e}thermin}, M., {et~al.} 2018, \aap, 620, A152

\bibitem[{{Fudamoto} {et~al.}(2020){Fudamoto}, {Oesch}, {Magnelli},
  {Schinnerer}, {Liu}, {Lang}, {Jim{\'e}nez-Andrade}, {Groves}, {Leslie}, \&
  {Sargent}}]{Fudamoto2020}
{Fudamoto}, Y., {Oesch}, P.~A., {Magnelli}, B., {et~al.} 2020, \mnras, 491,
  4724

\bibitem[{{Gordon} {et~al.}(2003){Gordon}, {Clayton}, {Misselt}, {Landolt}, \&
  {Wolff}}]{Gordon2003}
{Gordon}, K.~D., {Clayton}, G.~C., {Misselt}, K.~A., {Landolt}, A.~U., \&
  {Wolff}, M.~J. 2003, \apj, 594, 279

\bibitem[{Grogin {et~al.}(2011)Grogin, Kocevski, Faber, Ferguson, Koekemoer,
  Riess, Acquaviva, Alexander, Almaini, Ashby, Barden, Bell, Bournaud, Brown,
  Caputi, Casertano, Cassata, Castellano, Challis, Chary, Cheung, Cirasuolo,
  Conselice, Cooray, Croton, Daddi, Dahlen, Dav{\'{e}}, de~Mello, Dekel,
  Dickinson, Dolch, Donley, Dunlop, Dutton, Elbaz, Fazio, Filippenko,
  Finkelstein, Fontana, Gardner, Garnavich, Gawiser, Giavalisco, Grazian, Guo,
  Hathi, Häussler, Hopkins, Huang, Huang, Jha, Kartaltepe, Kirshner, Koo, Lai,
  Lee, Li, Lotz, Lucas, Madau, McCarthy, McGrath, McIntosh, McLure, Mobasher,
  Moustakas, Mozena, Nandra, Newman, Niemi, Noeske, Papovich, Pentericci, Pope,
  Primack, Rajan, Ravindranath, Reddy, Renzini, Rix, Robaina, Rodney, Rosario,
  Rosati, Salimbeni, Scarlata, Siana, Simard, Smidt, Somerville, Spinrad,
  Straughn, Strolger, Telford, Teplitz, Trump, van~der Wel, Villforth,
  Wechsler, Weiner, Wiklind, Wild, Wilson, Wuyts, Yan, \& Yun}]{Grogin2011}
Grogin, N.~A., Kocevski, D.~D., Faber, S.~M., {et~al.} 2011, The Astrophysical
  Journal Supplement Series, 197, 35

\bibitem[{{Gruppioni} {et~al.}(2020){Gruppioni}, {B{\'e}thermin}, {Loiacono},
  {Le F{\`e}vre}, {Capak}, {Cassata}, {Faisst}, {Schaerer}, {Silverman}, {Yan},
  {Bardelli}, {Boquien}, {Carraro}, {Cimatti}, {Dessauges-Zavadsky}, {Ginolfi},
  {Fujimoto}, {Hathi}, {Jones}, {Khusanova}, {Koekemoer}, {Lagache}, {Lemaux},
  {Oesch}, {Pozzi}, {Riechers}, {Rodighiero}, {Romano}, {Talia}, {Vallini},
  {Vergani}, {Zamorani}, \& {Zucca}}]{Gruppioni2020}
{Gruppioni}, C., {B{\'e}thermin}, M., {Loiacono}, F., {et~al.} 2020, \aap, 643,
  A8

\bibitem[{{Hauser} \& {Dwek}(2001)}]{Hauser2001}
{Hauser}, M.~G. \& {Dwek}, E. 2001, \araa, 39, 249

\bibitem[{{Holwerda} {et~al.}(2015){Holwerda}, {Bouwens}, {Oesch}, {Smit},
  {Illingworth}, \& {Labbe}}]{Holwerda2015}
{Holwerda}, B.~W., {Bouwens}, R., {Oesch}, P., {et~al.} 2015, \apj, 808, 6

\bibitem[{{Holwerda} {et~al.}(2020){Holwerda}, {Bridge}, {Steele}, {Kusmic},
  {Bradley}, {Livermore}, {Bernard}, \& {Jacques}}]{Holwerda2020}
{Holwerda}, B.~W., {Bridge}, J.~S., {Steele}, R.~L., {et~al.} 2020, \aj, 160,
  154

\bibitem[{{Jin} {et~al.}(2022){Jin}, {Sillassen}, {Magdis}, {Vijayan},
  {Brammer}, {Kokorev}, {Weaver}, {Gobat}, {Gim{\'e}nez-Arteaga}, {Valentino},
  {Brinch}, {G{\'o}mez-Guijarro}, {Shuntov}, {Toft}, {Greve}, \& {Blanquez
  Sese}}]{Jin2022}
{Jin}, S., {Sillassen}, N.~B., {Magdis}, G.~E., {et~al.} 2022, arXiv e-prints,
  arXiv:2212.09372

\bibitem[{{Kennicutt}(1998)}]{Kennicutt1998}
{Kennicutt}, Robert~C., J. 1998, \apj, 498, 541

\bibitem[{{Koekemoer} {et~al.}(2011){Koekemoer}, {Faber}, {Ferguson}, {Grogin},
  {Kocevski}, {Koo}, {Lai}, {Lotz}, {Lucas}, {McGrath}, {Ogaz}, {Rajan},
  {Riess}, {Rodney}, {Strolger}, {Casertano}, {Castellano}, {Dahlen},
  {Dickinson}, {Dolch}, {Fontana}, {Giavalisco}, {Grazian}, {Guo}, {Hathi},
  {Huang}, {van der Wel}, {Yan}, {Acquaviva}, {Alexander}, {Almaini}, {Ashby},
  {Barden}, {Bell}, {Bournaud}, {Brown}, {Caputi}, {Cassata}, {Challis},
  {Chary}, {Cheung}, {Cirasuolo}, {Conselice}, {Roshan Cooray}, {Croton},
  {Daddi}, {Dav{\'e}}, {de Mello}, {de Ravel}, {Dekel}, {Donley}, {Dunlop},
  {Dutton}, {Elbaz}, {Fazio}, {Filippenko}, {Finkelstein}, {Frazer}, {Gardner},
  {Garnavich}, {Gawiser}, {Gruetzbauch}, {Hartley}, {H{\"a}ussler},
  {Herrington}, {Hopkins}, {Huang}, {Jha}, {Johnson}, {Kartaltepe},
  {Khostovan}, {Kirshner}, {Lani}, {Lee}, {Li}, {Madau}, {McCarthy},
  {McIntosh}, {McLure}, {McPartland}, {Mobasher}, {Moreira}, {Mortlock},
  {Moustakas}, {Mozena}, {Nandra}, {Newman}, {Nielsen}, {Niemi}, {Noeske},
  {Papovich}, {Pentericci}, {Pope}, {Primack}, {Ravindranath}, {Reddy},
  {Renzini}, {Rix}, {Robaina}, {Rosario}, {Rosati}, {Salimbeni}, {Scarlata},
  {Siana}, {Simard}, {Smidt}, {Snyder}, {Somerville}, {Spinrad}, {Straughn},
  {Telford}, {Teplitz}, {Trump}, {Vargas}, {Villforth}, {Wagner}, {Wandro},
  {Wechsler}, {Weiner}, {Wiklind}, {Wild}, {Wilson}, {Wuyts}, \&
  {Yun}}]{Koekemoer2011}
{Koekemoer}, A.~M., {Faber}, S.~M., {Ferguson}, H.~C., {et~al.} 2011, \apjs,
  197, 36

\bibitem[{{Koushan} {et~al.}(2021){Koushan}, {Driver}, {Bellstedt}, {Davies},
  {Robotham}, {Lagos}, {Hashemizadeh}, {Obreschkow}, {Thorne}, {Bremer},
  {Holwerda}, {Hopkins}, {Jarvis}, {Siudek}, \& {Windhorst}}]{Koushan2021}
{Koushan}, S., {Driver}, S.~P., {Bellstedt}, S., {et~al.} 2021, \mnras, 503,
  2033

\bibitem[{{Kramer} {et~al.}(2022){Kramer}, {Carleton}, {Cohen}, {Jansen},
  {Windhorst}, {Grogin}, {Koekemoer}, {MacKenty}, \& {Pirzkal}}]{Kramer2022}
{Kramer}, D.~M., {Carleton}, T., {Cohen}, S.~H., {et~al.} 2022, \apjl, 940, L15

\bibitem[{{Leja} {et~al.}(2017){Leja}, {Johnson}, {Conroy}, {van Dokkum}, \&
  {Byler}}]{Leja2017}
{Leja}, J., {Johnson}, B.~D., {Conroy}, C., {van Dokkum}, P.~G., \& {Byler}, N.
  2017, \apj, 837, 170

\bibitem[{{Madau} \& {Dickinson}(2014)}]{Madau2014}
{Madau}, P. \& {Dickinson}, M. 2014, \araa, 52, 415

\bibitem[{{McKee} {et~al.}(2015){McKee}, {Parravano}, \&
  {Hollenbach}}]{McKee2015}
{McKee}, C.~F., {Parravano}, A., \& {Hollenbach}, D.~J. 2015, \apj, 814, 13

\bibitem[{{McLure} {et~al.}(2018){McLure}, {Dunlop}, {Cullen}, {Bourne},
  {Best}, {Khochfar}, {Bowler}, {Biggs}, {Geach}, {Scott}, {Micha{\l}owski},
  {Rujopakarn}, {van Kampen}, {Kirkpatrick}, \& {Pope}}]{McLure2018}
{McLure}, R.~J., {Dunlop}, J.~S., {Cullen}, F., {et~al.} 2018, \mnras, 476,
  3991

\bibitem[{{Momcheva} {et~al.}(2016){Momcheva}, {Brammer}, {van Dokkum},
  {Skelton}, {Whitaker}, {Nelson}, {Fumagalli}, {Maseda}, {Leja}, {Franx},
  {Rix}, {Bezanson}, {Da Cunha}, {Dickey}, {F{\"o}rster Schreiber},
  {Illingworth}, {Kriek}, {Labb{\'e}}, {Ulf Lange}, {Lundgren}, {Magee},
  {Marchesini}, {Oesch}, {Pacifici}, {Patel}, {Price}, {Tal}, {Wake}, {van der
  Wel}, \& {Wuyts}}]{Momcheva2016}
{Momcheva}, I.~G., {Brammer}, G.~B., {van Dokkum}, P.~G., {et~al.} 2016, \apjs,
  225, 27

\bibitem[{{Naidu} {et~al.}(2022){Naidu}, {Oesch}, {van Dokkum}, {Nelson},
  {Suess}, {Whitaker}, {Allen}, {Bezanson}, {Bouwens}, {Brammer}, {Conroy},
  {Illingworth}, {Labbe}, {Leja}, {Leonova}, {Matthee}, {Price}, {Setton},
  {Strait}, {Stefanon}, {Tacchella}, {Toft}, {Weaver}, \& {Weibel}}]{Naidu2022}
{Naidu}, R.~P., {Oesch}, P.~A., {van Dokkum}, P., {et~al.} 2022, arXiv
  e-prints, arXiv:2207.09434

\bibitem[{{Noeske} {et~al.}(2007){Noeske}, {Weiner}, {Faber}, {Papovich},
  {Koo}, {Somerville}, {Bundy}, {Conselice}, {Newman}, {Schiminovich}, {Le
  Floc'h}, {Coil}, {Rieke}, {Lotz}, {Primack}, {Barmby}, {Cooper}, {Davis},
  {Ellis}, {Fazio}, {Guhathakurta}, {Huang}, {Kassin}, {Martin}, {Phillips},
  {Rich}, {Small}, {Willmer}, \& {Wilson}}]{Noeske2007}
{Noeske}, K.~G., {Weiner}, B.~J., {Faber}, S.~M., {et~al.} 2007, \apjl, 660,
  L43

\bibitem[{{Novak} {et~al.}(2017){Novak}, {Smol{\v{c}}i{\'c}}, {Delhaize},
  {Delvecchio}, {Zamorani}, {Baran}, {Bondi}, {Capak}, {Carilli}, {Ciliegi},
  {Civano}, {Ilbert}, {Karim}, {Laigle}, {Le F{\`e}vre}, {Marchesi},
  {McCracken}, {Miettinen}, {Salvato}, {Sargent}, {Schinnerer}, \&
  {Tasca}}]{Novak2017}
{Novak}, M., {Smol{\v{c}}i{\'c}}, V., {Delhaize}, J., {et~al.} 2017, \aap, 602,
  A5

\bibitem[{{Oesch} {et~al.}(2018){Oesch}, {Bouwens}, {Illingworth}, {Labb{\'e}},
  \& {Stefanon}}]{Oesch2018}
{Oesch}, P.~A., {Bouwens}, R.~J., {Illingworth}, G.~D., {Labb{\'e}}, I., \&
  {Stefanon}, M. 2018, \apj, 855, 105

\bibitem[{{Oke} \& {Gunn}(1983)}]{Oke1983}
{Oke}, J.~B. \& {Gunn}, J.~E. 1983, \apj, 266, 713

\bibitem[{{Pannella} {et~al.}(2015){Pannella}, {Elbaz}, {Daddi}, {Dickinson},
  {Hwang}, {Schreiber}, {Strazzullo}, {Aussel}, {Bethermin}, {Buat},
  {Charmandaris}, {Cibinel}, {Juneau}, {Ivison}, {Le Borgne}, {Le Floc'h},
  {Leiton}, {Lin}, {Magdis}, {Morrison}, {Mullaney}, {Onodera}, {Renzini},
  {Salim}, {Sargent}, {Scott}, {Shu}, \& {Wang}}]{Pannella2015}
{Pannella}, M., {Elbaz}, D., {Daddi}, E., {et~al.} 2015, \apj, 807, 141

\bibitem[{{P{\'e}rez-Gonz{\'a}lez} {et~al.}(2022){P{\'e}rez-Gonz{\'a}lez},
  {Barro}, {Annunziatella}, {Costantin}, {Garc{\'\i}a-Argum{\'a}nez},
  {McGrath}, {M{\'e}rida}, {Zavala}, {Arrabal Haro}, {Bagley}, {Backhaus},
  {Behroozi}, {Bell}, {Buat}, {Calabr{\`o}}, {Casey}, {Cleri}, {Coogan},
  {Cooper}, {Cooray}, {Dekel}, {Dickinson}, {Elbaz}, {Ferguson}, {Finkelstein},
  {Fontana}, {Franco}, {Gardner}, {Giavalisco}, {G{\'o}mez-Guijarro},
  {Grazian}, {Grogin}, {Guo}, {Jogee}, {Kartaltepe}, {Kewley}, {Kirkpatrick},
  {Kocevski}, {Koekemoer}, {Long}, {Lotz}, {Lucas}, {Papovich}, {Pirzkal},
  {Ravindranath}, {Somerville}, {Tacchella}, {Trump}, {Wang}, {Wilkins},
  {Wuyts}, {Yang}, \& {Yung}}]{PerezGonzales2023}
{P{\'e}rez-Gonz{\'a}lez}, P.~G., {Barro}, G., {Annunziatella}, M., {et~al.}
  2022, arXiv e-prints, arXiv:2211.00045

\bibitem[{{P{\'e}rez-Gonz{\'a}lez} {et~al.}(2023){P{\'e}rez-Gonz{\'a}lez},
  {Costantin}, {Langeroodi}, {Rinaldi}, {Annunziatella}, {Ilbert}, {Colina},
  {Noorgaard-Nielsen}, {Greve}, {Ostlin}, {Wright}, {Alonso-Herrero},
  {{\'A}lvarez-M{\'a}rquez}, {Caputi}, {Eckart}, {Le F{\`e}vre}, {Labiano},
  {Garc{\'\i}a-Mar{\'\i}n}, {Hjorth}, {Kendrew}, {Pye}, {Tikkanen}, {van der
  Werf}, {Walter}, {Ward}, {Bosman}, {Gillman}, {Garc{\'\i}a-Argum{\'a}nez}, \&
  {Mar{\'\i}a M{\'e}rida}}]{PerezGonzales2023b}
{P{\'e}rez-Gonz{\'a}lez}, P.~G., {Costantin}, L., {Langeroodi}, D., {et~al.}
  2023, arXiv e-prints, arXiv:2302.02429

\bibitem[{{Puglisi} {et~al.}(2017){Puglisi}, {Daddi}, {Renzini}, {Rodighiero},
  {Silverman}, {Kashino}, {Rodr{\'\i}guez-Mu{\~n}oz}, {Mancini}, {Mainieri},
  {Man}, {Franceschini}, {Valentino}, {Calabr{\`o}}, {Jin}, {Darvish}, {Maier},
  {Kartaltepe}, \& {Sanders}}]{Puglisi2017}
{Puglisi}, A., {Daddi}, E., {Renzini}, A., {et~al.} 2017, \apjl, 838, L18

\bibitem[{{Rodighiero} {et~al.}(2023){Rodighiero}, {Bisigello}, {Iani},
  {Marasco}, {Grazian}, {Sinigaglia}, {Cassata}, \&
  {Gruppioni}}]{Rodighiero2023}
{Rodighiero}, G., {Bisigello}, L., {Iani}, E., {et~al.} 2023, \mnras, 518, L19

\bibitem[{{Rodighiero} {et~al.}(2011){Rodighiero}, {Daddi}, {Baronchelli},
  {Cimatti}, {Renzini}, {Aussel}, {Popesso}, {Lutz}, {Andreani}, {Berta},
  {Cava}, {Elbaz}, {Feltre}, {Fontana}, {F{\"o}rster Schreiber},
  {Franceschini}, {Genzel}, {Grazian}, {Gruppioni}, {Ilbert}, {Le Floch},
  {Magdis}, {Magliocchetti}, {Magnelli}, {Maiolino}, {McCracken}, {Nordon},
  {Poglitsch}, {Santini}, {Pozzi}, {Riguccini}, {Tacconi}, {Wuyts}, \&
  {Zamorani}}]{Rodighiero2011}
{Rodighiero}, G., {Daddi}, E., {Baronchelli}, I., {et~al.} 2011, \apjl, 739,
  L40

\bibitem[{{Rowan-Robinson} {et~al.}(2016){Rowan-Robinson}, {Oliver}, {Wang},
  {Farrah}, {Clements}, {Gruppioni}, {Marchetti}, {Rigopoulou}, \&
  {Vaccari}}]{Rowanrobinson2016}
{Rowan-Robinson}, M., {Oliver}, S., {Wang}, L., {et~al.} 2016, \mnras, 461,
  1100

\bibitem[{{Santini} {et~al.}(2017){Santini}, {Fontana}, {Castellano}, {Di
  Criscienzo}, {Merlin}, {Amorin}, {Cullen}, {Daddi}, {Dickinson}, {Dunlop},
  {Grazian}, {Lamastra}, {McLure}, {Micha{\l}owski}, {Pentericci}, \&
  {Shu}}]{Santini2017}
{Santini}, P., {Fontana}, A., {Castellano}, M., {et~al.} 2017, \apj, 847, 76

\bibitem[{{Scalo} \& {Slavsky}(1980)}]{Scalo1980}
{Scalo}, J.~M. \& {Slavsky}, D.~B. 1980, \apjl, 239, L73

\bibitem[{{Shapley} {et~al.}(2022){Shapley}, {Sanders}, {Salim}, {Reddy},
  {Kriek}, {Mobasher}, {Coil}, {Siana}, {Price}, {Shivaei}, {Dunlop}, {McLure},
  \& {Cullen}}]{Shapley2022}
{Shapley}, A.~E., {Sanders}, R.~L., {Salim}, S., {et~al.} 2022, \apj, 926, 145

\bibitem[{{Simpson} {et~al.}(2014){Simpson}, {Swinbank}, {Smail}, {Alexander},
  {Brandt}, {Bertoldi}, {de Breuck}, {Chapman}, {Coppin}, {da Cunha},
  {Danielson}, {Dannerbauer}, {Greve}, {Hodge}, {Ivison}, {Karim}, {Knudsen},
  {Poggianti}, {Schinnerer}, {Thomson}, {Walter}, {Wardlow}, {Wei{\ss}}, \&
  {van der Werf}}]{Simpson2014}
{Simpson}, J.~M., {Swinbank}, A.~M., {Smail}, I., {et~al.} 2014, \apj, 788, 125

\bibitem[{{Speagle} {et~al.}(2014){Speagle}, {Steinhardt}, {Capak}, \&
  {Silverman}}]{Speagle2014}
{Speagle}, J.~S., {Steinhardt}, C.~L., {Capak}, P.~L., \& {Silverman}, J.~D.
  2014, \apjs, 214, 15

\bibitem[{{Steinhardt} {et~al.}(2014){Steinhardt}, {Speagle}, {Capak},
  {Silverman}, {Carollo}, {Dunlop}, {Hashimoto}, {Hsieh}, {Ilbert}, {Le Fevre},
  {Le Floc'h}, {Lee}, {Lin}, {Lin}, {Masters}, {McCracken}, {Nagao}, {Petric},
  {Salvato}, {Sanders}, {Scoville}, {Sheth}, {Strauss}, \&
  {Taniguchi}}]{Steinhardt2014}
{Steinhardt}, C.~L., {Speagle}, J.~S., {Capak}, P., {et~al.} 2014, \apjl, 791,
  L25

\bibitem[{{Talia} {et~al.}(2021){Talia}, {Cimatti}, {Giulietti}, {Zamorani},
  {Bethermin}, {Faisst}, {Le F{\`e}vre}, \& {Smol{\c{c}}i{\'c}}}]{Talia2021}
{Talia}, M., {Cimatti}, A., {Giulietti}, M., {et~al.} 2021, \apj, 909, 23

\bibitem[{{Treu} {et~al.}(2022){Treu}, {Roberts-Borsani}, {Bradac}, {Brammer},
  {Fontana}, {Henry}, {Mason}, {Morishita}, {Pentericci}, {Wang}, {Acebron},
  {Bagley}, {Bergamini}, {Belfiori}, {Bonchi}, {Boyett}, {Boutsia},
  {Calabr{\'o}}, {Caminha}, {Castellano}, {Dressler}, {Glazebrook}, {Grillo},
  {Jacobs}, {Jones}, {Kelly}, {Leethochawalit}, {Malkan}, {Marchesini},
  {Mascia}, {Mercurio}, {Merlin}, {Nanayakkara}, {Nonino}, {Paris},
  {Poggianti}, {Rosati}, {Santini}, {Scarlata}, {Shipley}, {Strait}, {Trenti},
  {Tubthong}, {Vanzella}, {Vulcani}, \& {Yang}}]{Treu2022}
{Treu}, T., {Roberts-Borsani}, G., {Bradac}, M., {et~al.} 2022, \apj, 935, 110

\bibitem[{{van der Wel} {et~al.}(2014){van der Wel}, {Franx}, {van Dokkum},
  {Skelton}, {Momcheva}, {Whitaker}, {Brammer}, {Bell}, {Rix}, {Wuyts},
  {Ferguson}, {Holden}, {Barro}, {Koekemoer}, {Chang}, {McGrath},
  {H{\"a}ussler}, {Dekel}, {Behroozi}, {Fumagalli}, {Leja}, {Lundgren},
  {Maseda}, {Nelson}, {Wake}, {Patel}, {Labb{\'e}}, {Faber}, {Grogin}, \&
  {Kocevski}}]{Wel2014}
{van der Wel}, A., {Franx}, M., {van Dokkum}, P.~G., {et~al.} 2014, \apj, 788,
  28

\bibitem[{{Wang} {et~al.}(2019){Wang}, {Schreiber}, {Elbaz}, {Yoshimura},
  {Kohno}, {Shu}, {Yamaguchi}, {Pannella}, {Franco}, {Huang}, {Lim}, \&
  {Wang}}]{Wang2019}
{Wang}, T., {Schreiber}, C., {Elbaz}, D., {et~al.} 2019, \nat, 572, 211

\bibitem[{{Whitaker} {et~al.}(2017){Whitaker}, {Pope}, {Cybulski}, {Casey},
  {Popping}, \& {Yun}}]{Whitaker2017}
{Whitaker}, K.~E., {Pope}, A., {Cybulski}, R., {et~al.} 2017, \apj, 850, 208

\bibitem[{{Whitaker} {et~al.}(2012){Whitaker}, {van Dokkum}, {Brammer}, \&
  {Franx}}]{Whitaker2012}
{Whitaker}, K.~E., {van Dokkum}, P.~G., {Brammer}, G., \& {Franx}, M. 2012,
  \apjl, 754, L29

\bibitem[{{Wilkins} {et~al.}(2014){Wilkins}, {Stanway}, \&
  {Bremer}}]{Wilkins2014}
{Wilkins}, S.~M., {Stanway}, E.~R., \& {Bremer}, M.~N. 2014, \mnras, 439, 1038

\bibitem[{{Xiao} {et~al.}(2023){Xiao}, {Elbaz}, {G{\'o}mez-Guijarro}, {Leroy},
  {Bing}, {Daddi}, {Magnelli}, {Franco}, {Zhou}, {Dickinson}, {Wang},
  {Rujopakarn}, {Magdis}, {Treister}, {Inami}, {Demarco}, {Sargent}, {Shu},
  {Kartaltepe}, {Alexander}, {B{\'e}thermin}, {Bournaud}, {Ciesla}, {Ferguson},
  {Finkelstein}, {Giavalisco}, {Gu}, {Iono}, {Juneau}, {Lagache}, {Leiton},
  {Messias}, {Motohara}, {Mullaney}, {Nagar}, {Pannella}, {Papovich}, {Pope},
  {Schreiber}, \& {Silverman}}]{Xiao2023}
{Xiao}, M.~Y., {Elbaz}, D., {G{\'o}mez-Guijarro}, C., {et~al.} 2023, \aap, 672,
  A18

\bibitem[{{Yan} {et~al.}(2023{\natexlab{a}}){Yan}, {Cohen}, {Windhorst},
  {Jansen}, {Ma}, {Beacom}, {Ling}, {Cheng}, {Huang}, {Grogin}, {Willner},
  {Yun}, {Hammel}, {Milam}, {Conselice}, {Driver}, {Frye}, {Marshall},
  {Koekemoer}, {Willmer}, {Robotham}, {D'Silva}, {Summers}, {Lim},
  {Harrington}, {Ferreira}, {Diego}, {Pirzkal}, {Wilkins}, {Wang}, {Hathi},
  {Zitrin}, {Bhatawdekar}, {Adams}, {Furtak}, {Maksym}, {Rutkowski}, \&
  {Fazio}}]{Yan2023}
{Yan}, H., {Cohen}, S.~H., {Windhorst}, R.~A., {et~al.} 2023{\natexlab{a}},
  \apjl, 942, L8

\bibitem[{{Yan} {et~al.}(2023{\natexlab{b}}){Yan}, {Ma}, {Ling}, {Cheng}, \&
  {Huang}}]{Yan2023a}
{Yan}, H., {Ma}, Z., {Ling}, C., {Cheng}, C., \& {Huang}, J.-S.
  2023{\natexlab{b}}, \apjl, 942, L9

\bibitem[{{Zavala} {et~al.}(2022){Zavala}, {Buat}, {Casey}, {Burgarella},
  {Finkelstein}, {Bagley}, {Ciesla}, {Daddi}, {Dickinson}, {Ferguson},
  {Franco}, {Jim'enez-Andrade}, {Kartaltepe}, {Koekemoer}, {Le Bail}, {Murphy},
  {Papovich}, {Tacchella}, {Wilkins}, {Aretxaga}, {Behroozi}, {Champagne},
  {Fontana}, {Giavalisco}, {Grazian}, {Grogin}, {Kewley}, {Kocevski},
  {Kirkpatrick}, {Lotz}, {Pentericci}, {Perez-Gonzalez}, {Pirzkal},
  {Ravindranath}, {Somerville}, {Trump}, {Yang}, {Yung}, {Almaini}, {Amorin},
  {Annunziatella}, {Arrabal Haro}, {Backhaus}, {Barro}, {Bell}, {Bhatawdekar},
  {Bisigello}, {Buitrago}, {Calabro}, {Castellano}, {Chavez Ortiz},
  {Chworowsky}, {Cleri}, {Cohen}, {Cole}, {Cooke}, {Cooper}, {Cooray},
  {Costantin}, {Cox}, {Croton}, {Dave}, {de la Vega}, {Dekel}, {Elbaz},
  {Estrada-Carpenter}, {Fern{\'a}ndez}, {Finkelstein}, {Freundlich},
  {Fujimoto}, {Garc{\'\i}a-Argum{\'a}nez}, {Gardner}, {Gawiser},
  {G{\'o}mez-Guijarro}, {Guo}, {Hamilton}, {Hathi}, {Holwerda}, {Hirschmann},
  {Huertas-Company}, {Hutchison}, {Iyer}, {Jaskot}, {Jha}, {Jogee}, {Juneau},
  {Jung}, {Kassin}, {Kurczynski}, {Larson}, {Leung}, {Long}, {Lucas},
  {Magnelli}, {Mantha}, {Matharu}, {McGrath}, {McIntosh}, {Medrano}, {Merlin},
  {Mobasher}, {Morales}, {Newman}, {Nicholls}, {Pandya}, {Rafelski}, {Ronayne},
  {Rose}, {Ryan}, {Santini}, {Seill{\'e}}, {Shah}, {Shen}, {Simons}, {Snyder},
  {Stanway}, {Straughn}, {Teplitz}, {Vanderhoof}, {Vega-Ferrero}, {Wang},
  {Weiner}, {Willmer}, \& {Wuyts}}]{Zavala2022}
{Zavala}, J.~A., {Buat}, V., {Casey}, C.~M., {et~al.} 2022, arXiv e-prints,
  arXiv:2208.01816

\end{thebibliography}
%

\begin{appendix} 
\section{Comparison with aperture photometry}\label{sec:fixedpaerture}
As reported before, the sample selection and the derivation of the physical properties are based on the Kron aperture photometry, however, given the faintness of our sources, the photometric extraction could strongly influence our results. In this Appendix we analyse the reliability of our results repeating the analysis by deriving the flux on a 0.2"-radius aperture, corrected to total flux. \par
\begin{figure}
    \centering
    \includegraphics[trim={15 20 20 20},clip,width=0.99\linewidth,keepaspectratio]{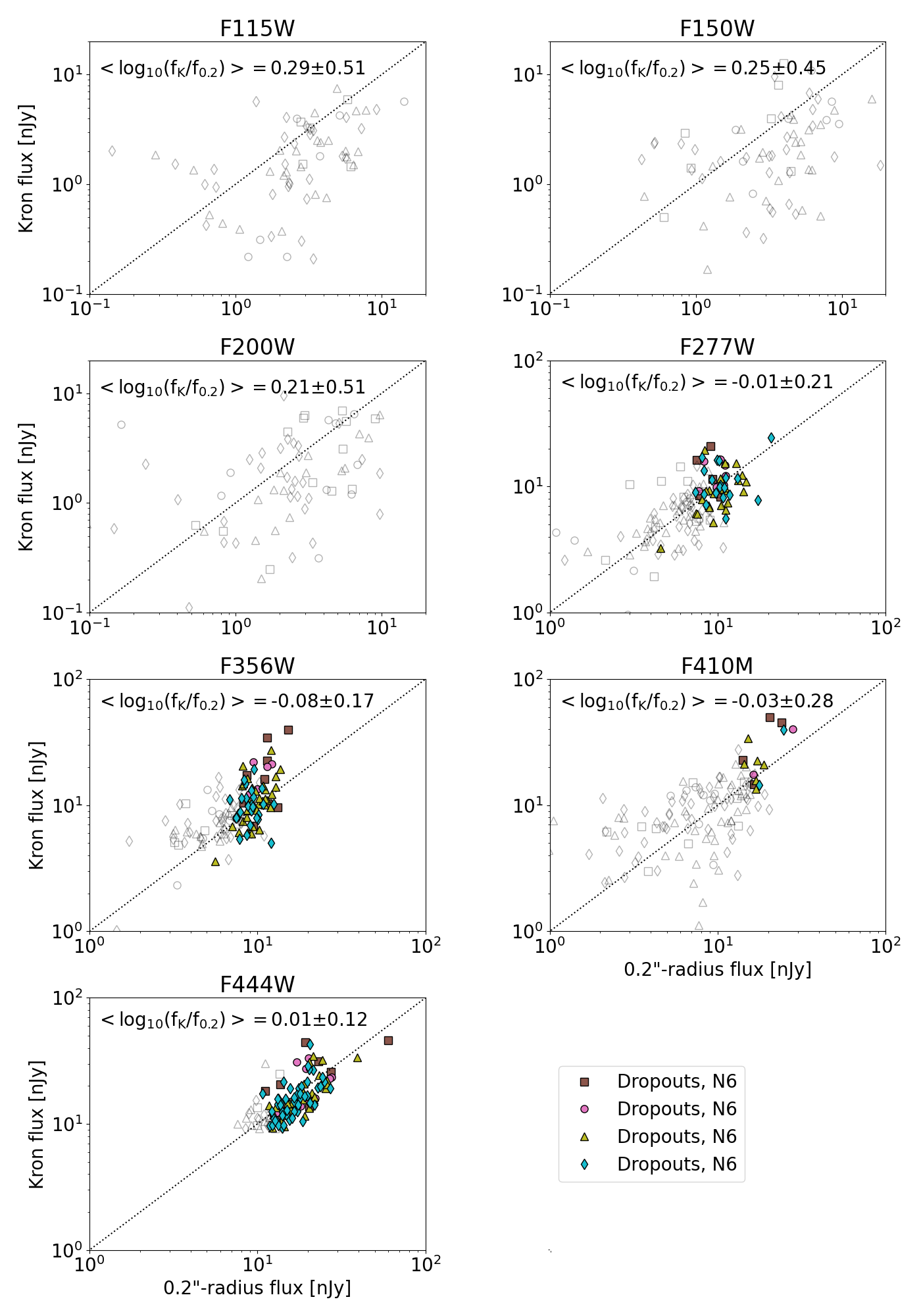}
    \caption{Comparison between the flux density obtained with a Kron aperture and a 0.2"-radius aperture, for JWST bands from F115W (upper left) to F444W (lower left). Empty symbols correspond to objects with S/N$<3$ in the considered filter photometry with at least one of the two apertures. The average logarithm of the ratios between Kron and 0.2"-radius apertures is reported in the top left of each panel. }
    \label{fig:Ap_kron_comparison}
\end{figure}

In Figure \ref{fig:Ap_kron_comparison} we compare the photometry derived with the two apertures. The ratios between fixed and Kron apertures is, on average, within 20$\%$ for filters at $\lambda>2\mu m$. It is larger for the filters at shorter wavelengths, where however all objects have a S/N$<$2. No clear difference is present between the different pointings. \par

\begin{figure}
    \centering
    \includegraphics[width=0.99\linewidth,keepaspectratio]{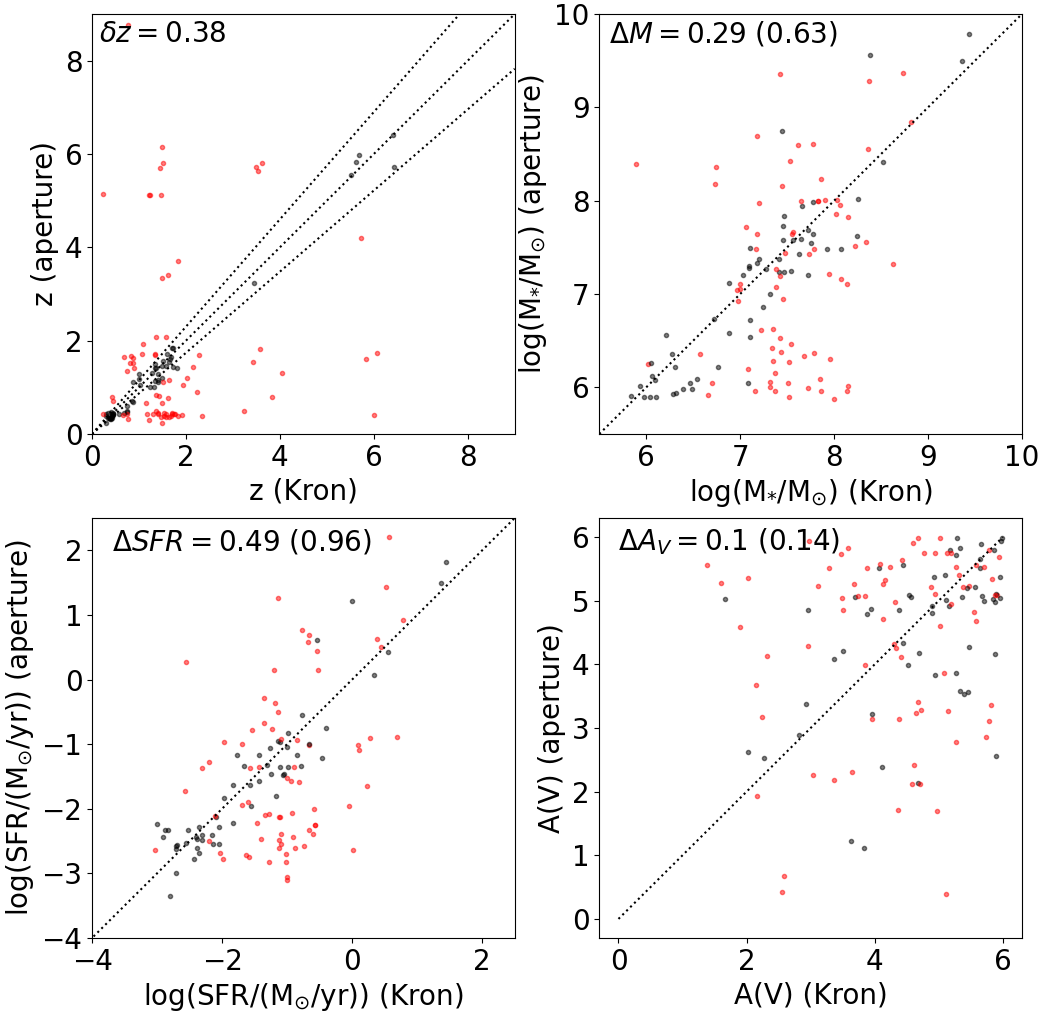}
    \caption{Comparison between the physical parameters derived using BAGPIPES based on photometry derived using a Kron aperture and a 0.2"-radius aperture. From top left to bottom left: redshift, stellar mass, SFR and A(V). Black dots show galaxies for which the redshift is well recovered, while red dots show redshift outliers, i.e. $\delta z=|z_{kron}-z_{0.2"}|/(1+z_{kron})>0.15$. On the top left of each panel we shown the absolute difference between the physical properties derived with Kron and 0.2" aperture, for galaxies with a good redshift and, in brackets, for the entire sample.}
    \label{fig:Ap_kron_comparison_phys}
\end{figure}

We then repeat the SED fitting described in Section \ref{sec:SEDfitting} using the 0.2"-radius aperture and the comparison between redshift, stellar mass, SFR, A(V) and age is reported in Figure \ref{fig:Ap_kron_comparison_phys}. Looking at the redshift, we have that 57$\%$ have a normalised redshift difference (i.e., $\delta z=|z_{kron}-z_{0.2"}|/(1+z_{kron})$) above 0.15 and 30$\%$ have a redshift difference above $|z_{kron}-z_{0.2"}|=1$. These are mainly galaxies for which a double or large peak is present in the probability redshift distribution, so small differences on the photometry results change the probability of the different peaks. For galaxies for which the redshift is consistent within the two photometric sets, the stellar mass is consistent within 0.29 dex, the SFR within 0.49 dex and the dust extinction within 0.10 mag. 

\begin{figure}
    \centering
    \includegraphics[clip,width=0.99\linewidth,keepaspectratio]{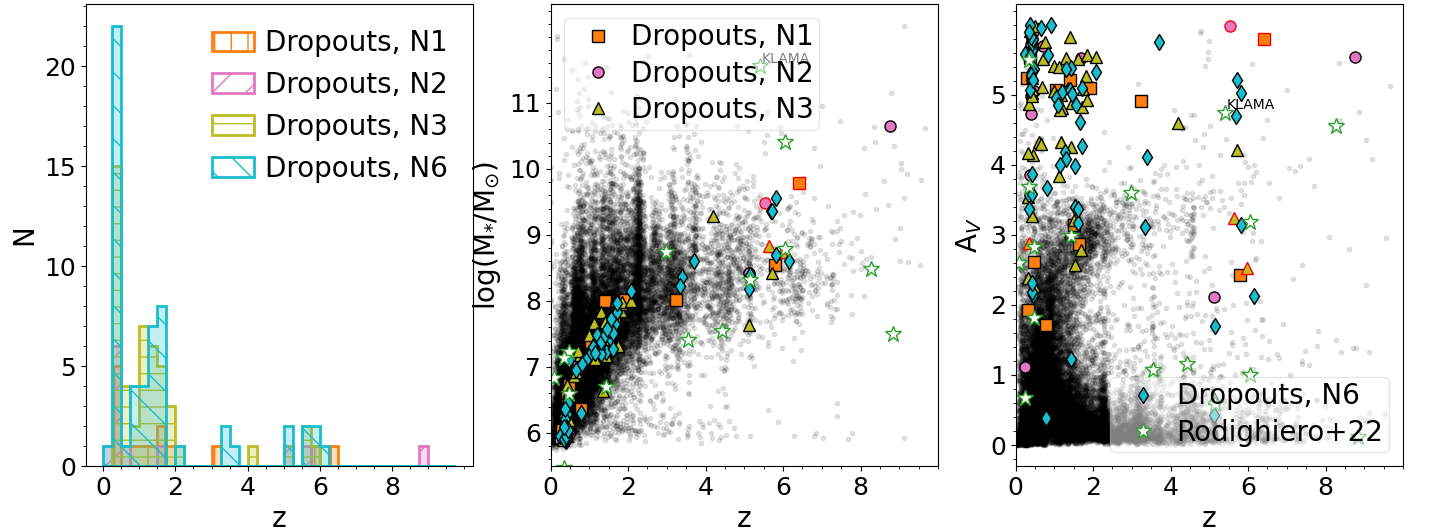}
    \caption{Same as figure \ref{fig:z_M_Av} but derived using 0.2"-radius aperture photometry.}
    \label{fig:z_ap}
\end{figure}

Even considering the differences in the physical properties of each single object derived changing from Kron-aperture to 0.2"-radius photometry, we have that the median properties of the F200W-sample remain similar and the conclusion of our work are unaffected. Indeed, the redshift distribution of the F200W-dropouts galaxies is similar to the one presented in the Section \ref{sec:results}, with 81$\%$ of them being at $z<2$ with a median stellar mass of 10$^{6.9}\,\rm M_{\odot}$ and a median dust extinction of 5.0 mag (Fig. \ref{fig:z_ap}). At the same time, the majority of the F200W-dropout galaxies are star-forming galaxies, with still 26$\%$ of SB at $z<2$, compared to 27$\%$ derived with the Kron photometry.

\begin{figure}
    \centering
    \includegraphics[width=0.99\linewidth,keepaspectratio]{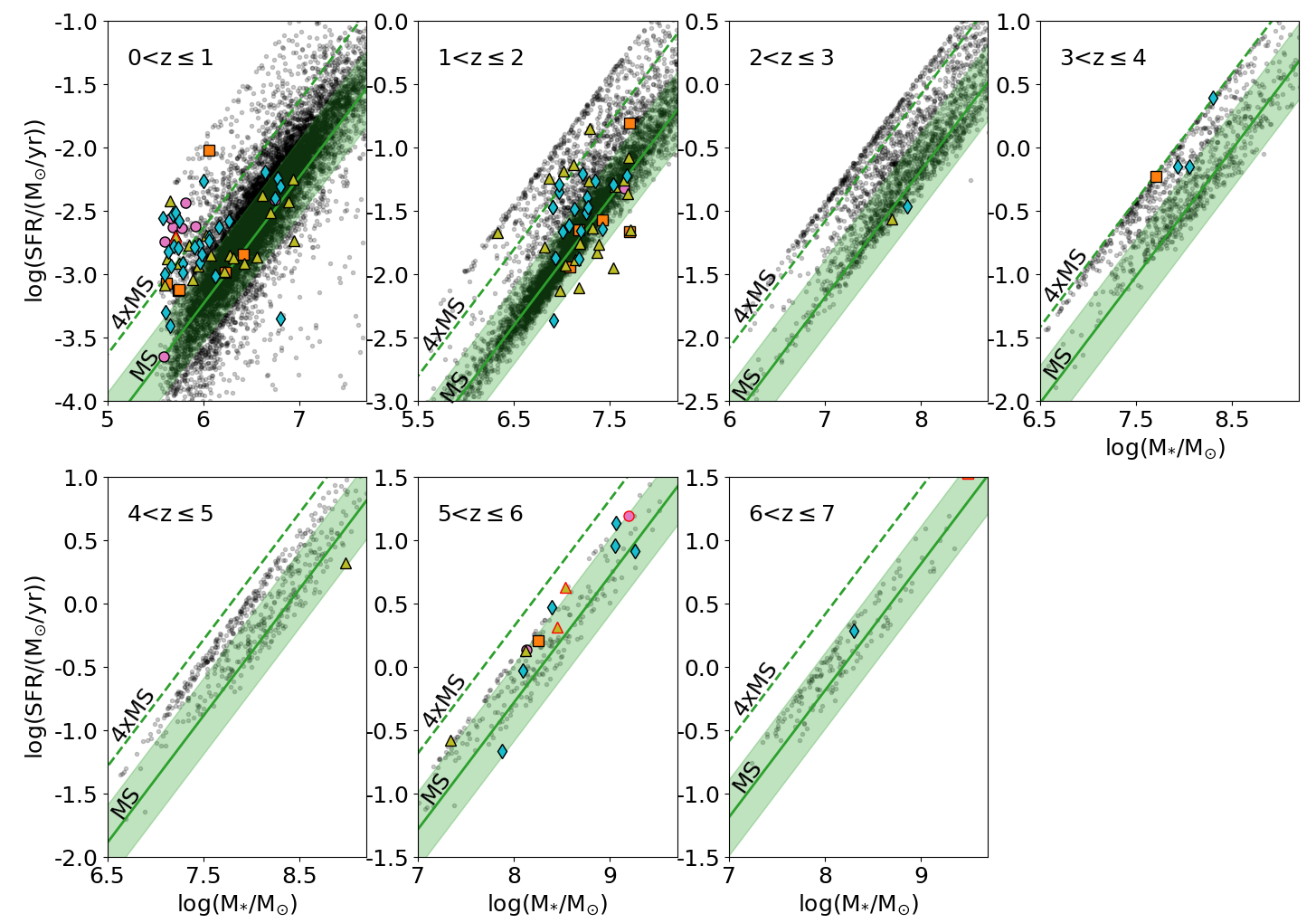}
    \caption{Same as previous figure but derived using 0.2"-radius aperture photometry.}
\end{figure}

We can therefore conclude that, when changing the aperture to calculate the fluxes from a Kron ellipse to a fixed 0.2"-radius circle aperture, there are difference, even large ones, between the physical properties of the objects in our F200W-dropout sample. However, the median properties of the sample are not largely affected by this fluxes changes, un-affecting the main findings of our work. 

\section{Testing the impact of the dust extinction law, PAH strength and star-formation history}\label{sec:SMC}
In this Appendix we investigate the impact of the chosen dust extinction law, i.e. \citet{Calzetti2000}, the PAH strength and star-formation history (SFH) on the results of our paper. For this reason we repeat the BAGPIPES run described in Section \ref{sec:SEDfitting}, but first assuming a Small Magellanic Cloud (SMC) extinction curve \citep{Gordon2003}, second fixing the dust mass associated to PAH to 10\% and third assuming an exponentially declining (i.e. SFR$\propto e^{-(t/\tau)}$) SFH. \par
\begin{figure}
    \centering
    \includegraphics[trim={30 40 40 50},clip,width=0.99\linewidth,keepaspectratio]{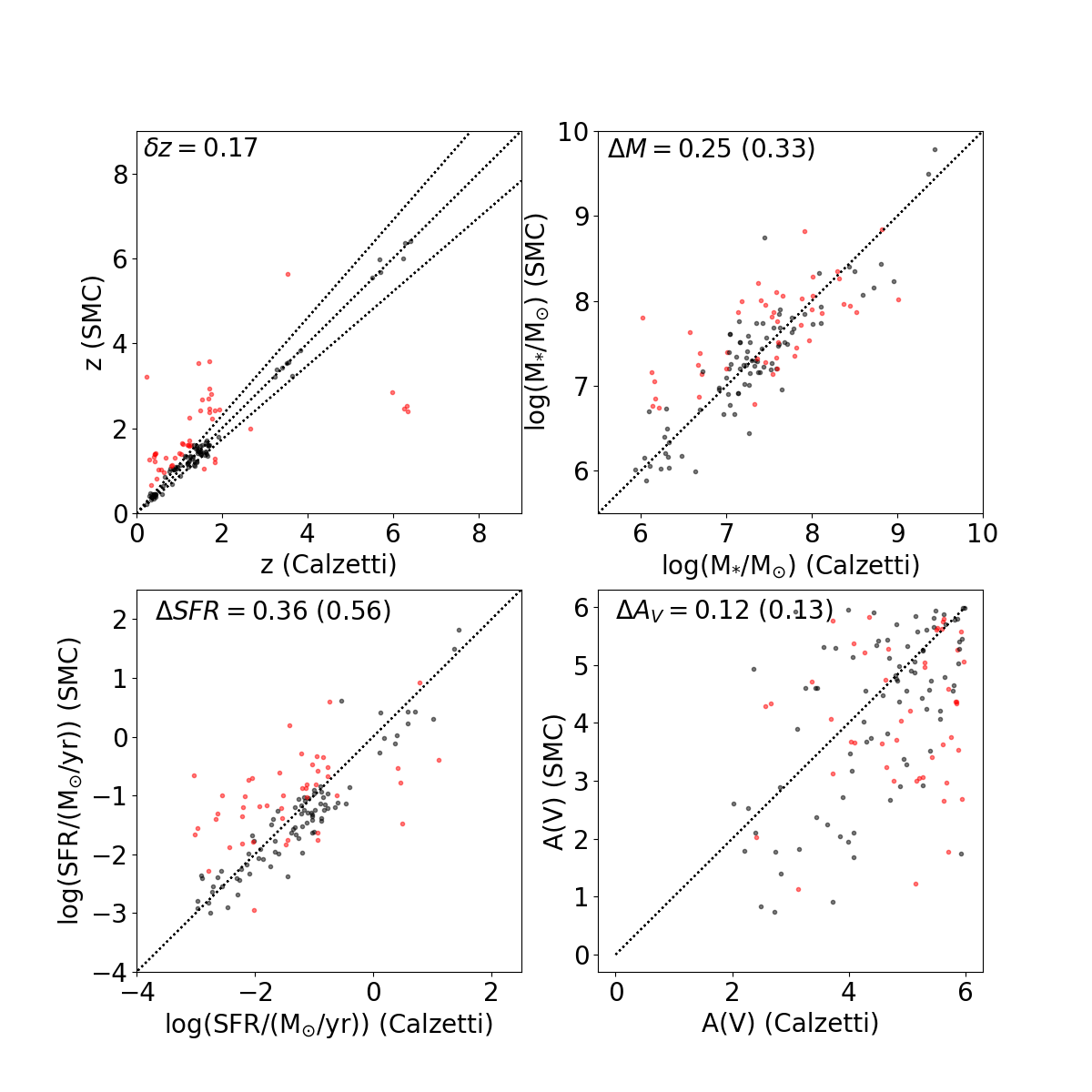}
    \caption{Same as Figure \ref{fig:Ap_kron_comparison_phys}, but comparing BAGPIPES assuming the dust extinction law by \citet{Calzetti2000} with the SMC one.}
    \label{fig:SMC_Calzetti}
\end{figure}
\begin{figure}
    \centering
    \includegraphics[trim={30 40 40 50},clip,width=0.99\linewidth,keepaspectratio]{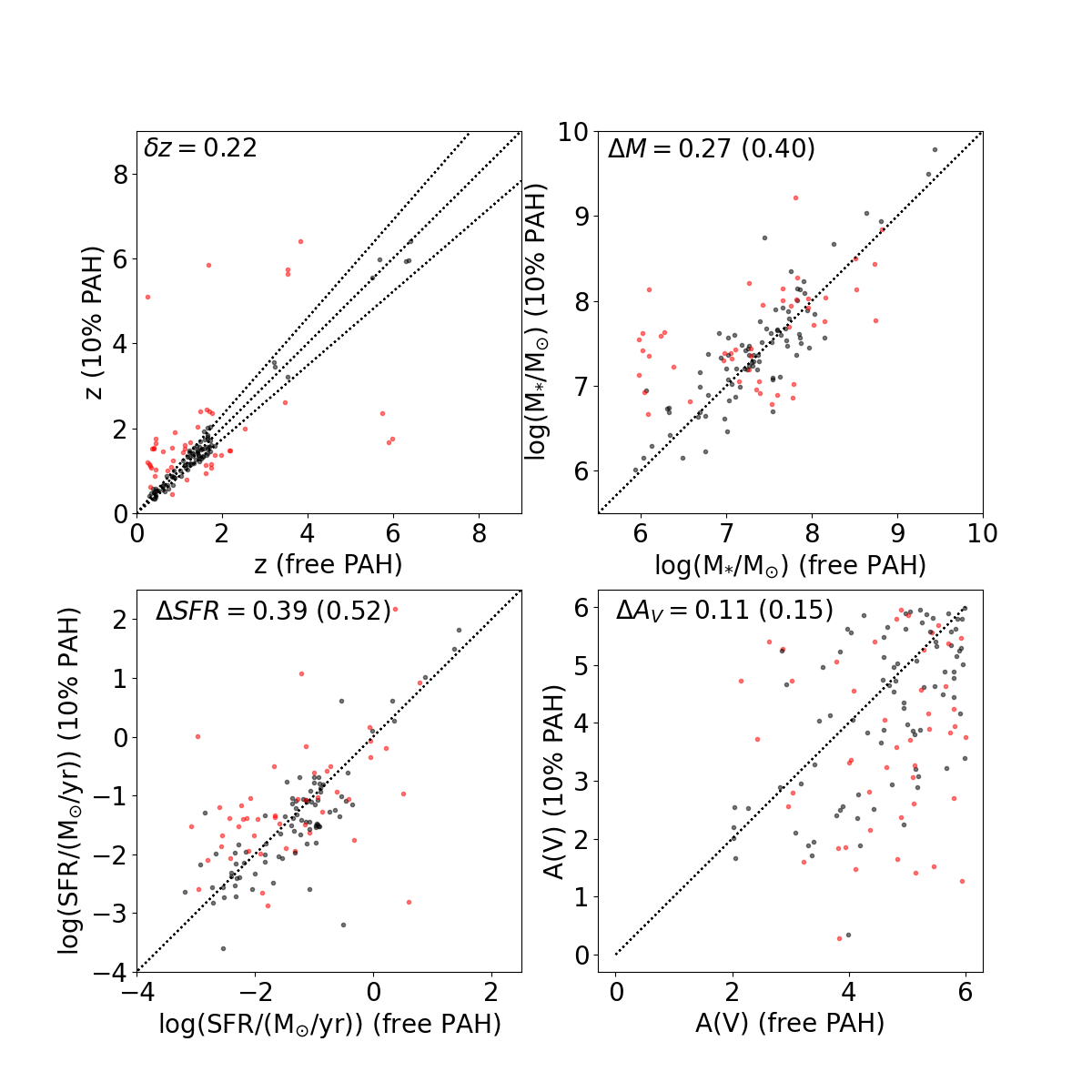}
    \caption{Same as Figure \ref{fig:Ap_kron_comparison_phys}, but letting the PAH free to vary or fixing the dust mass associated to them to 10\%.}
    \label{fig:PAH}
\end{figure}
\begin{figure}
    \centering
    \includegraphics[trim={30 40 40 50},clip,width=0.99\linewidth,keepaspectratio]{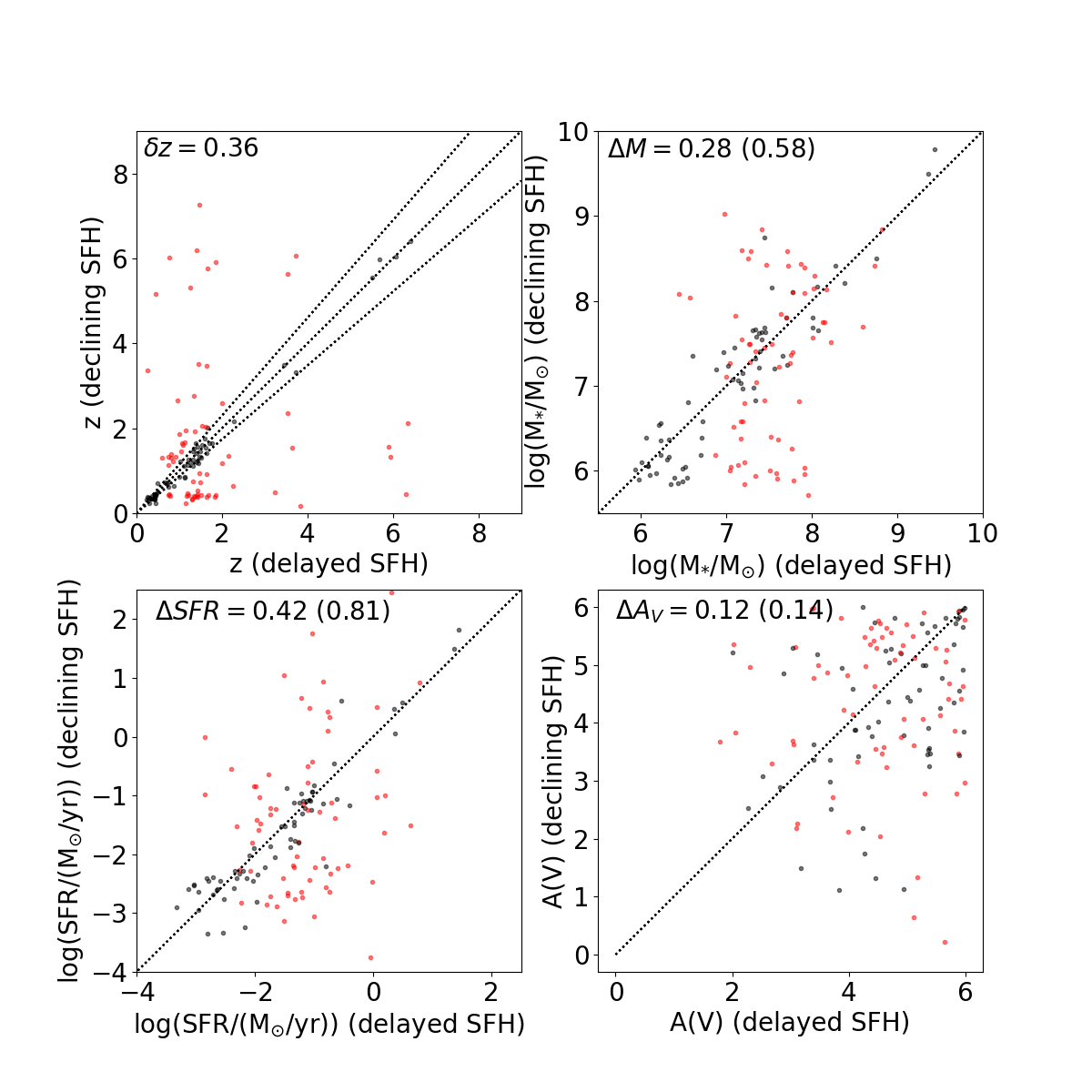}
    \caption{Same as Figure \ref{fig:Ap_kron_comparison_phys}, but comparing two different SFHs, i.e. delayed and declining.}
    \label{fig:SFH}
\end{figure}
\begin{figure}
    \centering
    \includegraphics[trim={30 40 40 50},clip,width=0.99\linewidth,keepaspectratio]{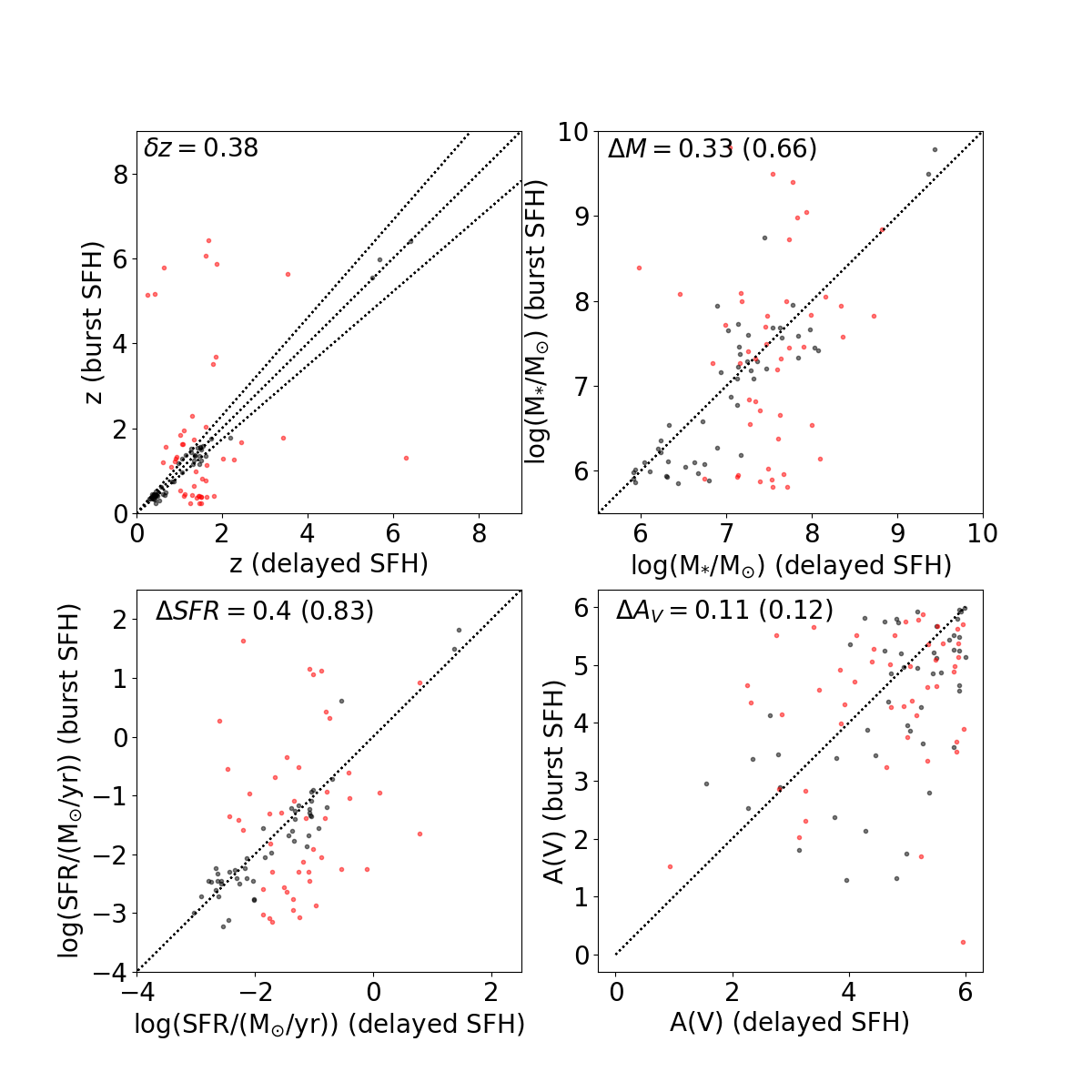}
    \caption{Same as Figure \ref{fig:Ap_kron_comparison_phys}, but comparing two different SFHs, i.e. delayed and a burst.}
    \label{fig:SFH2}
\end{figure}

We show the comparison of the derived redshift, stellar mass, SFR and A$_{V}$ in figure \ref{fig:SMC_Calzetti} for the test on the dust extinction law, in figure \ref{fig:PAH} for the test on the PAH strength and in figures \ref{fig:SFH} and \ref{fig:SFH2} for the test on the SFH. For 86/133 ($\sim65\%$) of the objects the change in the dust extinction law has a minor impact on the redshift estimation, as $\delta z=|z_{Calzetti}-z_{SMC}|/(1+z_{Calzetti})<0.15$. For these objects the other physical properties are also quite stable, with absolute differences in the stellar mass, SFR and A$_{V}$ of 0.25 dex, 0.36 dex and 0.12 mag, which are generally equal or smaller than the differences induced by changing from Kron to 0.2"-radius aperture photometry (see Appendix \ref{sec:fixedpaerture}). The median value of the optical dust extinction remains large, varying from 4.9 mag using a \citet{Calzetti2000} reddening law to 4.5 mag considering a SMC reddening law. At the same time, almost all galaxies, 130/133, are still very dusty sources as A$_{V}\geq$1. \par
Imposing a low mass content associated to PAH as a similar low impact on the results. Indeed, 85/133 objects have a redshift that change less than $\delta z=|z_{free}-z_{10\%}|/(1+z_{free})<0.15$. The median optical dust extinction is 4.2 mag, with only two objects with A$_{V}\leq$1. \par
Finally, we change the SFH considering separately an exponentially declining one and a burst one. The first has a larger impact on the redshift, with only 65/133 objects with a redshift change less than $\delta z=|z_{del}-z_{dec}|/(1+z_{del})<0.15$. However, the sample is still composed at 78$\%$ by galaxies with $z<2$ with median stellar mass of $10^{7.0}\,\rm M_{\odot}$ and median dust extinction $\rm A_{V}=4.7$. At the same time, 131/133 galaxies are still very dusty sources with A$_{V}\geq$1. The available data is not enough to discriminate between the two SFHs, as the difference in the $\chi_{red}^{2}$ is below one for all objects. \par
When we consider a single burst of star-formation, the fit is noticeably worst (i.e. $\Delta\chi^{2}_{red}>1$) than the case with a delayed SFH for 26\% of the sample and generally results on a larger $\chi^{2}$ for 86\% of the sample. Therefore, for at least part of the sample, a burst SFH is generally disfavoured by the data. In Figure \ref{fig:SFH2} we compare the physical parameters for the objects for which the two SFH produce comparably good fits, i.e., $\Delta\chi^{2}_{red}<1$. As for the previous case, there is a large scatter on the derived physical properties of each single objects, however the sample still remains dominated by $z<2$ galaxies, i.e. 84$\%$ of the sample even when considering a burst SFH. These galaxies at $z<2$ have a median dust extinction A$_{V}=4.8$ and median stellar mass $log(\rm M/M_{\odot})=7.0$, with only two of these galaxies having a A$_{V}<$1.  \par
Overall, we can conclude that the physical properties of each galaxy are quite uncertain, but the results presented in the paper are statistically robust.

\section{Possible brown dwarfs}\label{sec:BD}
In this appendix we report the photometry, cutouts, best models and redshift probability distributions of the two galaxies for which the brown dwarf templates correspond to a lower $\chi_{2}$ with respect to the galaxy ones. Both these galaxies are in pointing 2 (N2-2160 and N2-4539). The best brown dwarf fits of all two objects correspond to a model with gravity of $\rm 10^{5.5} cm\,s^{-2}$ and a metallicity of [Fe/H]=-0.5. The effective temperature are instead 1000 K and 1300 K  for the N2-2160 and N2-4539, respectively. At the same time, the normalisation of such models would place them at 2.7 and 4.4 pc from the Sun, respectively. \par
Looking instead at the fit with galaxy models, the N2-2160 has a main peak at $z\leq1$, considering both the flux derived with the Kron aperture and the one with a fixes 0.2"-radius aperture. The possible redshift of N2-4539 is more uncertain, as the two photometric sets give two different best solutions, i.e. $z=$0.7 and $z=$5.9. \par
Both objects are unresolved, but this is consistent both the brown-dwarf solution and the galaxy one, considering the stellar-mass-size relation by \citet{Wel2014} and their stellar masses (i.e.$10^{6}-10^{7} \rm M_{\odot}$). 
\begin{figure}
    \centering
    \includegraphics[trim={30 20 50 40},clip,width=0.99\linewidth,keepaspectratio]{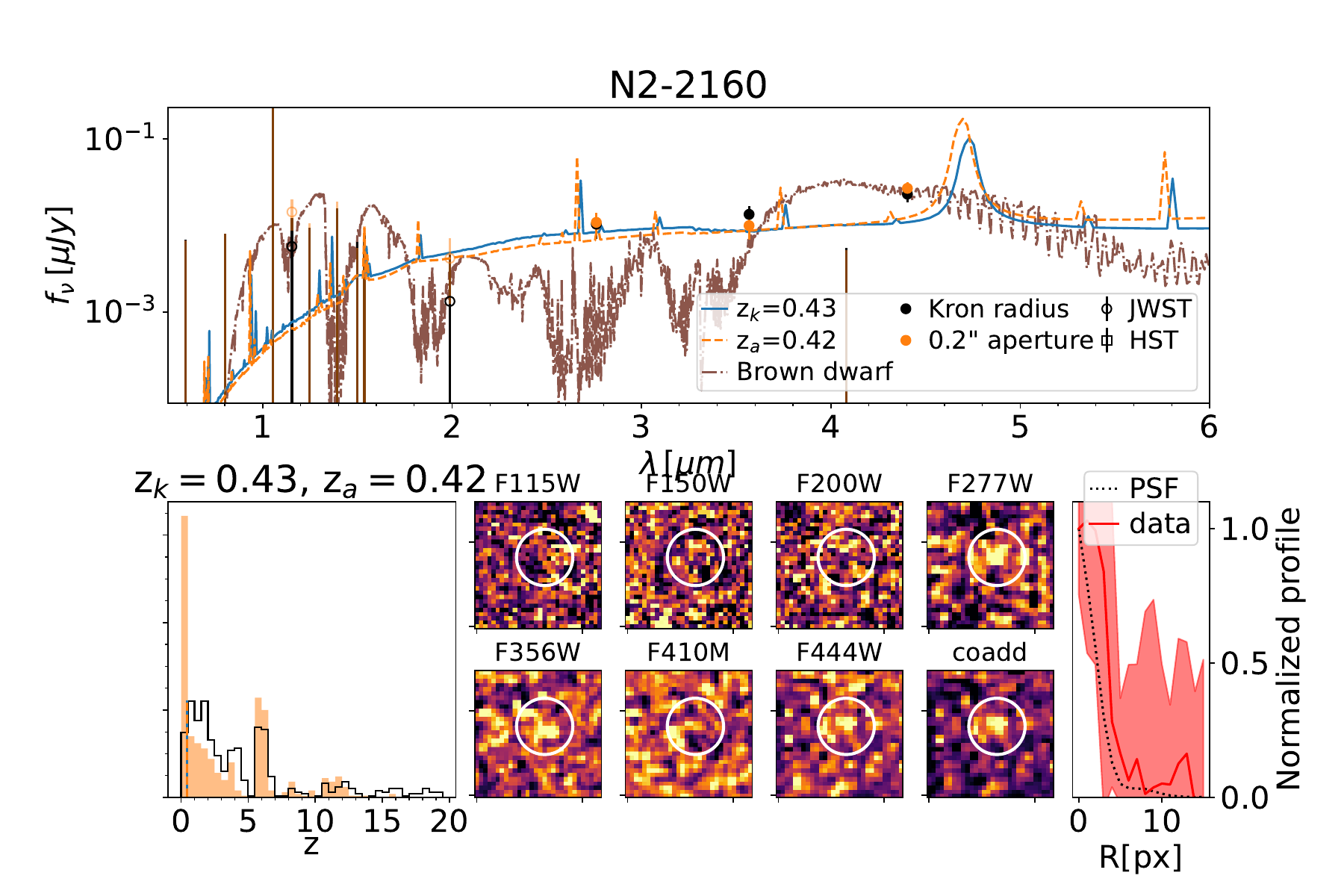}
    \includegraphics[trim={30 20 50 40},clip,width=0.99\linewidth,keepaspectratio]{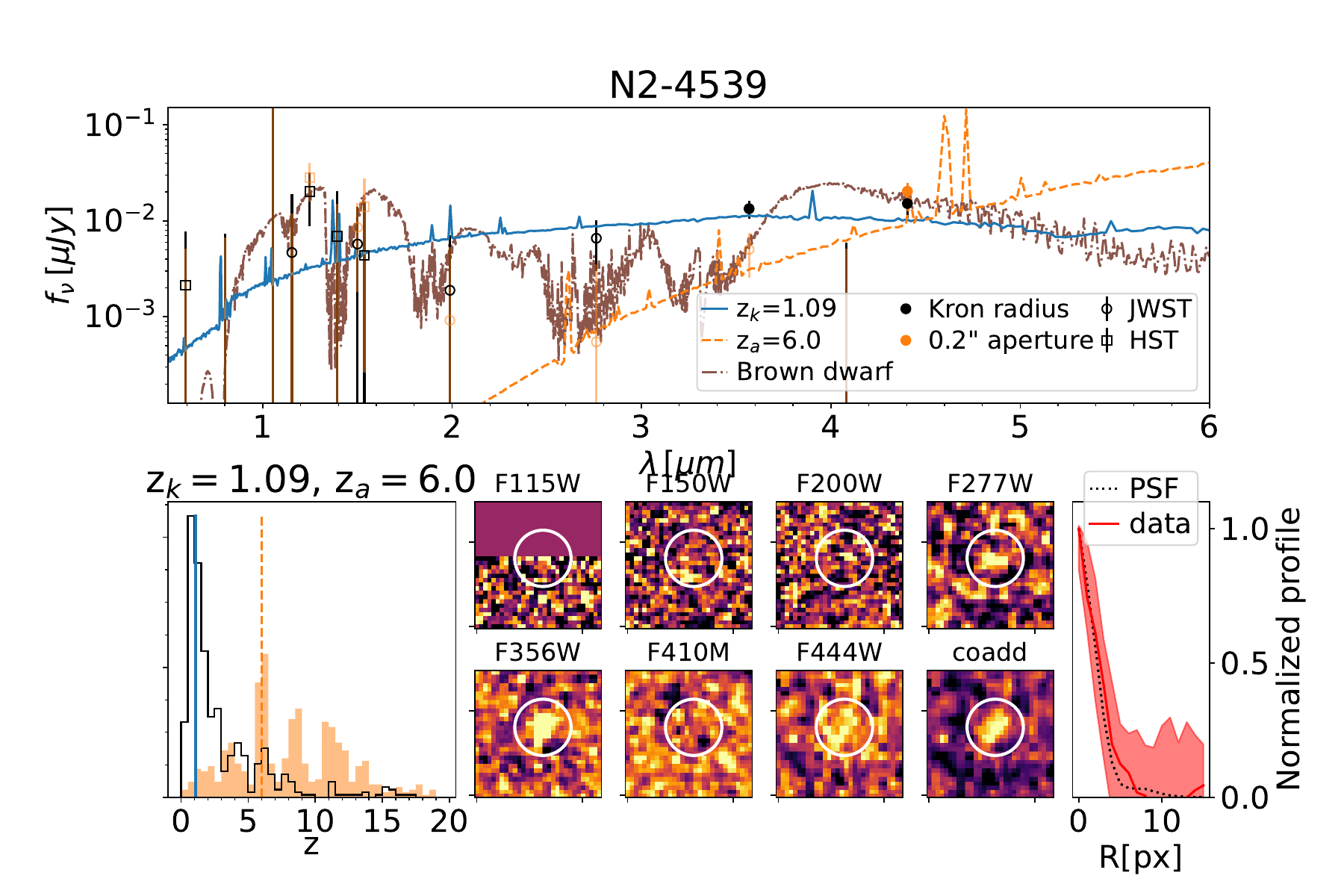}
    \caption{Same as Figure \ref{fig:SEDexample}, but for the three galaxies for which the brown-dwarf template (dashed-dotted brown lines) corresponded to a lower $\chi_{2}$ than the galaxy SEDs.}
\end{figure}

\section{SED fits and cutouts}\label{sec:allimages}
In this section we report the SED fits and cutouts images for all F200W-dropout images. The full list of images are available online.

\onecolumn
\begin{longfigure}{cc}    

    \includegraphics[trim={20 10 50 40},clip,width=0.44\linewidth,keepaspectratio]{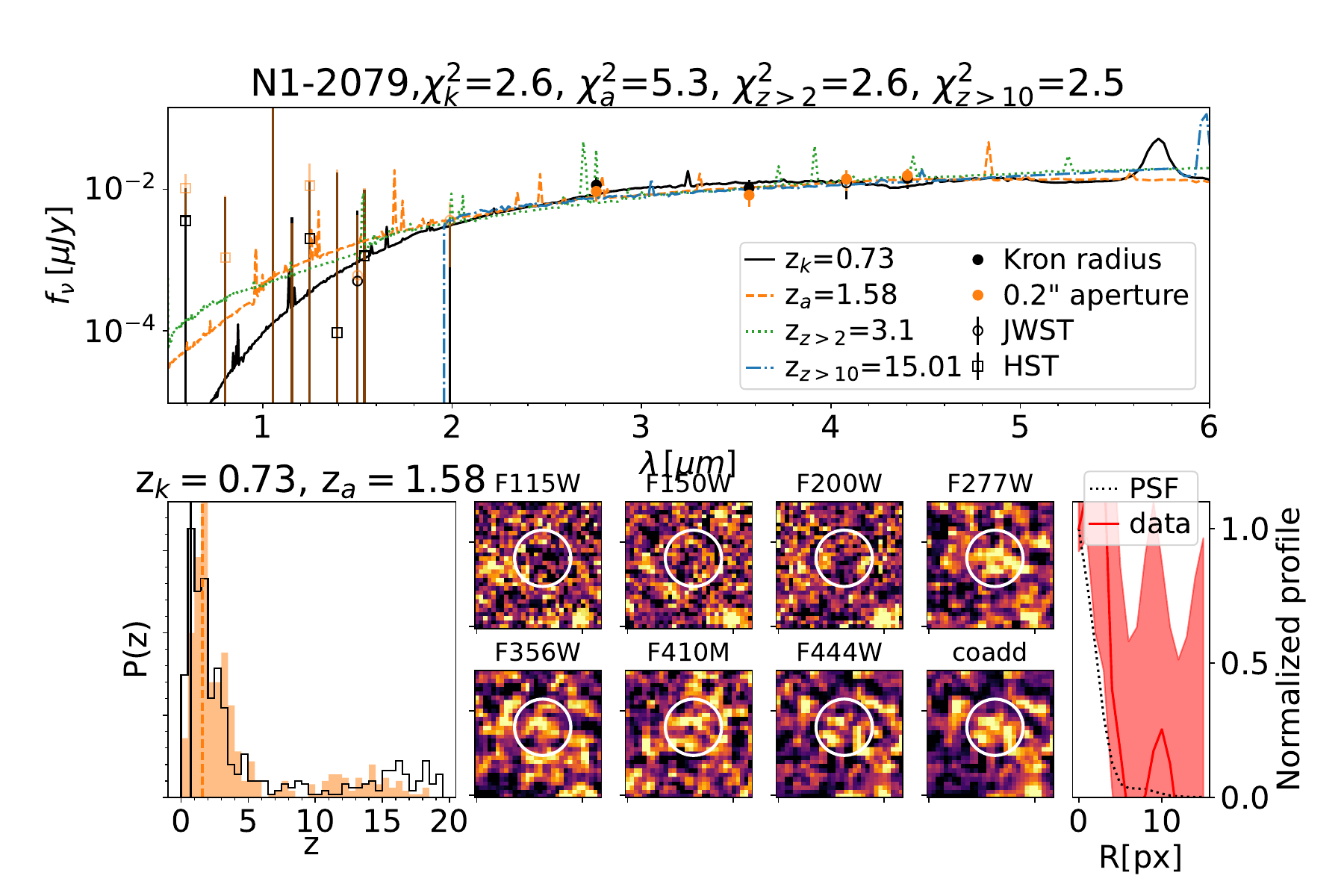} &
    \includegraphics[trim={20 10 50 40},clip,width=0.44\linewidth,keepaspectratio]{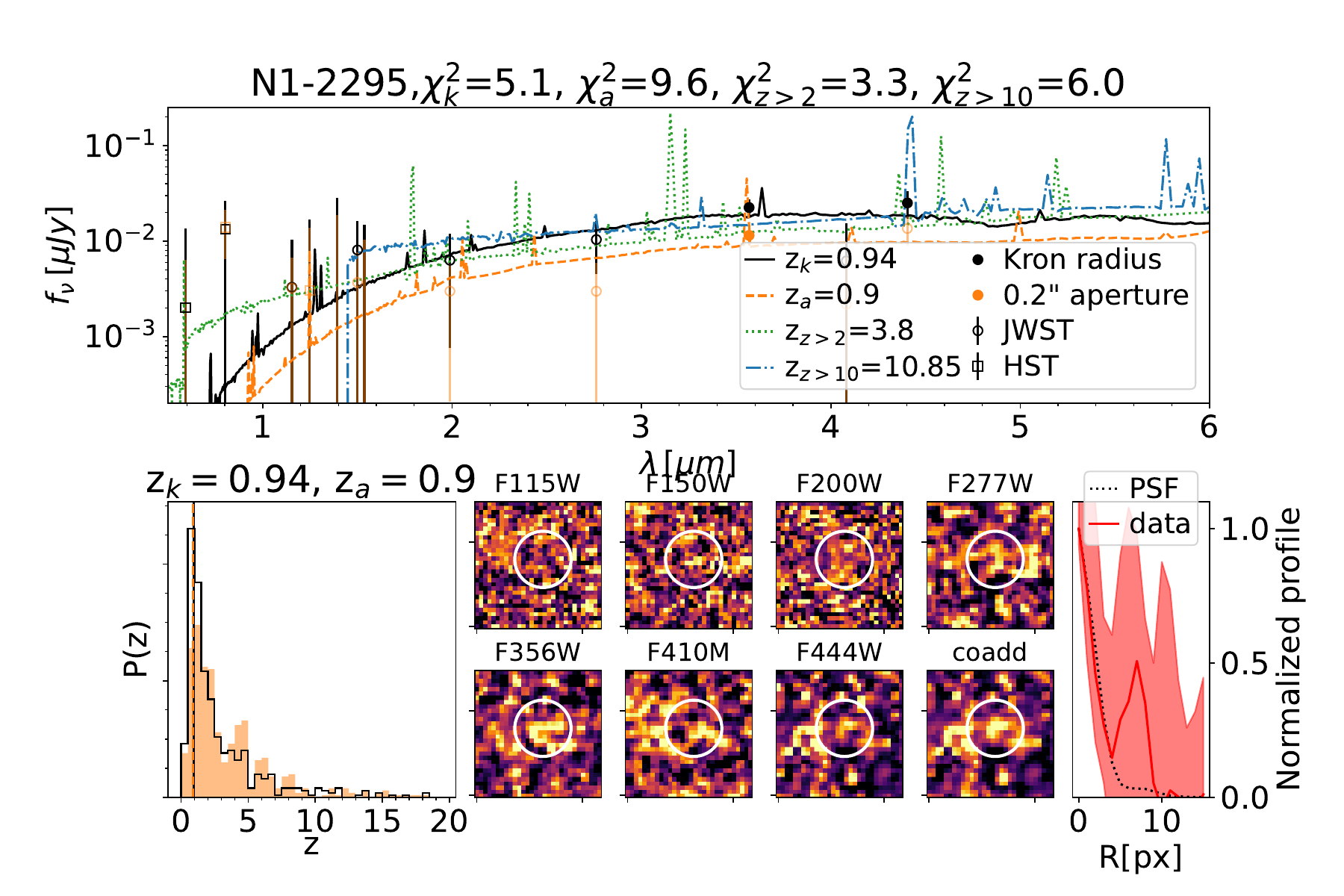}\\
    \includegraphics[trim={20 10 50 40},clip,width=0.44\linewidth,keepaspectratio]{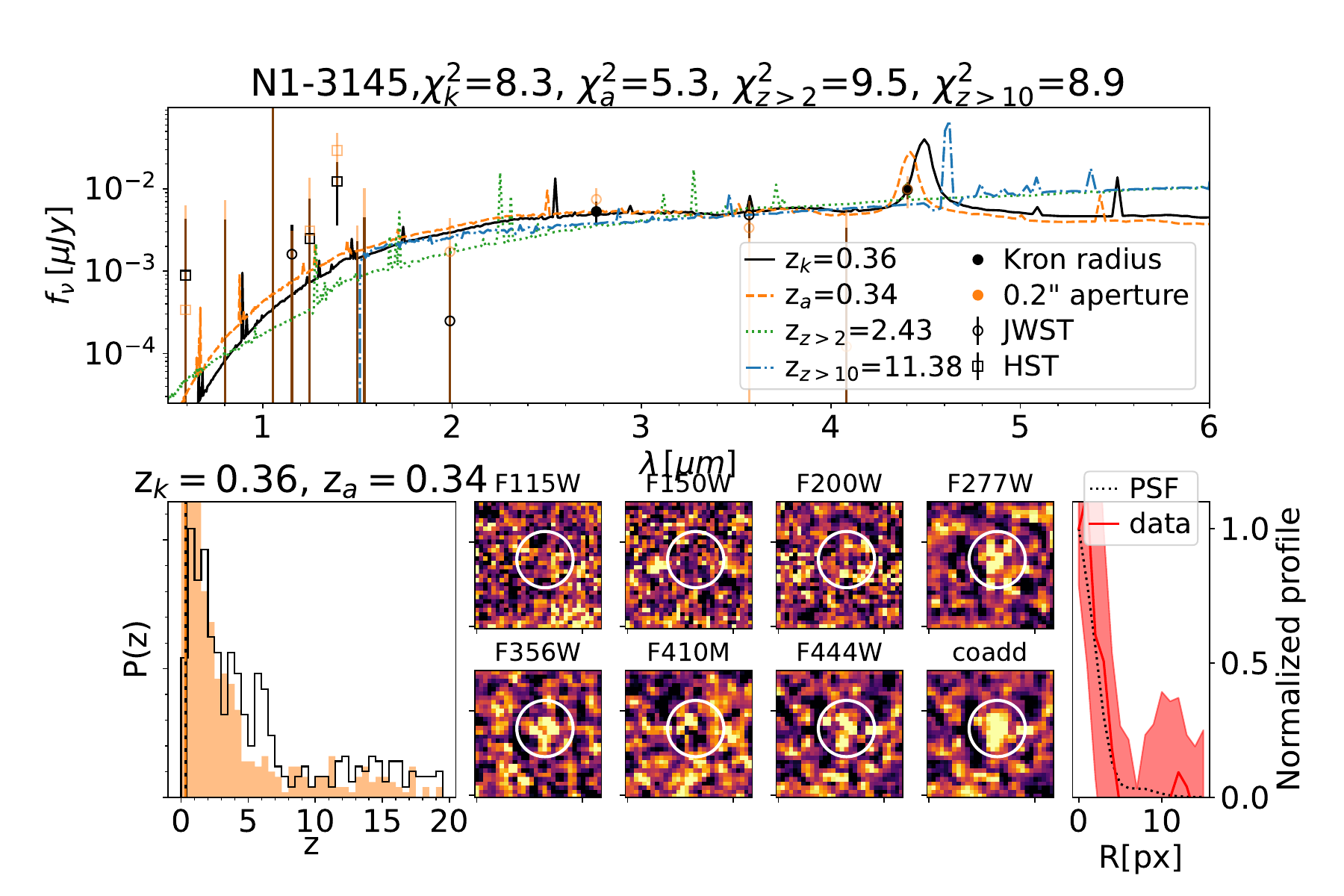} &
    \includegraphics[trim={20 10 50 40},clip,width=0.44\linewidth,keepaspectratio]{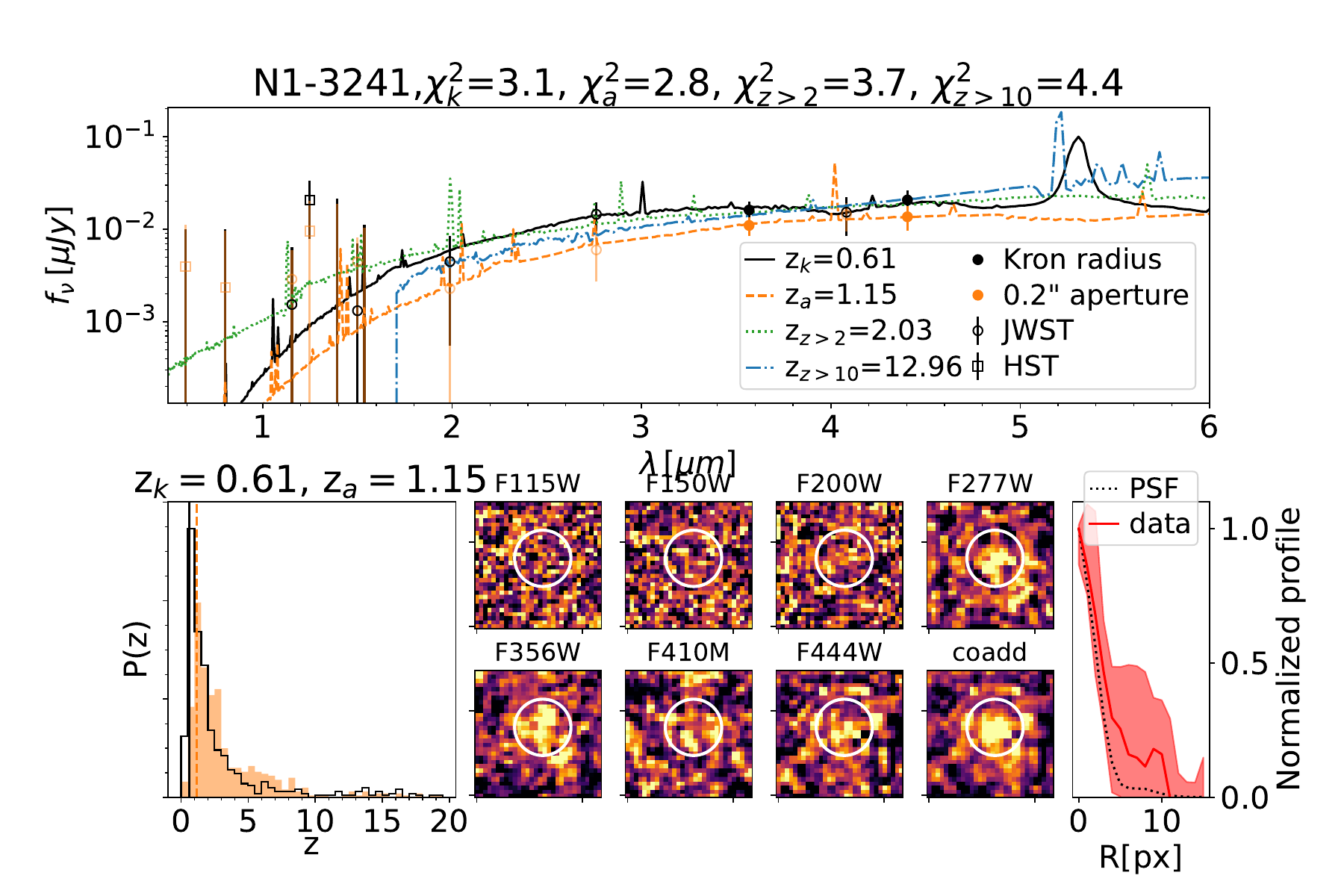}\\
    \includegraphics[trim={20 10 50 40},clip,width=0.44\linewidth,keepaspectratio]{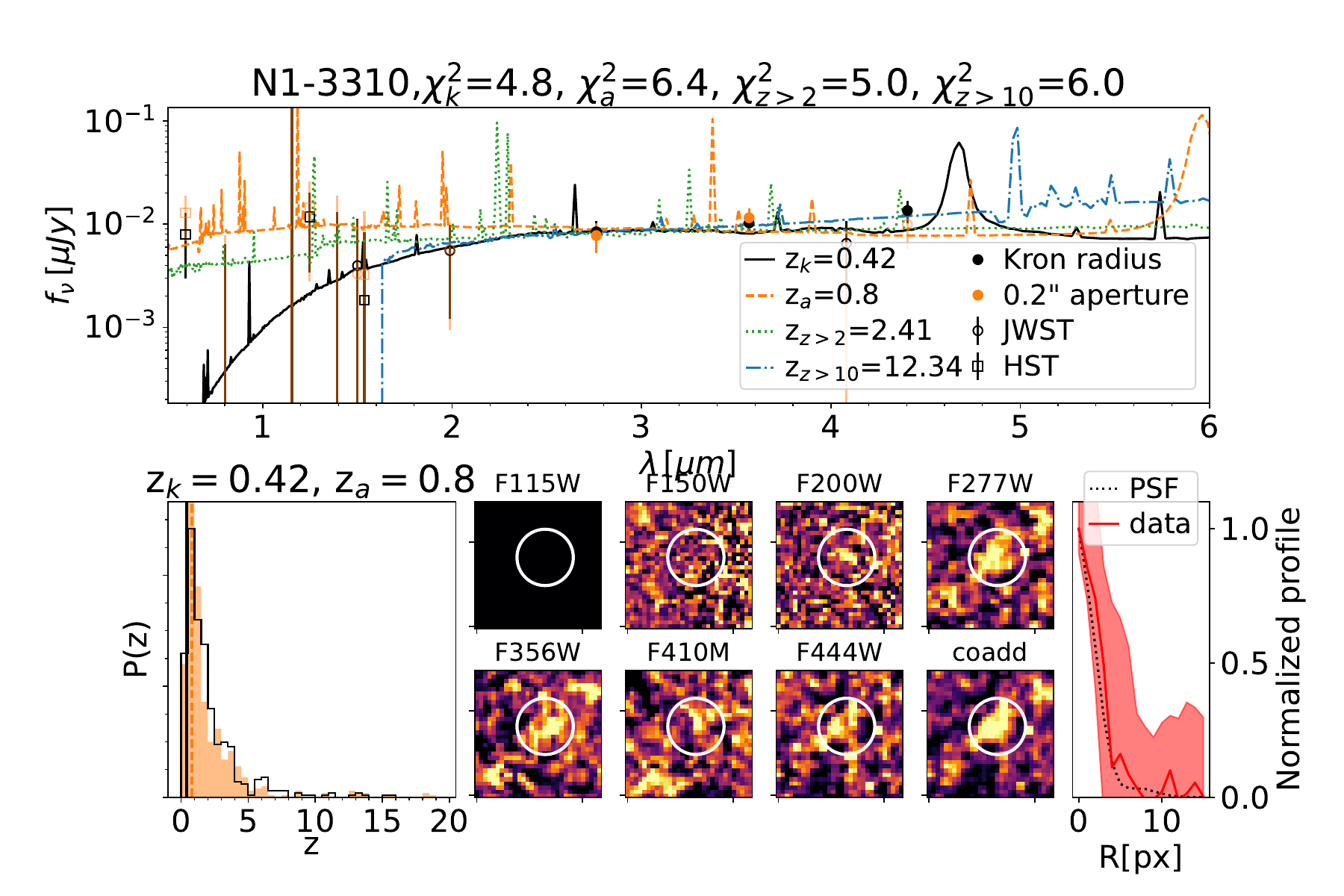} &
    \includegraphics[trim={20 10 50 40},clip,width=0.44\linewidth,keepaspectratio]{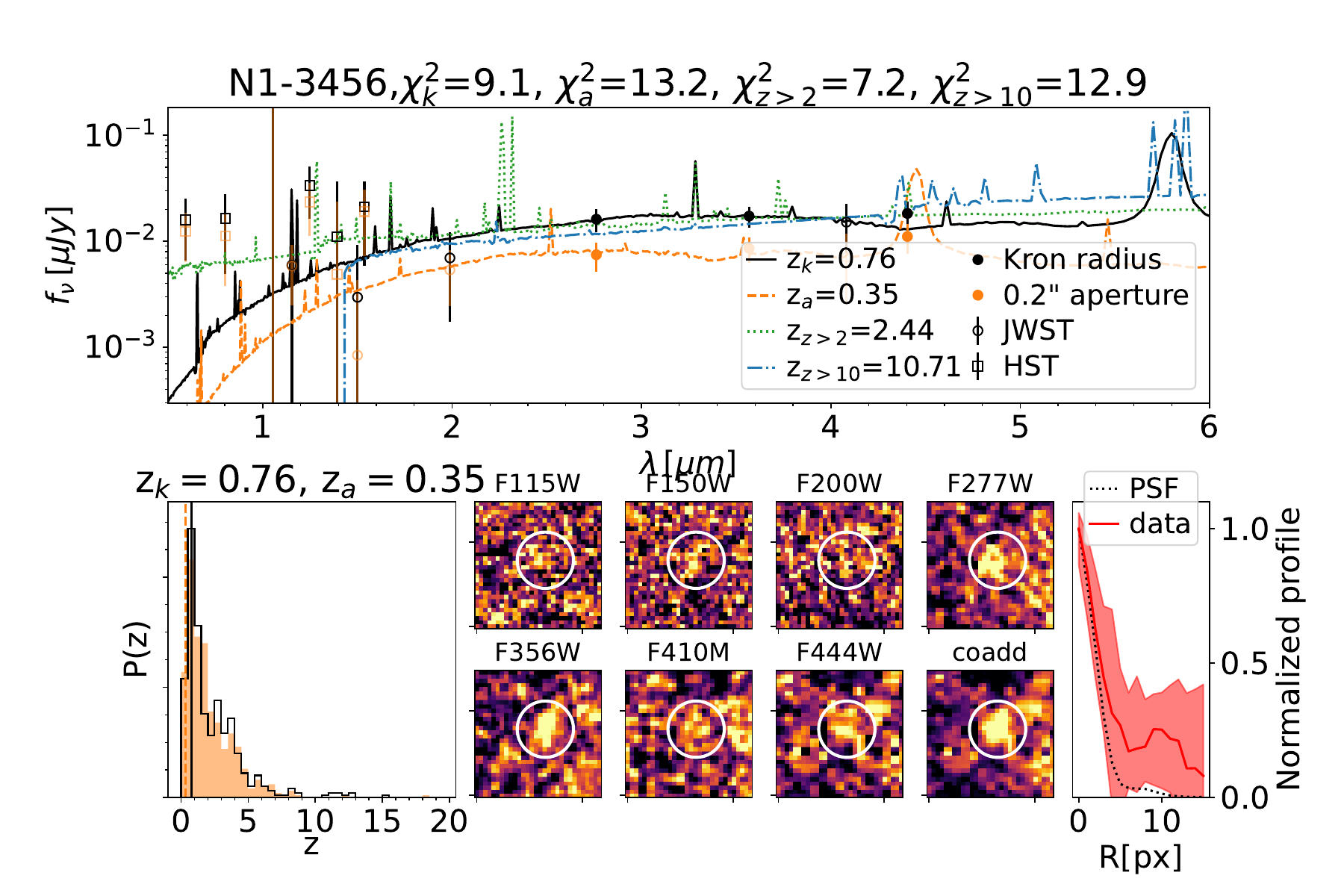}\\
    \includegraphics[trim={20 10 50 40},clip,width=0.44\linewidth,keepaspectratio]{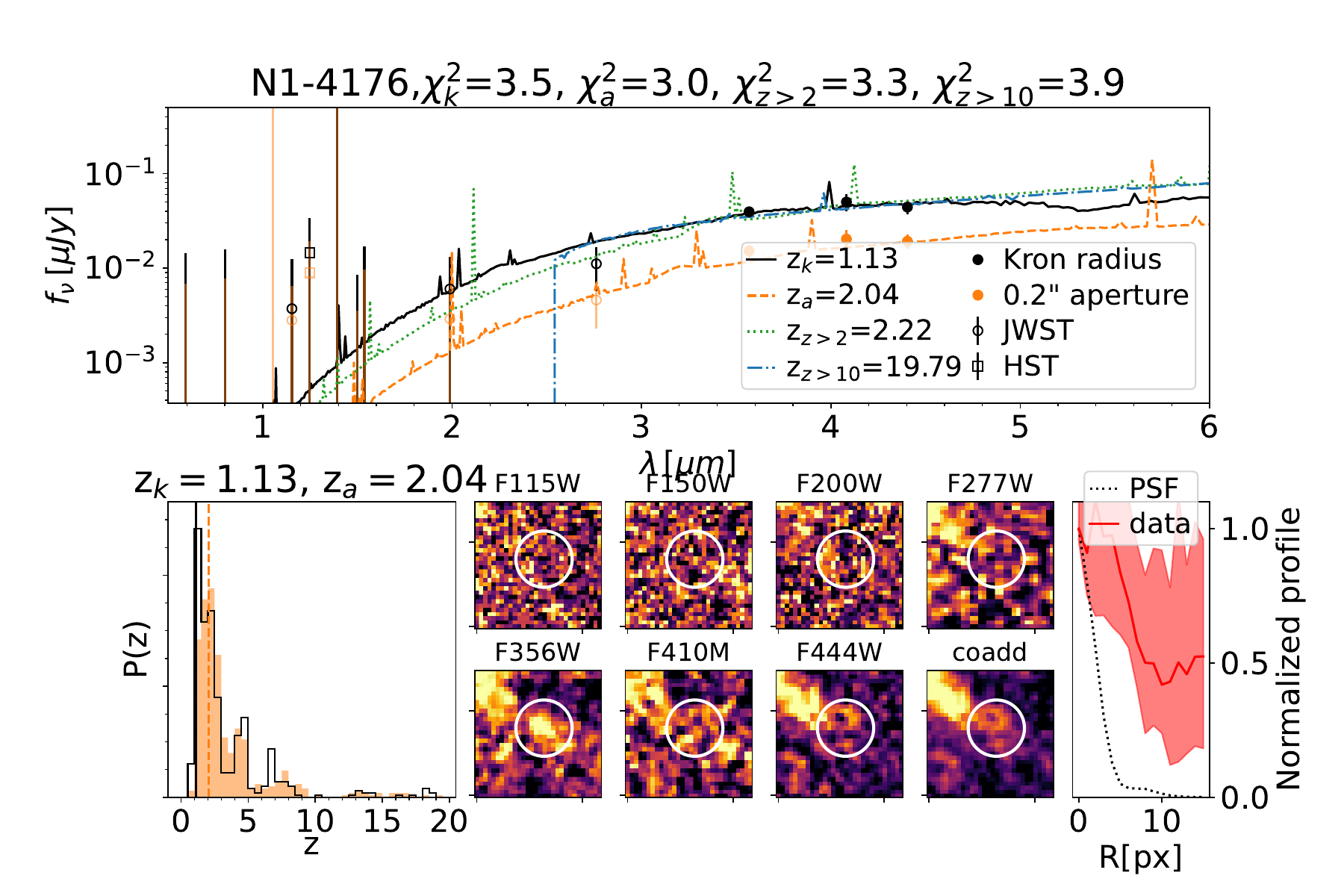} &
    \includegraphics[trim={20 10 50 40},clip,width=0.44\linewidth,keepaspectratio]{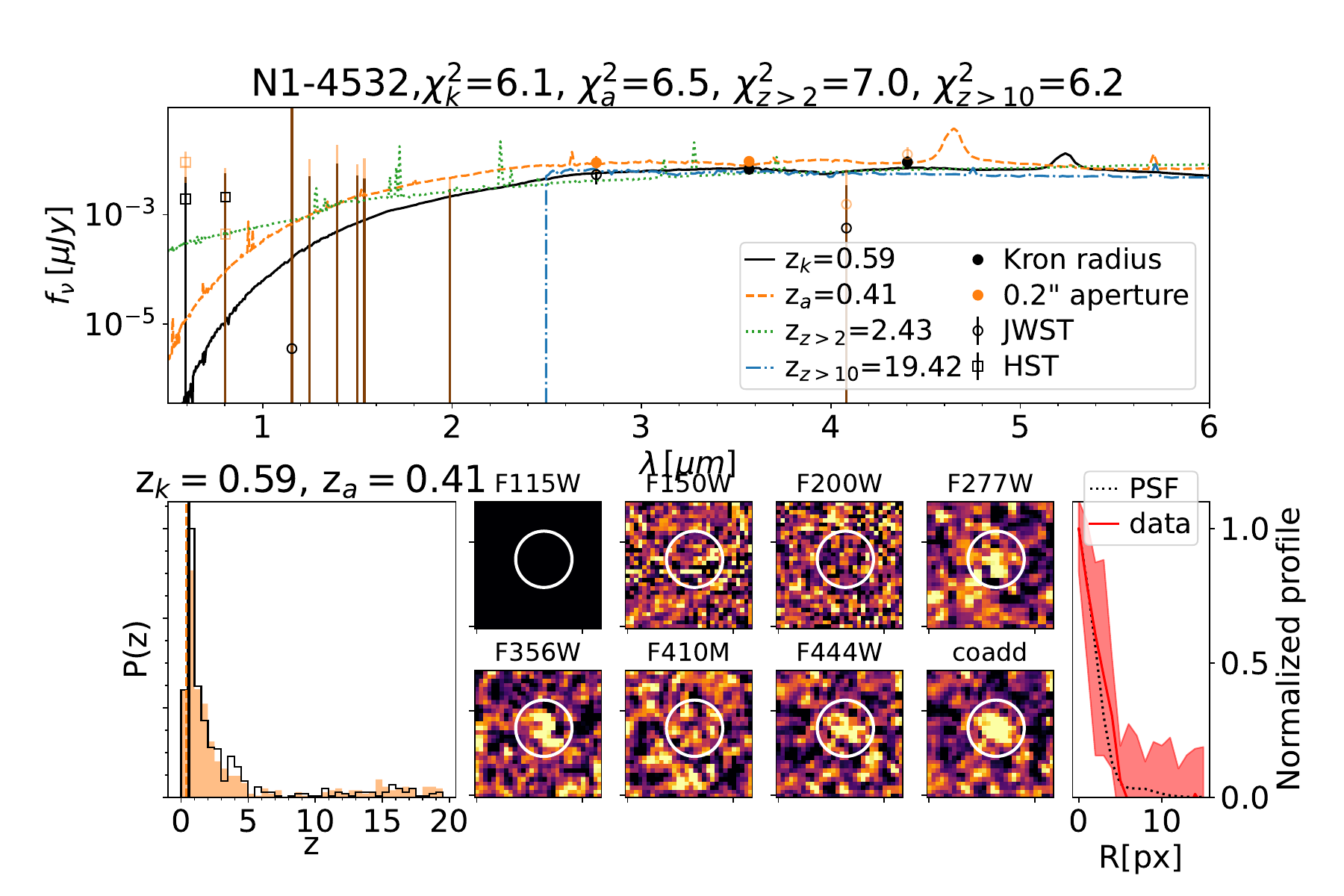}\\
 \caption{Same as Figure \ref{fig:SEDexample}, but for the F200W-dropout galaxies not previously shown.}\label{fig:longfig} \\
    \includegraphics[trim={20 10 50 40},clip,width=0.44\linewidth,keepaspectratio]{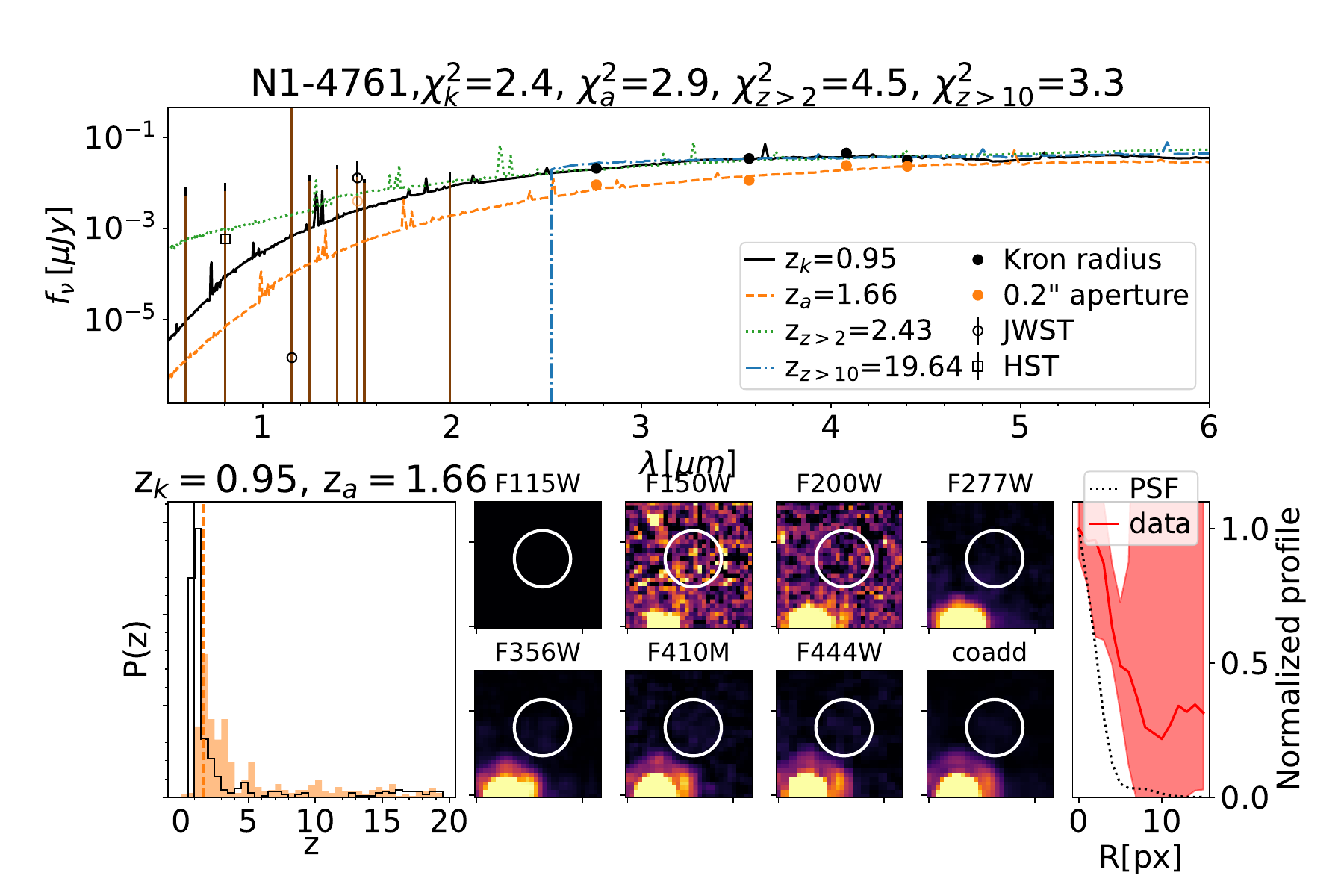} &
    \includegraphics[trim={20 10 50 40},clip,width=0.44\linewidth,keepaspectratio]{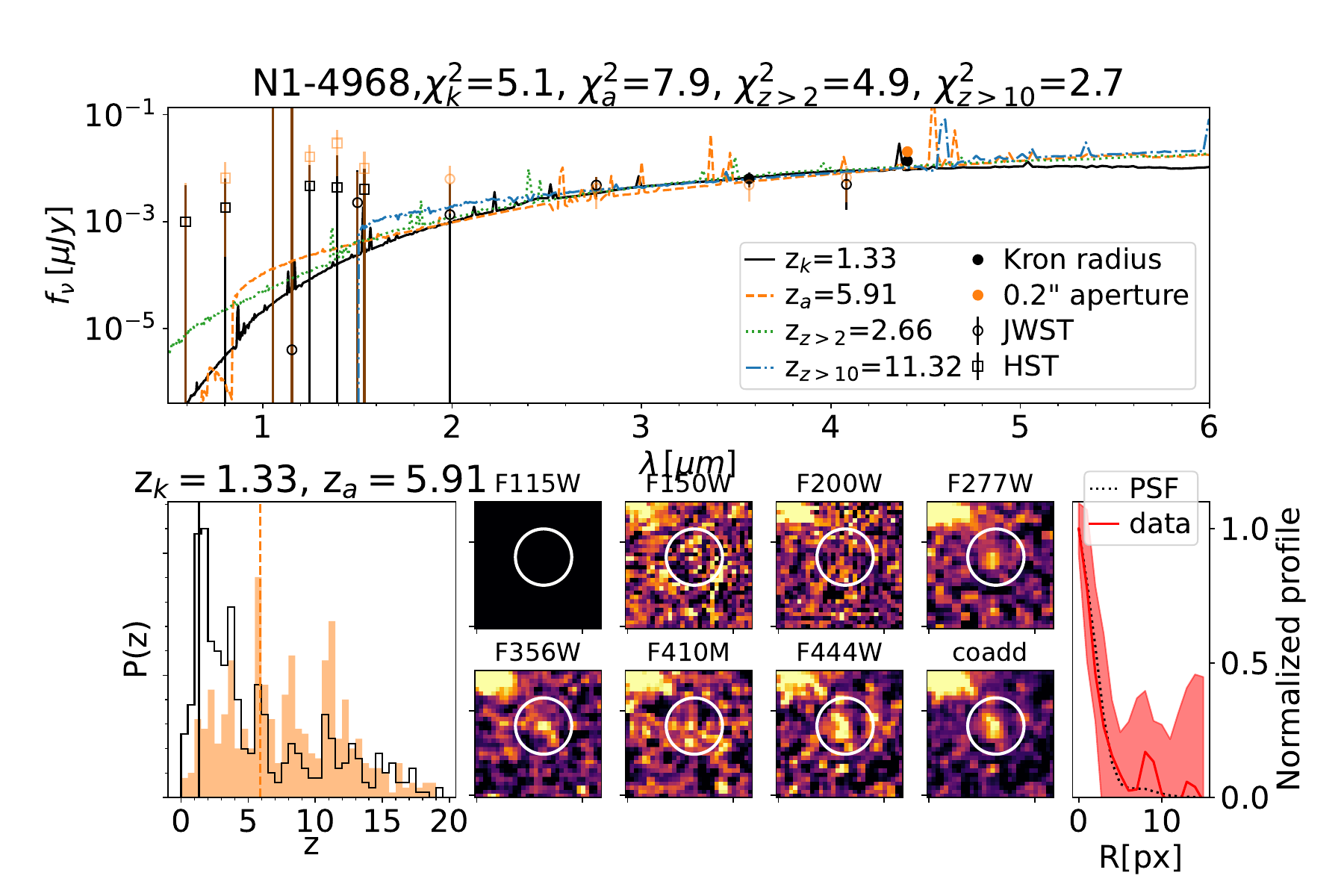} \\
    \includegraphics[trim={20 10 50 40},clip,width=0.44\linewidth,keepaspectratio]{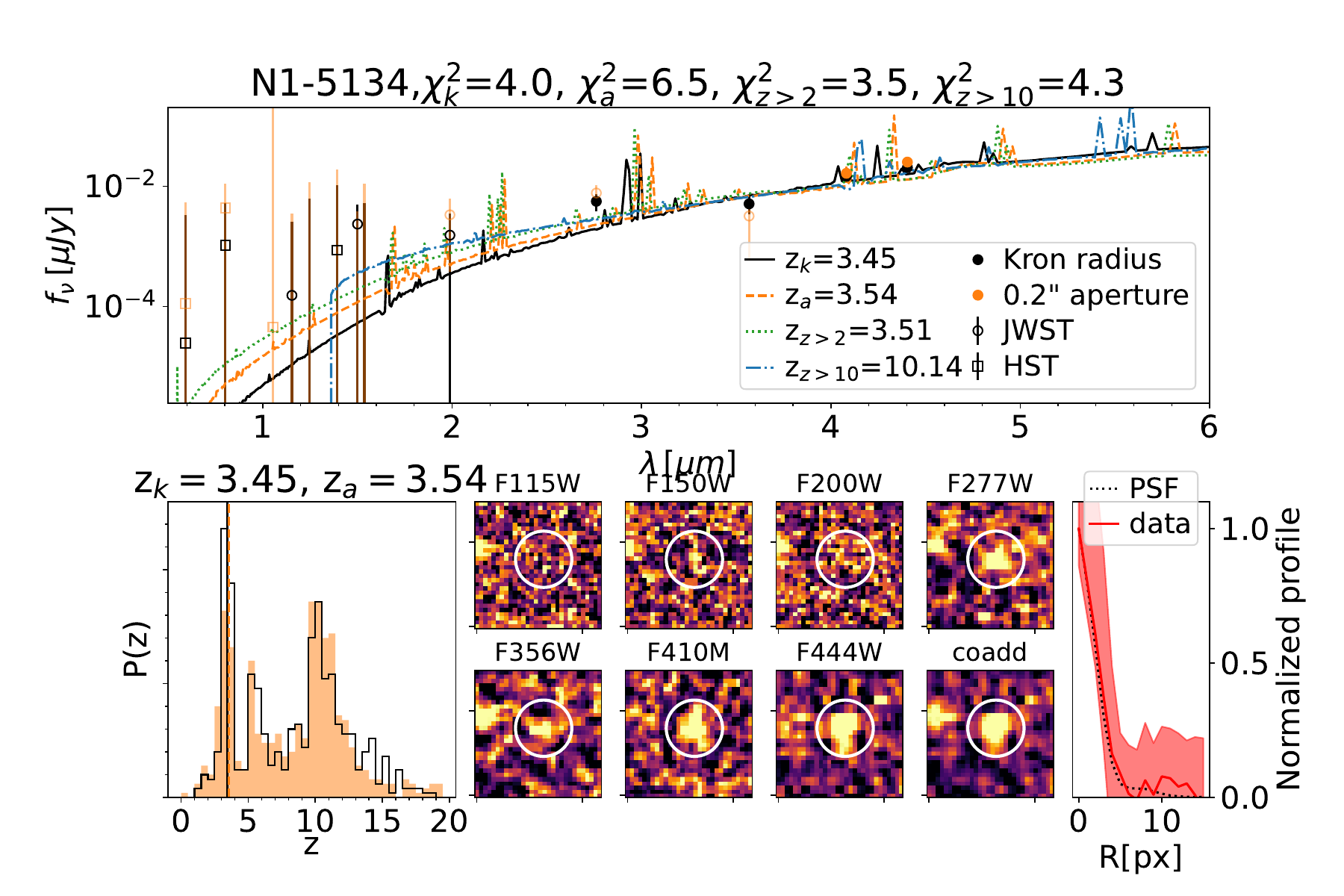} &
    \includegraphics[trim={20 10 50 40},clip,width=0.44\linewidth,keepaspectratio]{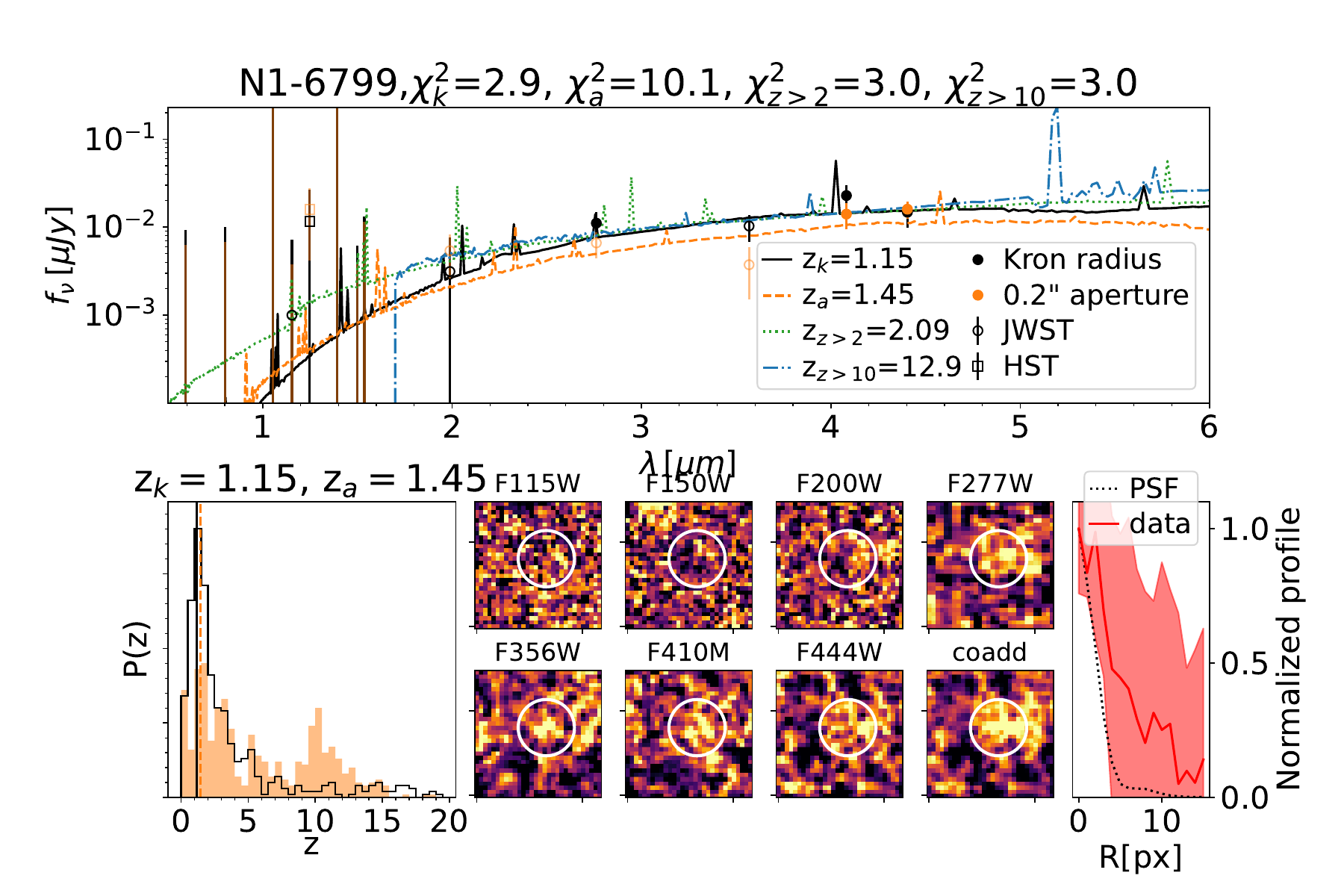} \\
    \includegraphics[trim={20 10 50 40},clip,width=0.44\linewidth,keepaspectratio]{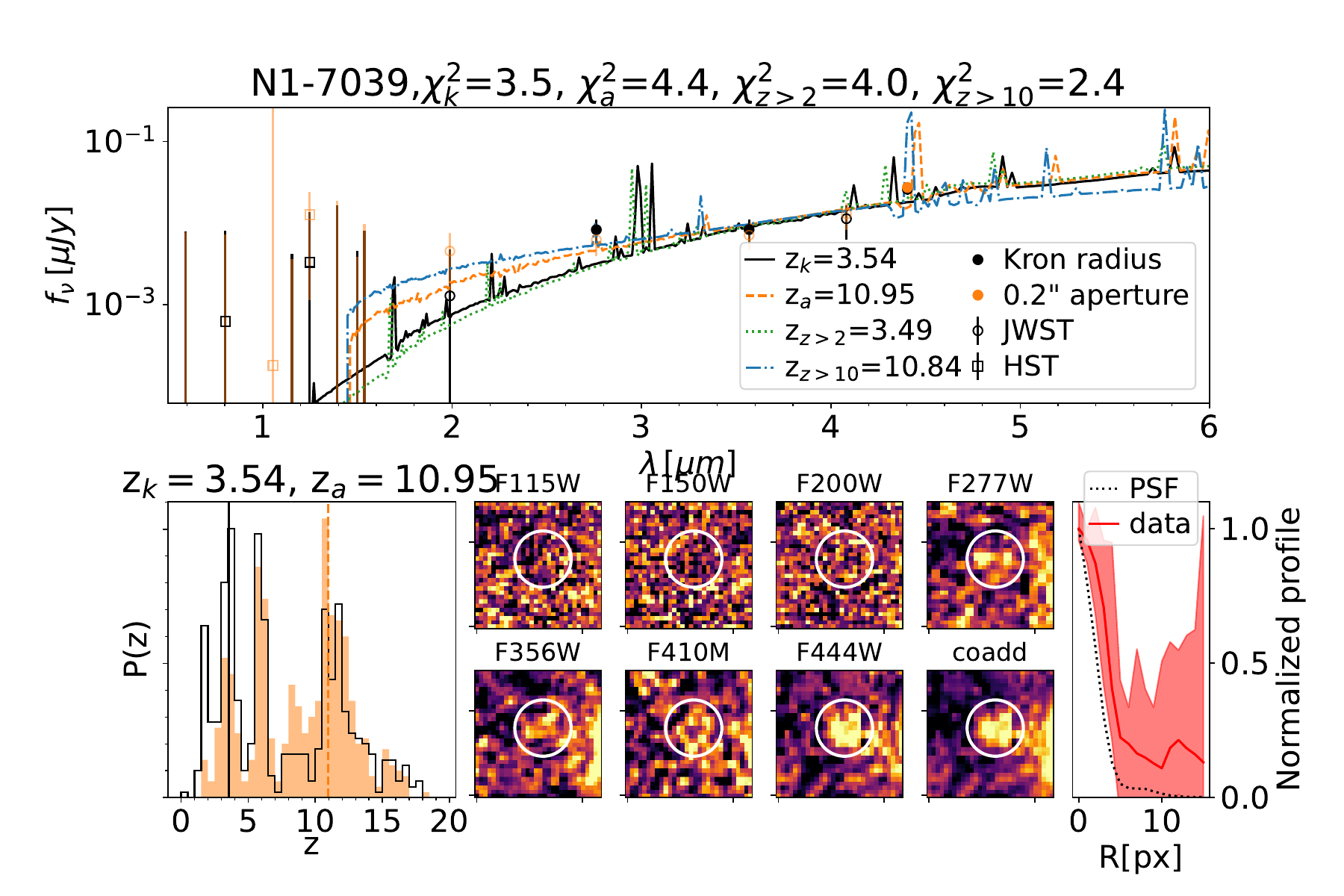} &
    \includegraphics[trim={20 10 50 40},clip,width=0.44\linewidth,keepaspectratio]{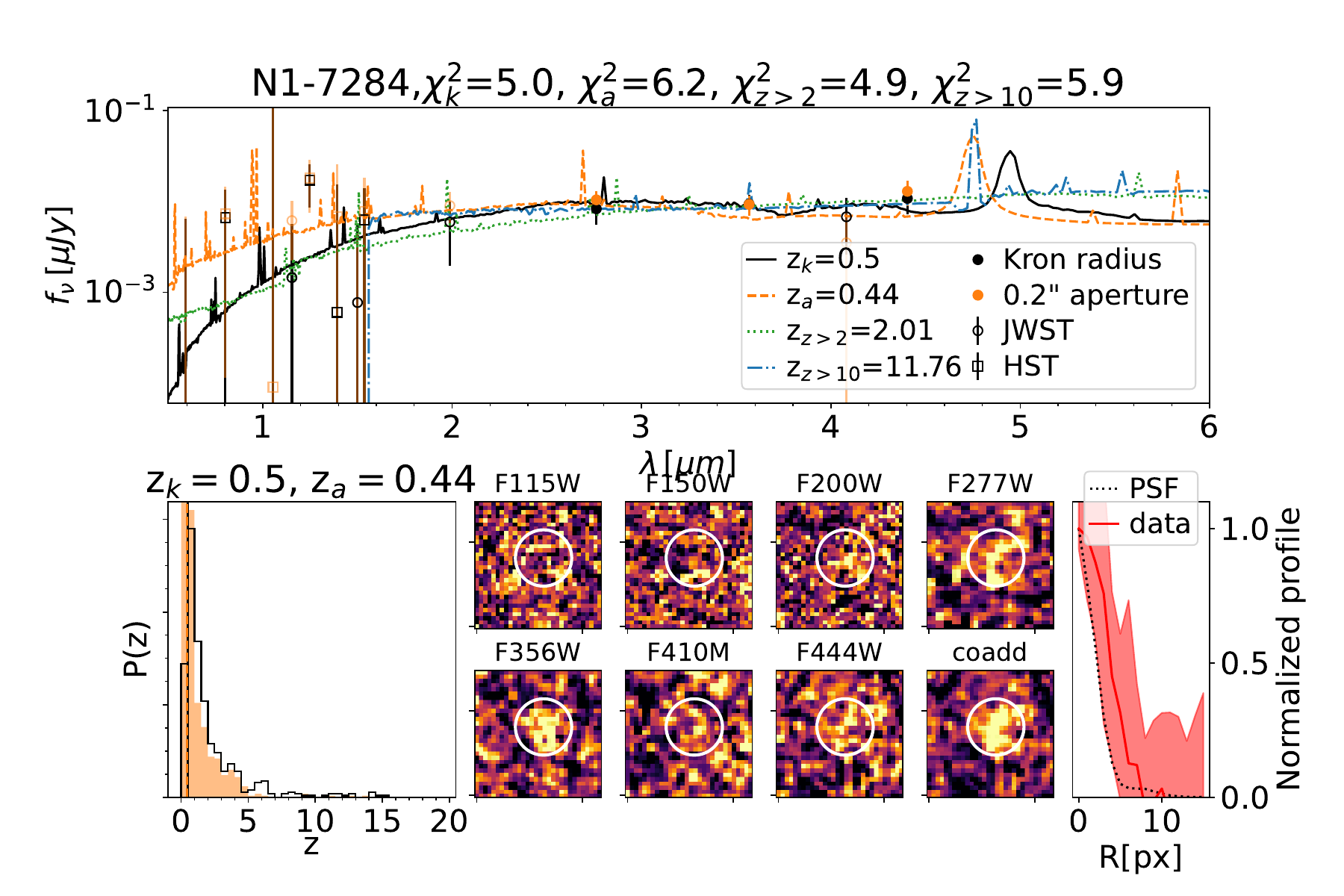} \\
    \includegraphics[trim={20 10 50 40},clip,width=0.44\linewidth,keepaspectratio]{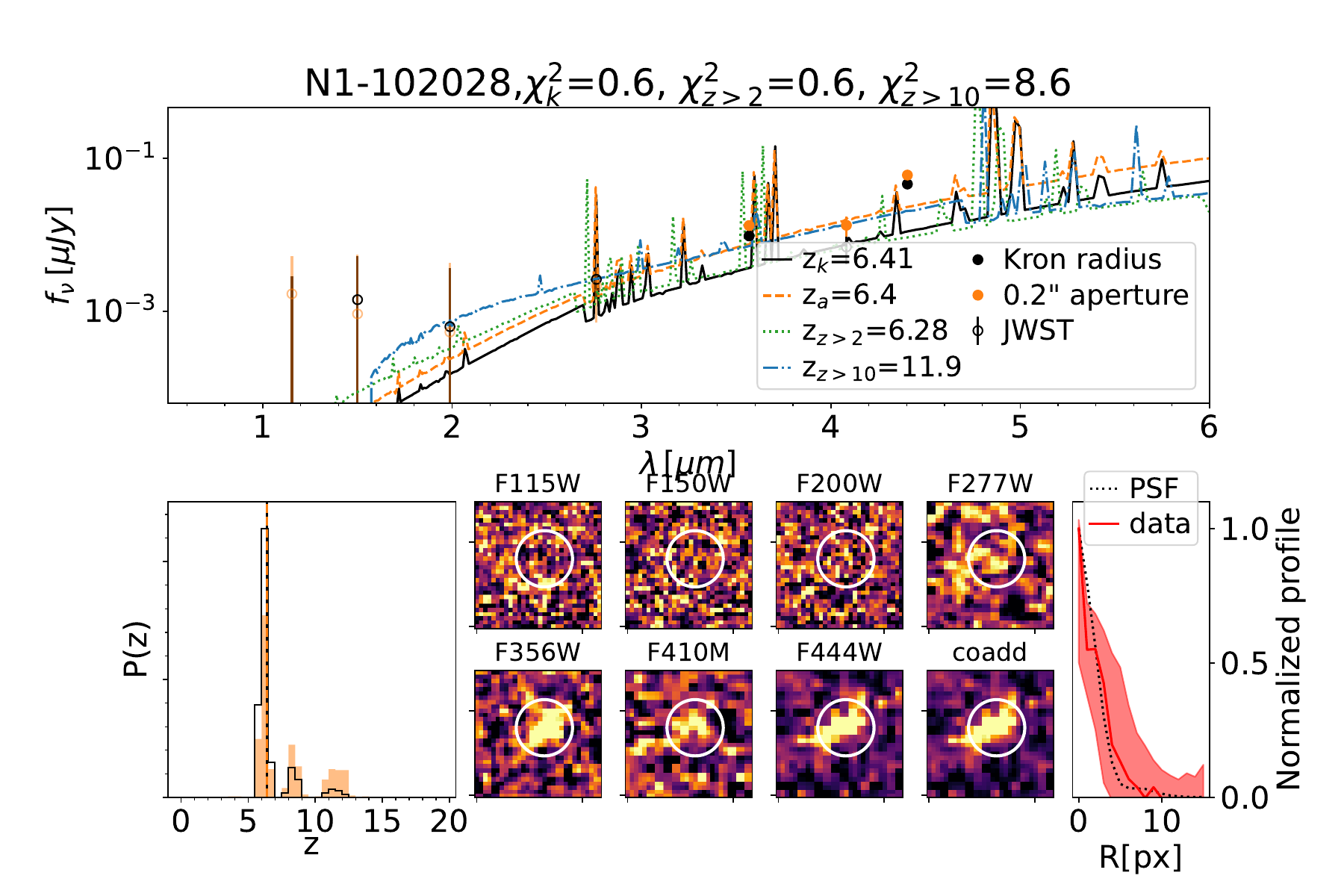} &
    \includegraphics[trim={20 10 50 40},clip,width=0.44\linewidth,keepaspectratio]{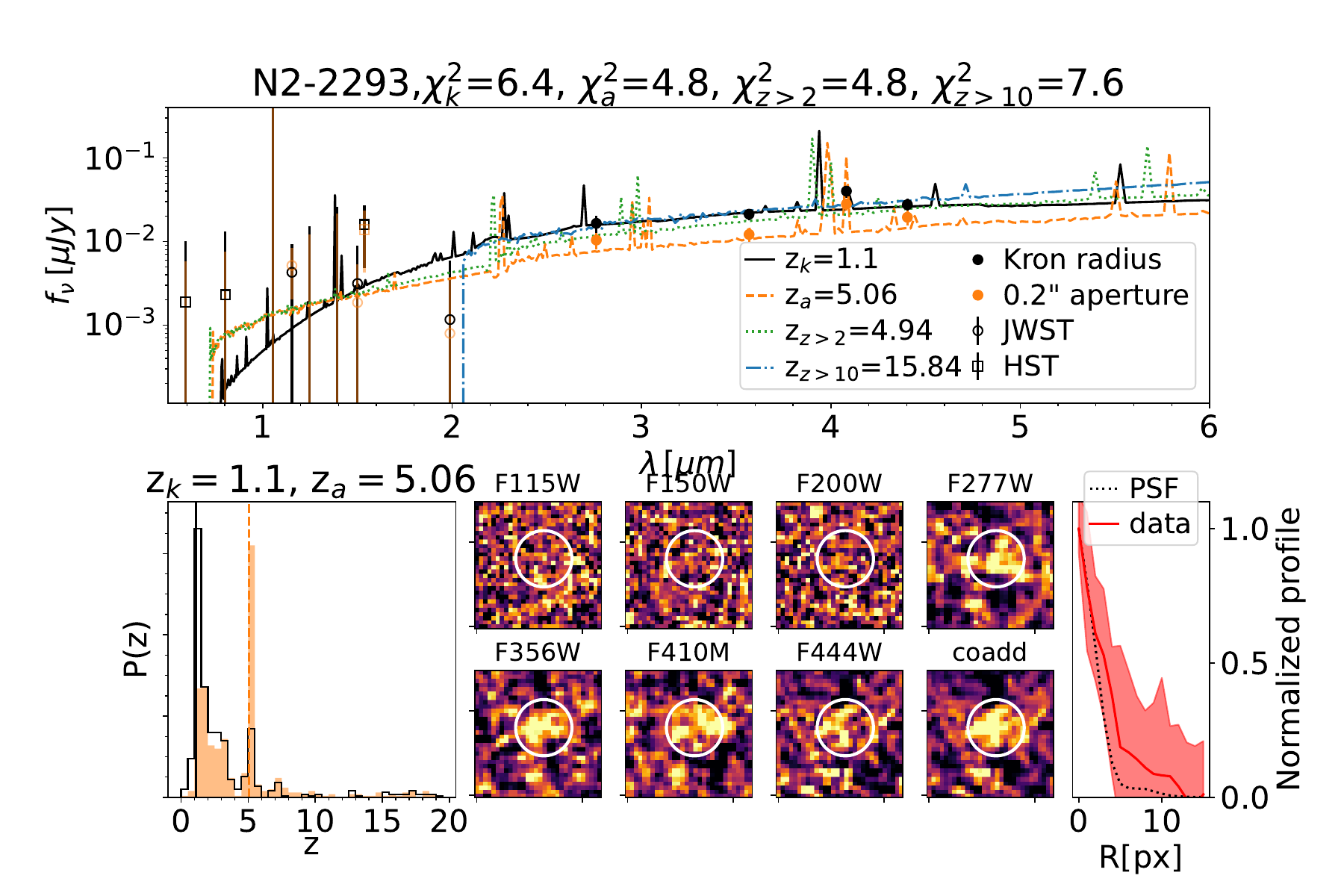} \\
    \caption{continued.}\\
    \includegraphics[trim={20 10 50 40},clip,width=0.44\linewidth,keepaspectratio]{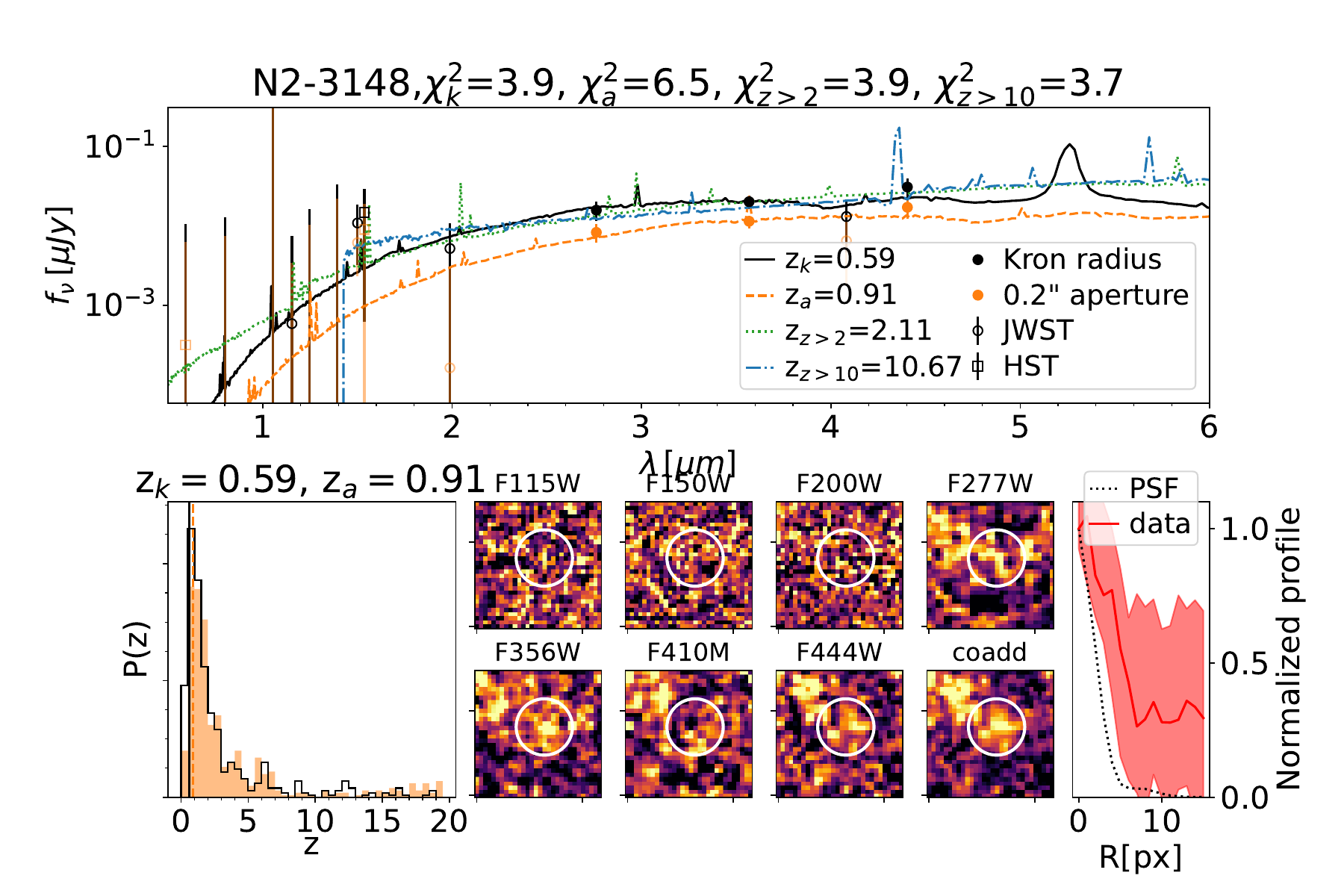} &
    \includegraphics[trim={20 10 50 40},clip,width=0.44\linewidth,keepaspectratio]{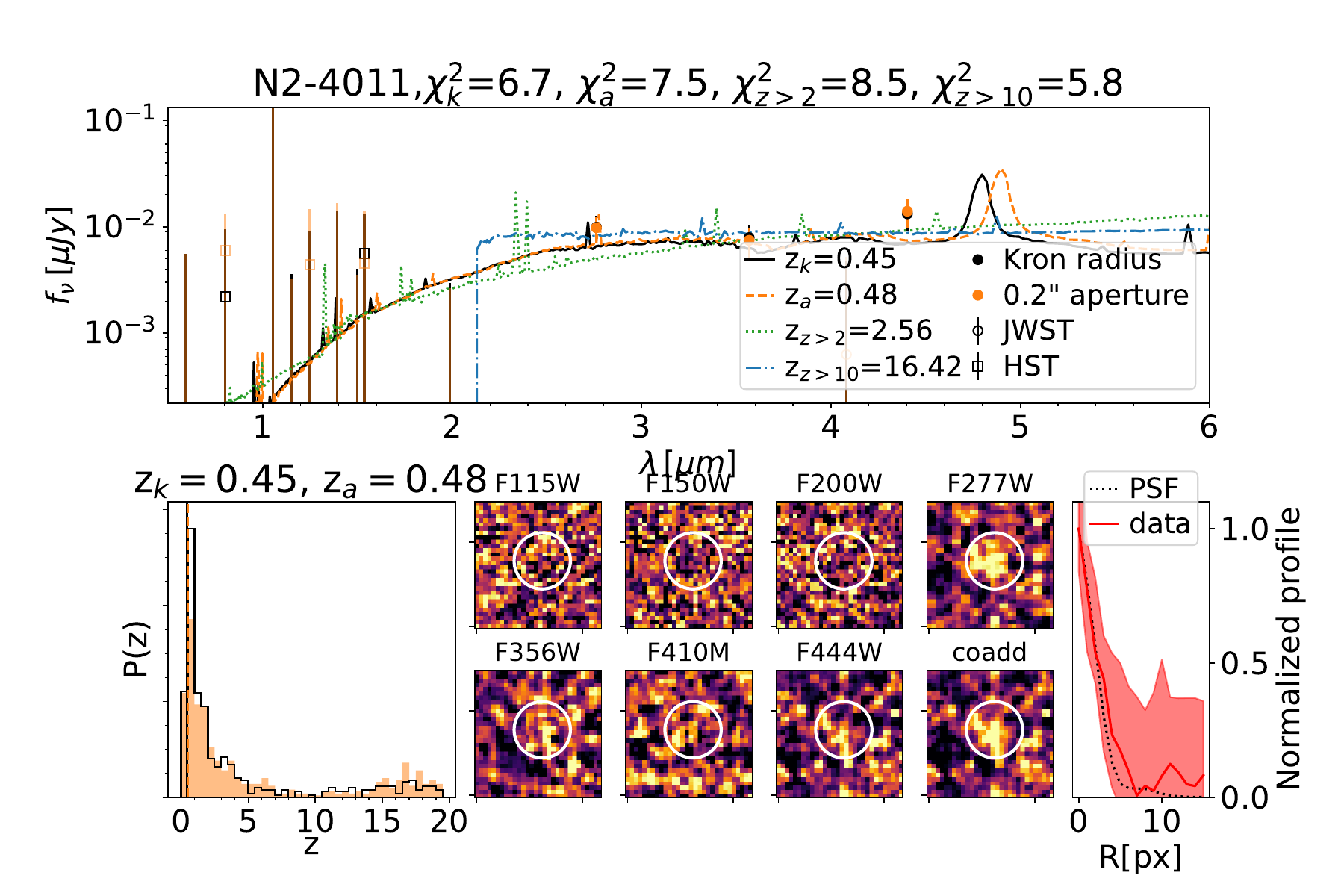} \\
    \includegraphics[trim={20 10 50 40},clip,width=0.44\linewidth,keepaspectratio]{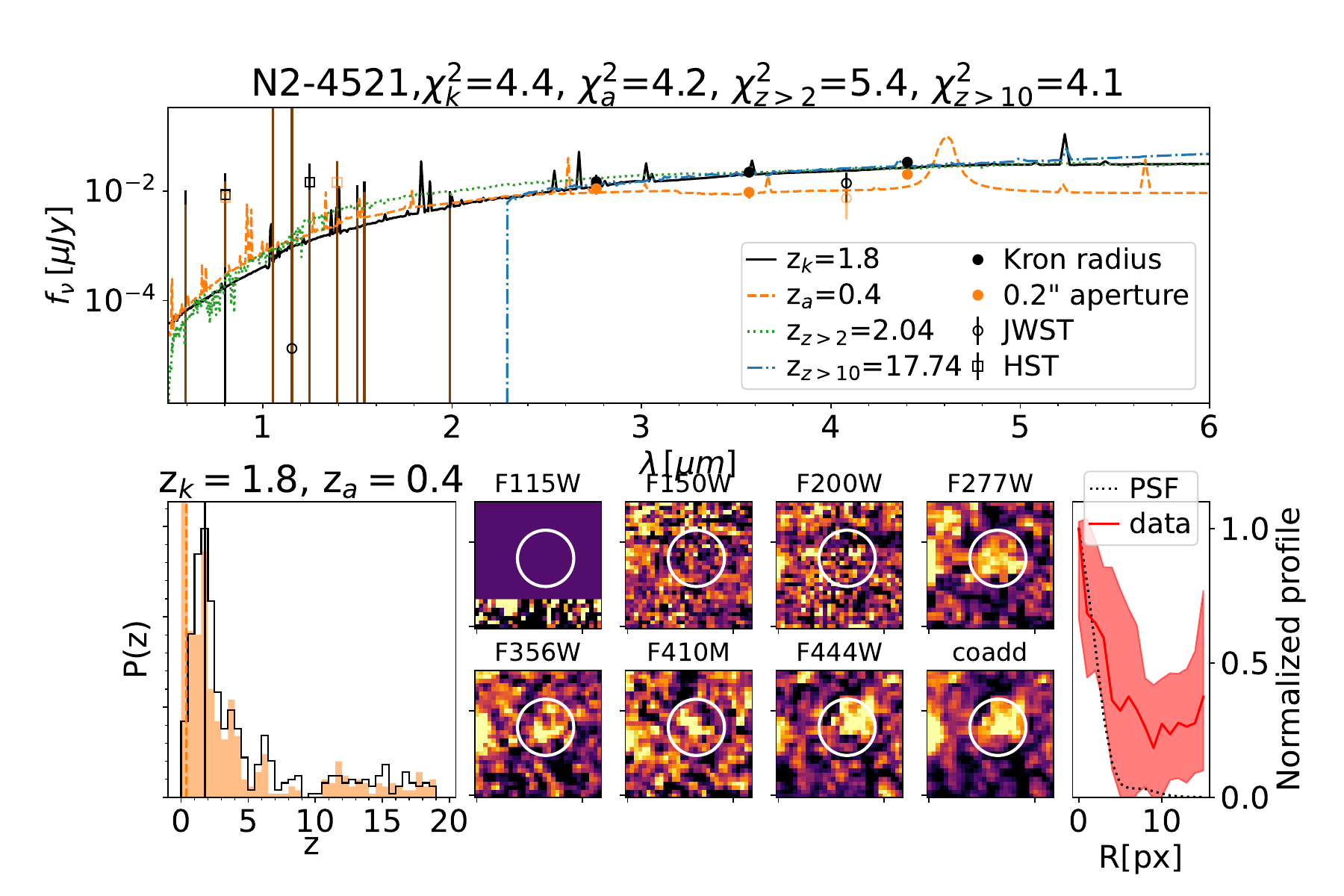} &
    \includegraphics[trim={20 10 50 40},clip,width=0.44\linewidth,keepaspectratio]{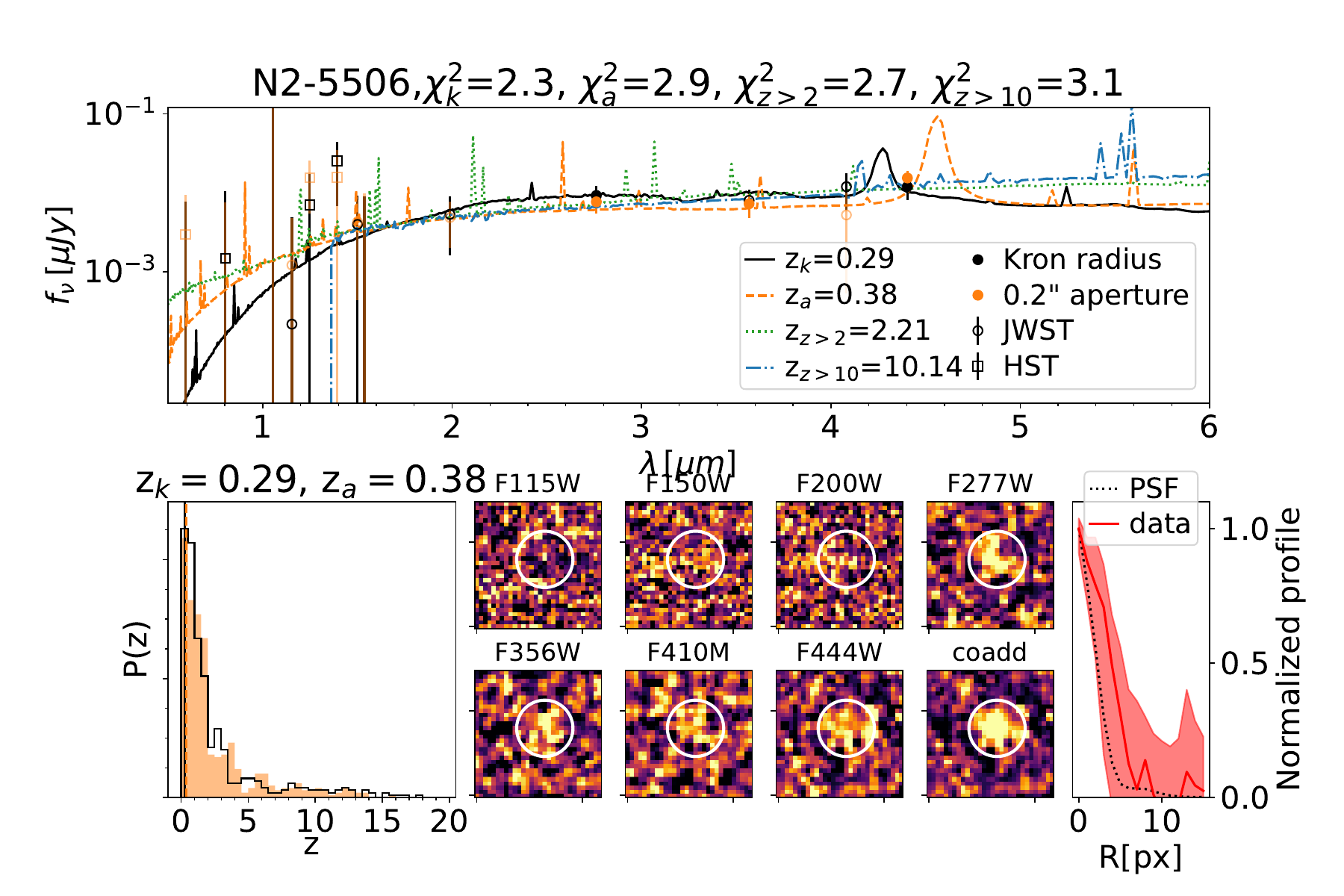} \\
    \includegraphics[trim={20 10 50 40},clip,width=0.44\linewidth,keepaspectratio]{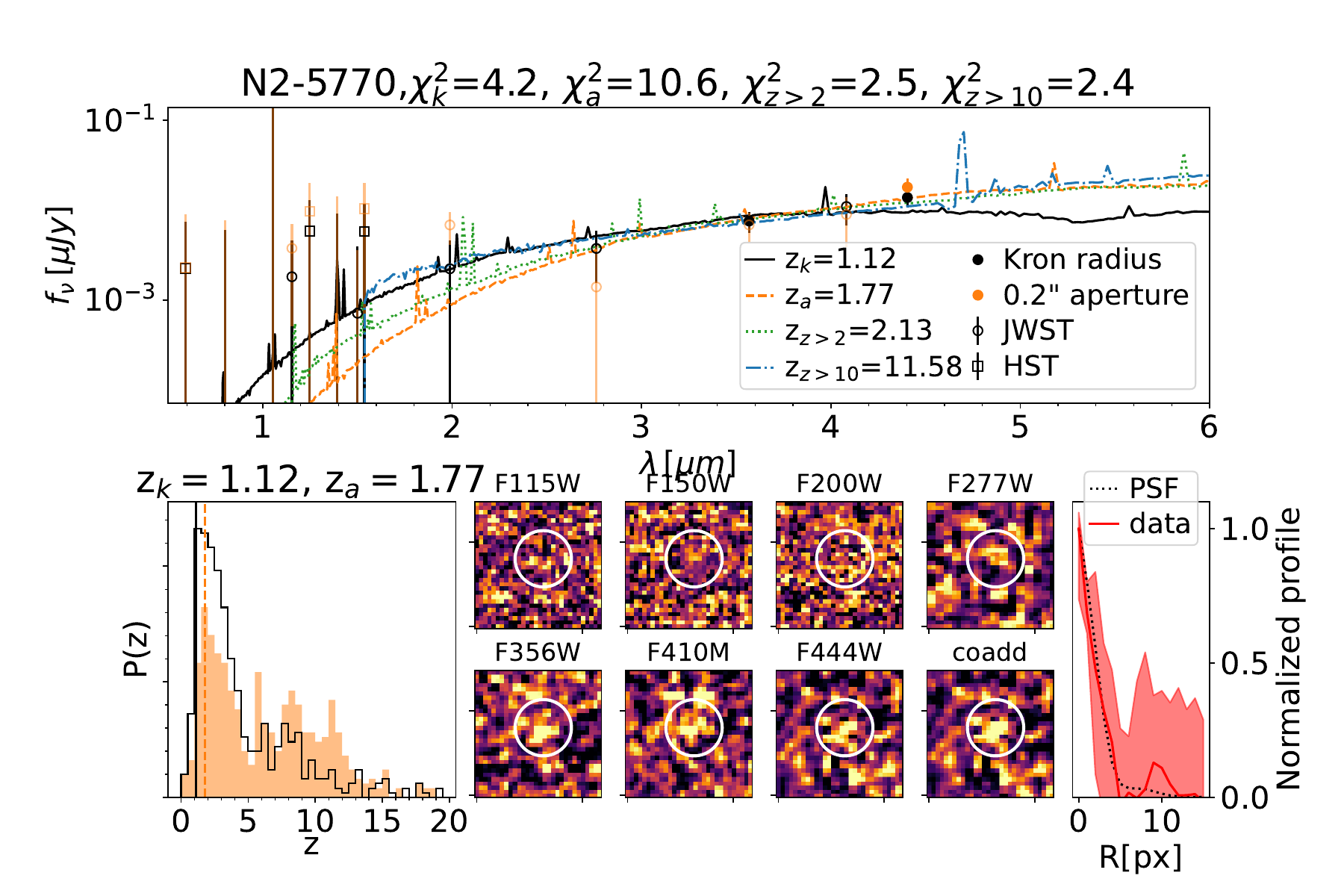} &
    \includegraphics[trim={20 10 50 40},clip,width=0.44\linewidth,keepaspectratio]{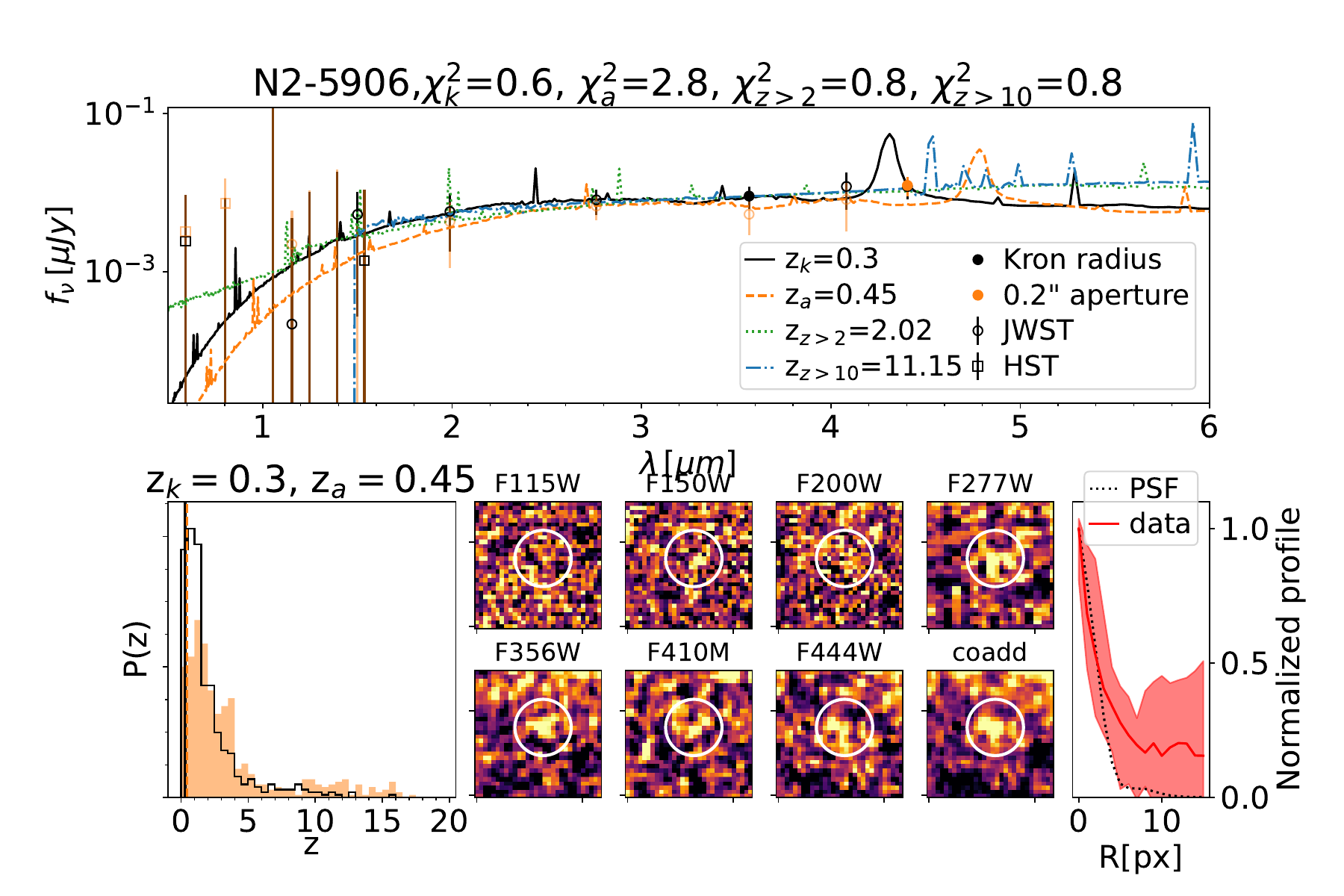} \\
    \includegraphics[trim={20 10 50 40},clip,width=0.44\linewidth,keepaspectratio]{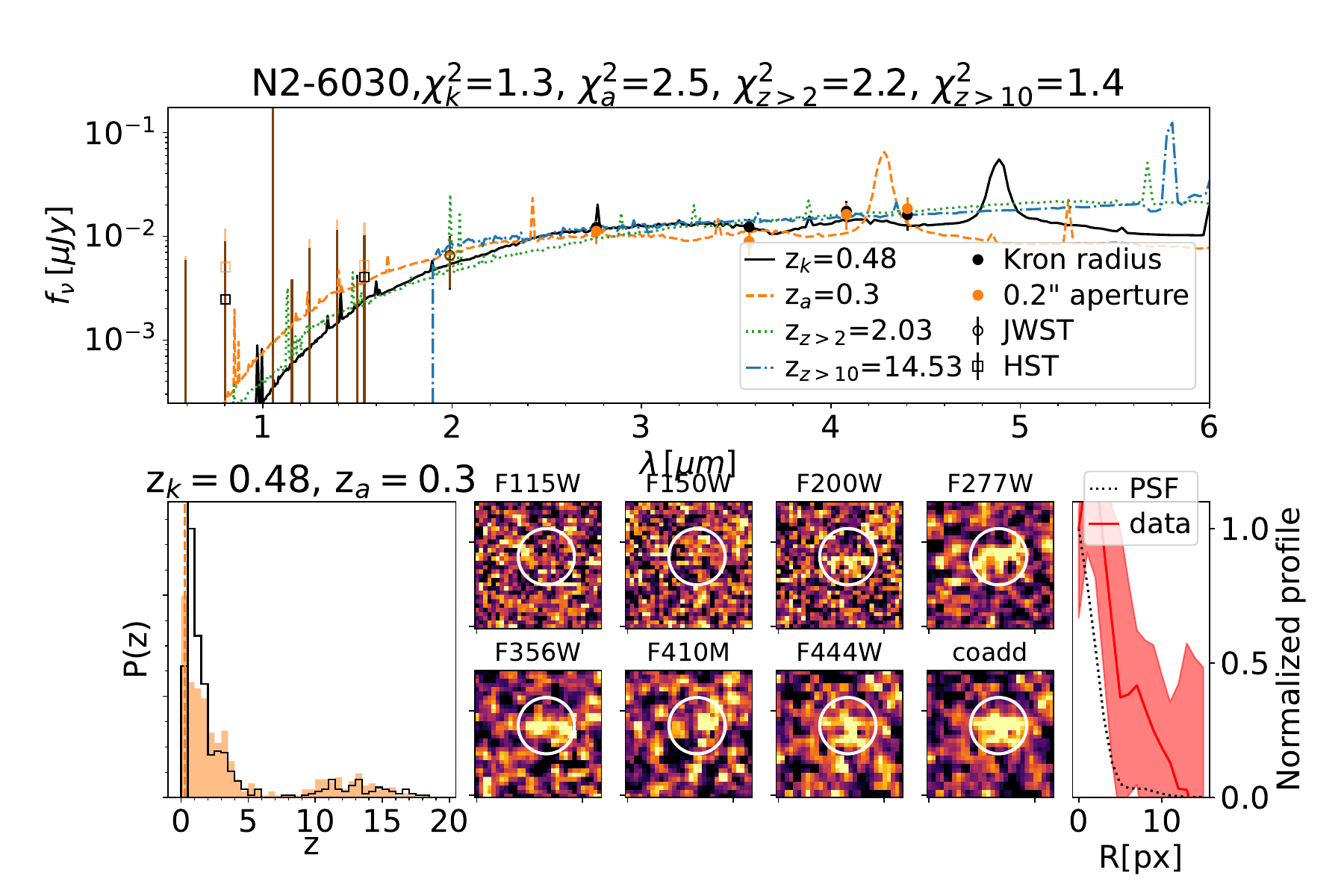} &
    \includegraphics[trim={20 10 50 40},clip,width=0.44\linewidth,keepaspectratio]{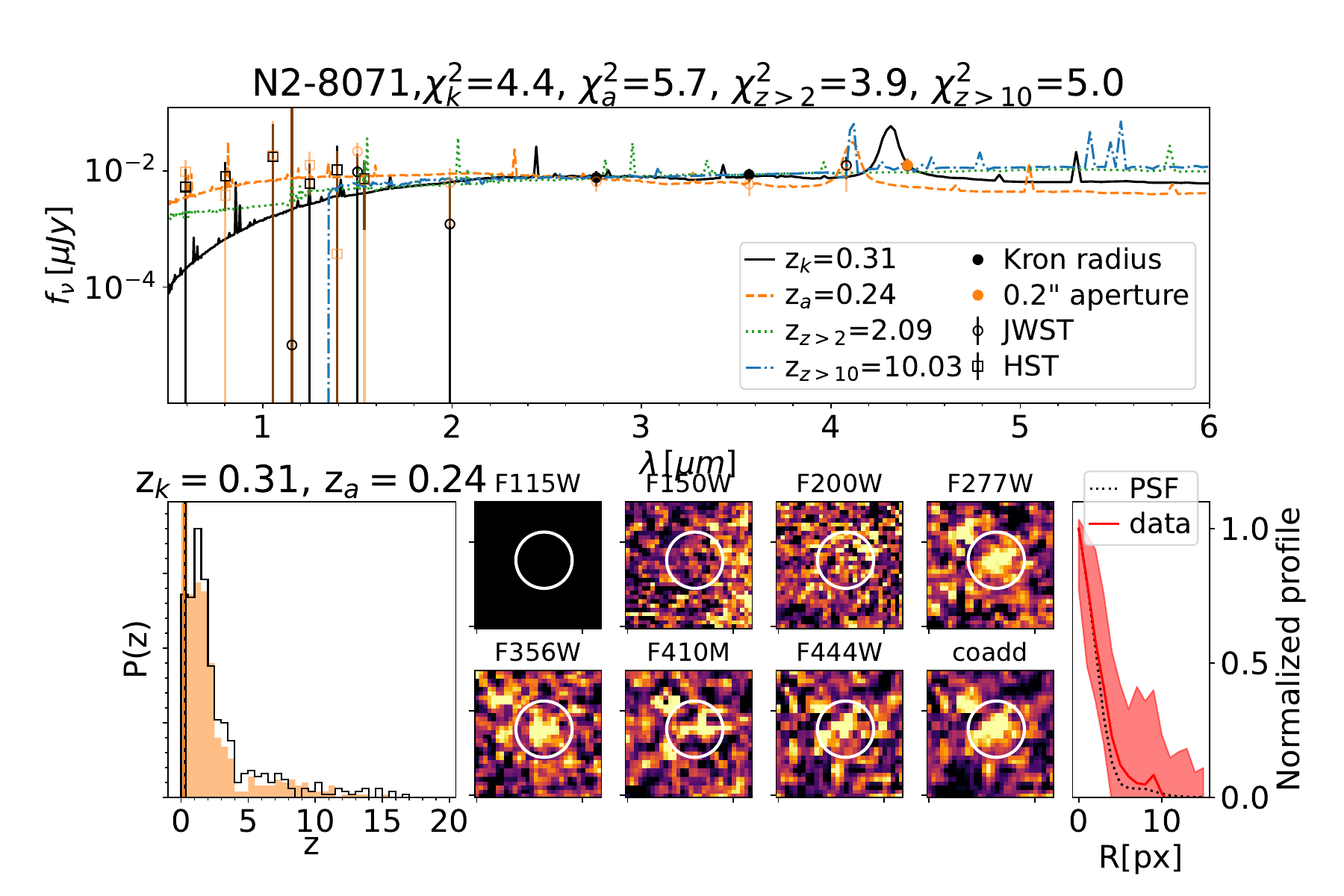} \\
  \caption{continued.}\\
    \includegraphics[trim={20 10 50 40},clip,width=0.44\linewidth,keepaspectratio]{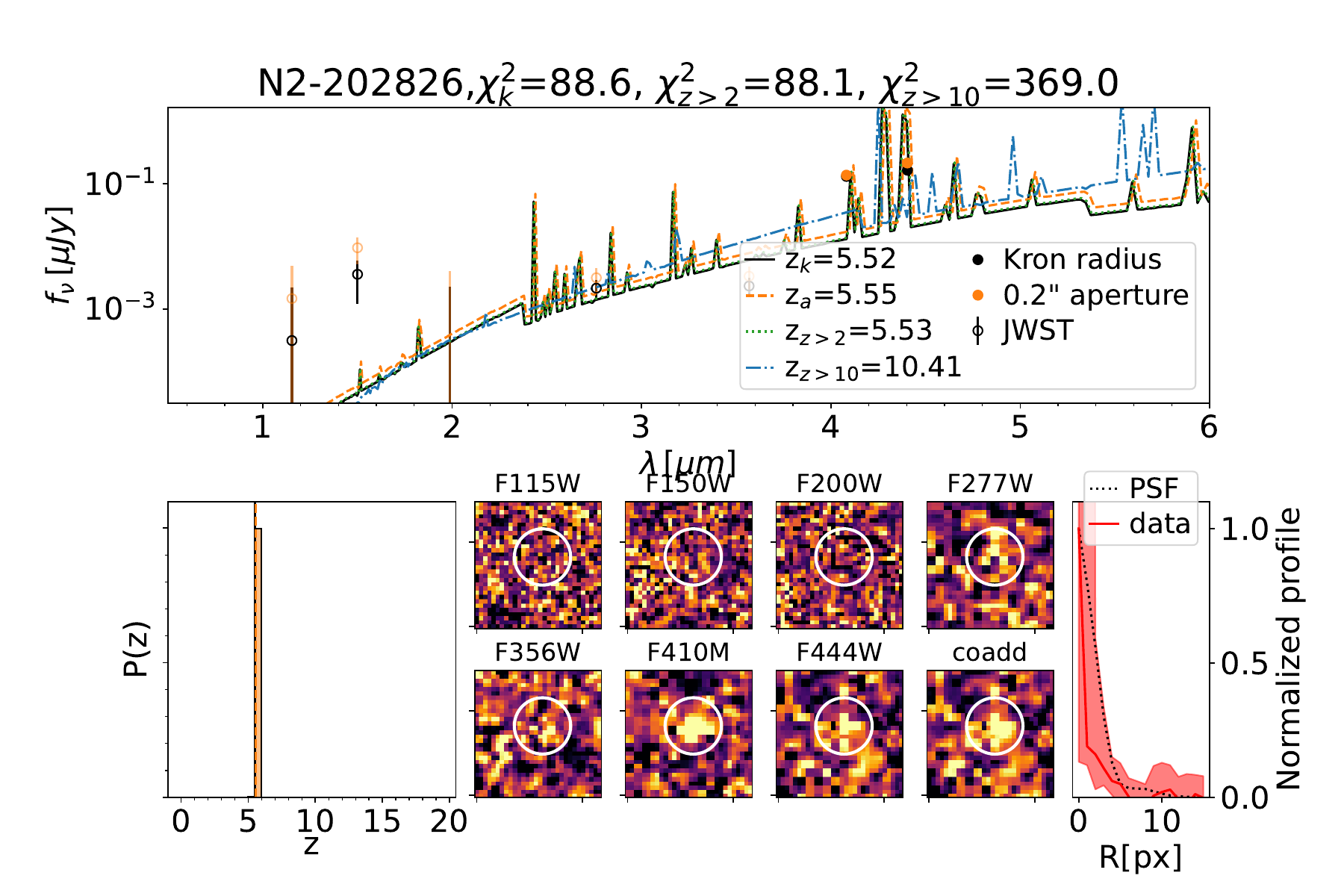} &
    \includegraphics[trim={20 10 50 40},clip,width=0.44\linewidth,keepaspectratio]{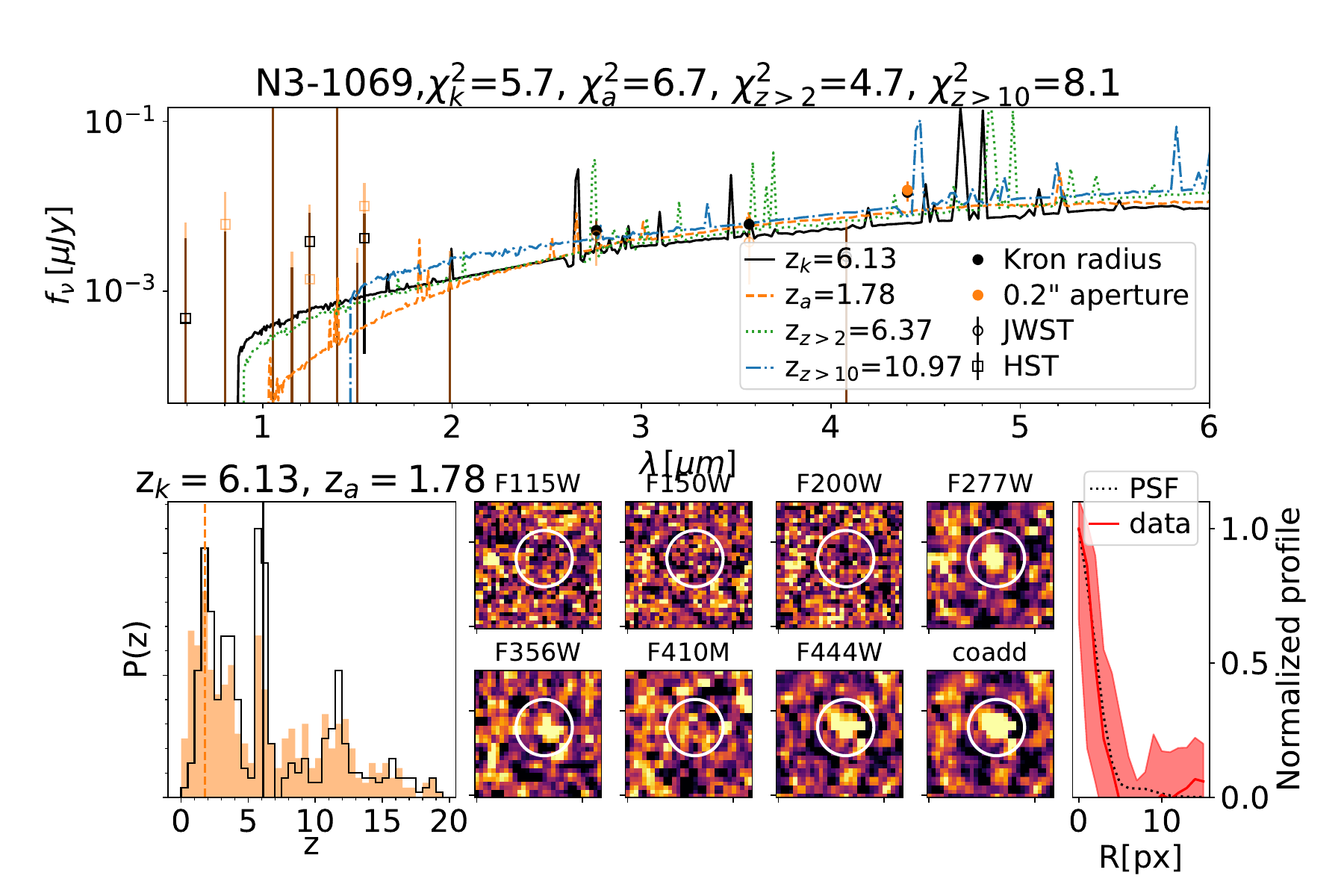} \\
    \includegraphics[trim={20 10 50 40},clip,width=0.44\linewidth,keepaspectratio]{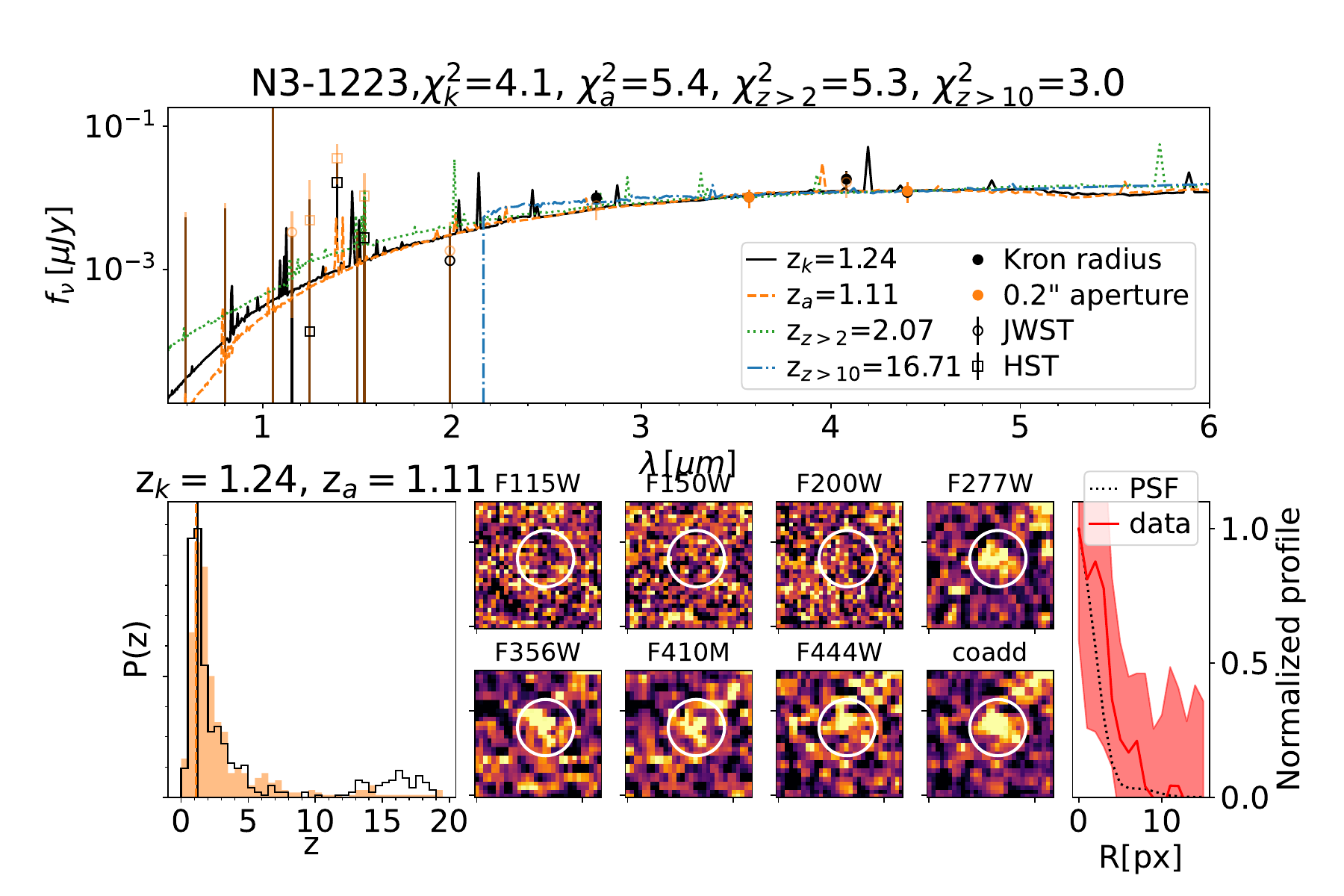} &
    \includegraphics[trim={20 10 50 40},clip,width=0.44\linewidth,keepaspectratio]{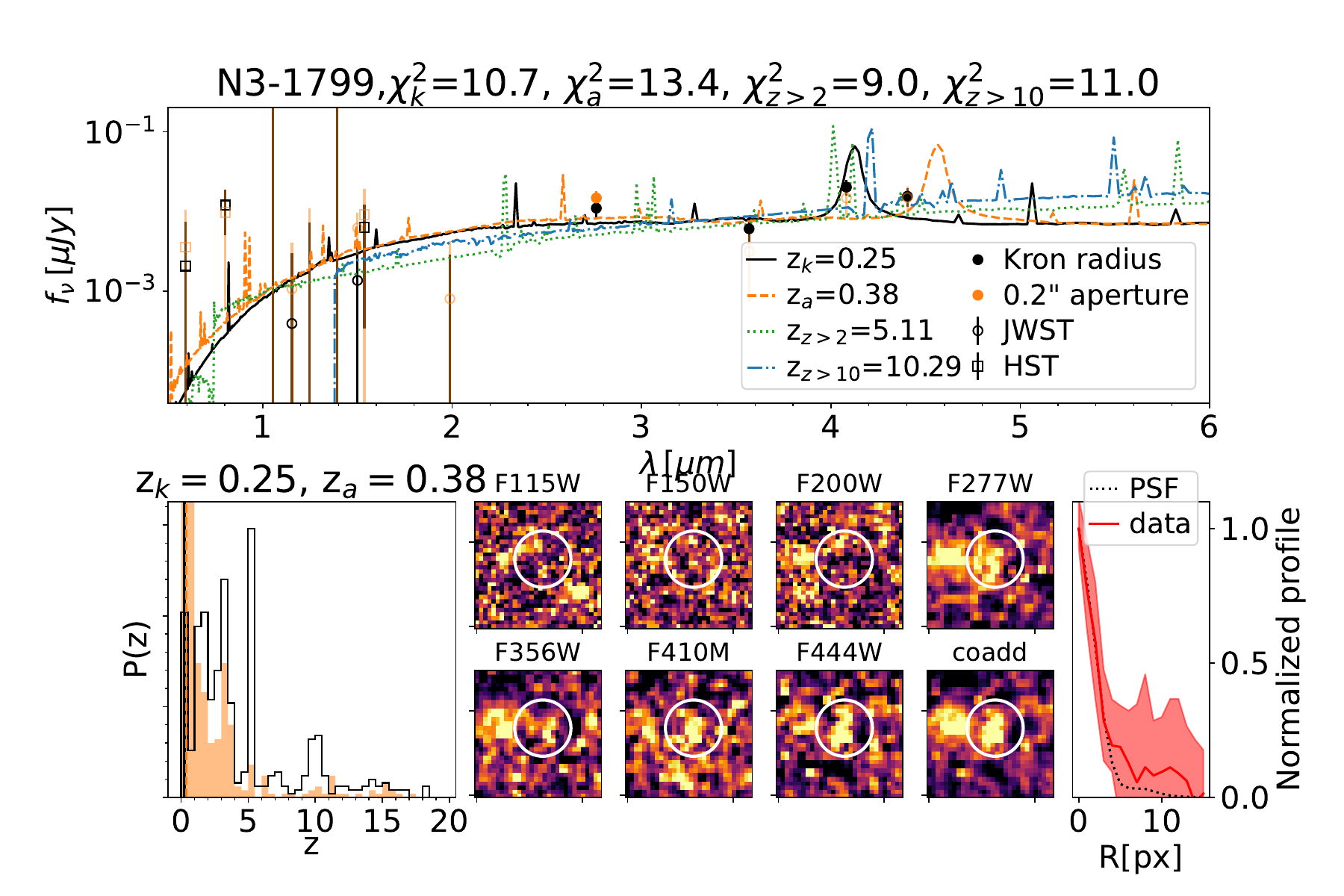} \\
    \includegraphics[trim={20 10 50 40},clip,width=0.44\linewidth,keepaspectratio]{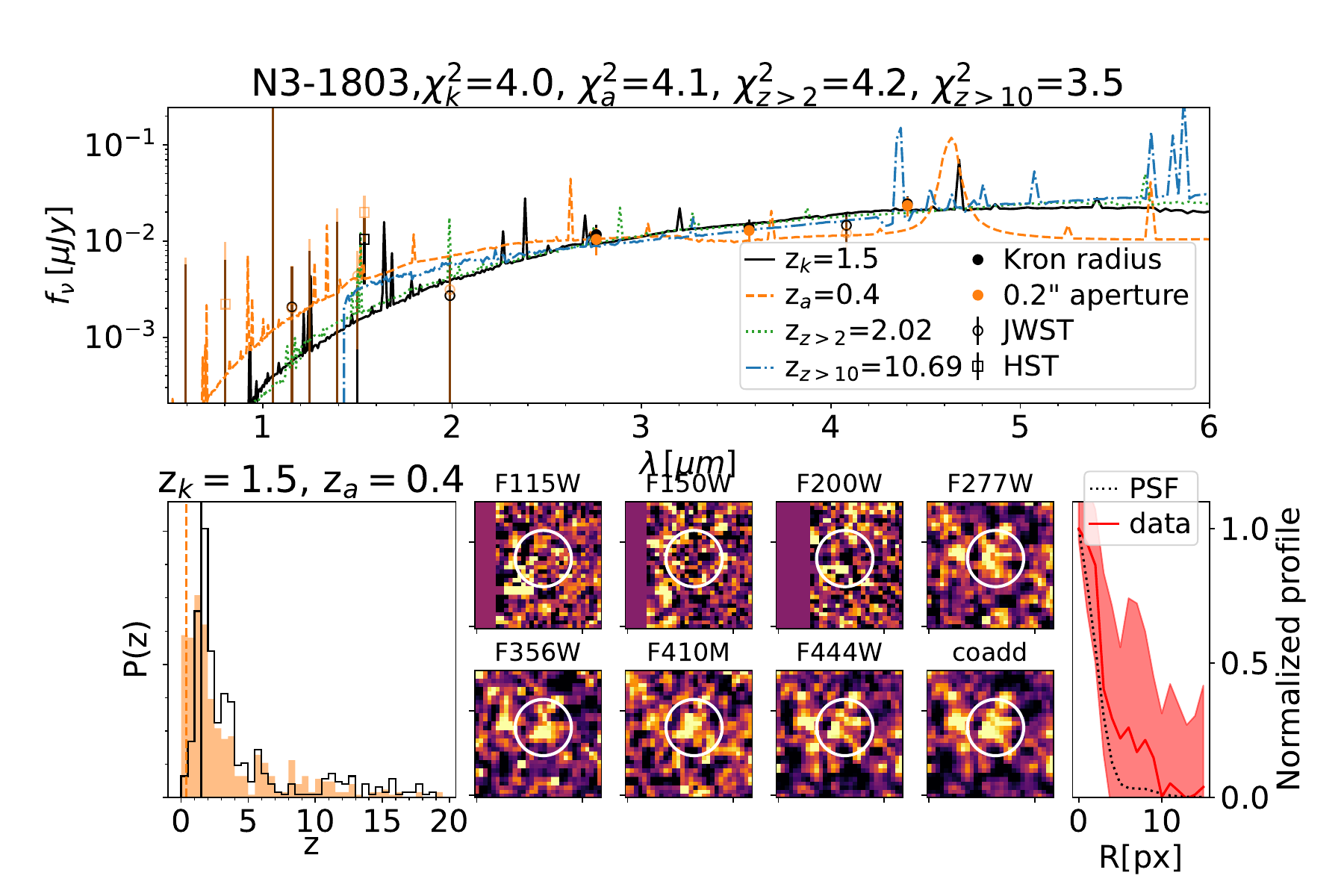} &
    \includegraphics[trim={20 10 50 40},clip,width=0.44\linewidth,keepaspectratio]{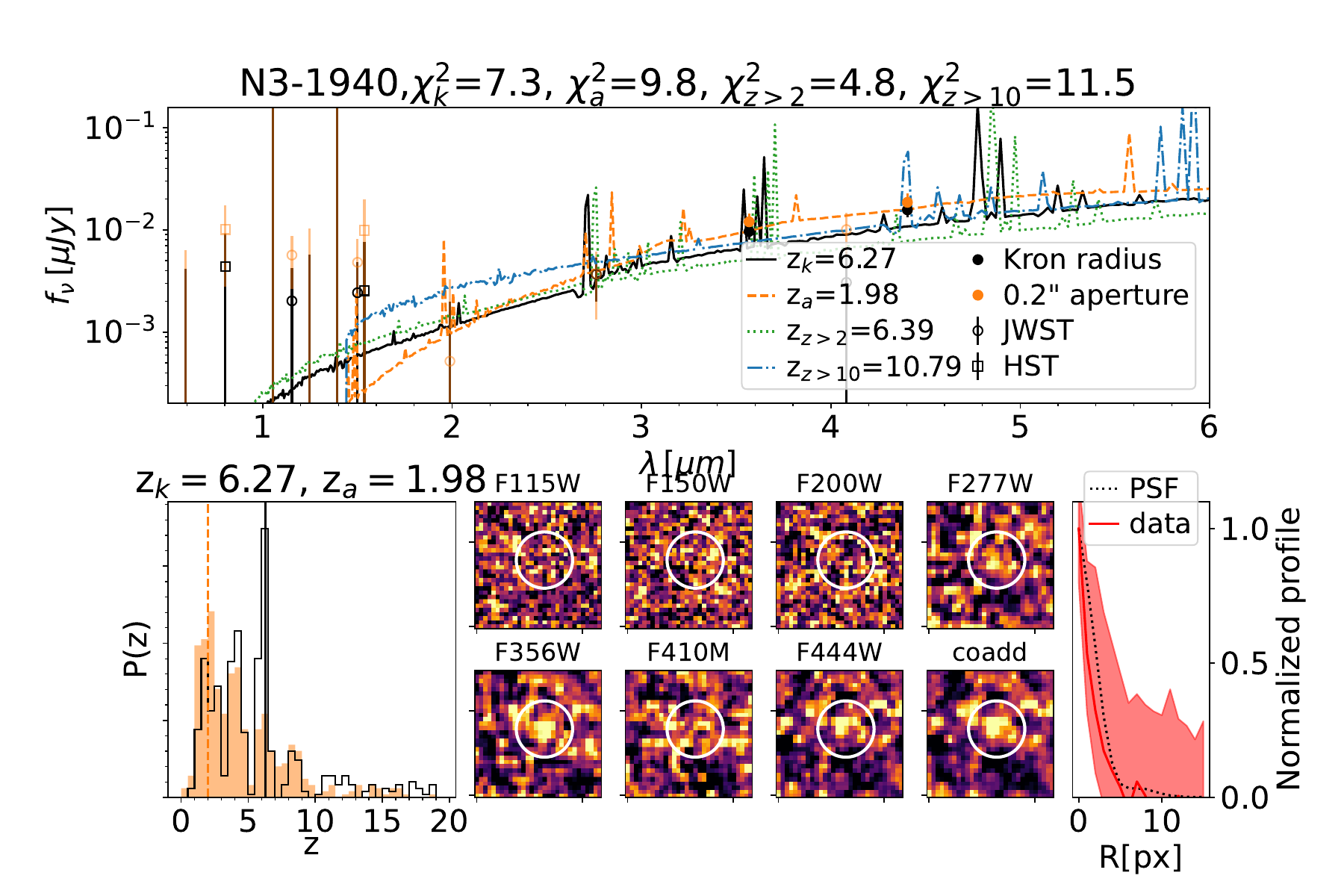} \\
    \includegraphics[trim={20 10 50 40},clip,width=0.44\linewidth,keepaspectratio]{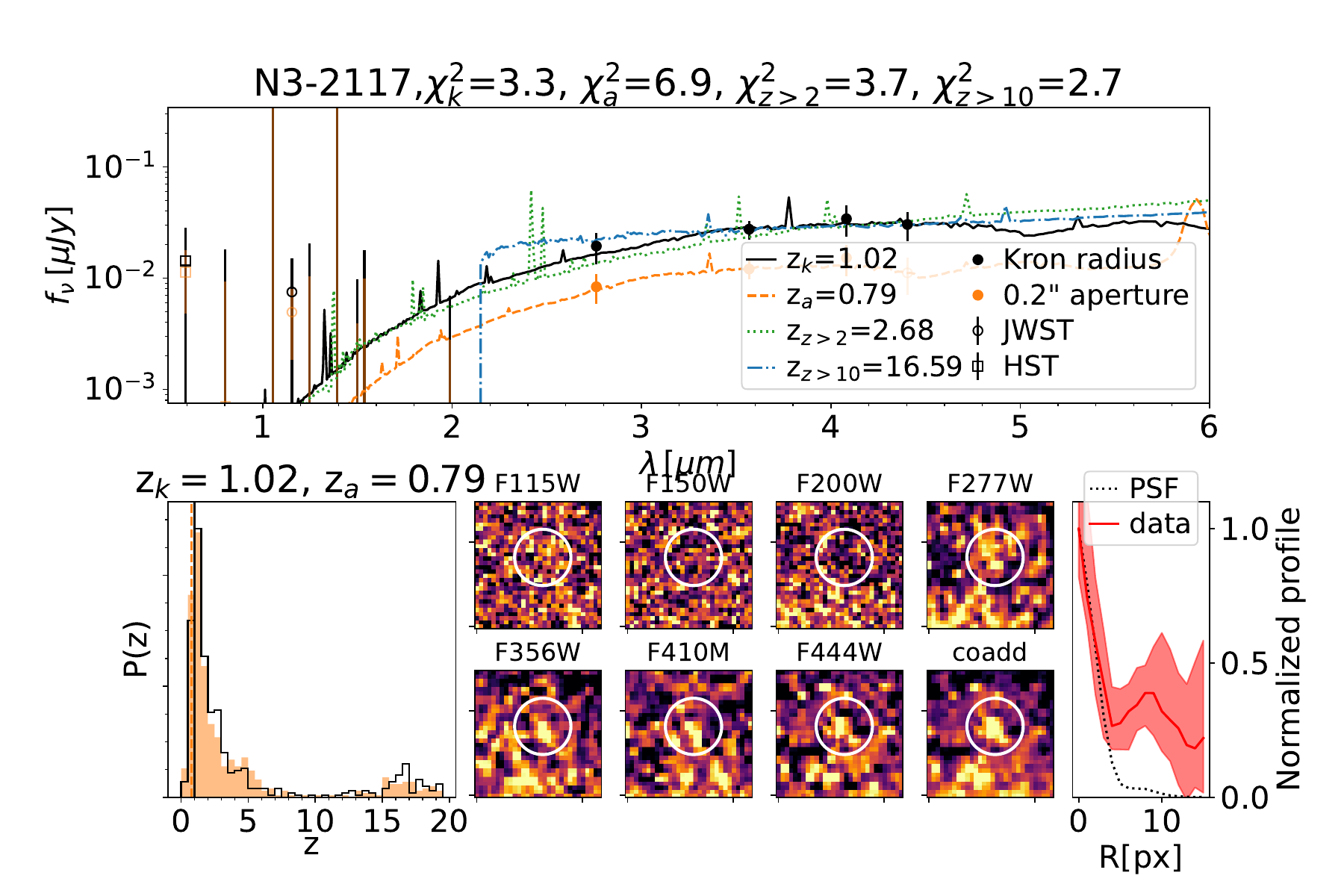} &
    \includegraphics[trim={20 10 50 40},clip,width=0.44\linewidth,keepaspectratio]{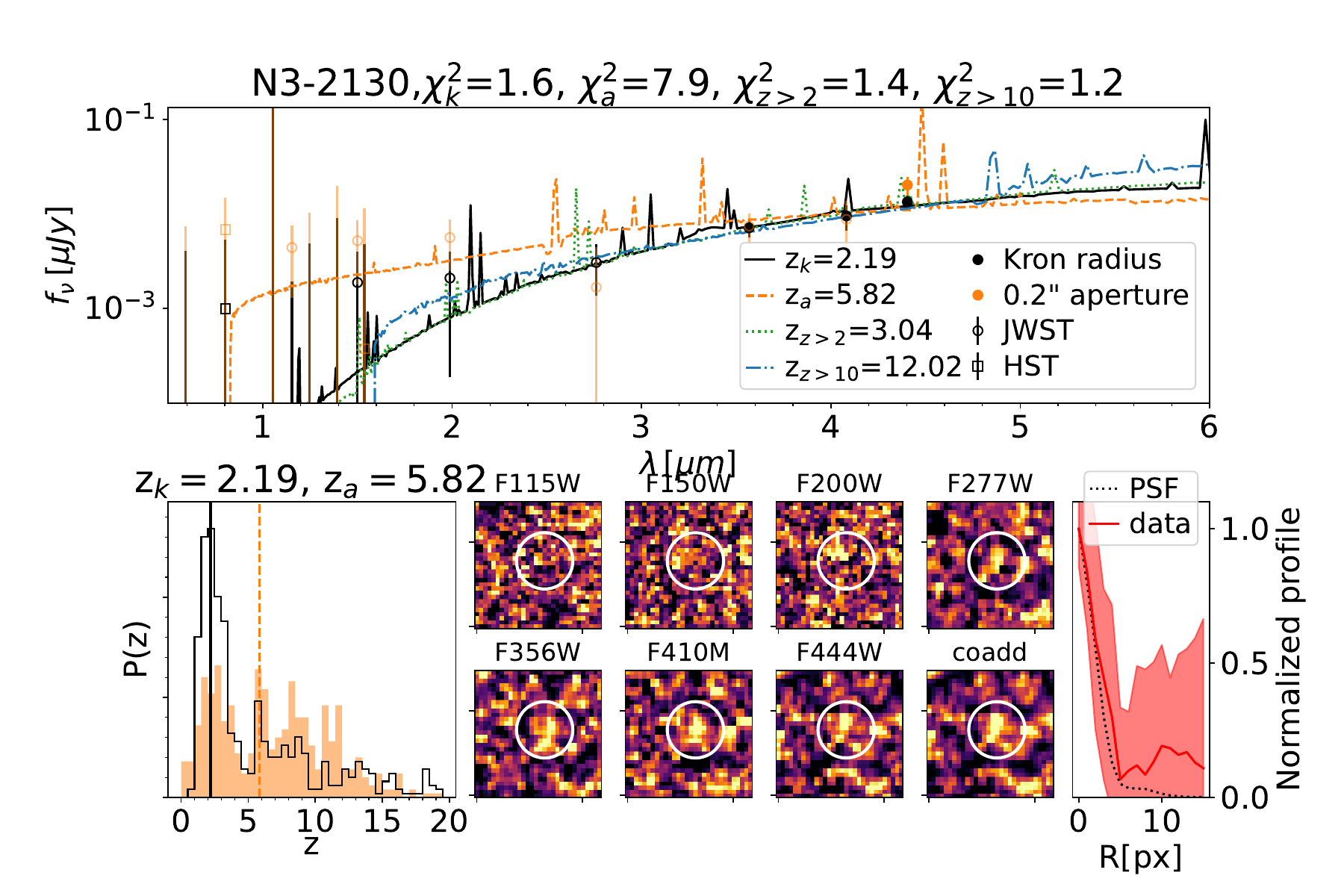} \\
  \caption{continued.}\\
    \includegraphics[trim={20 10 50 40},clip,width=0.44\linewidth,keepaspectratio]{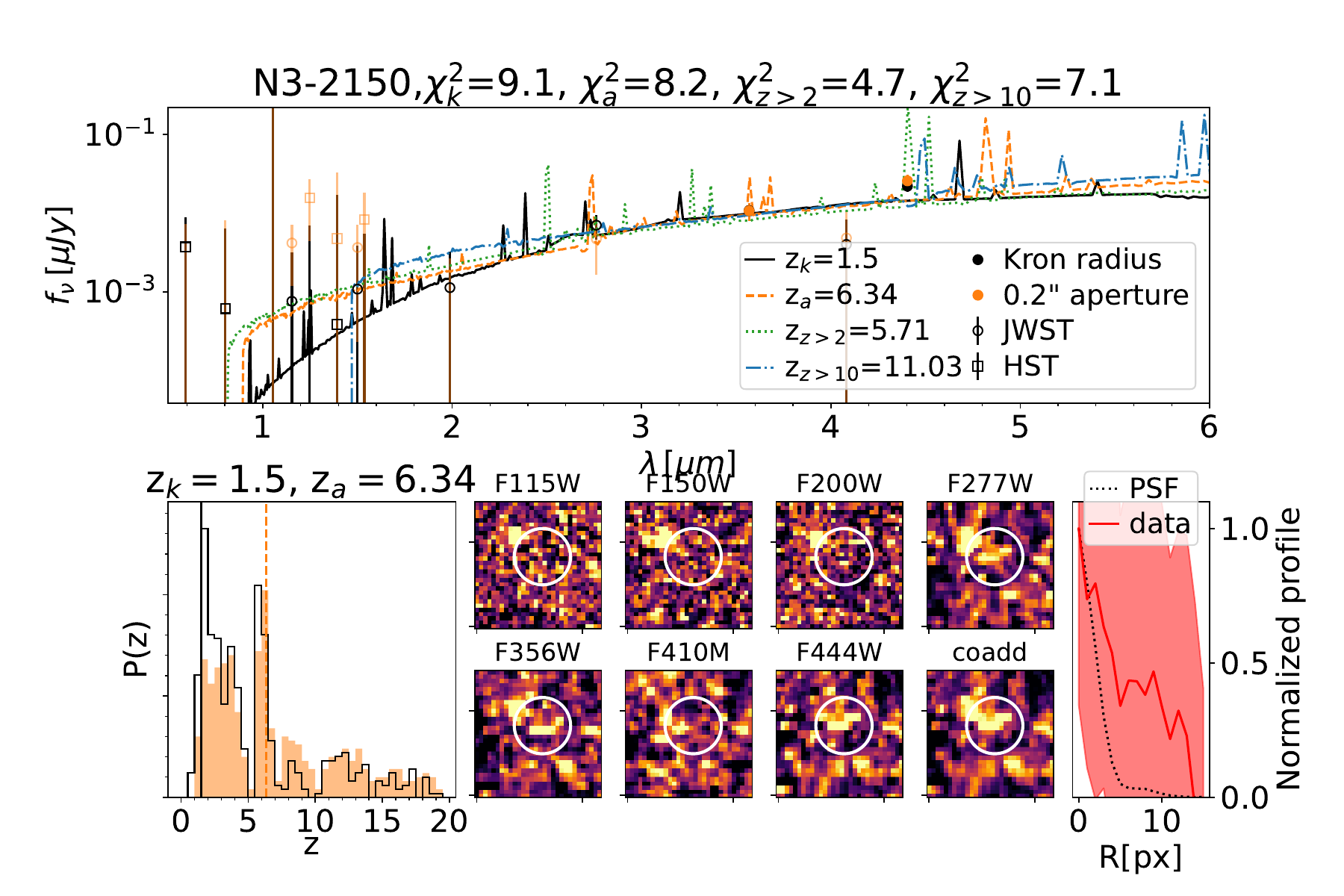} &
    \includegraphics[trim={20 10 50 40},clip,width=0.44\linewidth,keepaspectratio]{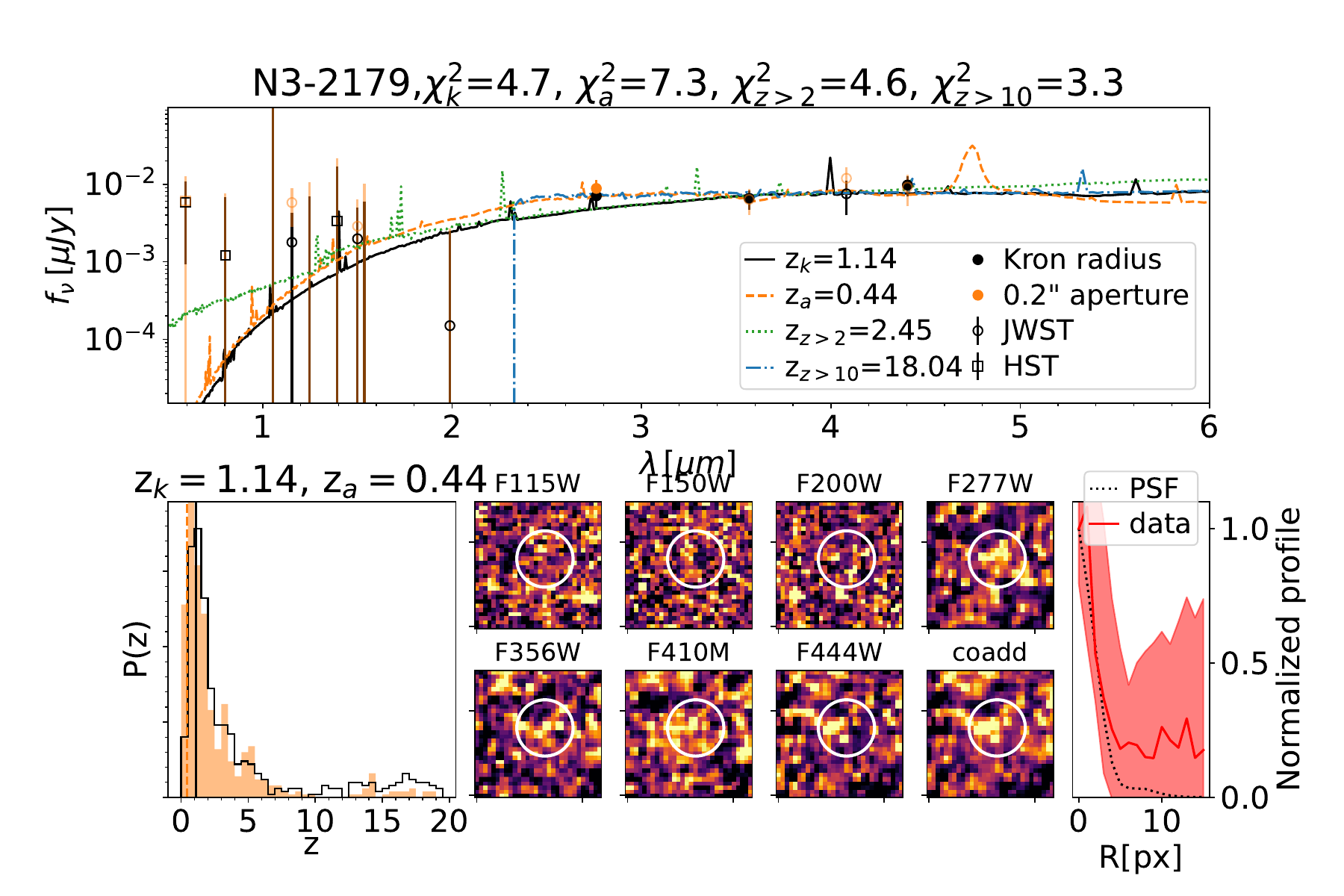} \\
    \includegraphics[trim={20 10 50 40},clip,width=0.44\linewidth,keepaspectratio]{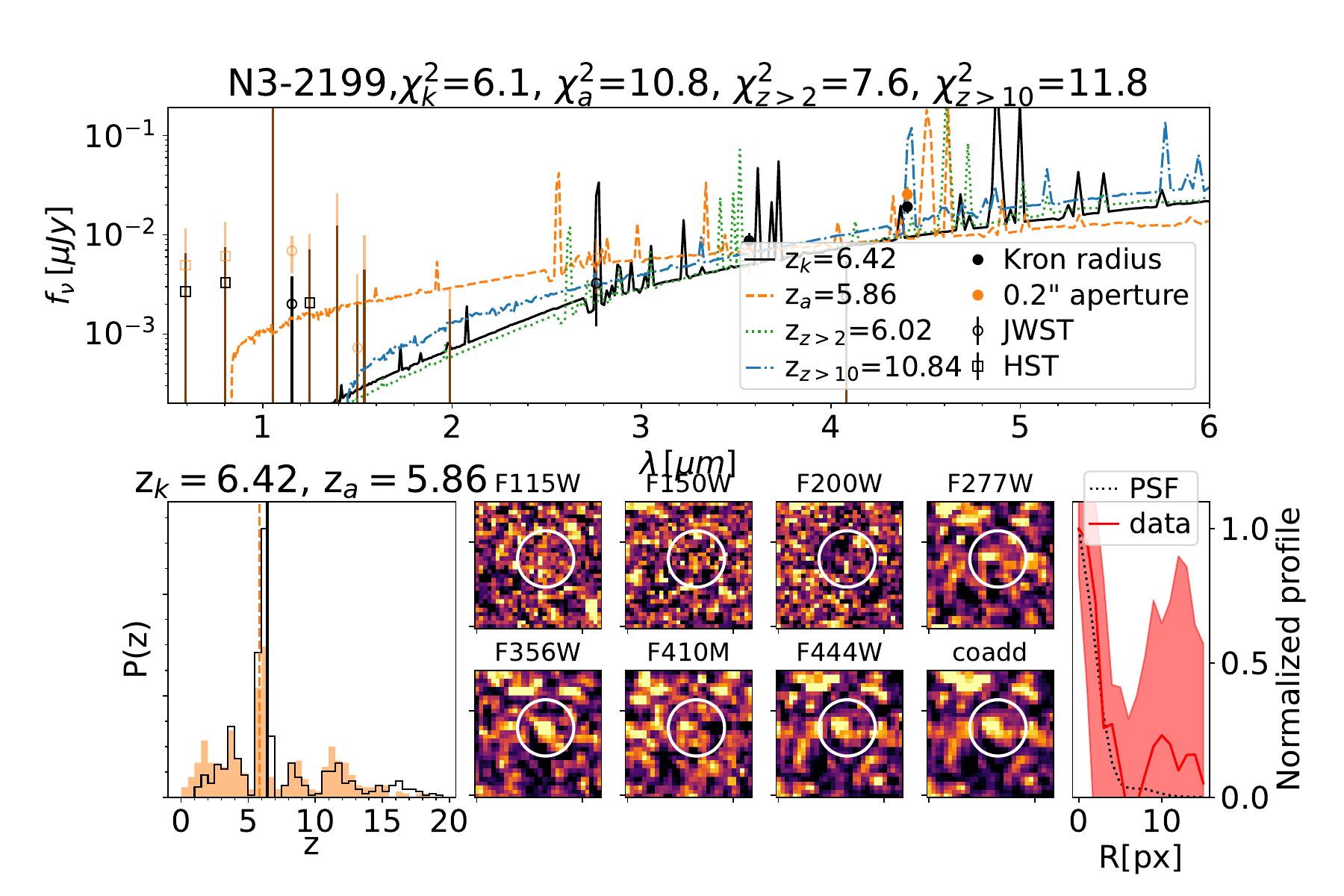} &
    \includegraphics[trim={20 10 50 40},clip,width=0.44\linewidth,keepaspectratio]{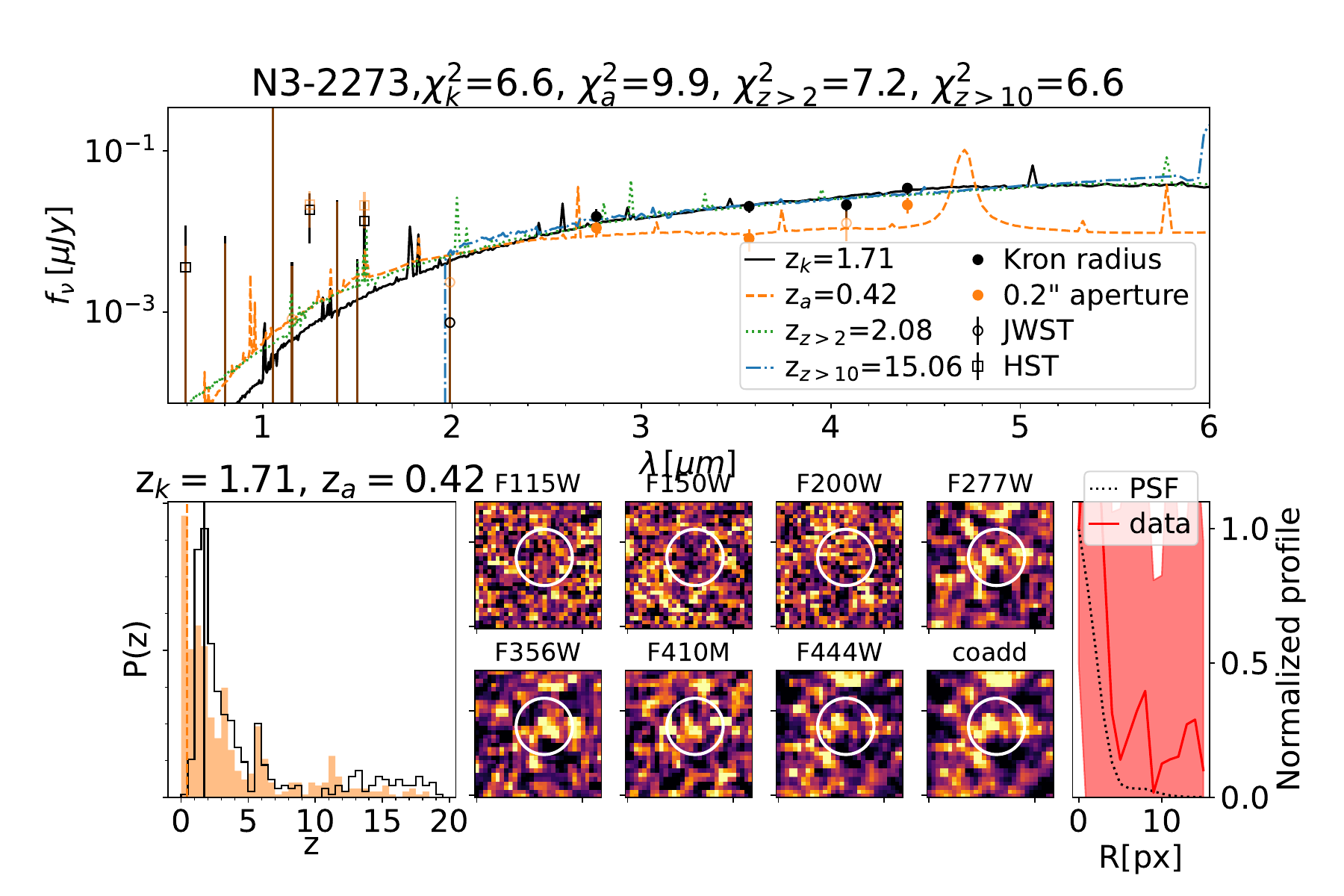} \\
    \includegraphics[trim={20 10 50 40},clip,width=0.44\linewidth,keepaspectratio]{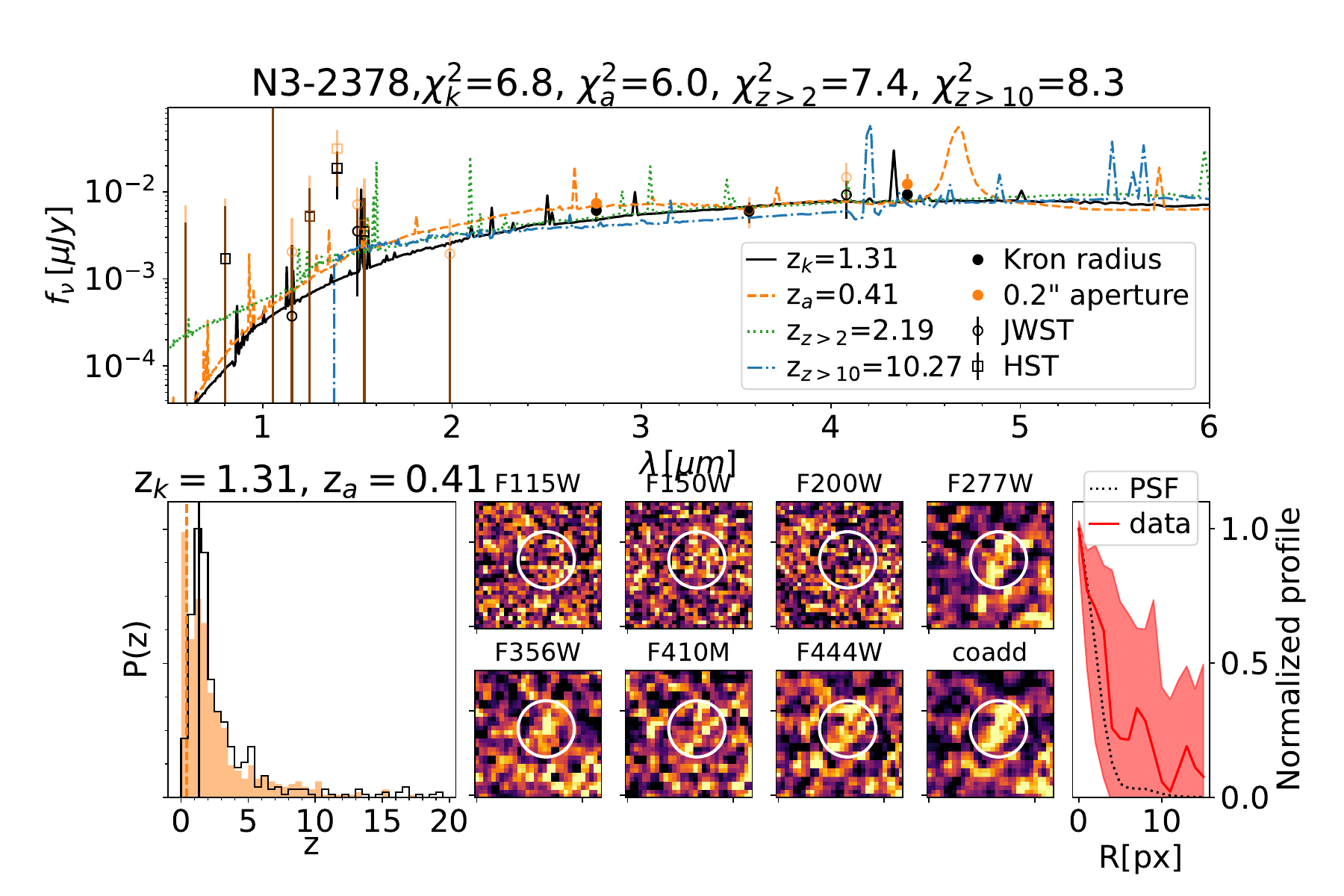} &
    \includegraphics[trim={20 10 50 40},clip,width=0.44\linewidth,keepaspectratio]{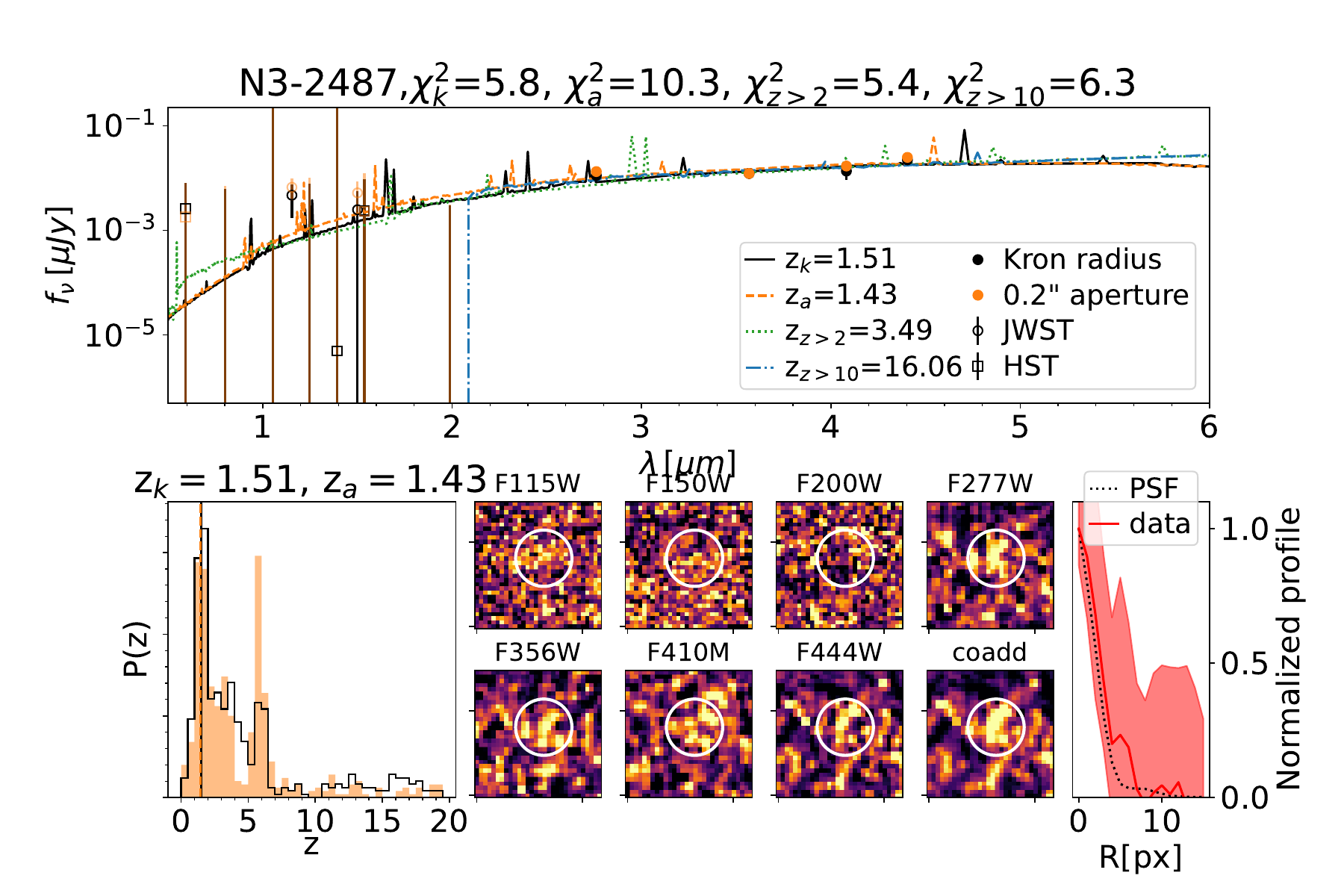} \\
    \includegraphics[trim={20 10 50 40},clip,width=0.44\linewidth,keepaspectratio]{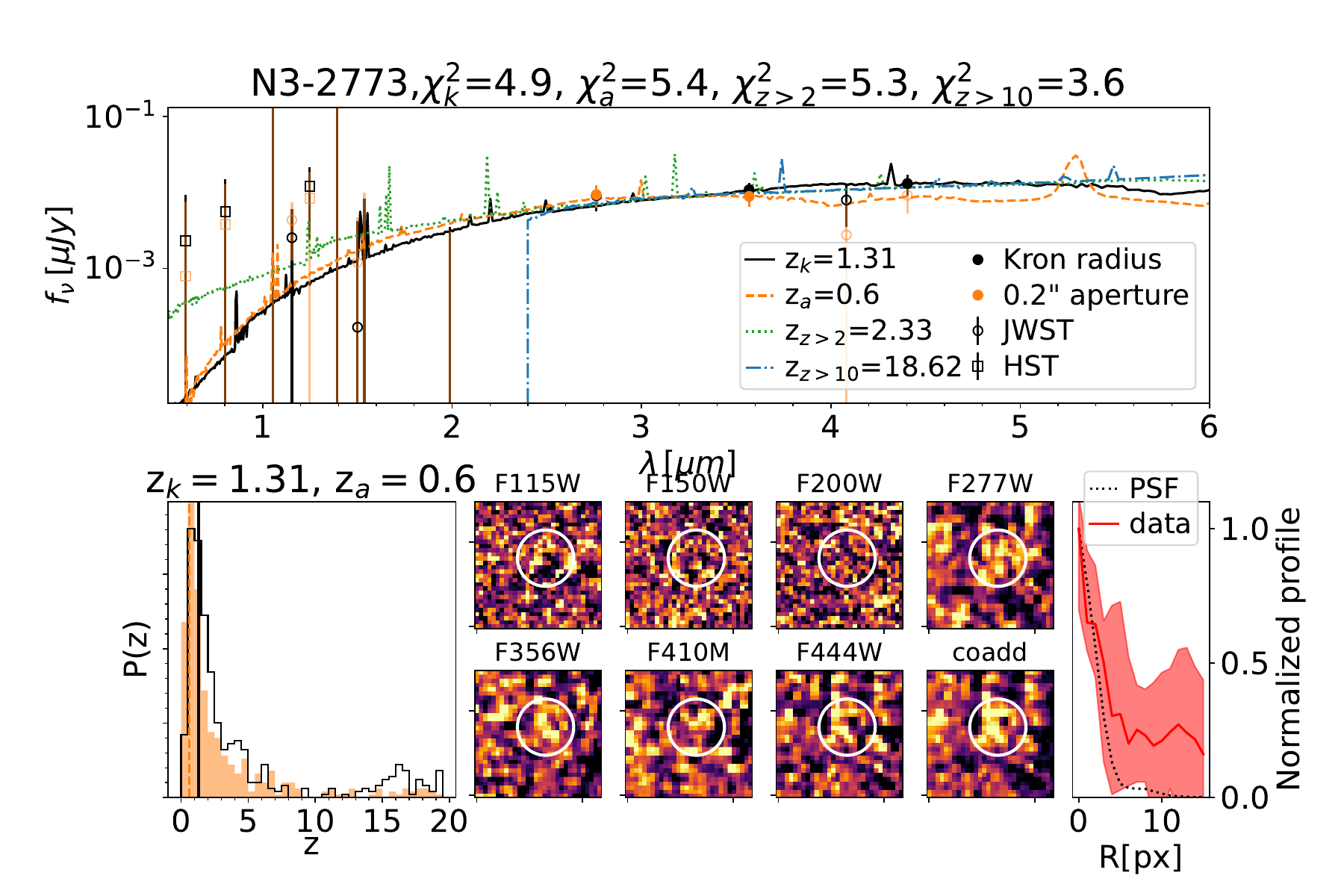} & 
    \includegraphics[trim={20 10 50 40},clip,width=0.44\linewidth,keepaspectratio]{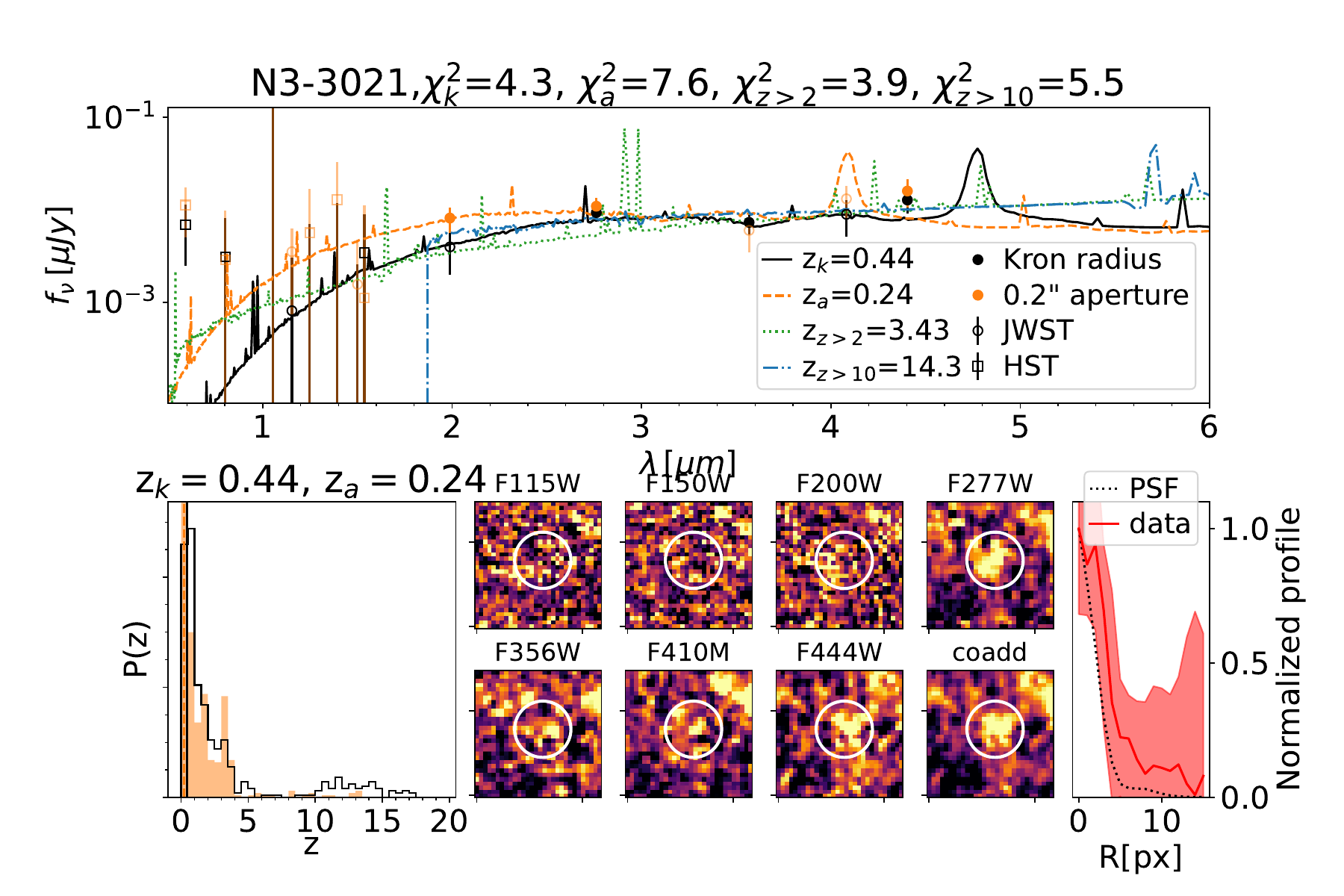} \\
  \caption{continued.}\\
    \includegraphics[trim={20 10 50 40},clip,width=0.44\linewidth,keepaspectratio]{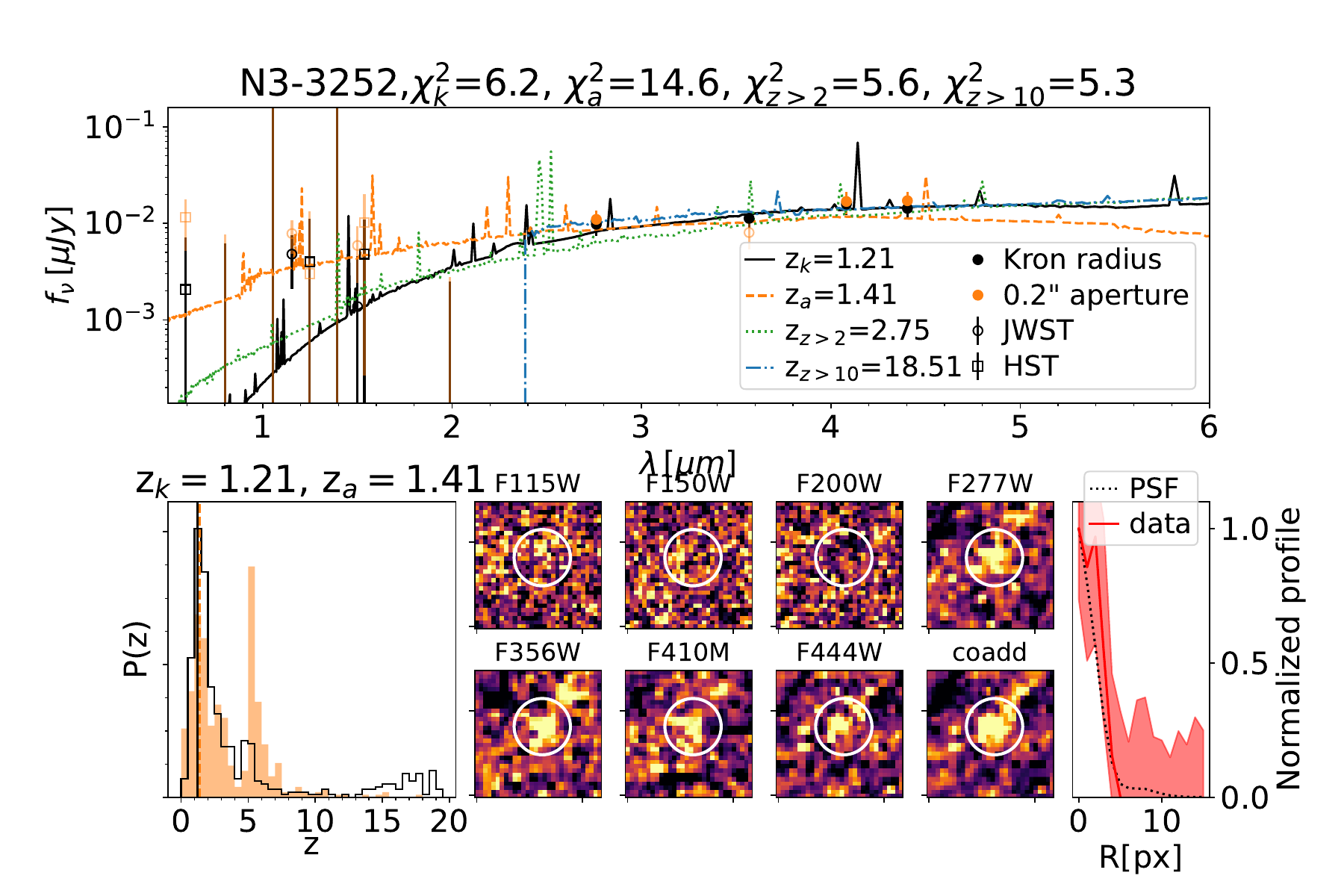} &
    \includegraphics[trim={20 10 50 40},clip,width=0.44\linewidth,keepaspectratio]{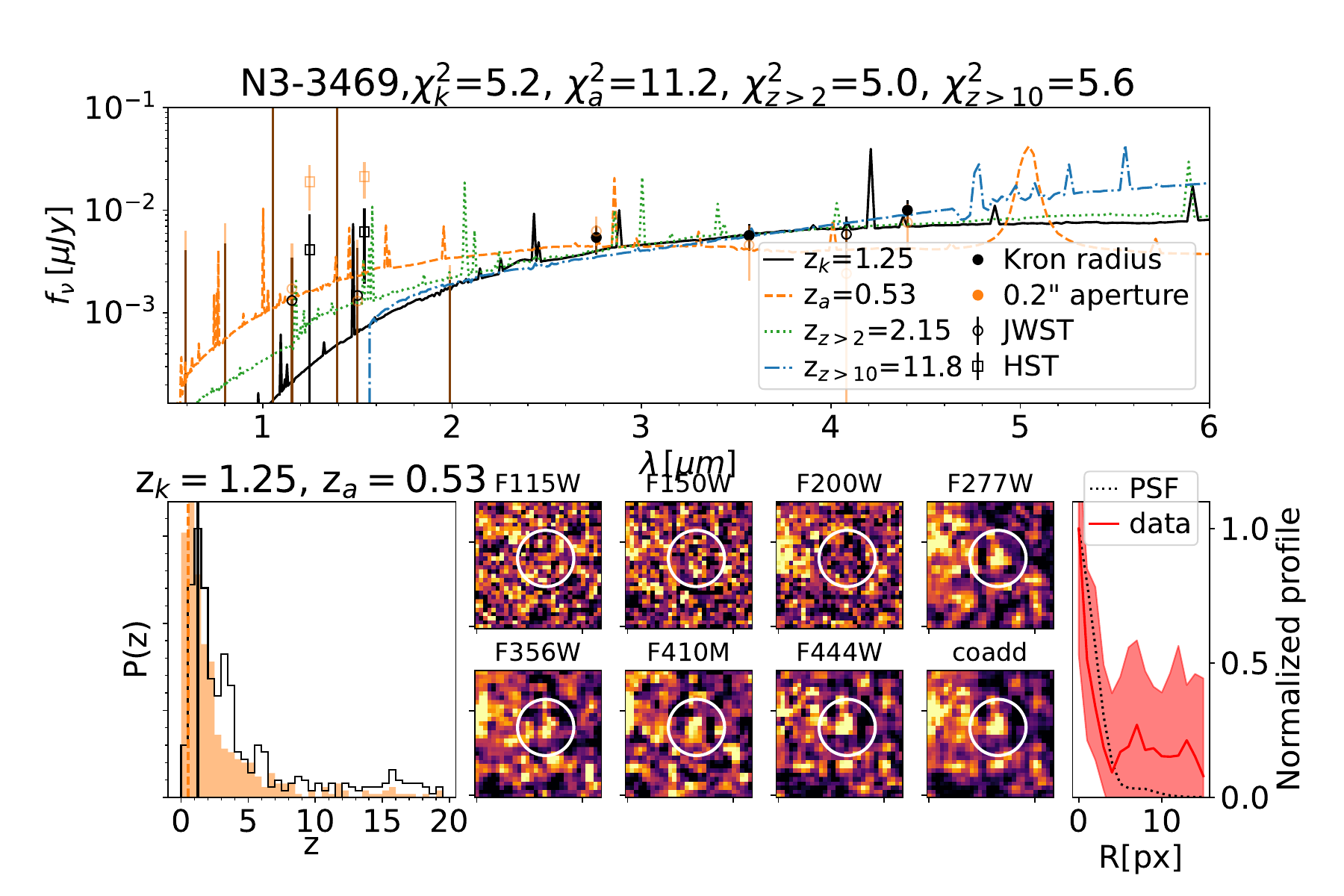} \\
    \includegraphics[trim={20 10 50 40},clip,width=0.44\linewidth,keepaspectratio]{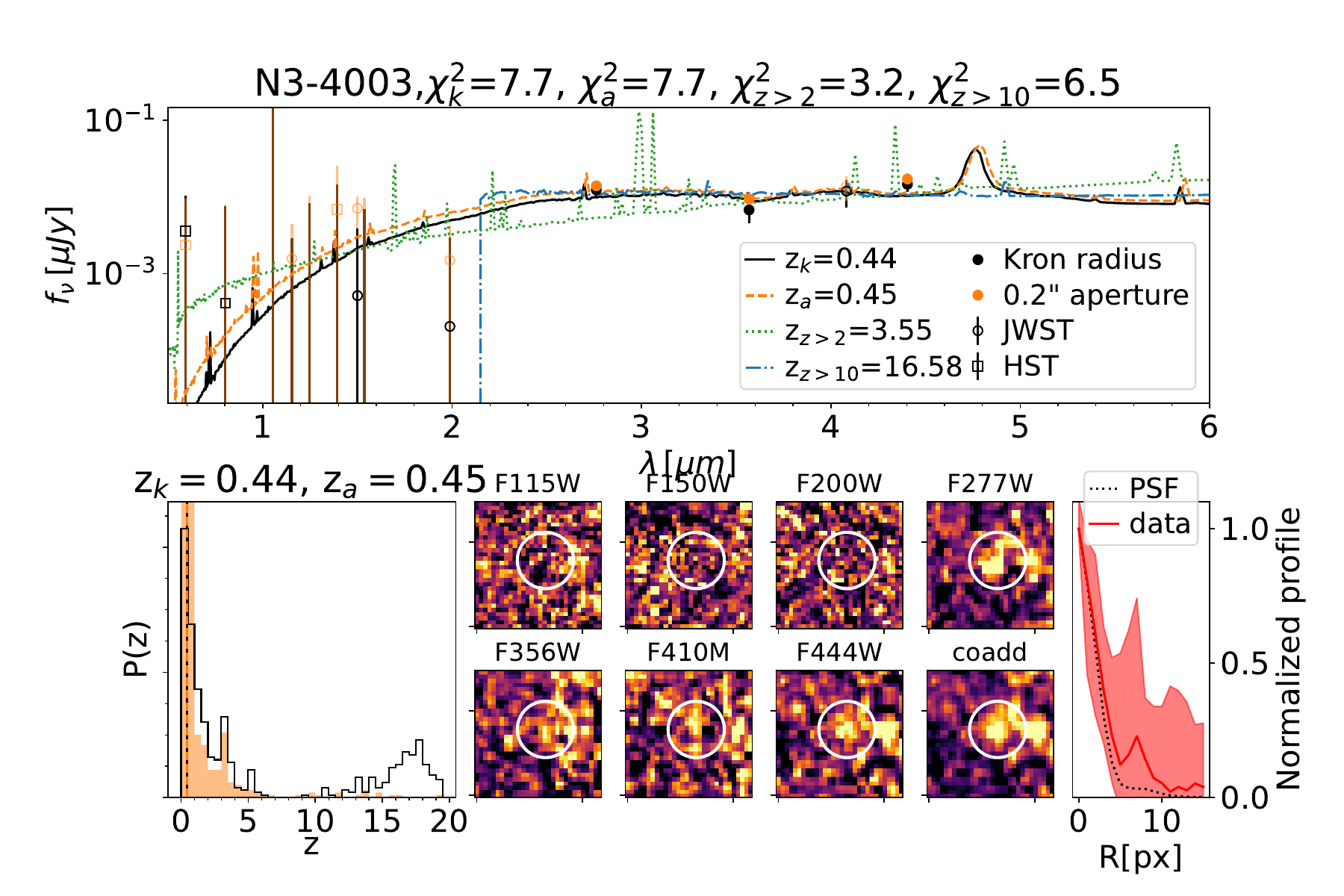} &
    \includegraphics[trim={20 10 50 40},clip,width=0.44\linewidth,keepaspectratio]{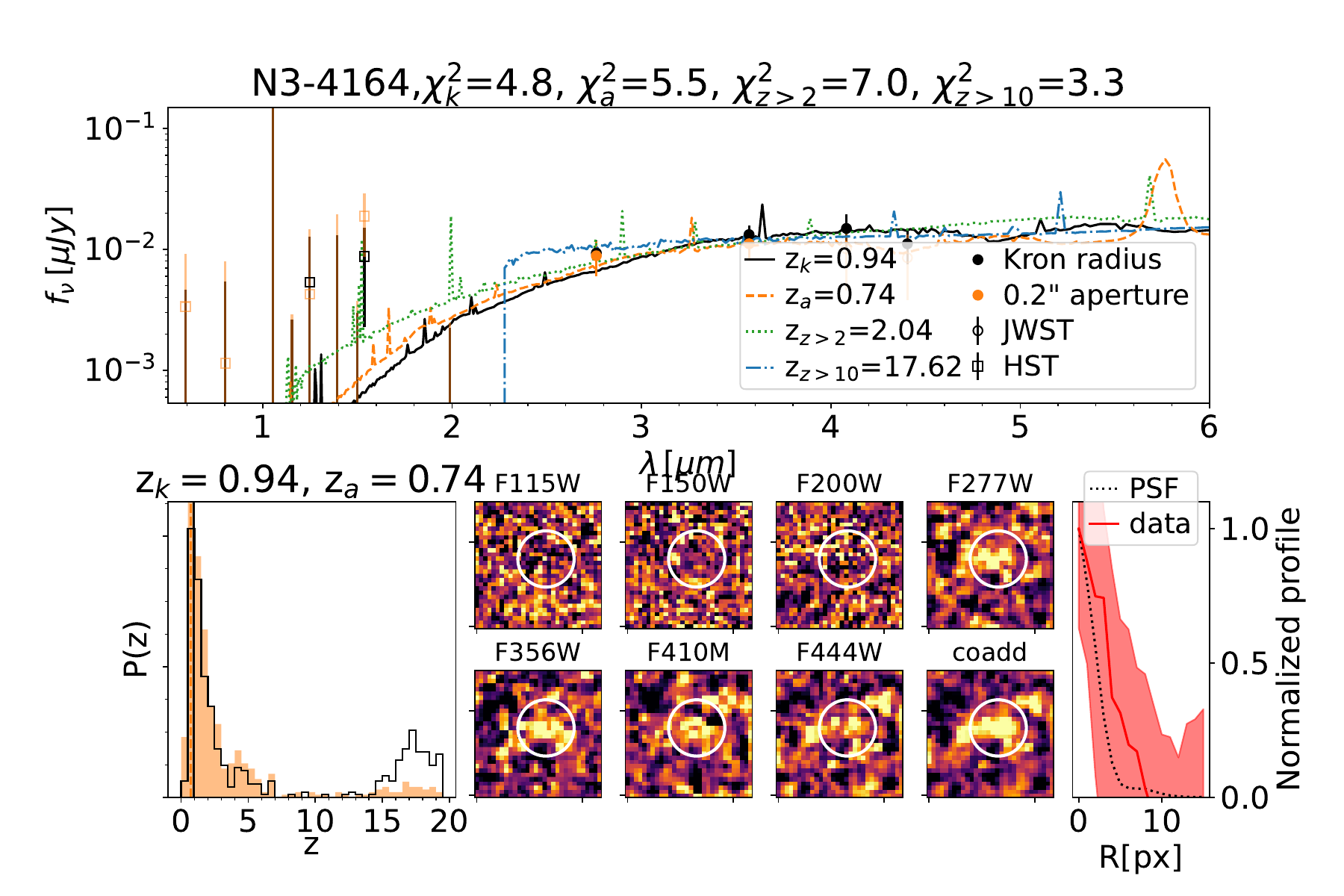} \\
    \includegraphics[trim={20 10 50 40},clip,width=0.44\linewidth,keepaspectratio]{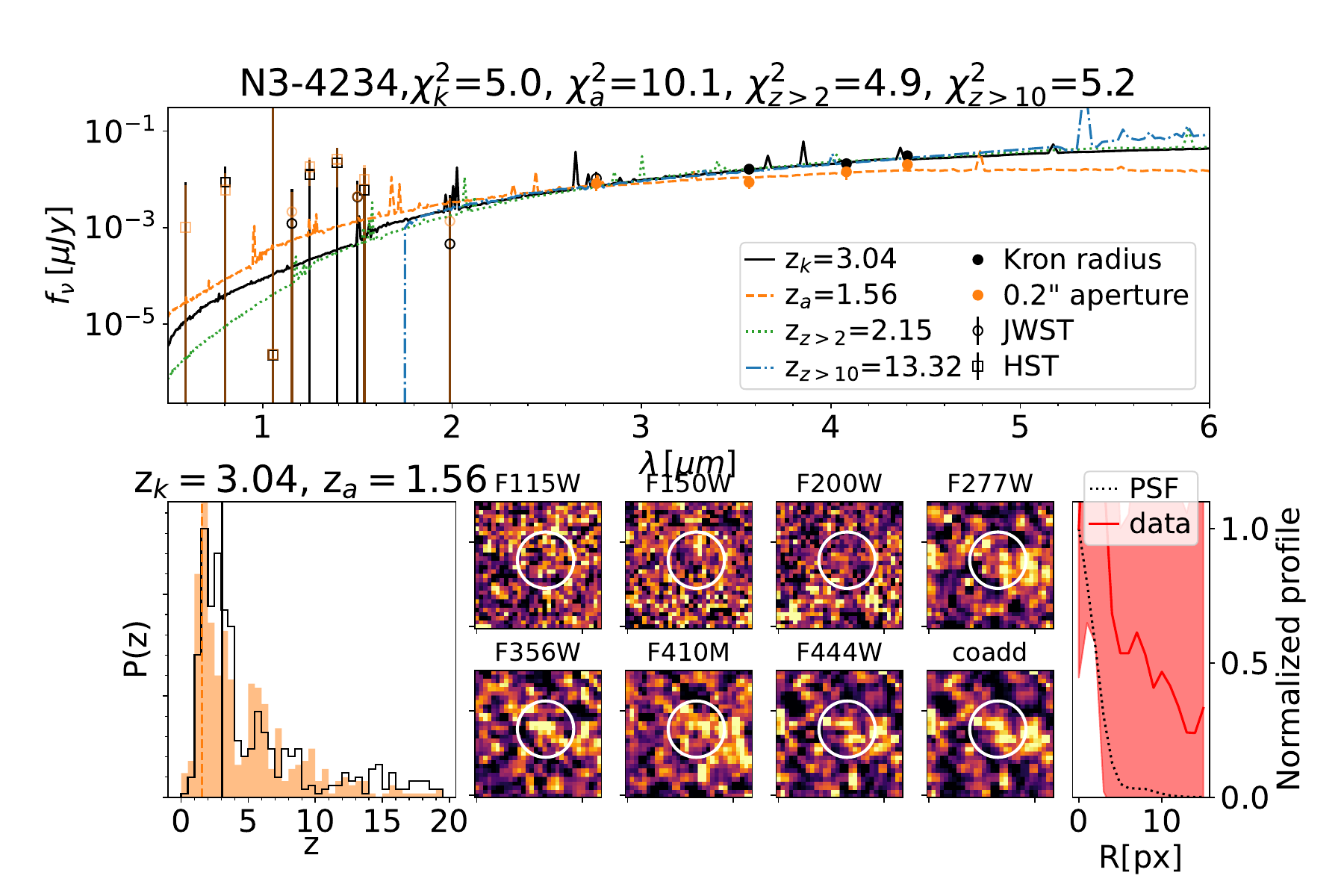} &
    \includegraphics[trim={20 10 50 40},clip,width=0.44\linewidth,keepaspectratio]{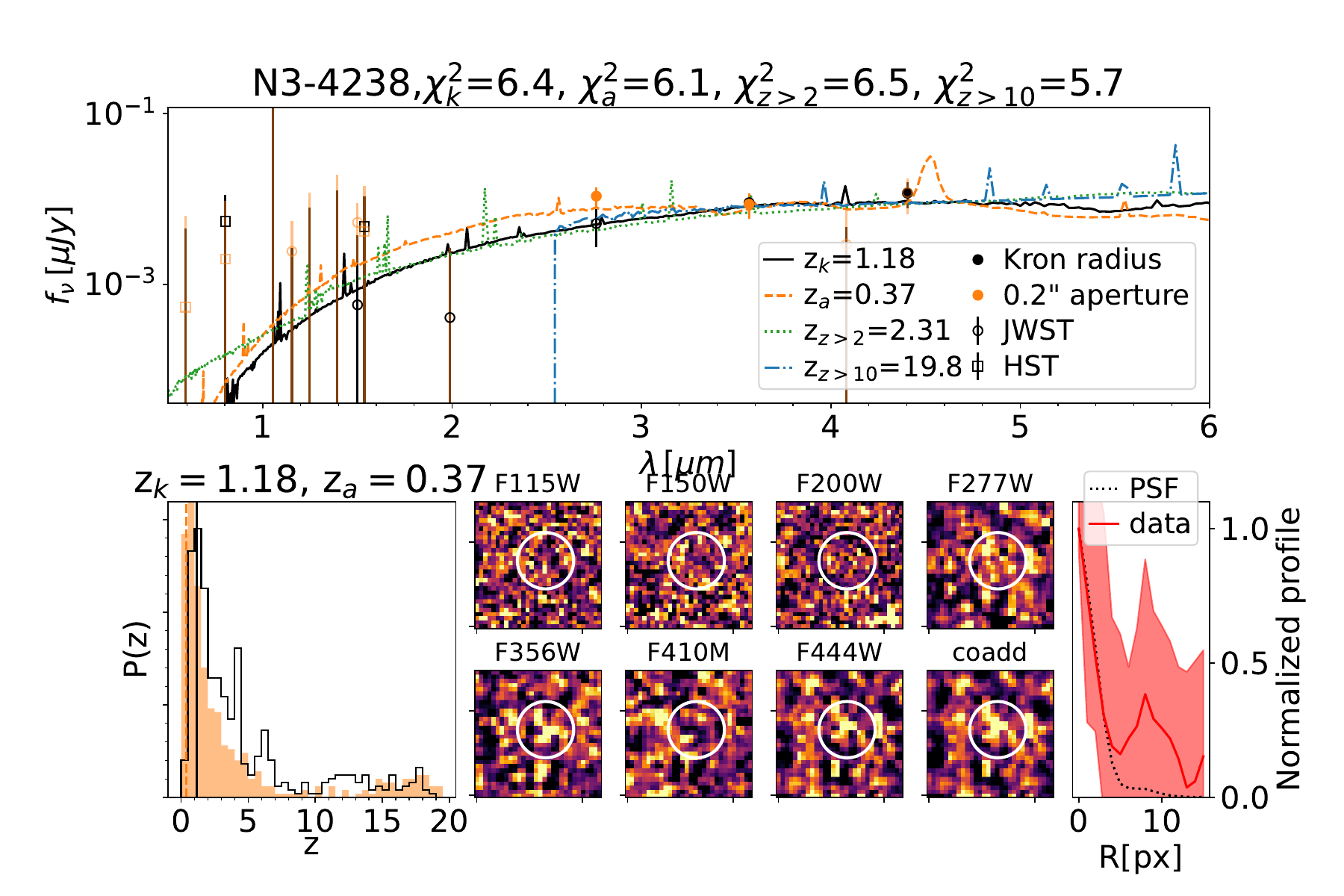}\\
    \includegraphics[trim={20 10 50 40},clip,width=0.44\linewidth,keepaspectratio]{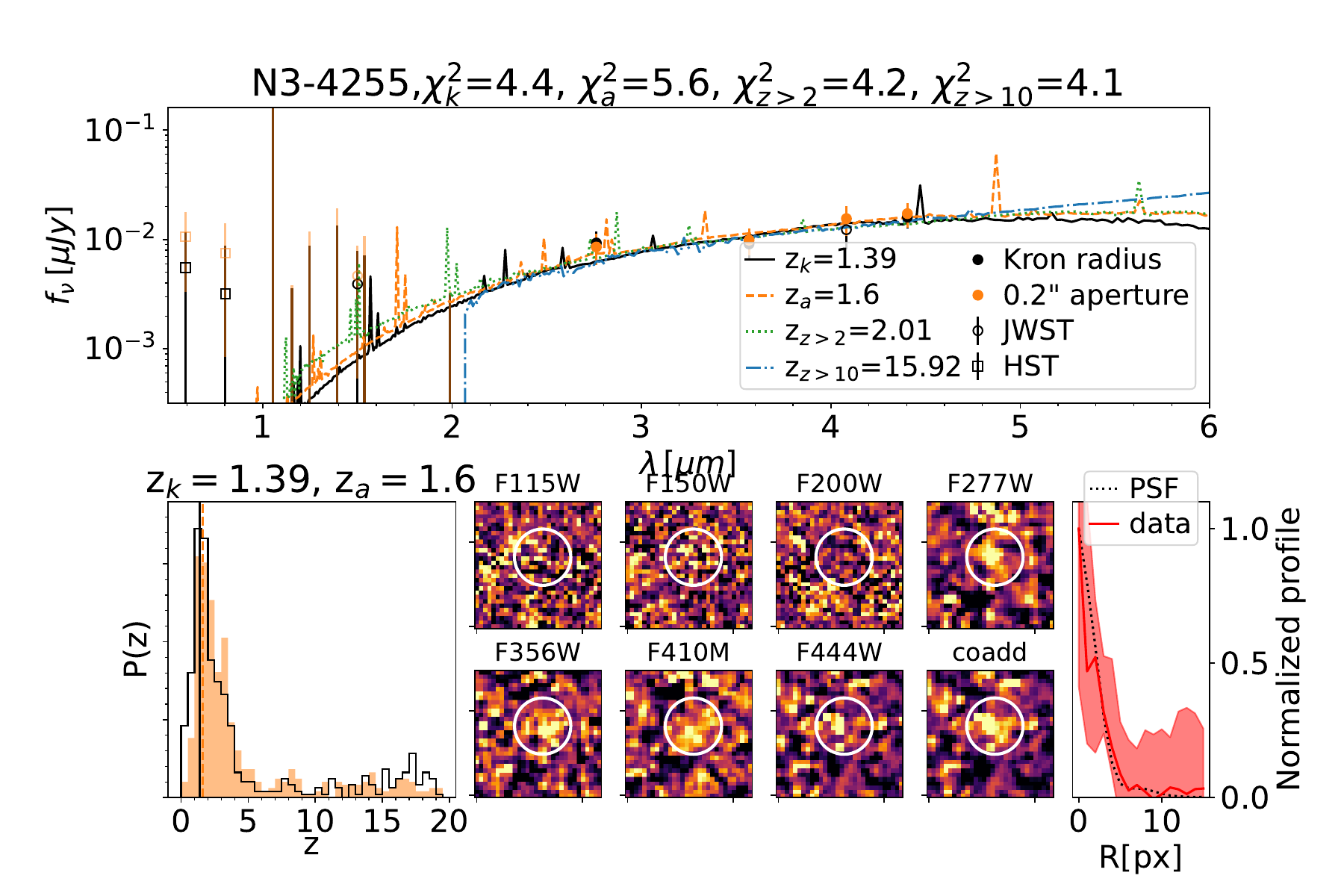}&
    \includegraphics[trim={20 10 50 40},clip,width=0.44\linewidth,keepaspectratio]{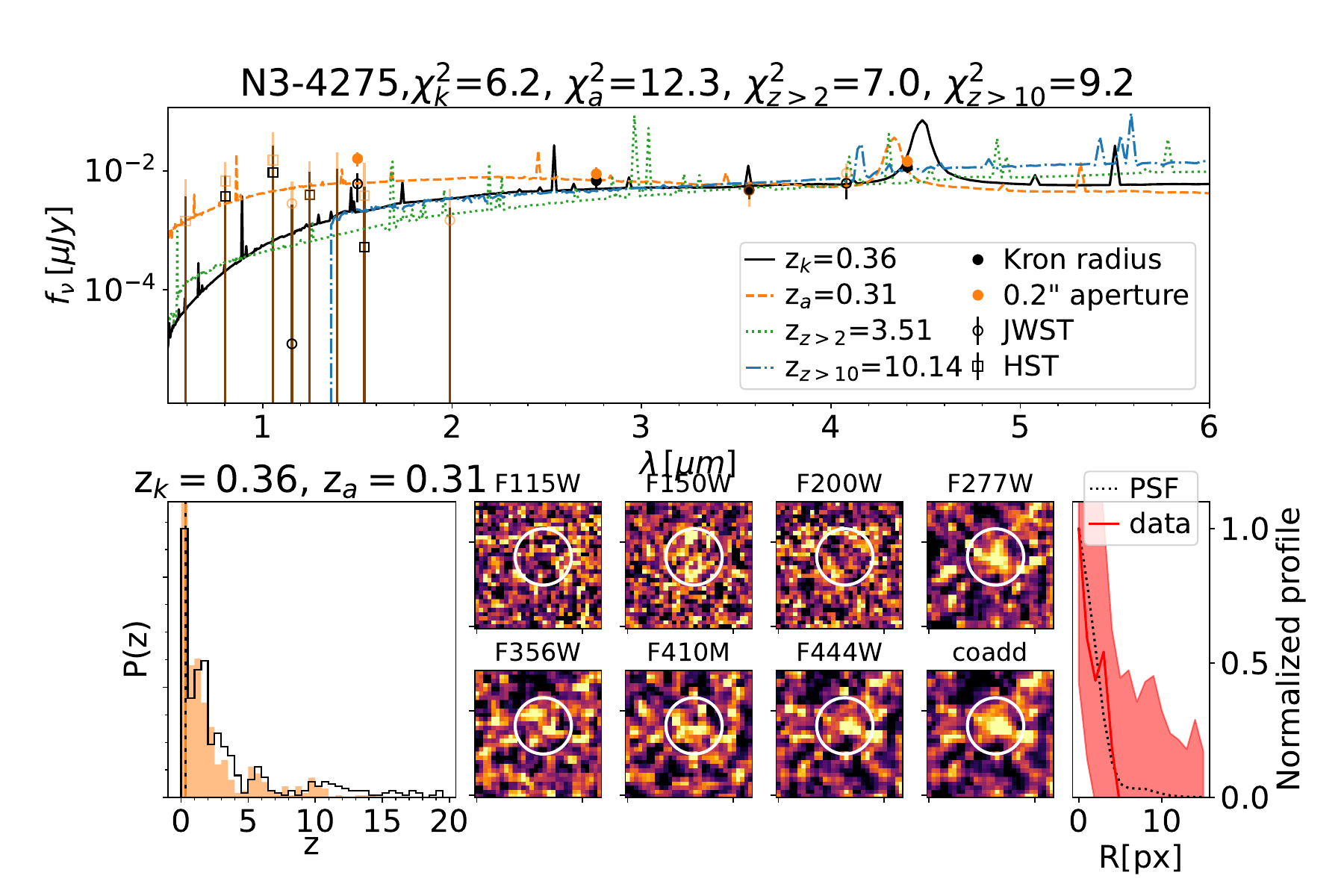}\\
  \caption{continued.}\\
    \includegraphics[trim={20 10 50 40},clip,width=0.44\linewidth,keepaspectratio]{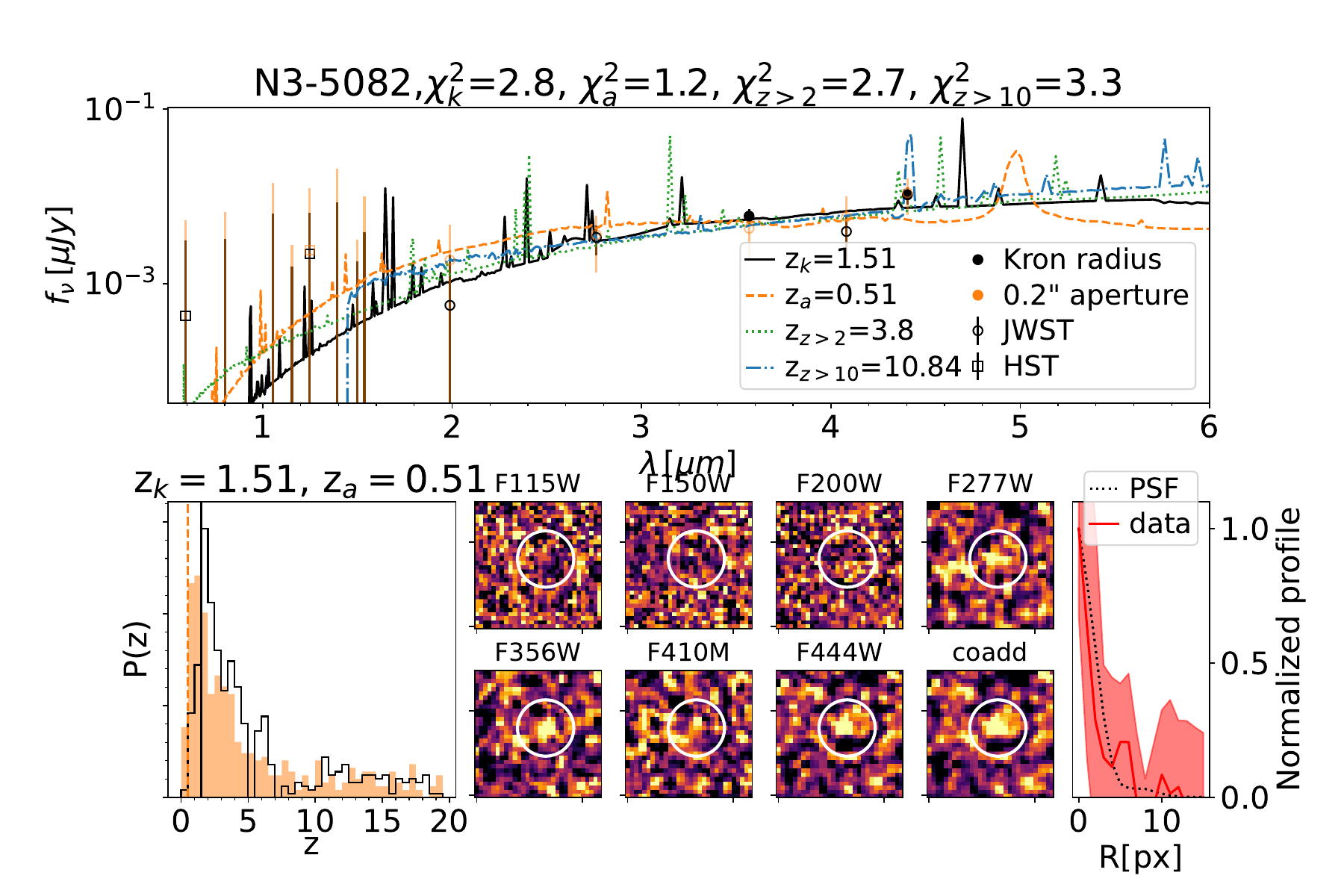}&
    \includegraphics[trim={20 10 50 40},clip,width=0.44\linewidth,keepaspectratio]{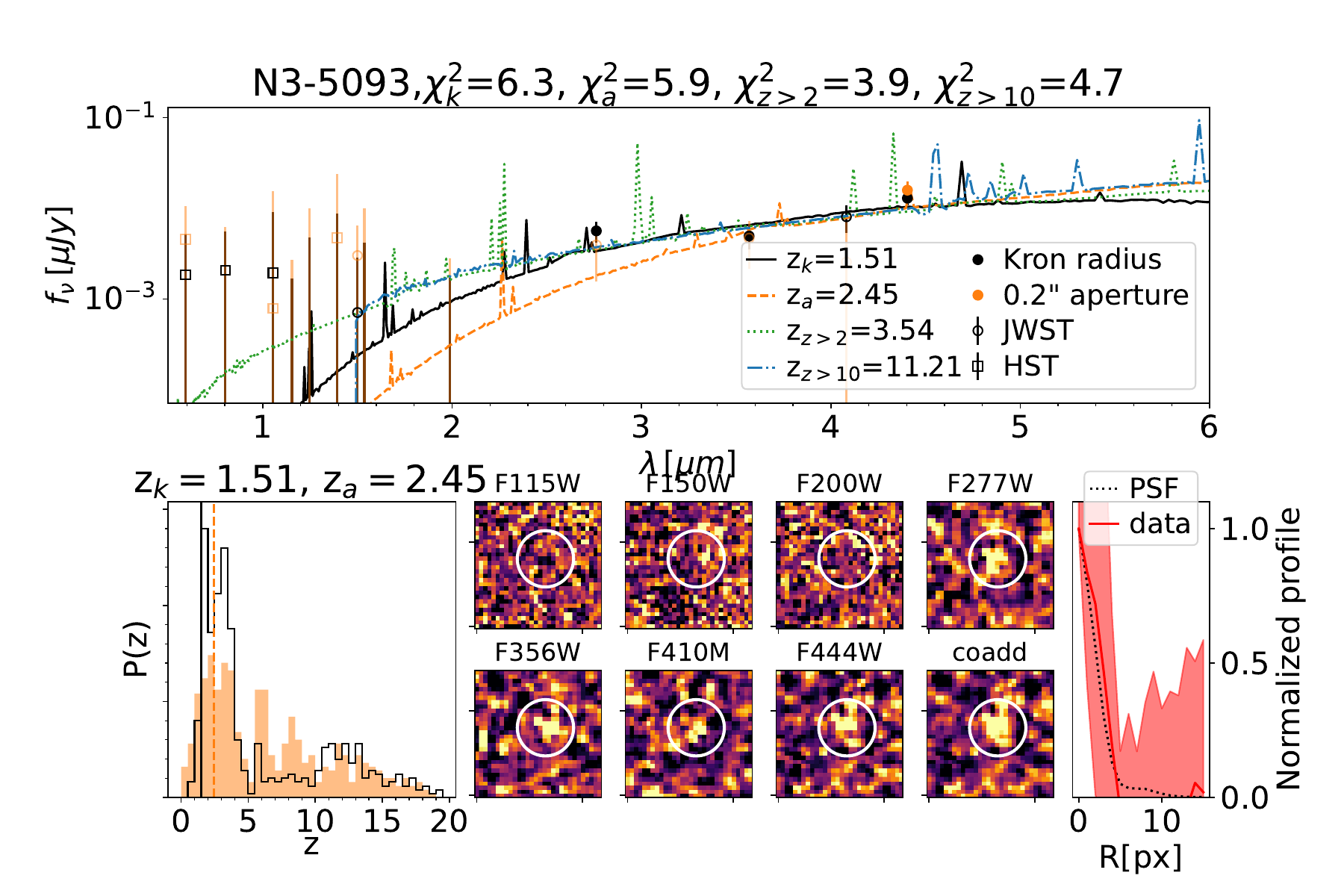}\\
    \includegraphics[trim={20 10 50 40},clip,width=0.44\linewidth,keepaspectratio]{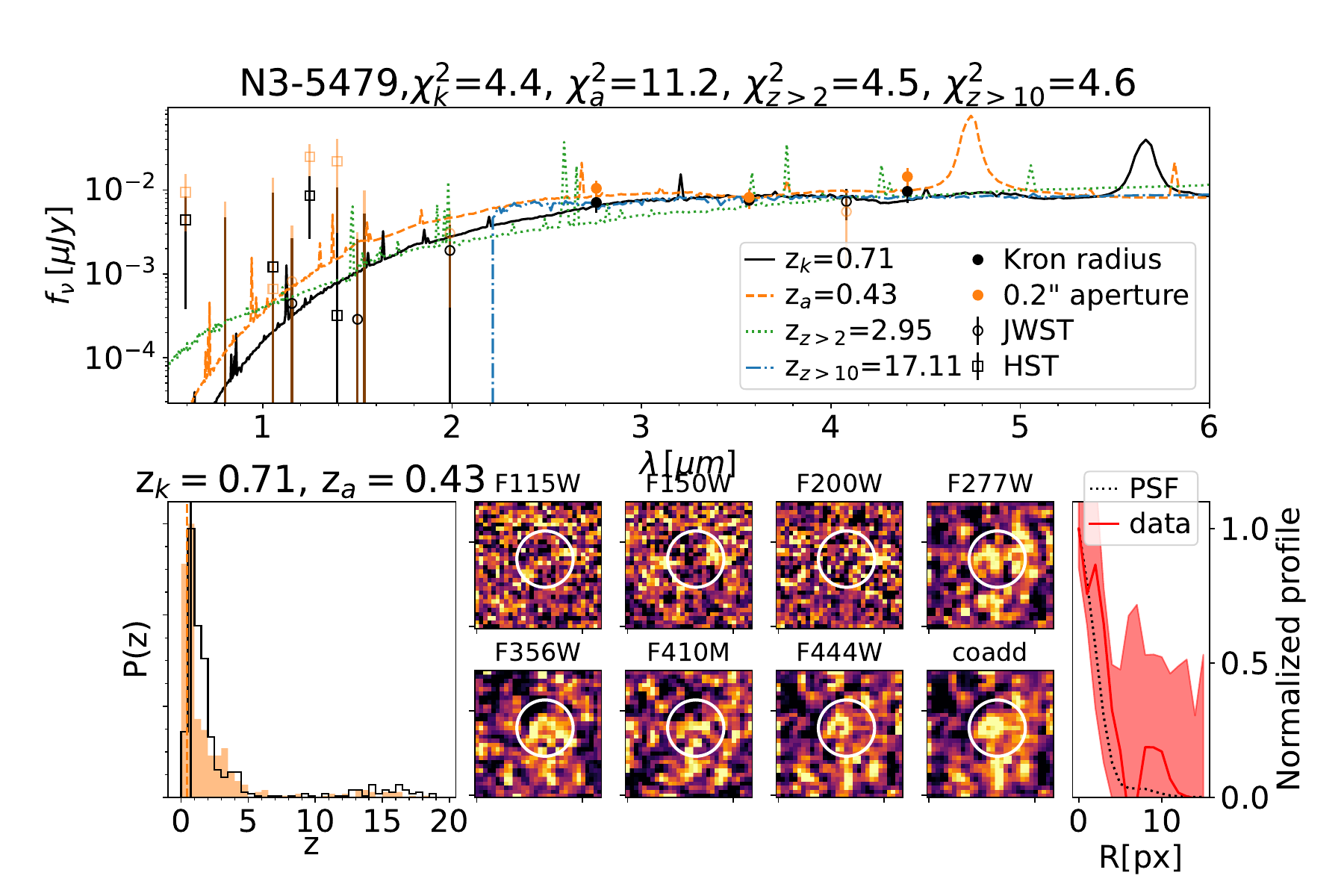}&
    \includegraphics[trim={20 10 50 40},clip,width=0.44\linewidth,keepaspectratio]{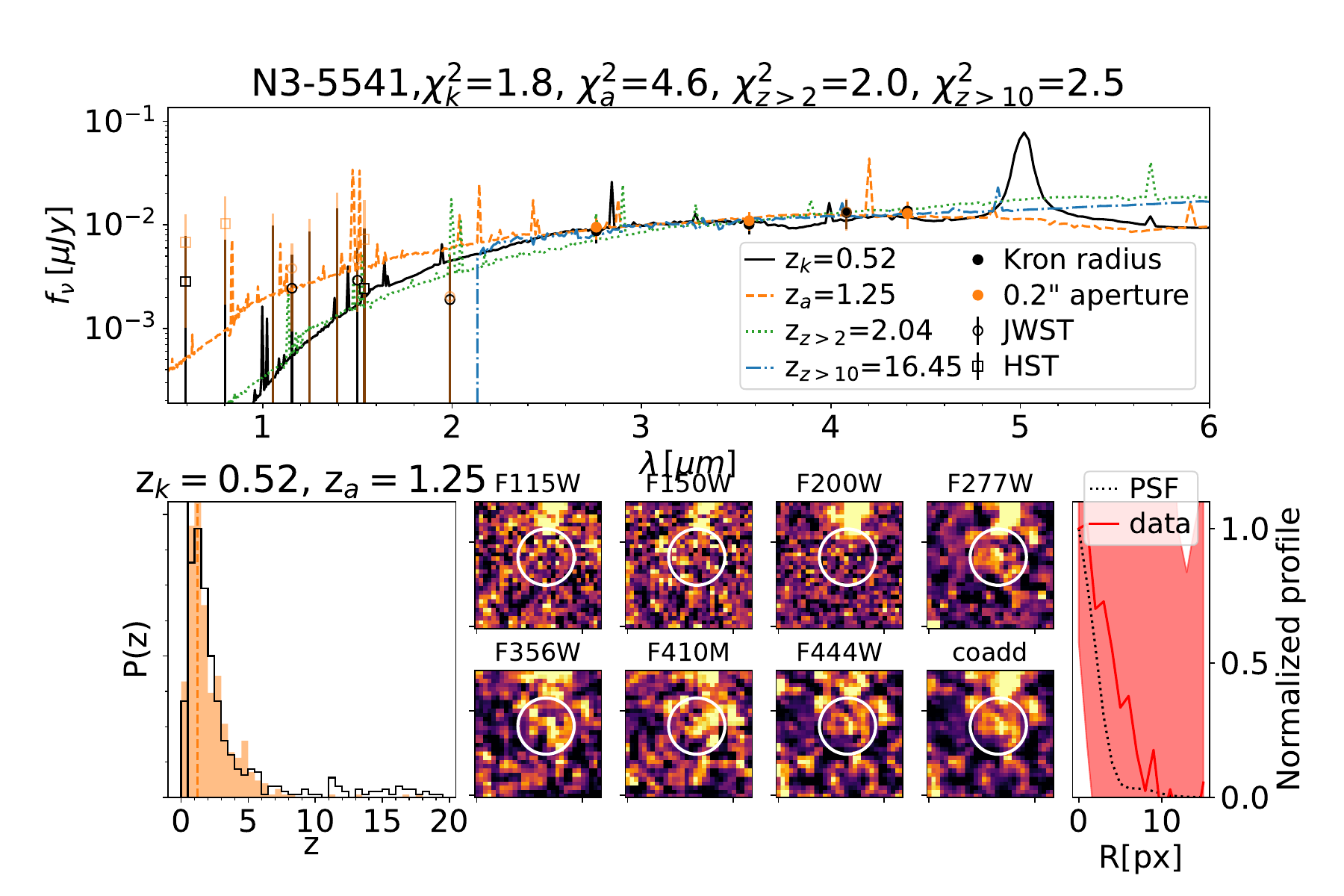}\\
    \includegraphics[trim={20 10 50 40},clip,width=0.44\linewidth,keepaspectratio]{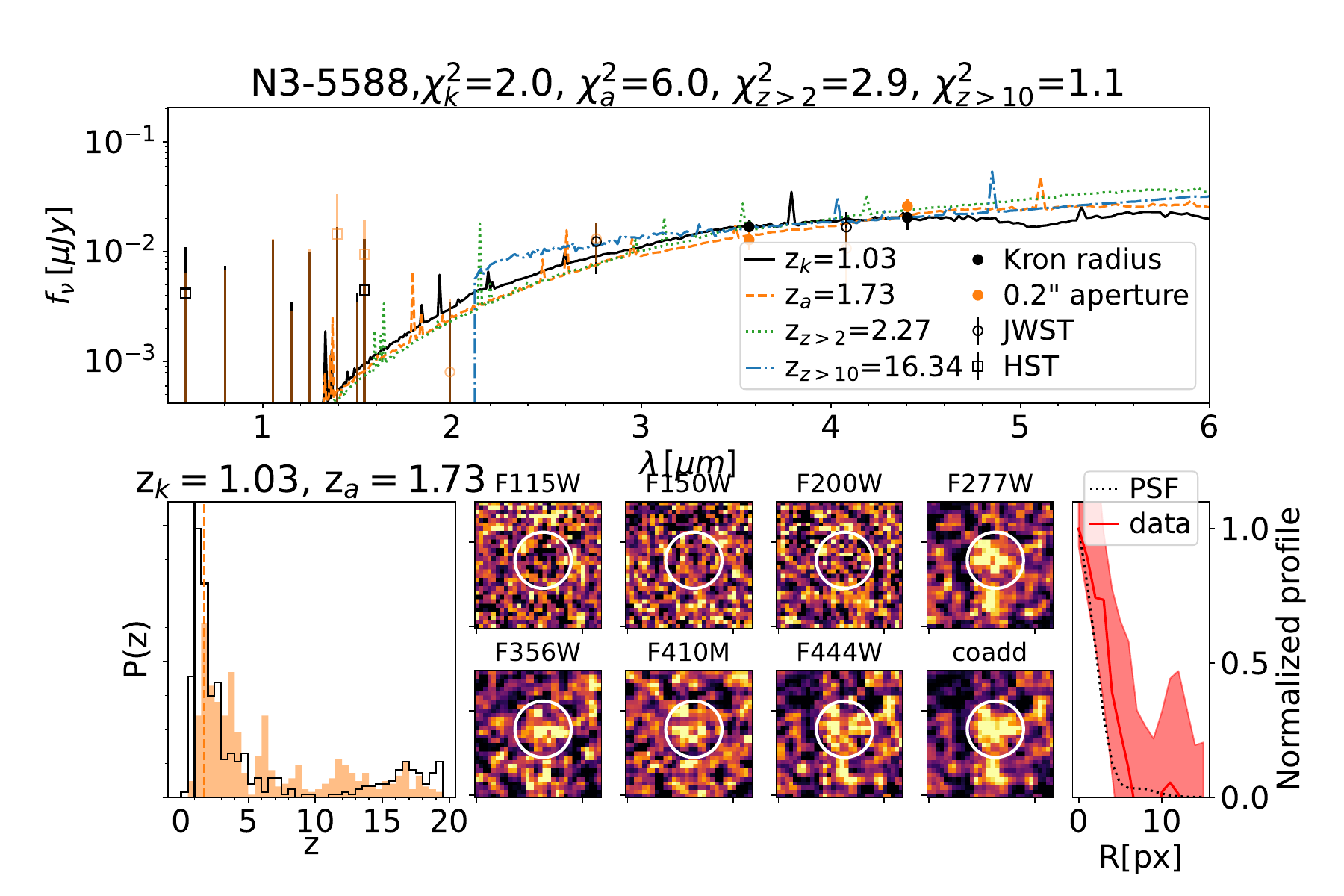}&
    \includegraphics[trim={20 10 50 40},clip,width=0.44\linewidth,keepaspectratio]{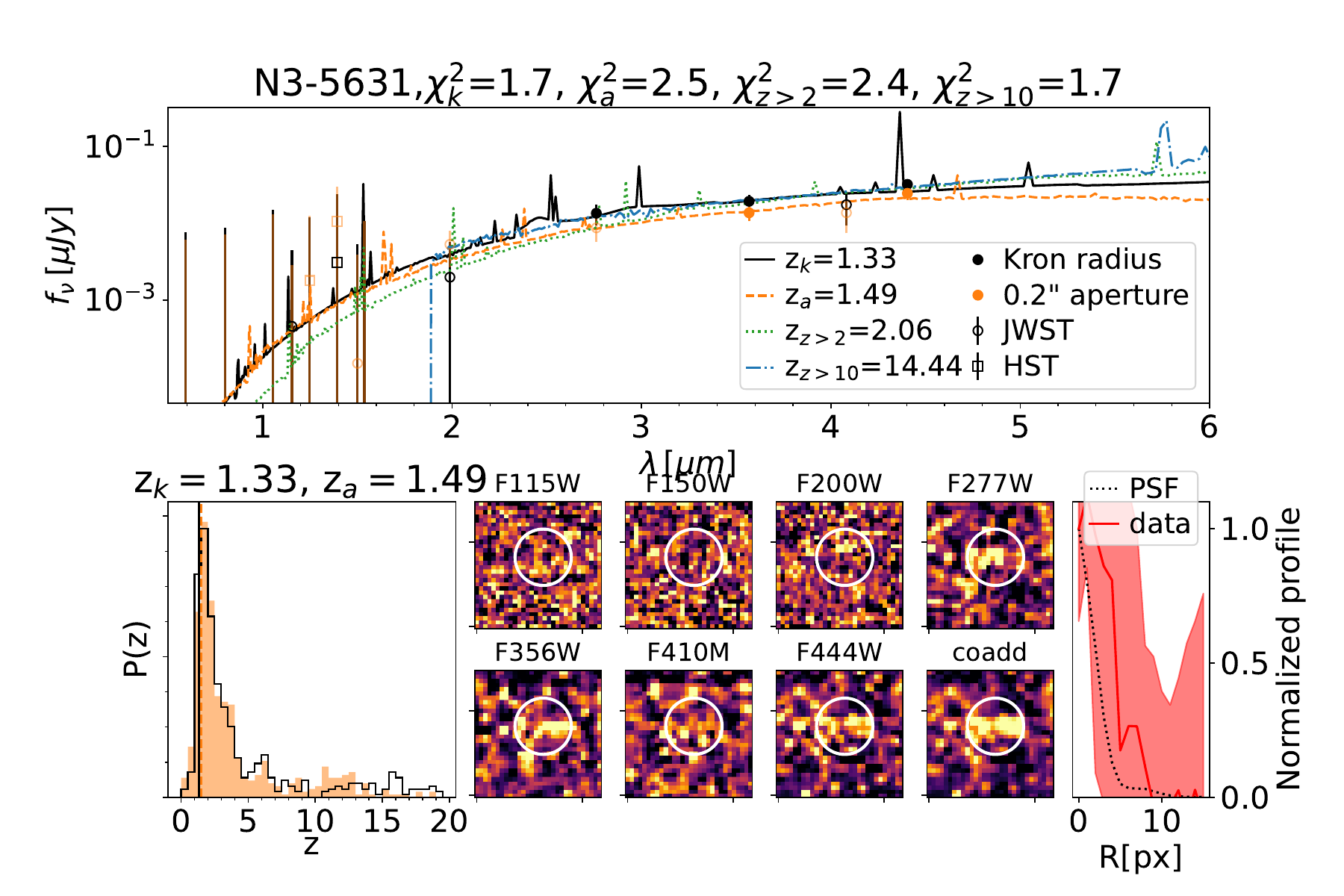}\\
    \includegraphics[trim={20 10 50 40},clip,width=0.44\linewidth,keepaspectratio]{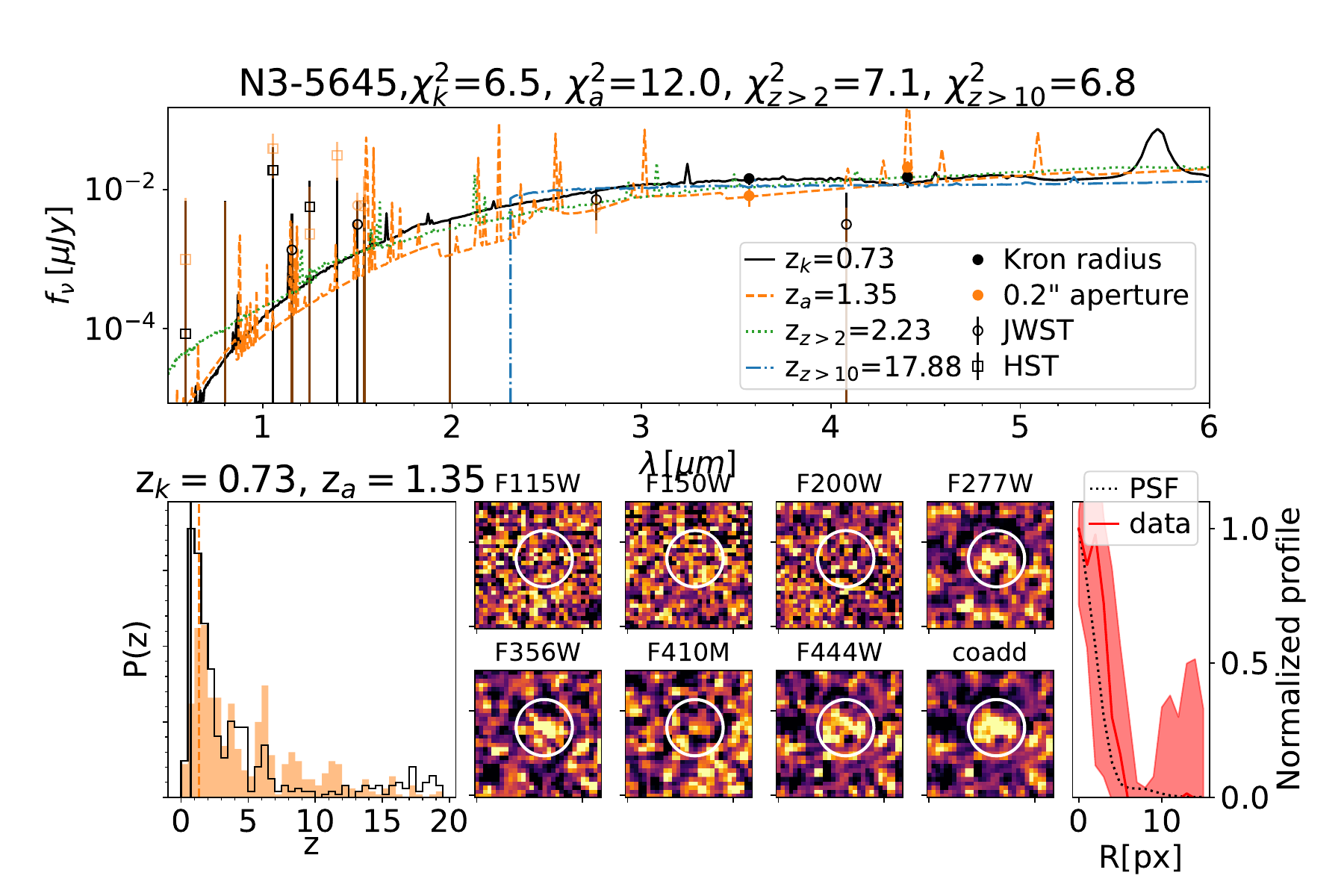}& 
    \includegraphics[trim={20 10 50 40},clip,width=0.44\linewidth,keepaspectratio]{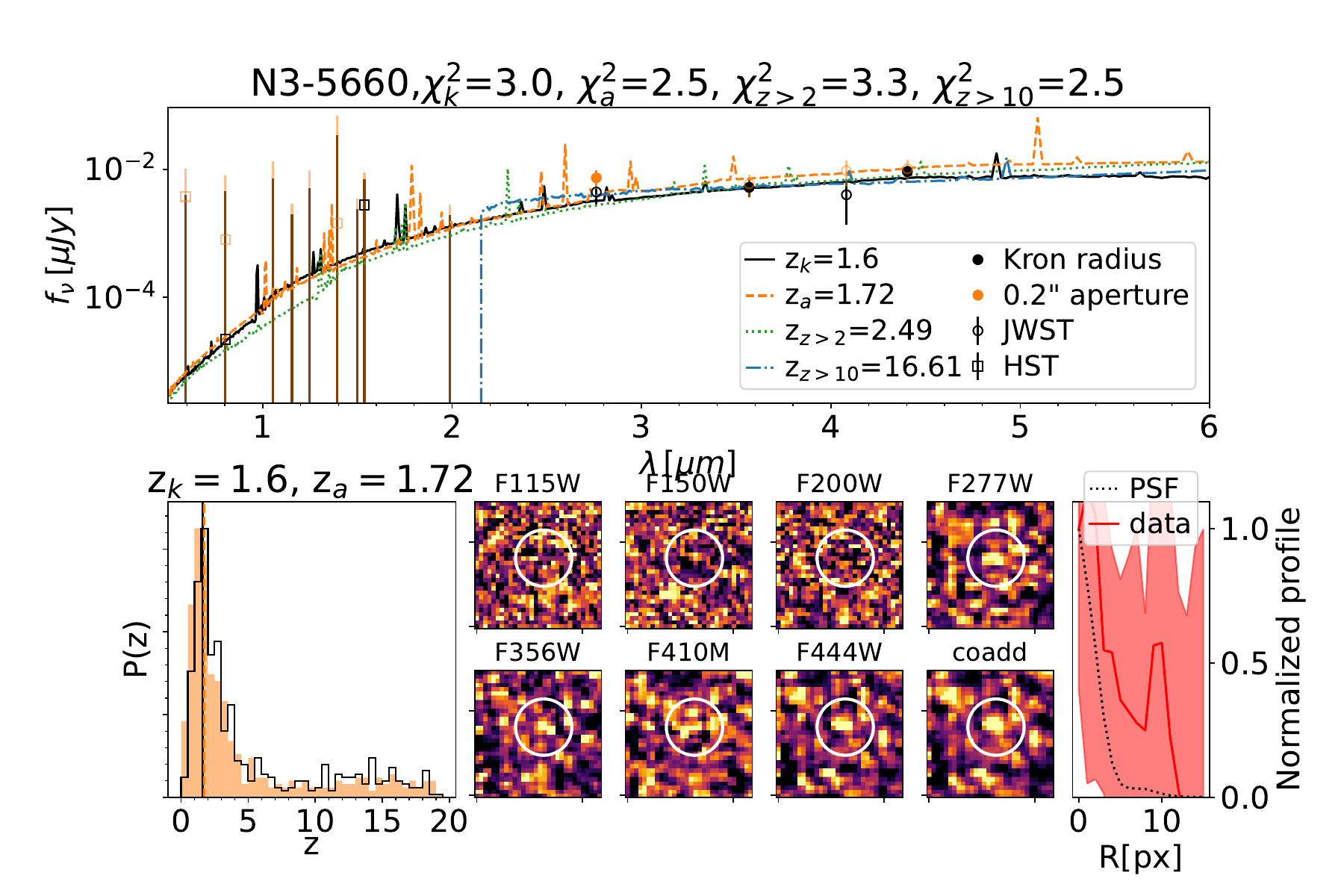}\\
  \caption{continued.}\\
    \includegraphics[trim={20 10 50 40},clip,width=0.44\linewidth,keepaspectratio]{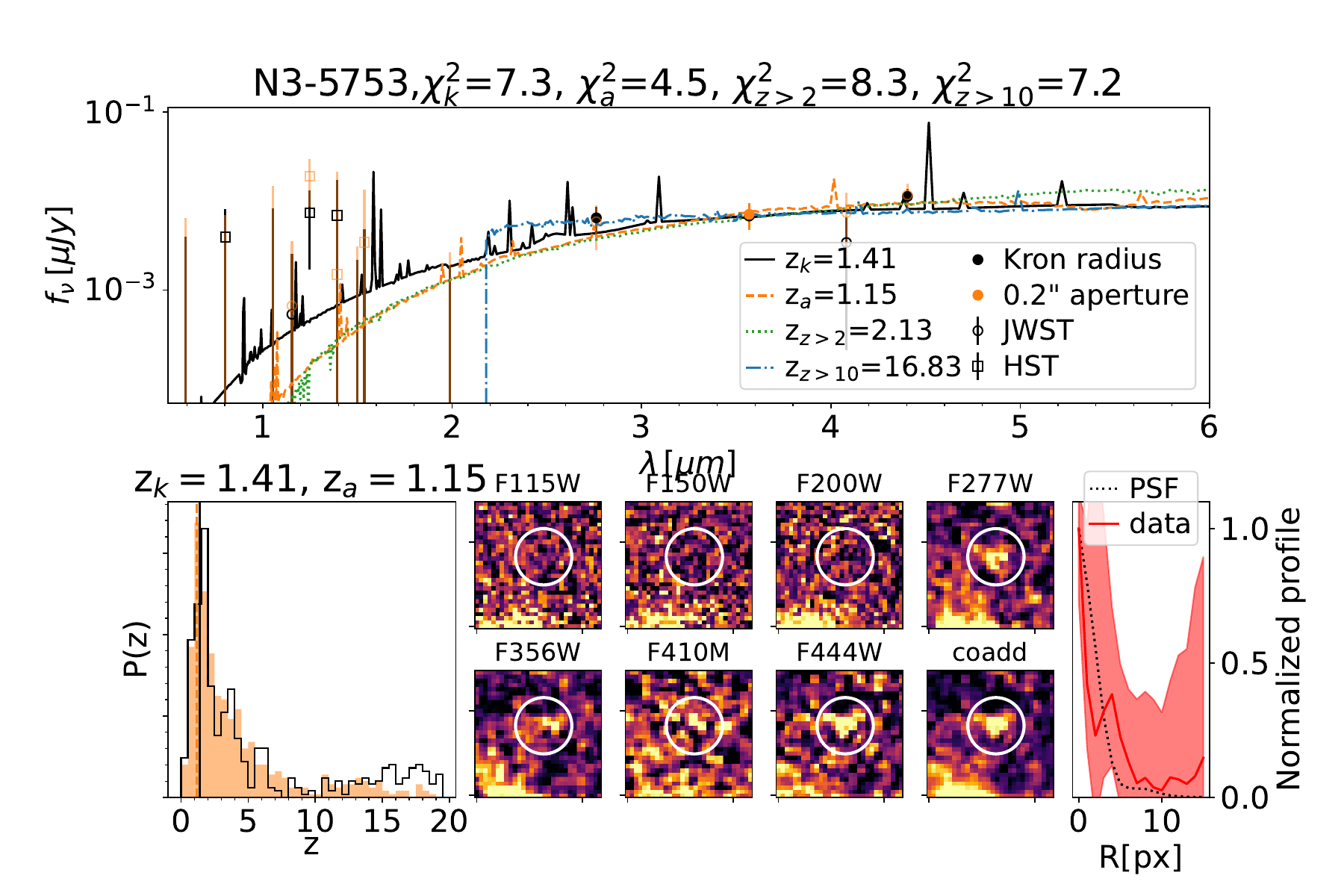}&
    \includegraphics[trim={20 10 50 40},clip,width=0.44\linewidth,keepaspectratio]{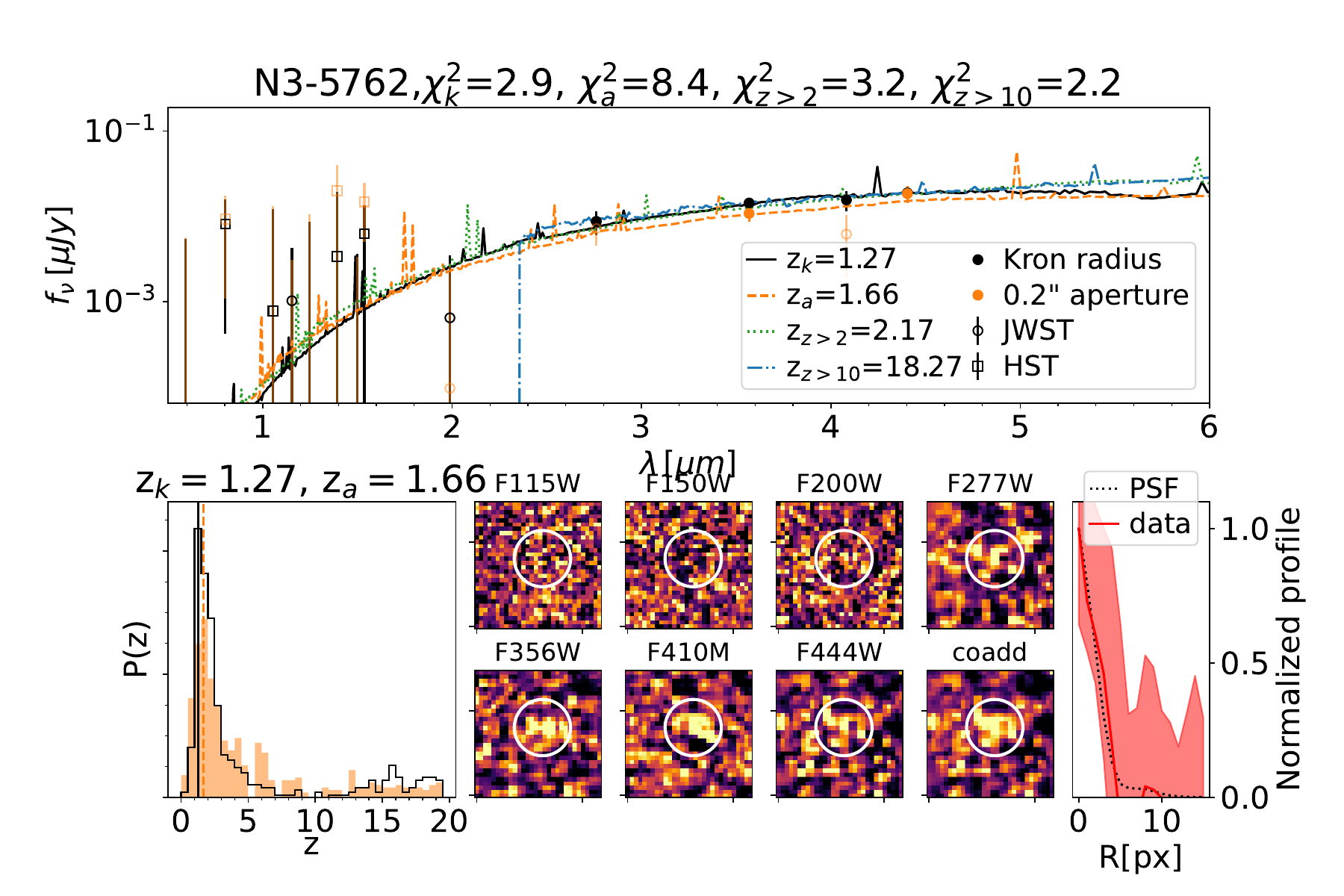}\\
    \includegraphics[trim={20 10 50 40},clip,width=0.44\linewidth,keepaspectratio]{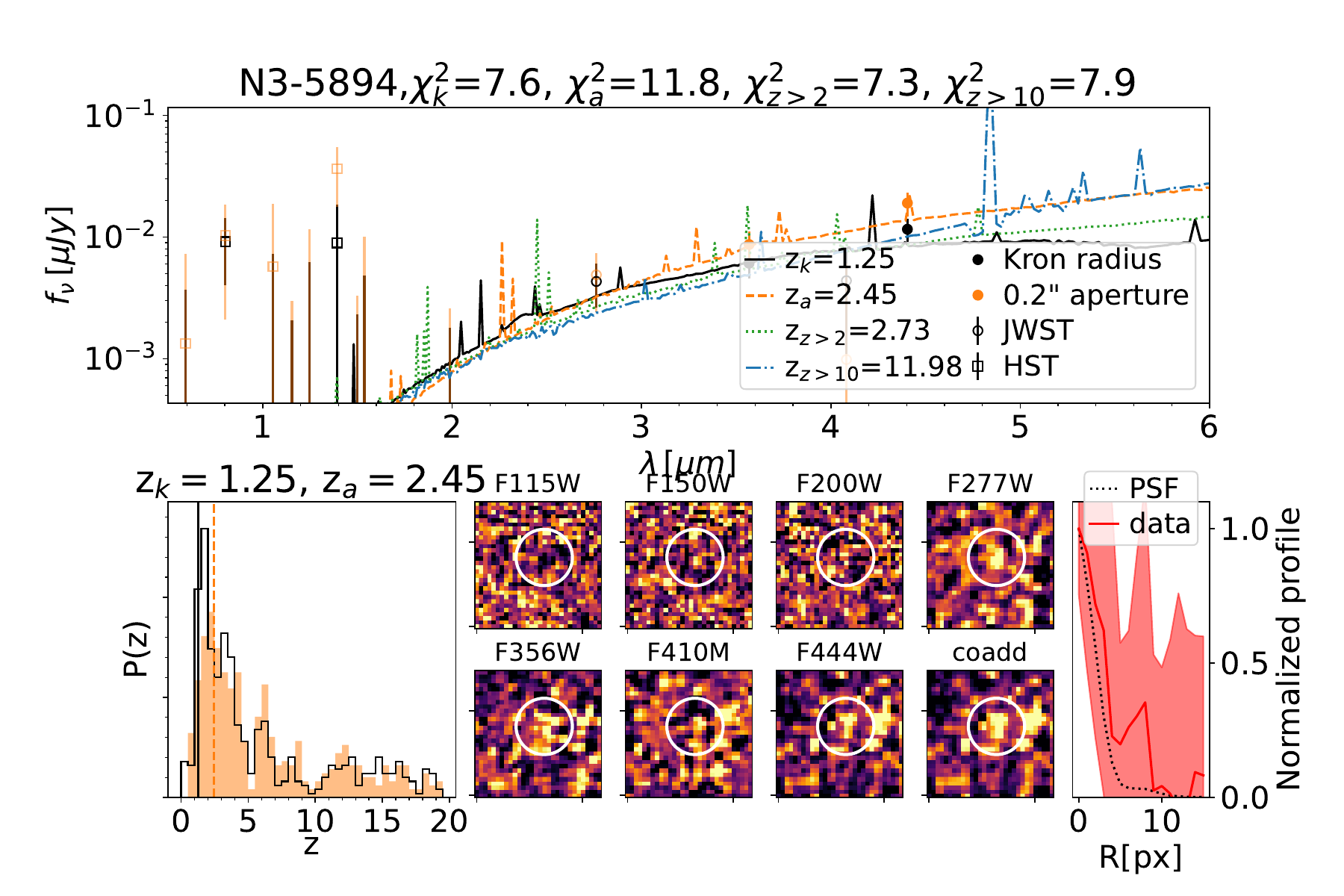}&
    \includegraphics[trim={20 10 50 40},clip,width=0.44\linewidth,keepaspectratio]{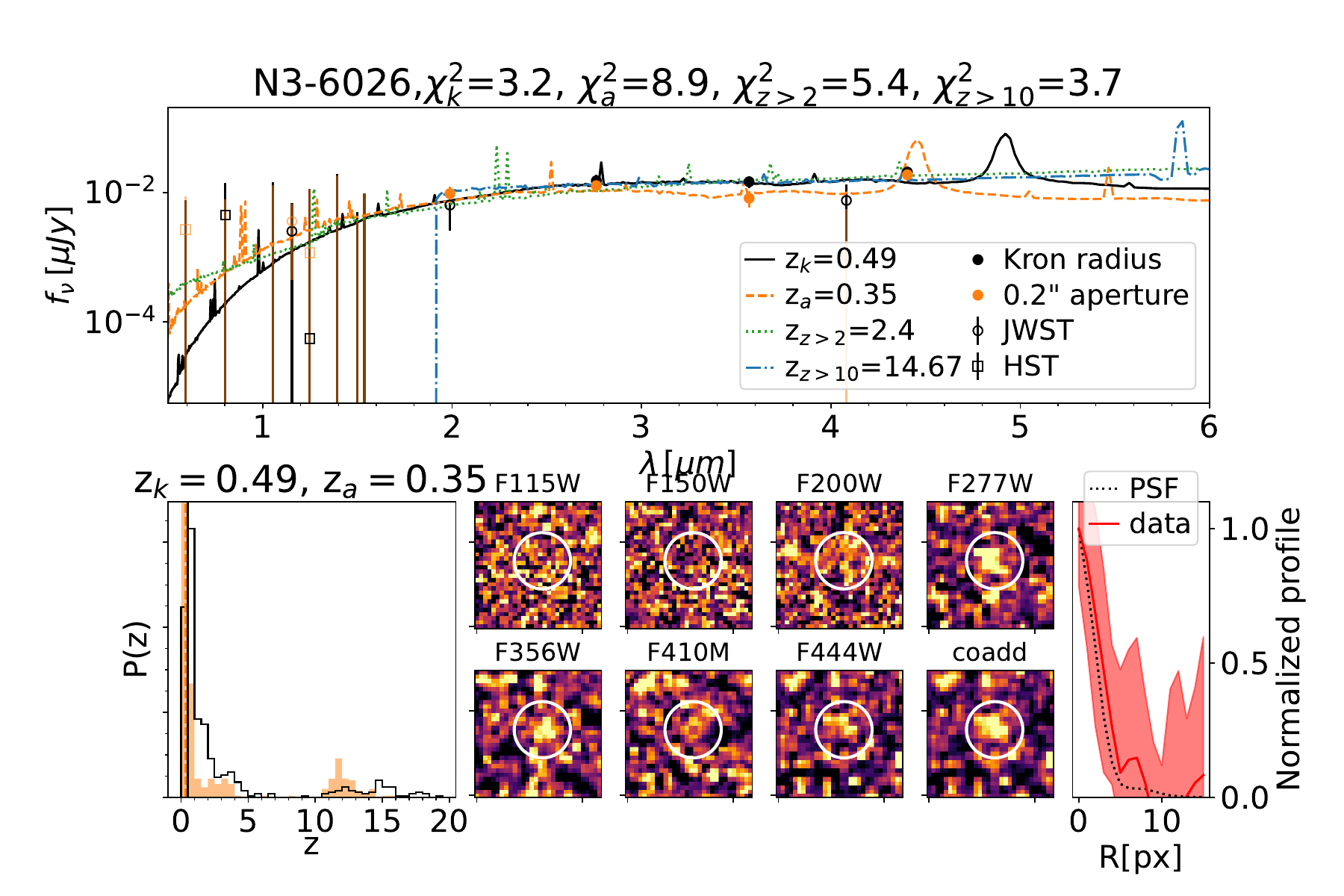}\\
    \includegraphics[trim={20 10 50 40},clip,width=0.44\linewidth,keepaspectratio]{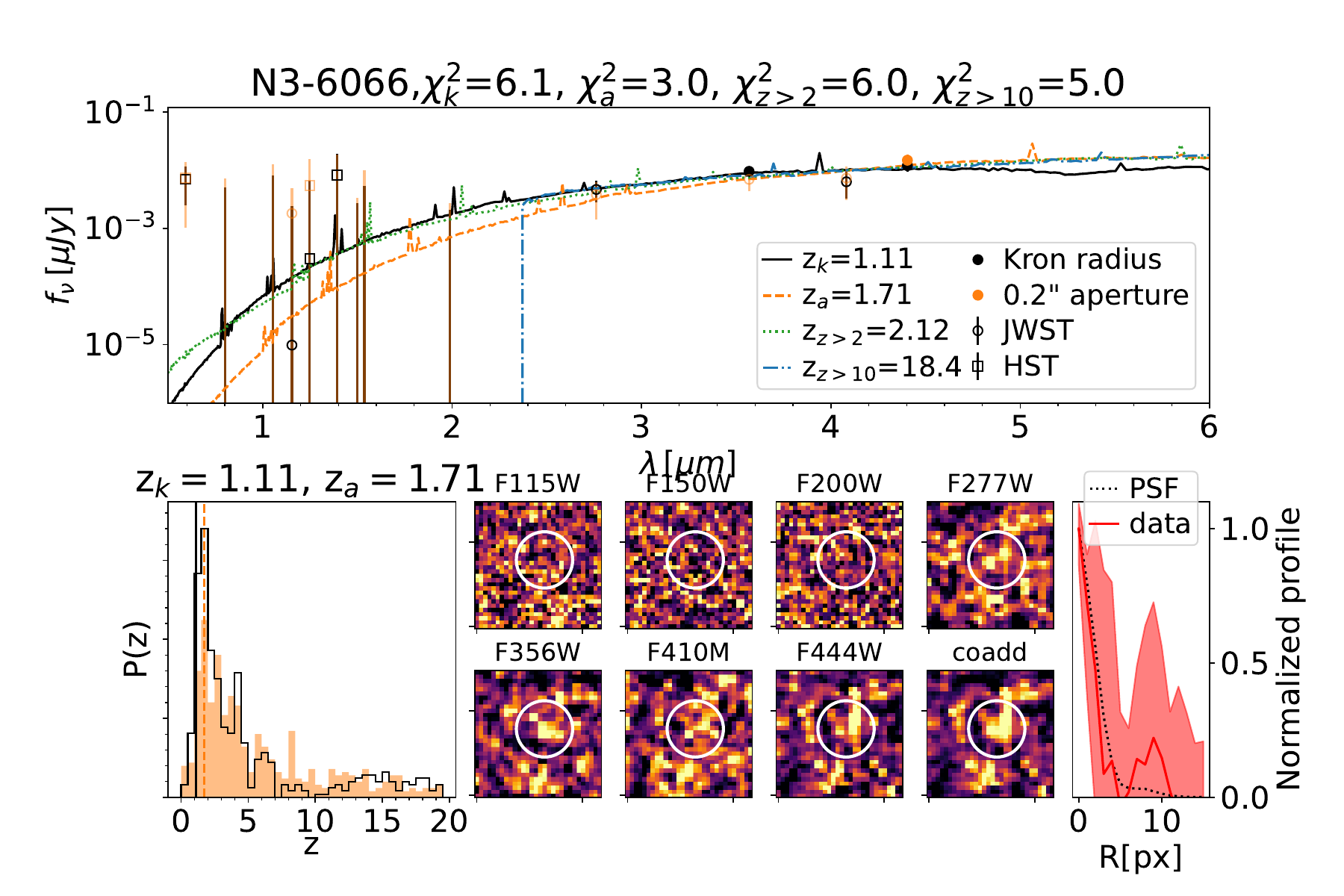}&
    \includegraphics[trim={20 10 50 40},clip,width=0.44\linewidth,keepaspectratio]{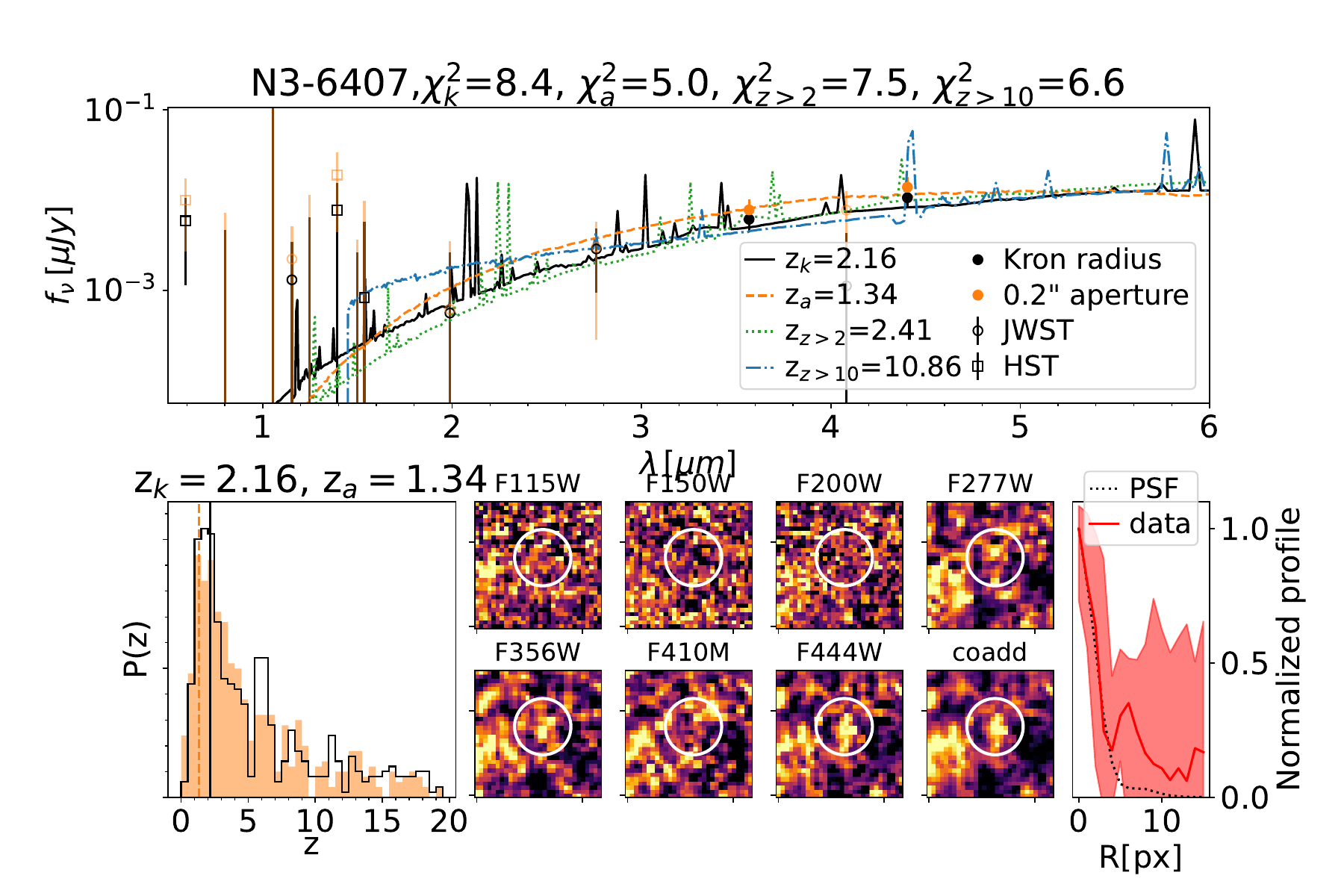}\\
    \includegraphics[trim={20 10 50 40},clip,width=0.44\linewidth,keepaspectratio]{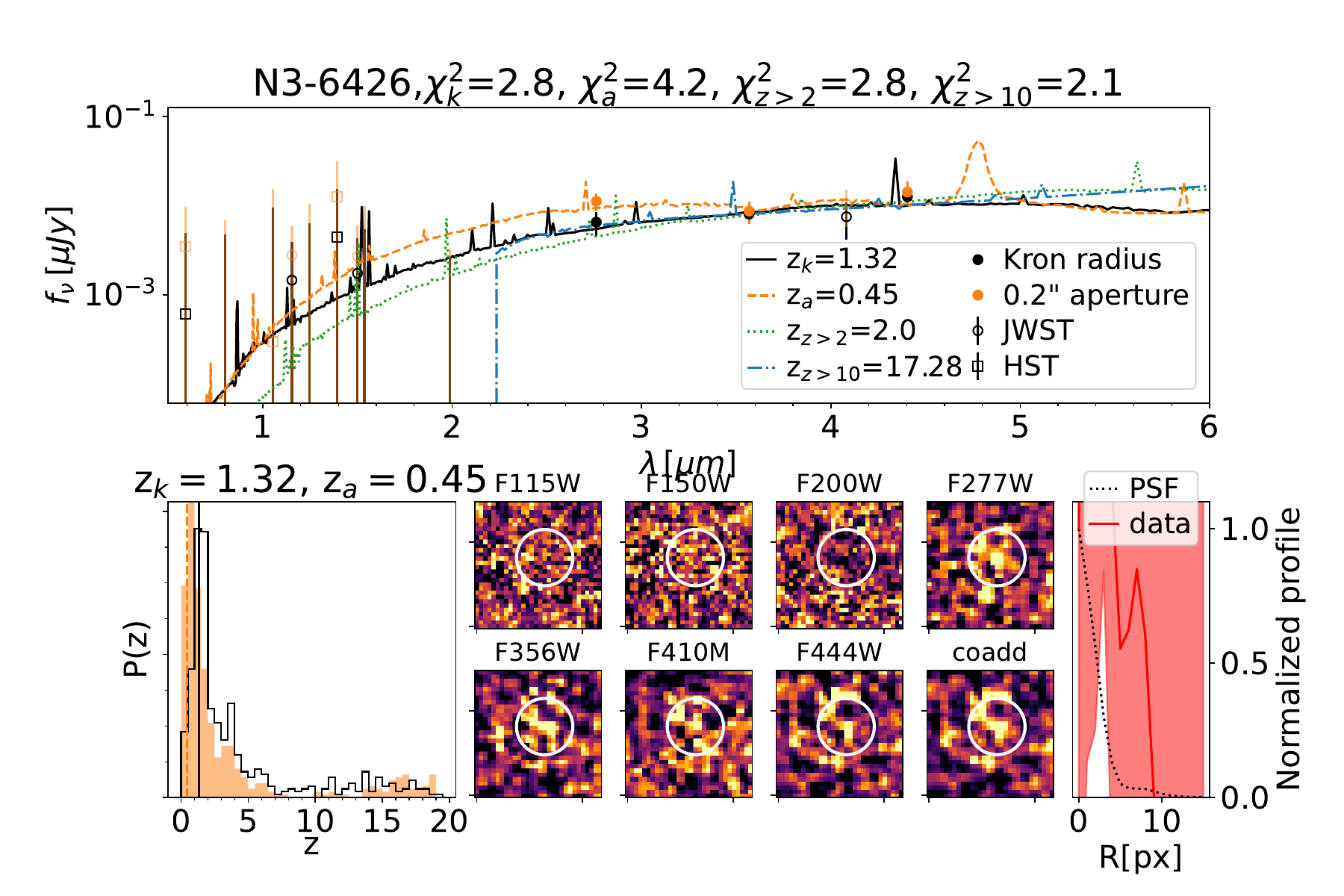}&
    \includegraphics[trim={20 10 50 40},clip,width=0.44\linewidth,keepaspectratio]{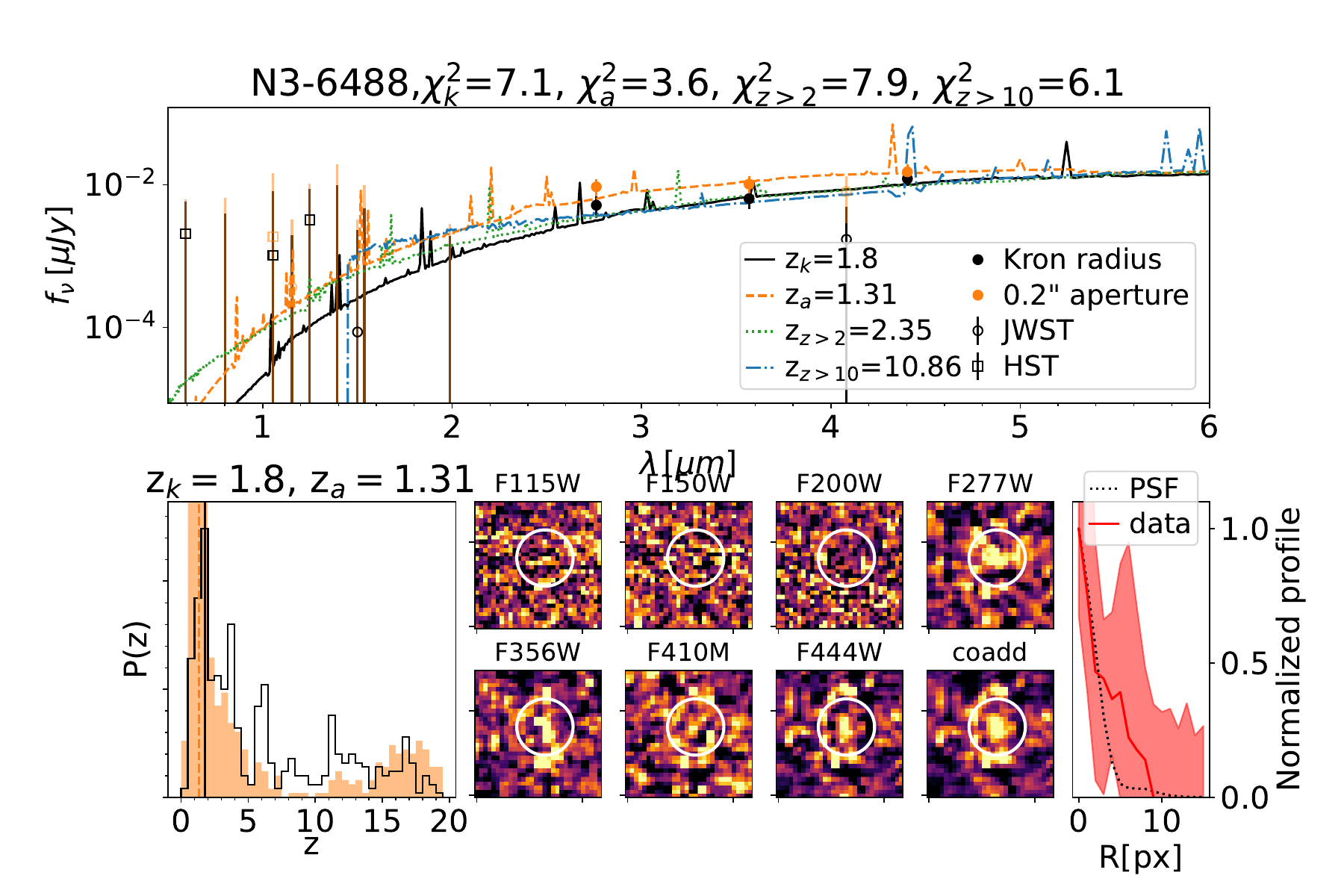}\\
  \caption{continued.}\\
    \includegraphics[trim={20 10 50 40},clip,width=0.44\linewidth,keepaspectratio]{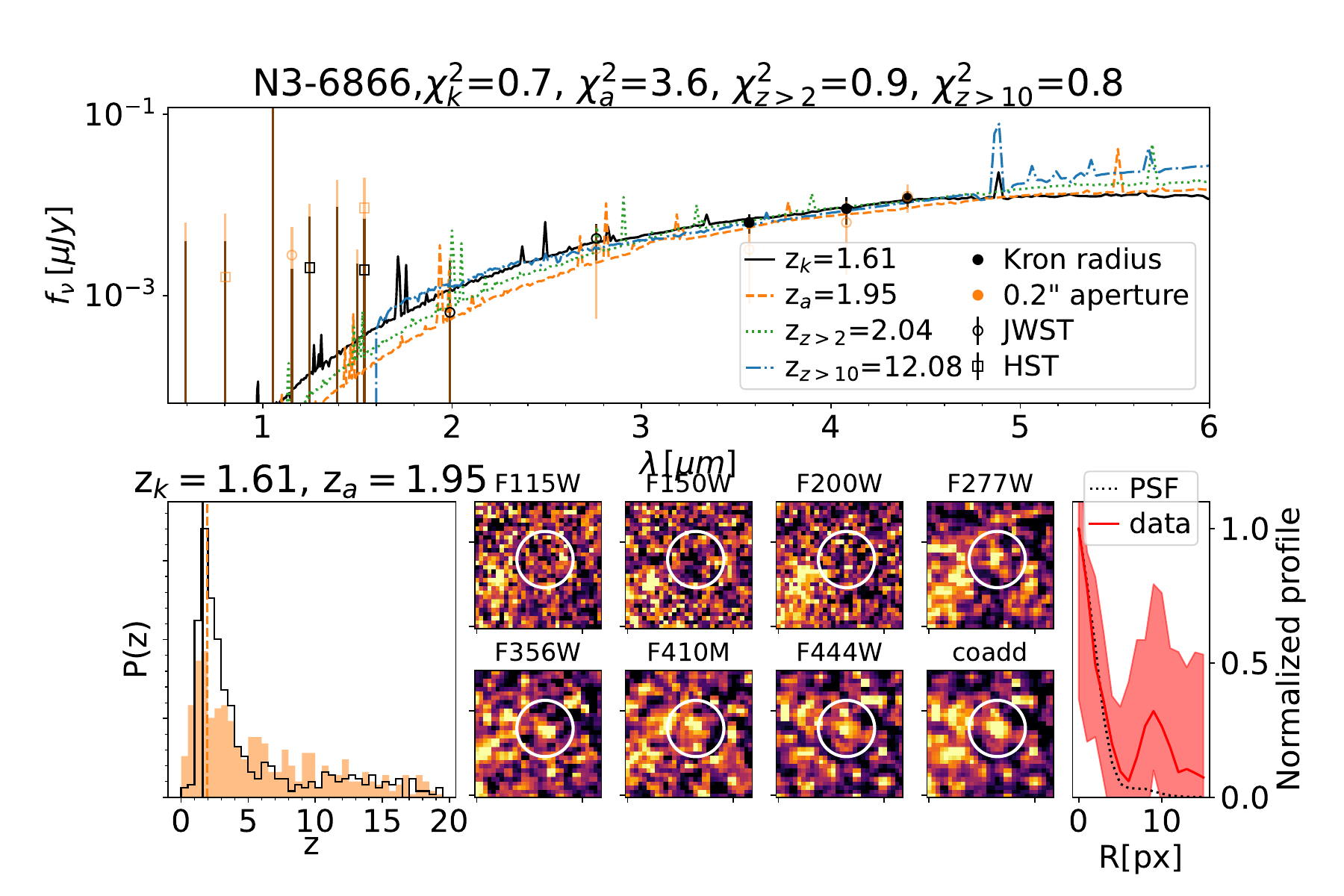}&
    \includegraphics[trim={20 10 50 40},clip,width=0.44\linewidth,keepaspectratio]{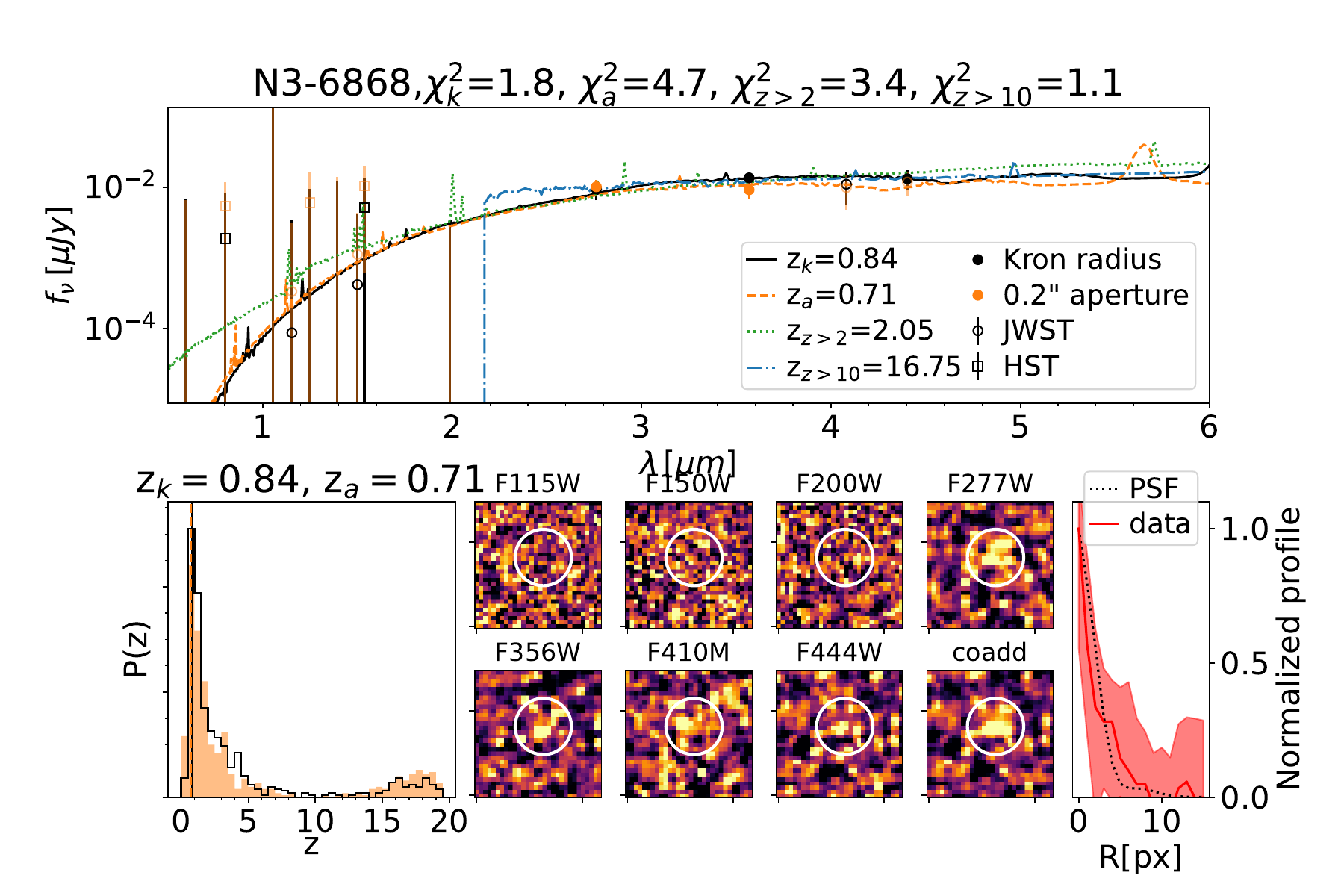}\\
    \includegraphics[trim={20 10 50 40},clip,width=0.44\linewidth,keepaspectratio]{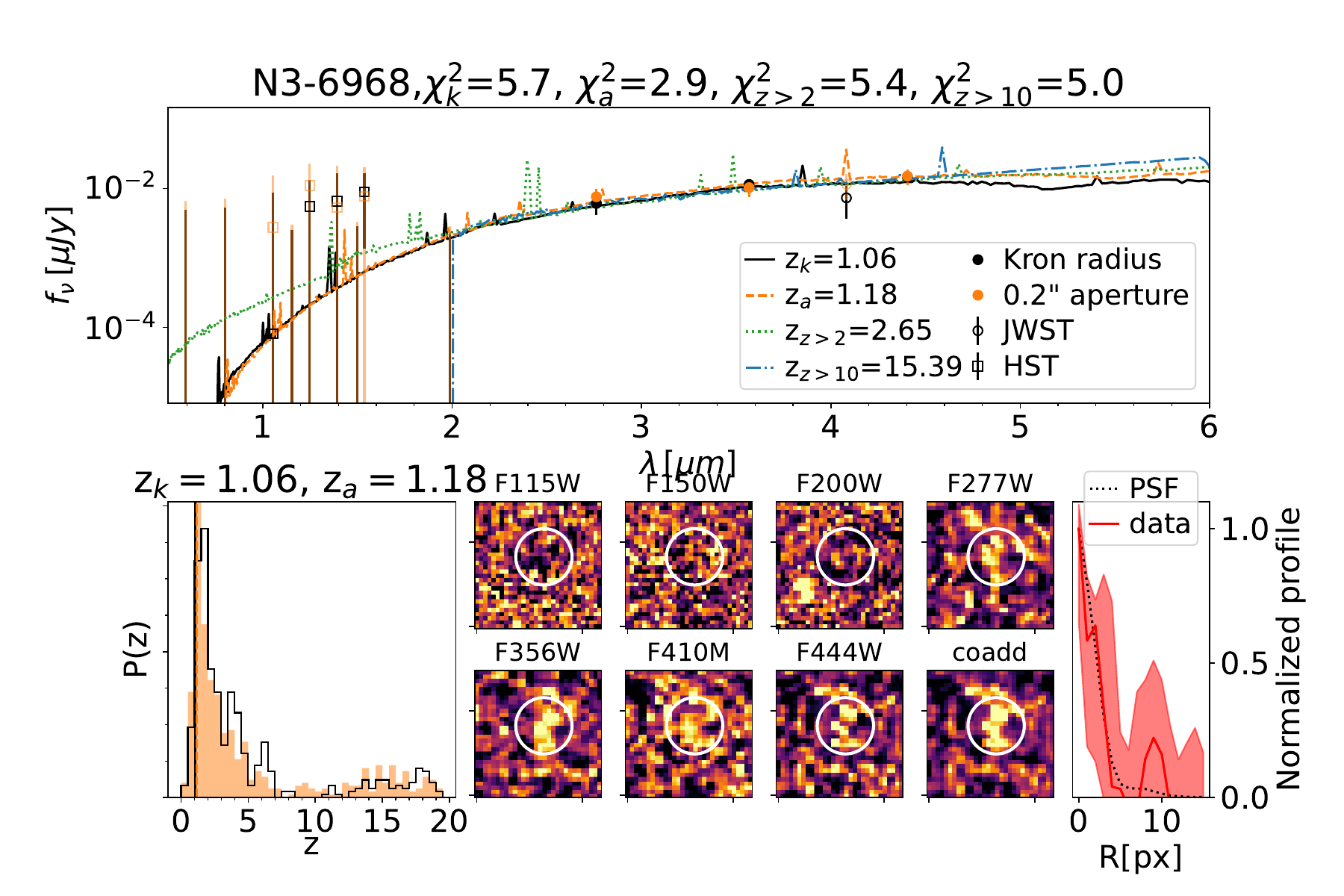}&
    \includegraphics[trim={20 10 50 40},clip,width=0.44\linewidth,keepaspectratio]{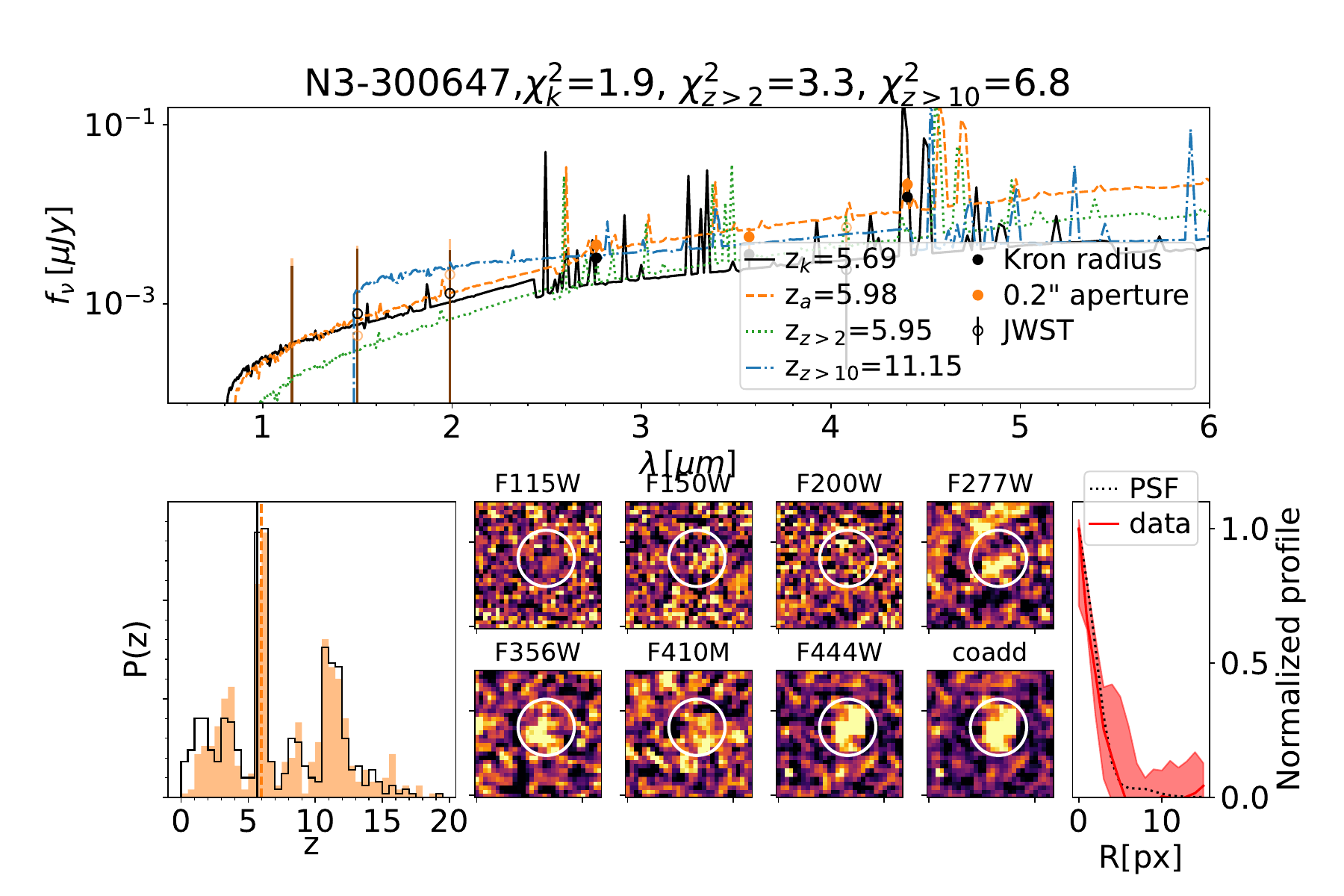} \\
    \includegraphics[trim={20 10 50 40},clip,width=0.44\linewidth,keepaspectratio]{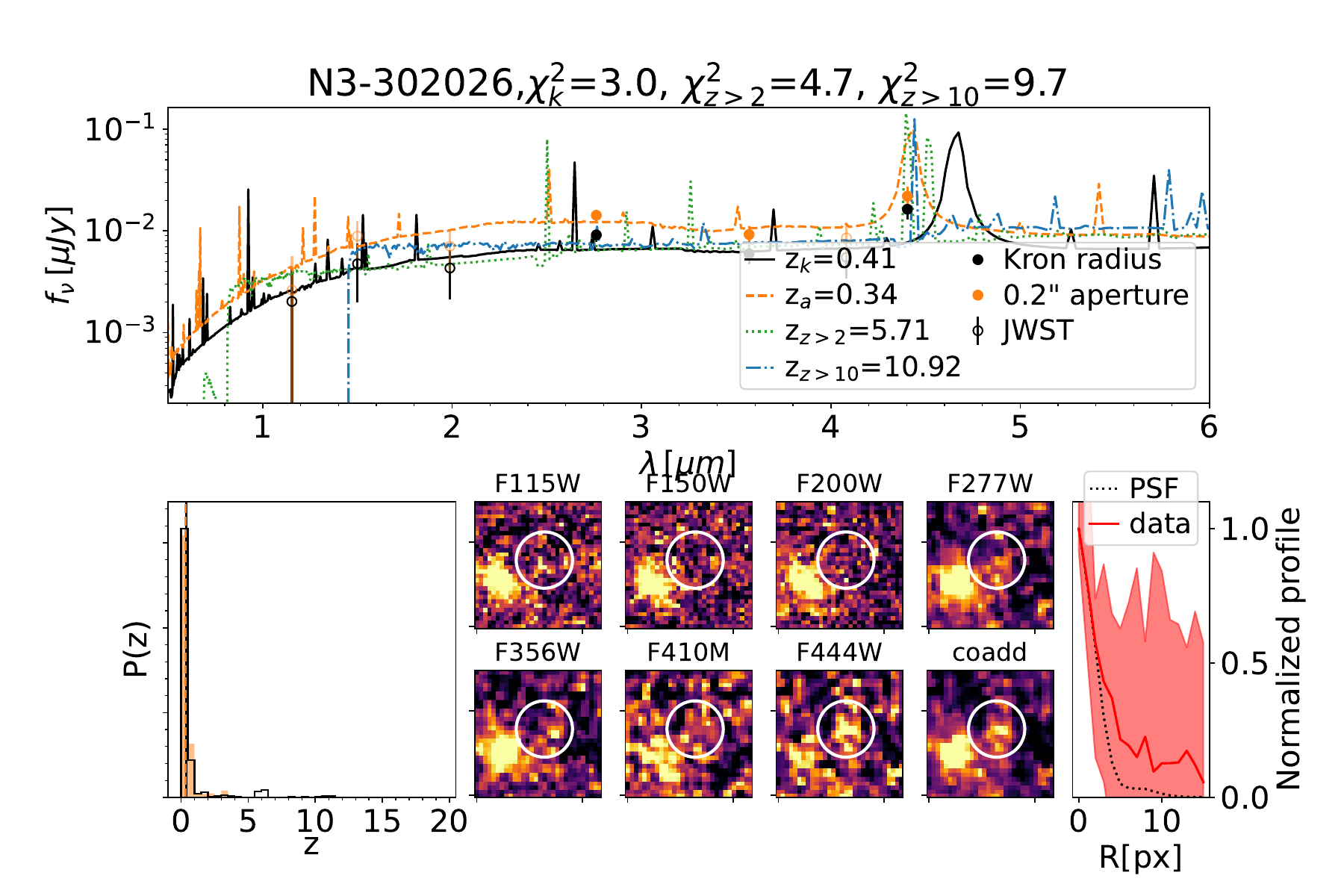} &
    \includegraphics[trim={20 10 50 40},clip,width=0.44\linewidth,keepaspectratio]{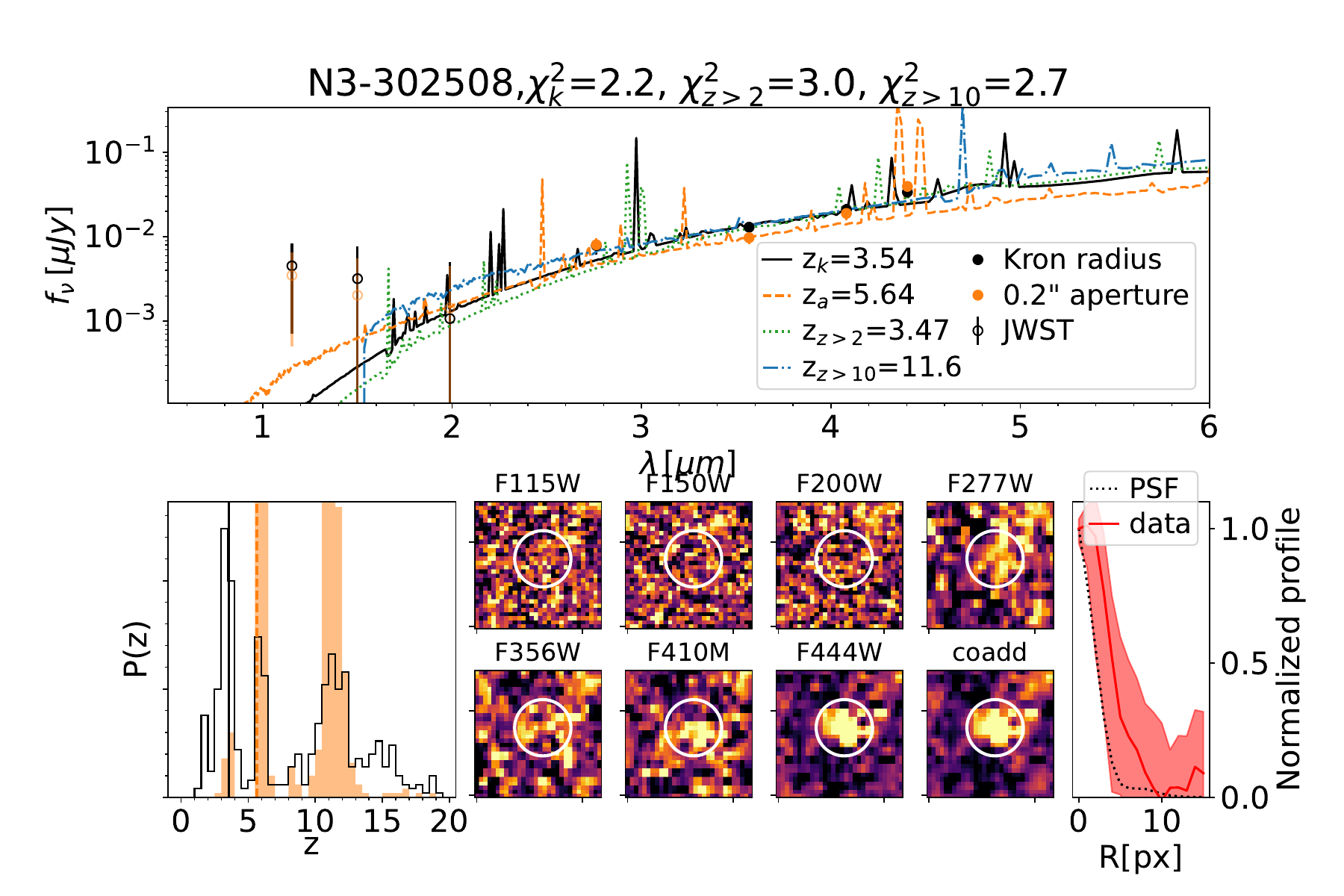}\\
    \includegraphics[trim={20 10 50 40},clip,width=0.44\linewidth,keepaspectratio]{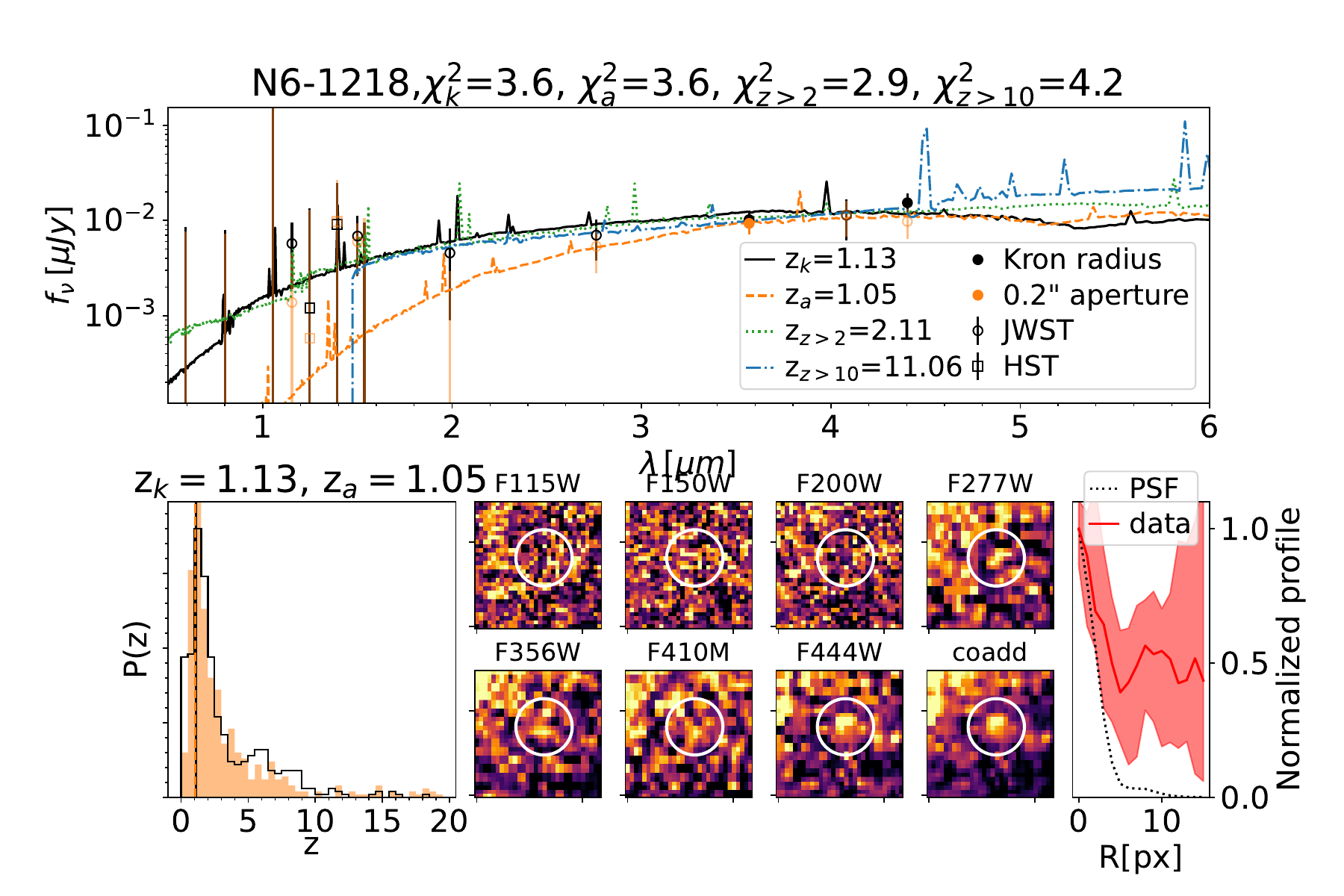}&
    \includegraphics[trim={20 10 50 40},clip,width=0.44\linewidth,keepaspectratio]{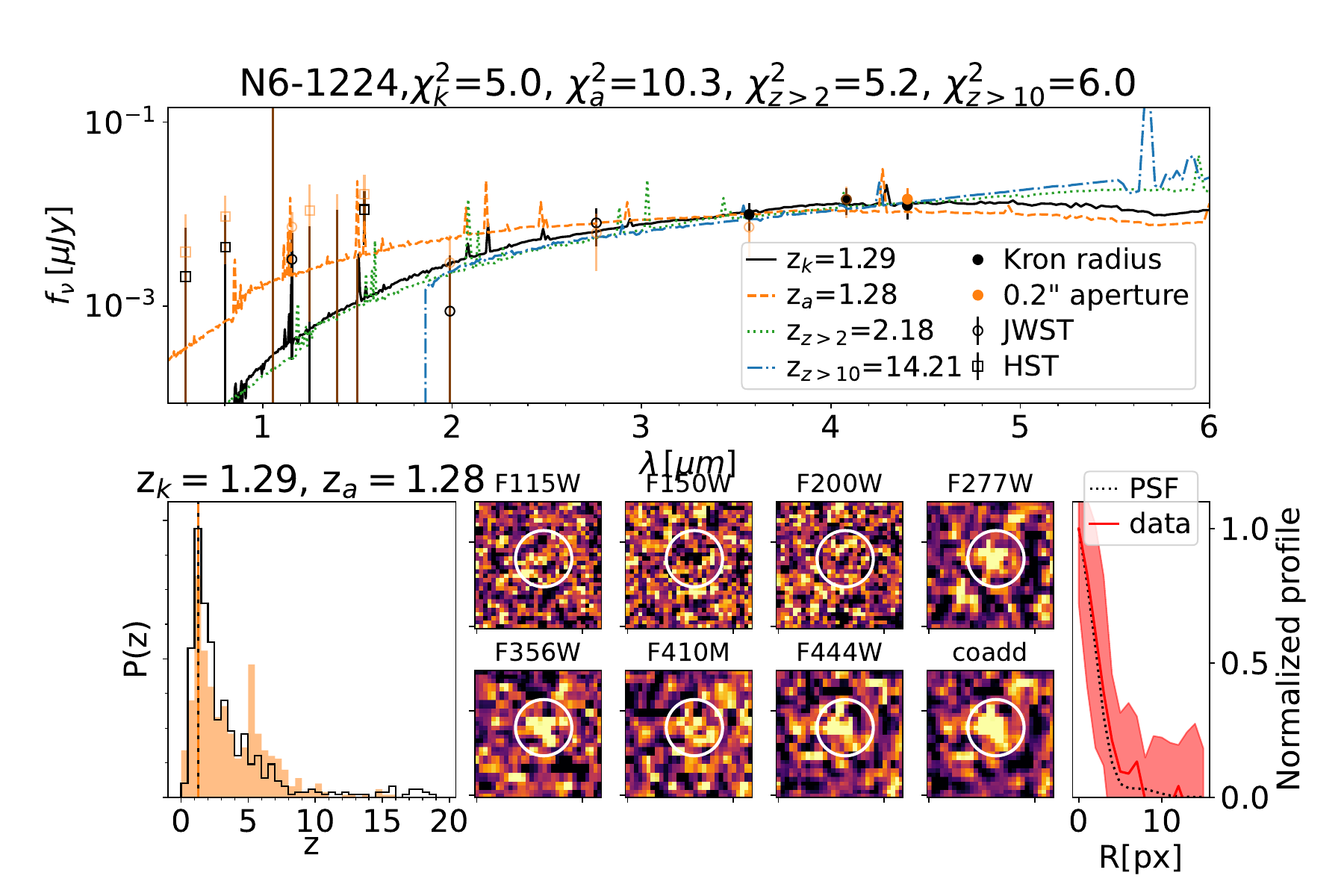}\\
  \caption{continued.}\\ 
    \includegraphics[trim={20 10 50 40},clip,width=0.44\linewidth,keepaspectratio]{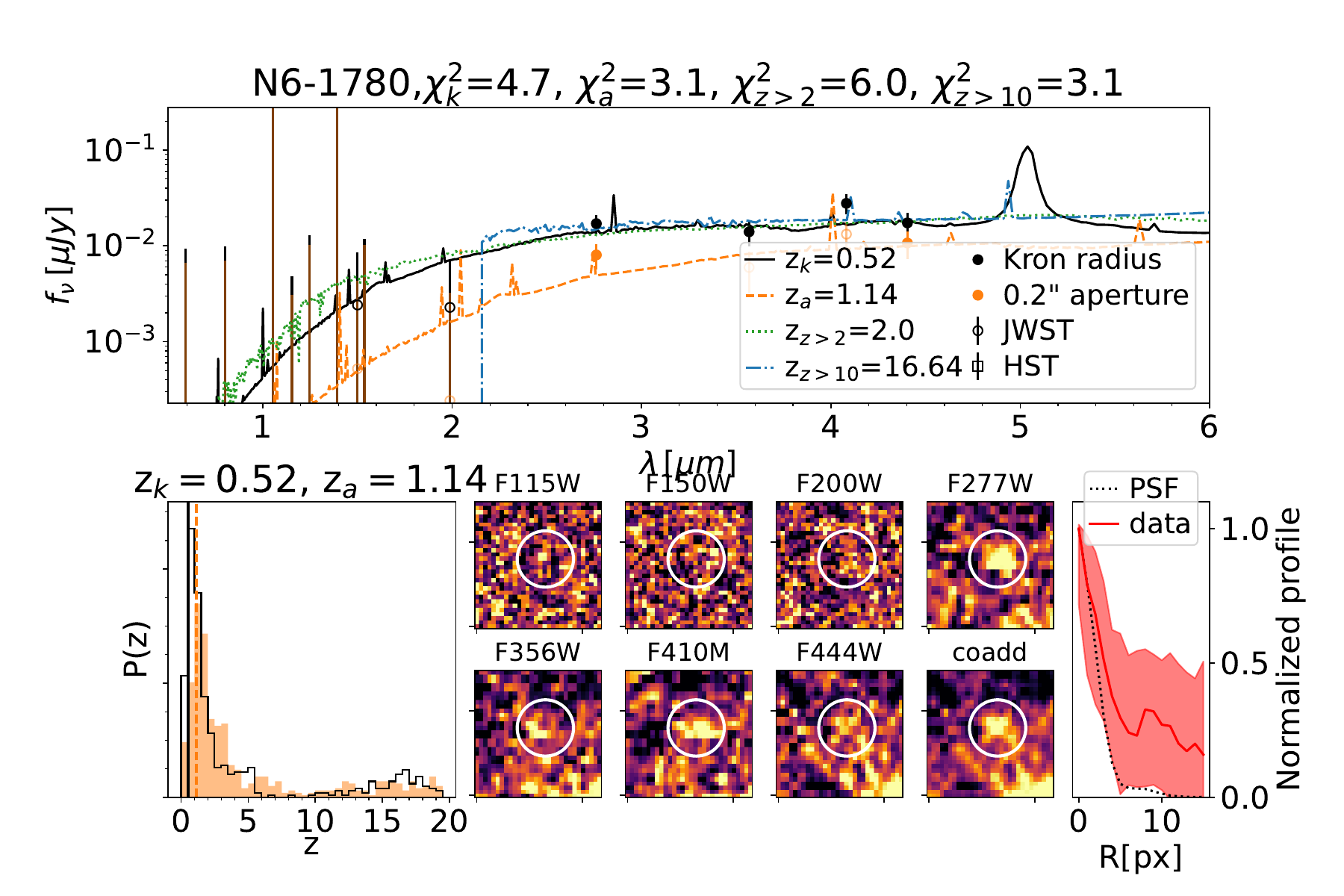}&
    \includegraphics[trim={20 10 50 40},clip,width=0.44\linewidth,keepaspectratio]{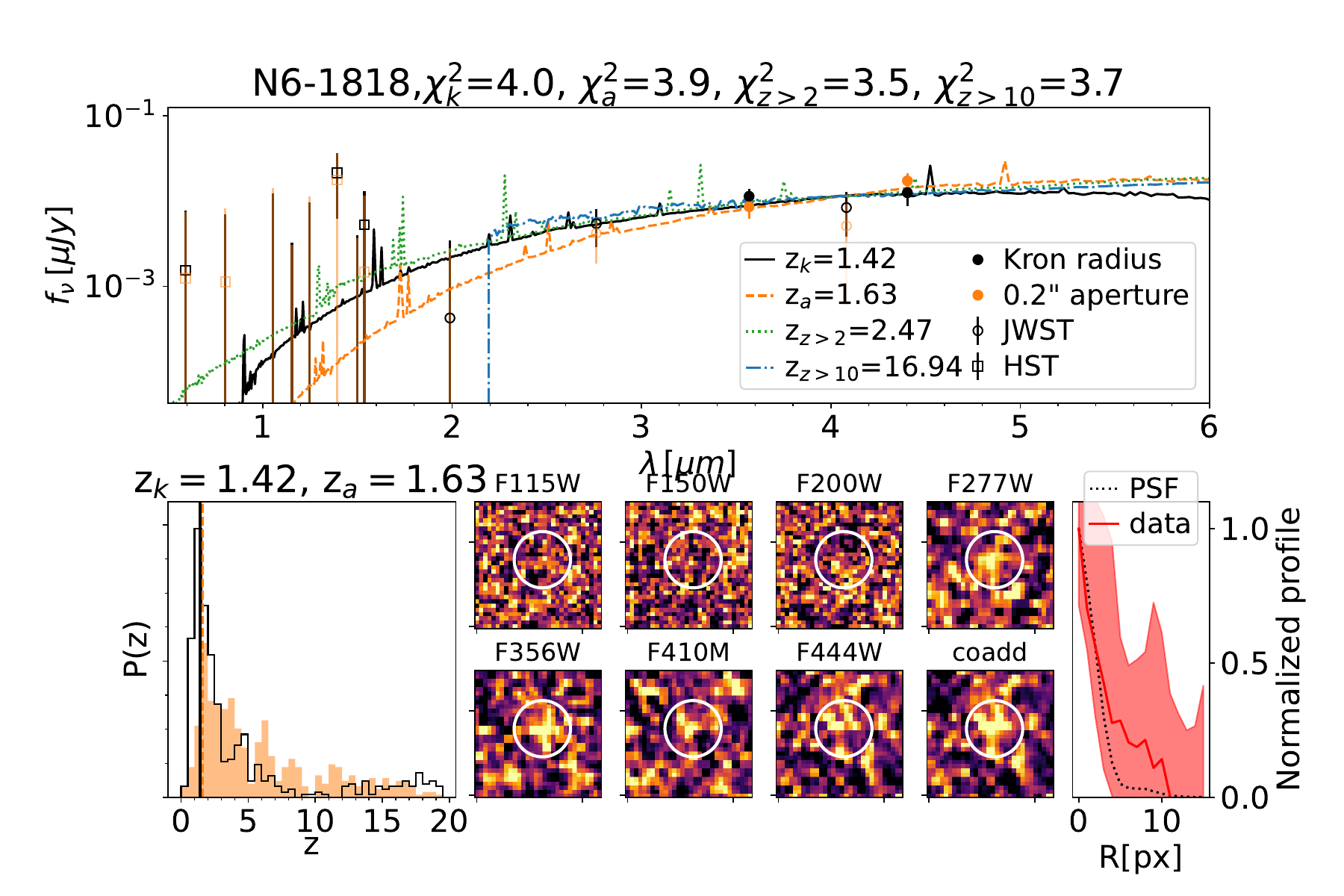}\\
    \includegraphics[trim={20 10 50 40},clip,width=0.44\linewidth,keepaspectratio]{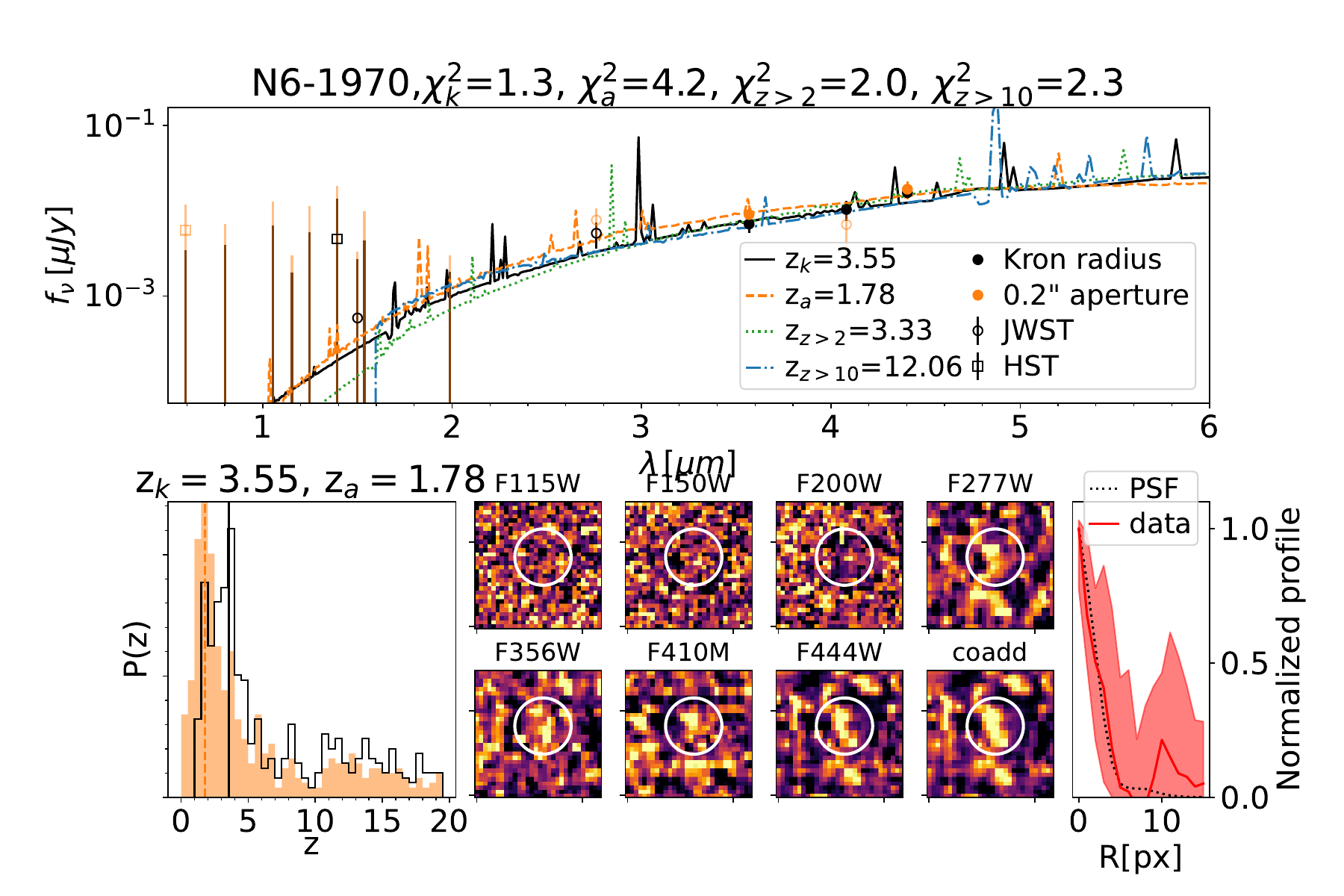}&
    \includegraphics[trim={20 10 50 40},clip,width=0.44\linewidth,keepaspectratio]{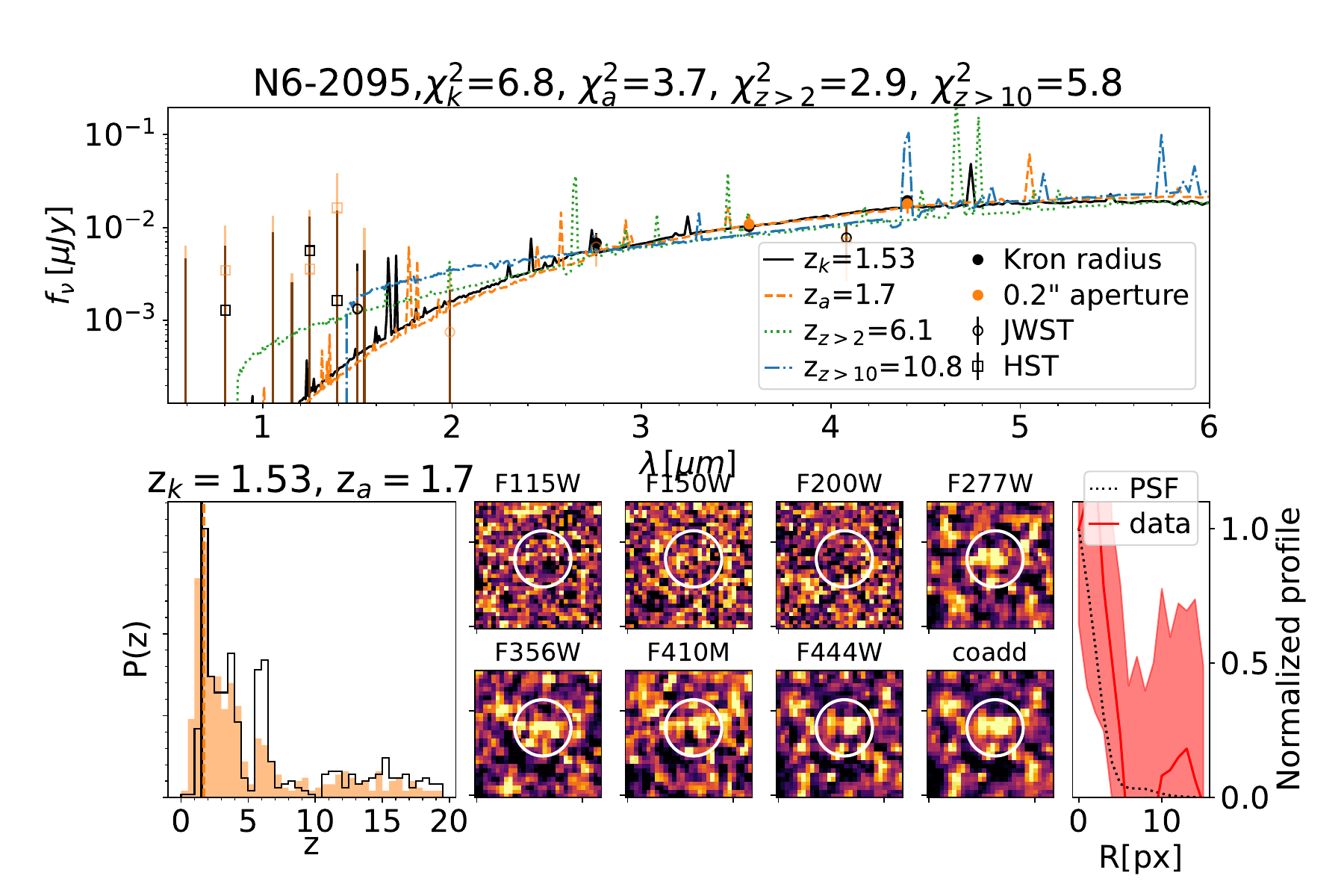}\\
    \includegraphics[trim={20 10 50 40},clip,width=0.44\linewidth,keepaspectratio]{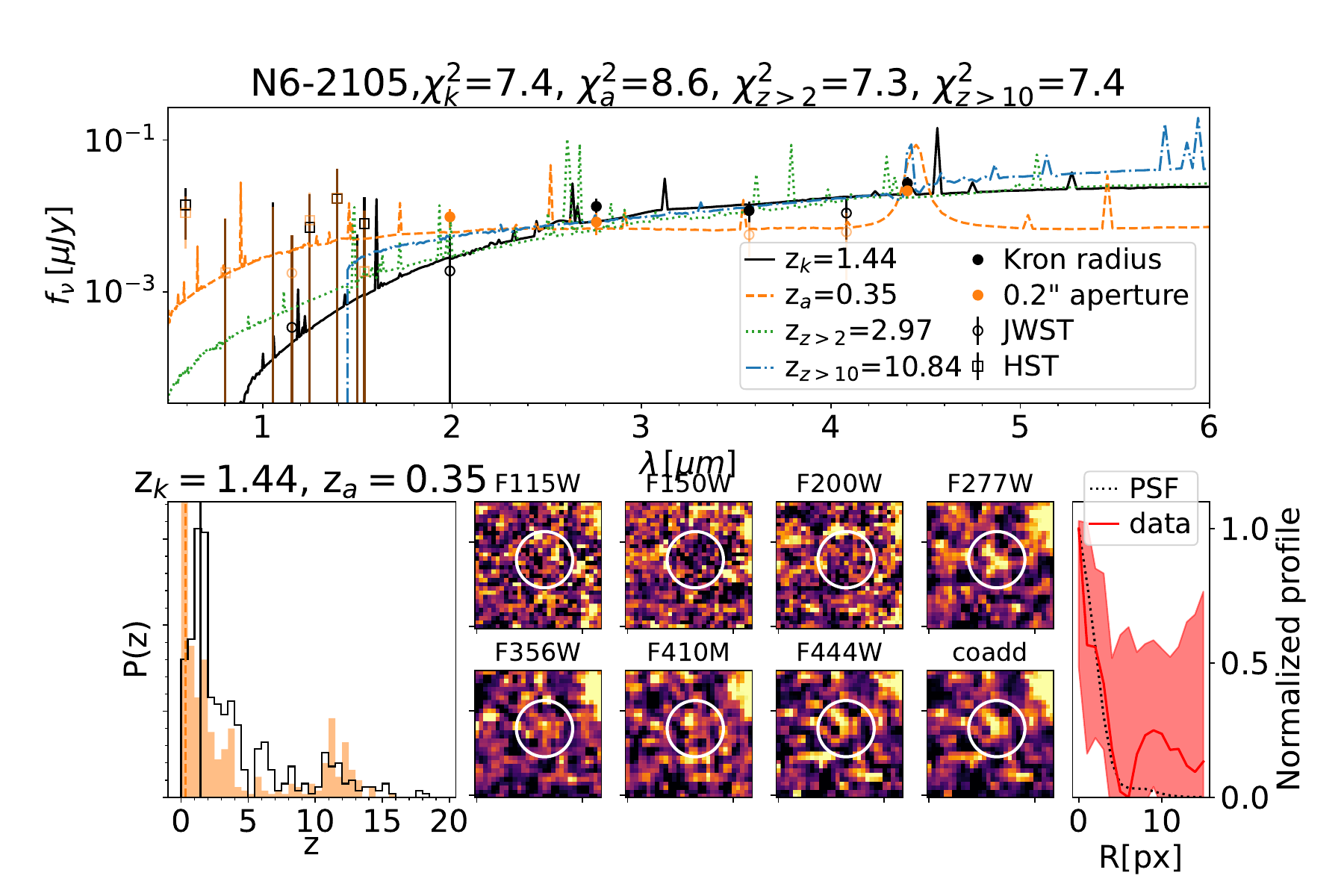}&
    \includegraphics[trim={20 10 50 40},clip,width=0.44\linewidth,keepaspectratio]{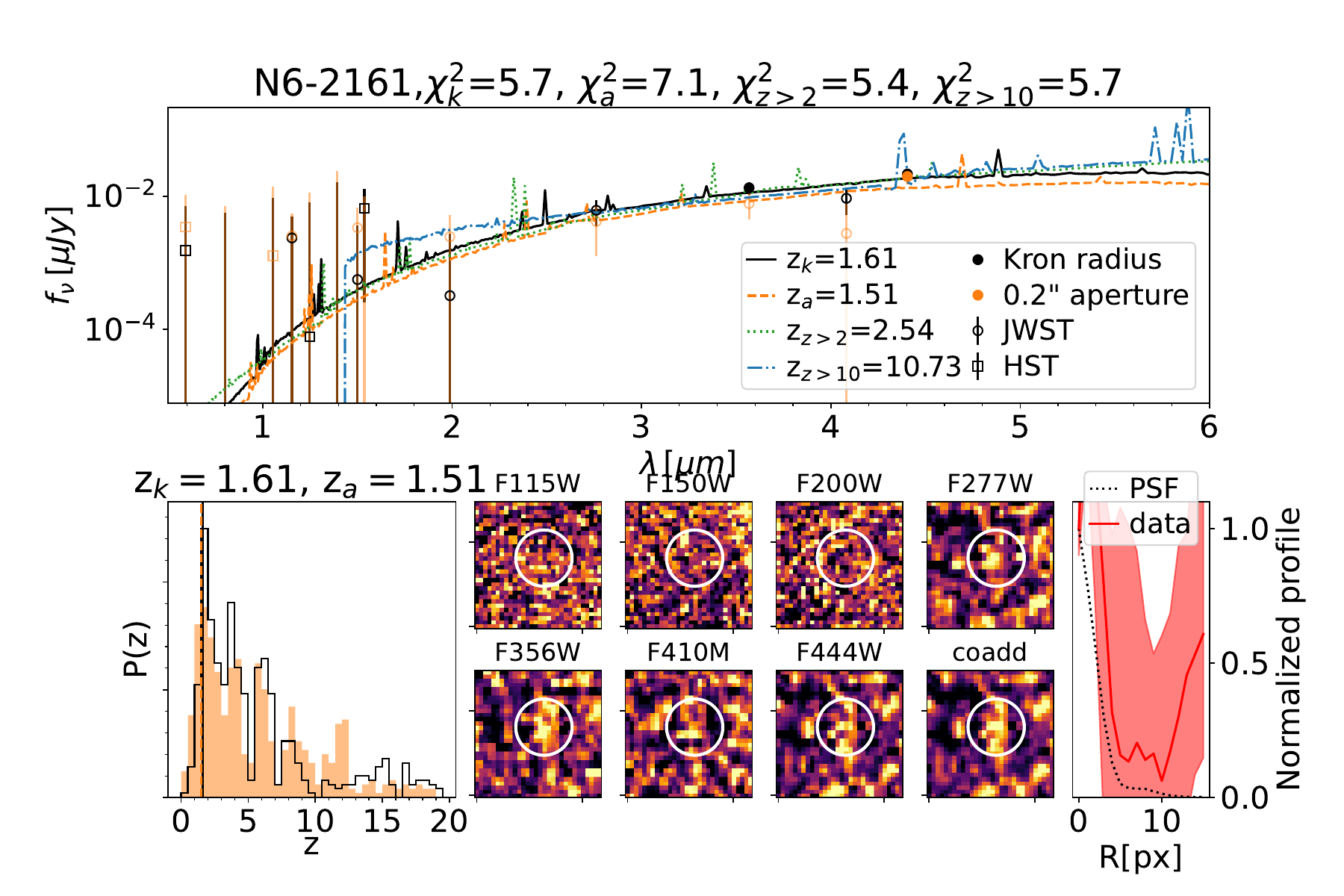}\\
    \includegraphics[trim={20 10 50 40},clip,width=0.44\linewidth,keepaspectratio]{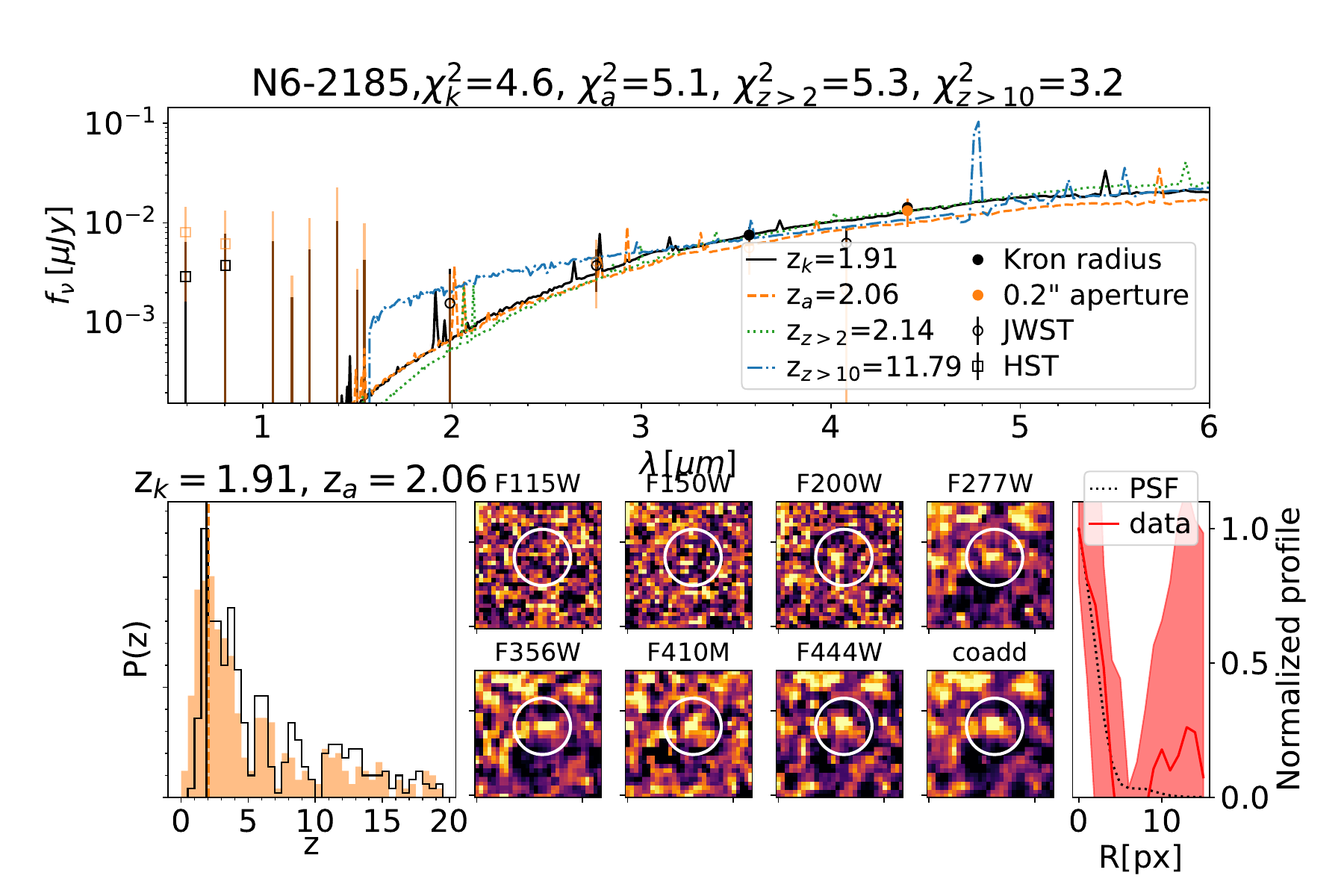}&
    \includegraphics[trim={20 10 50 40},clip,width=0.44\linewidth,keepaspectratio]{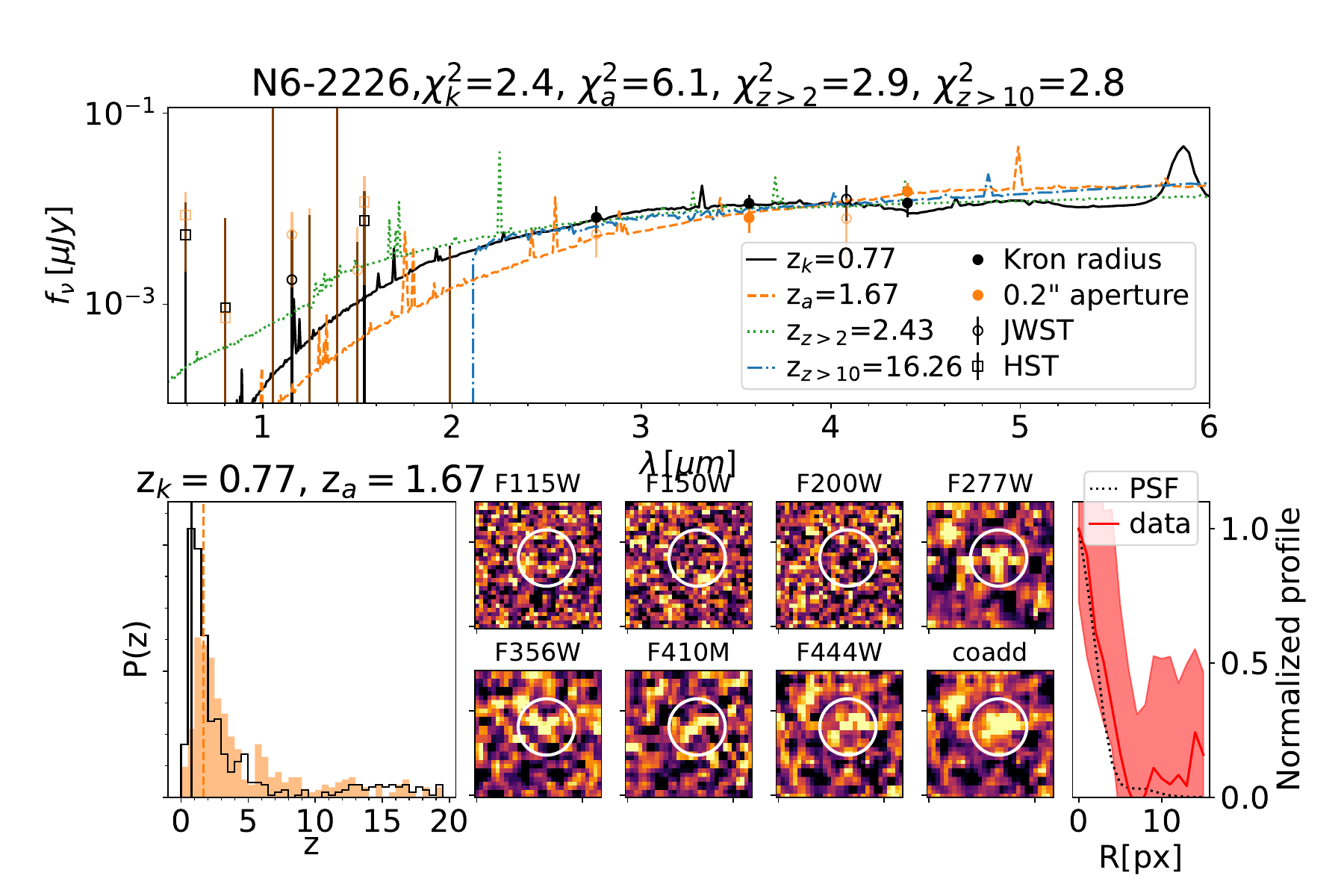}\\
      \caption{continued.}\\
    \includegraphics[trim={20 10 50 40},clip,width=0.44\linewidth,keepaspectratio]{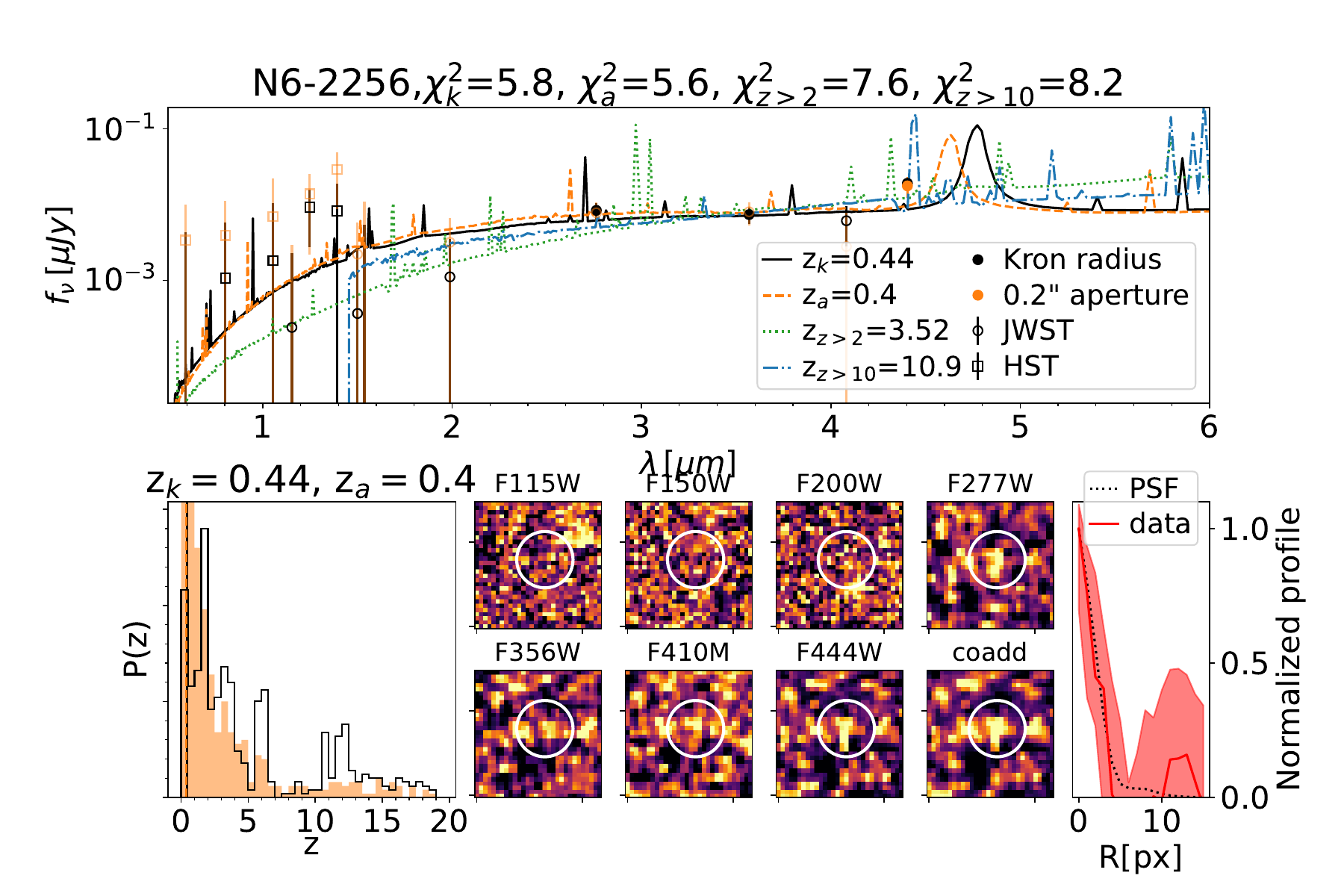}&
    \includegraphics[trim={20 10 50 40},clip,width=0.44\linewidth,keepaspectratio]{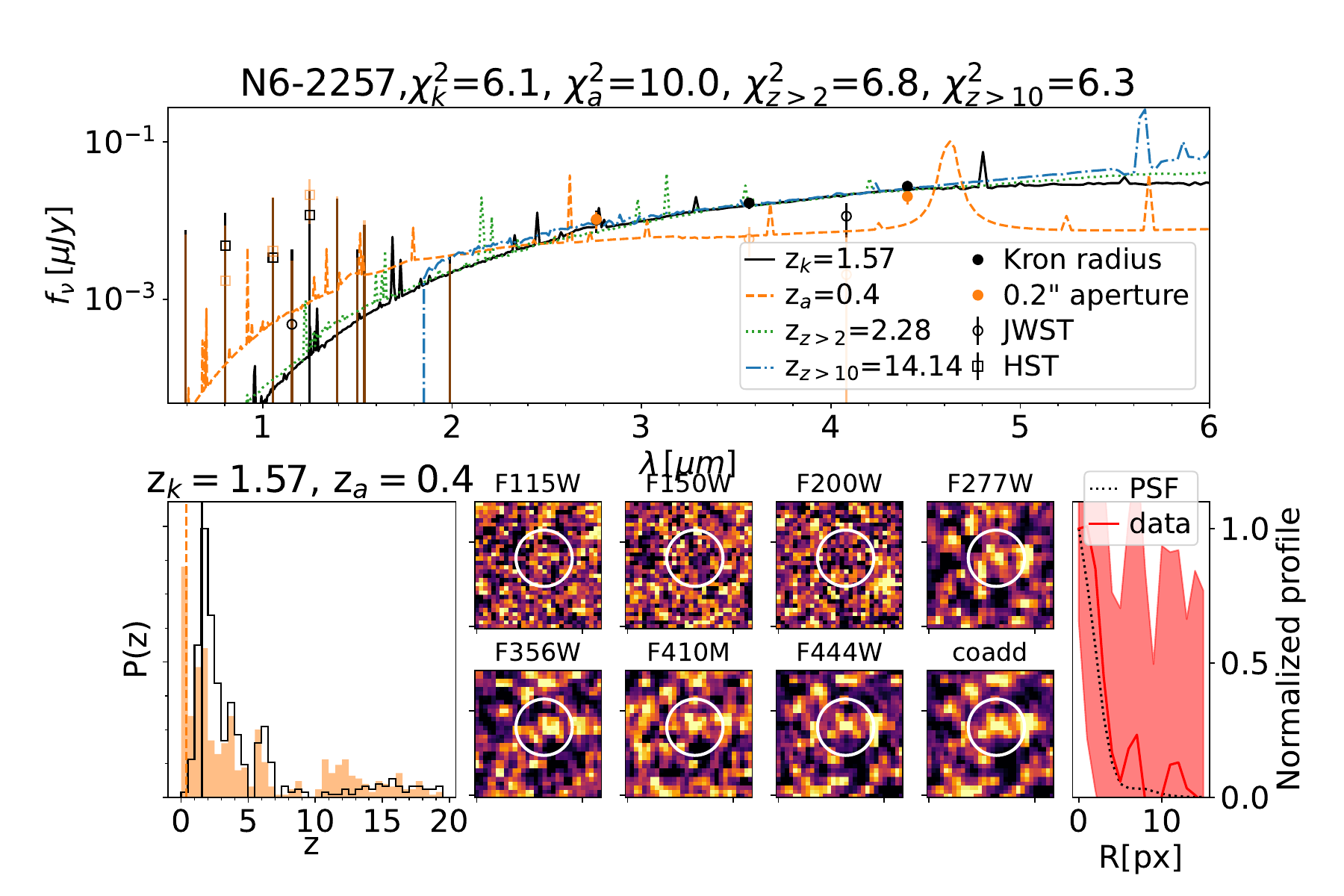}\\
    \includegraphics[trim={20 10 50 40},clip,width=0.44\linewidth,keepaspectratio]{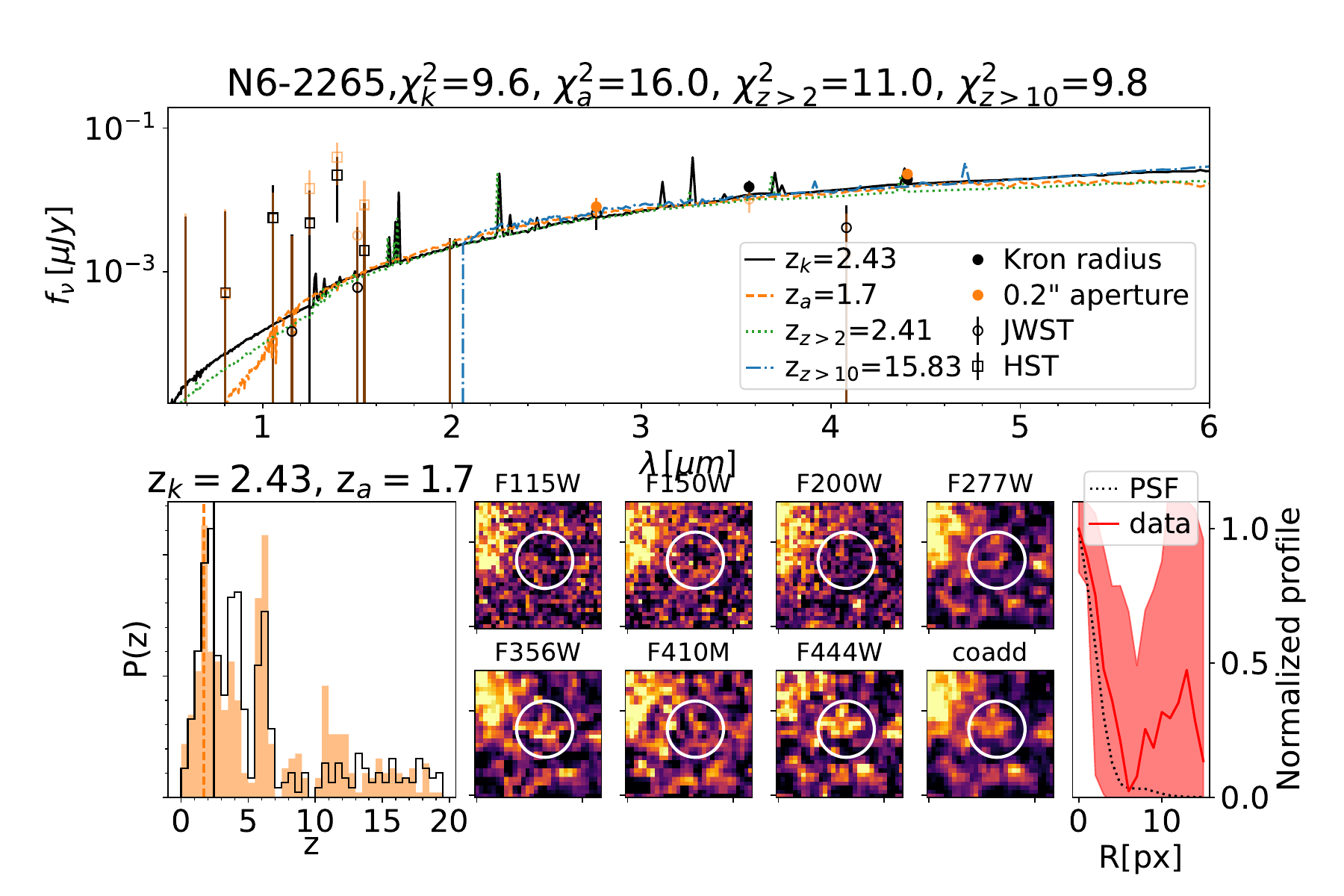}&
    \includegraphics[trim={20 10 50 40},clip,width=0.44\linewidth,keepaspectratio]{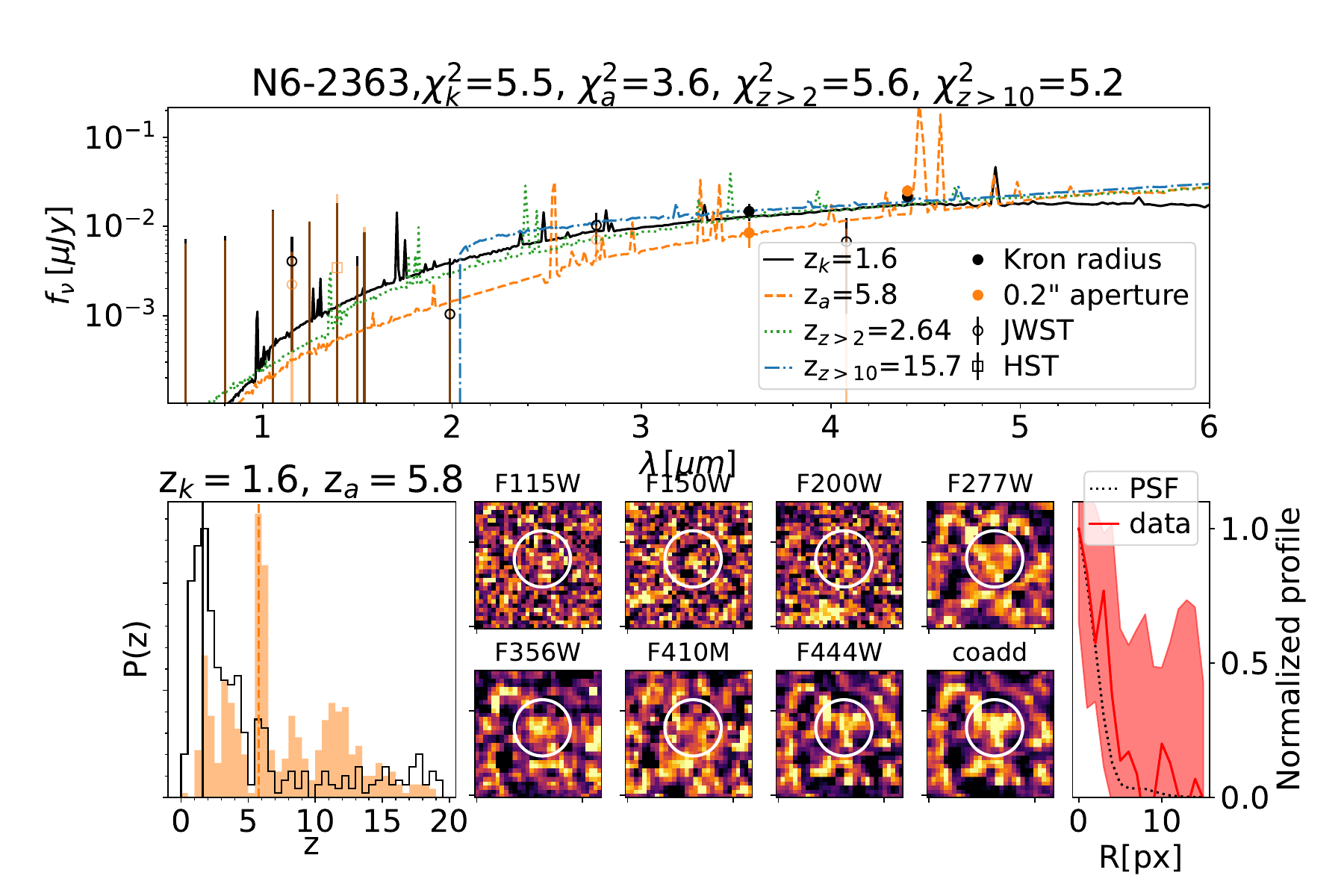}\\
    \includegraphics[trim={20 10 50 40},clip,width=0.44\linewidth,keepaspectratio]{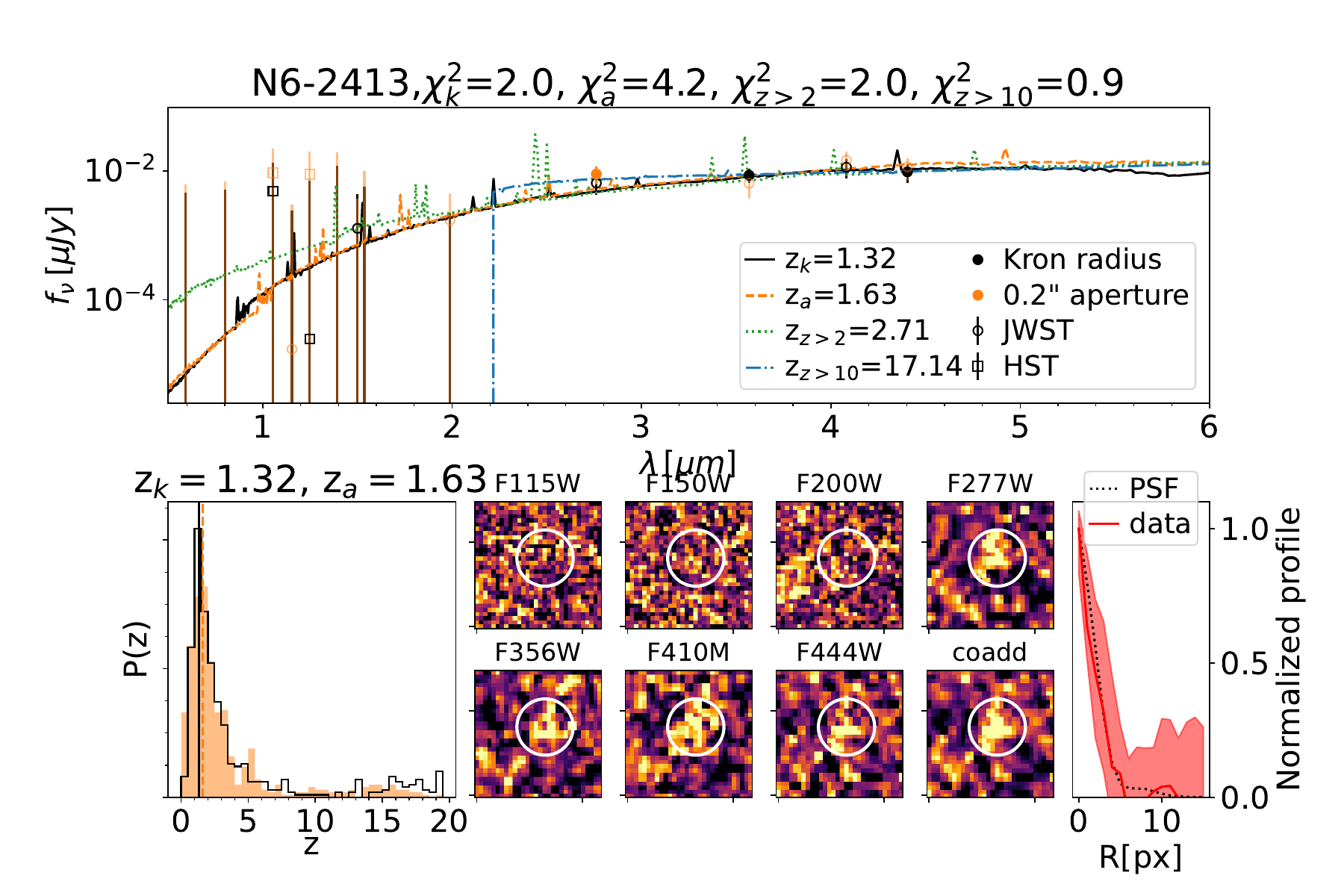}&
    \includegraphics[trim={20 10 50 40},clip,width=0.44\linewidth,keepaspectratio]{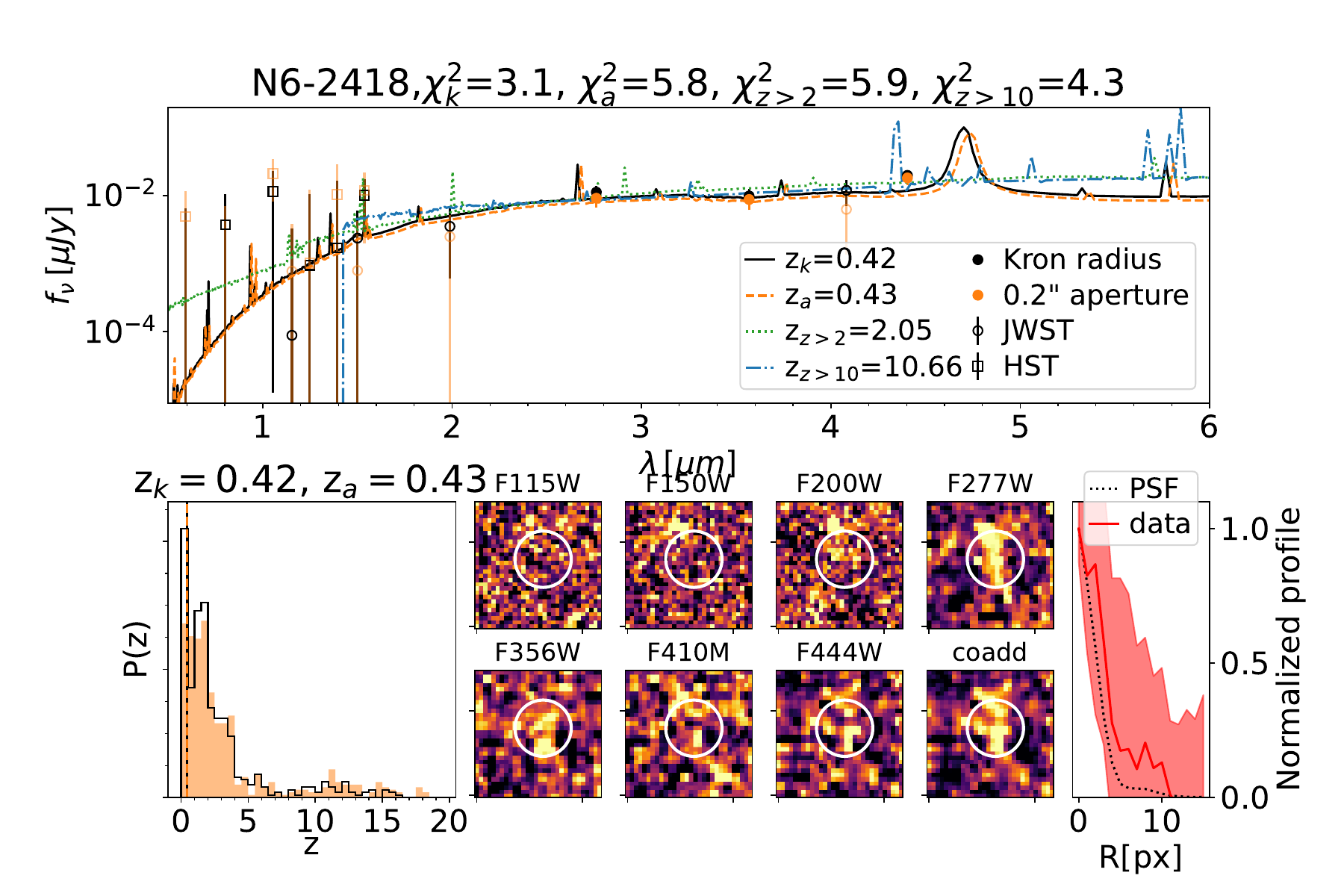}\\
    \includegraphics[trim={20 10 50 40},clip,width=0.44\linewidth,keepaspectratio]{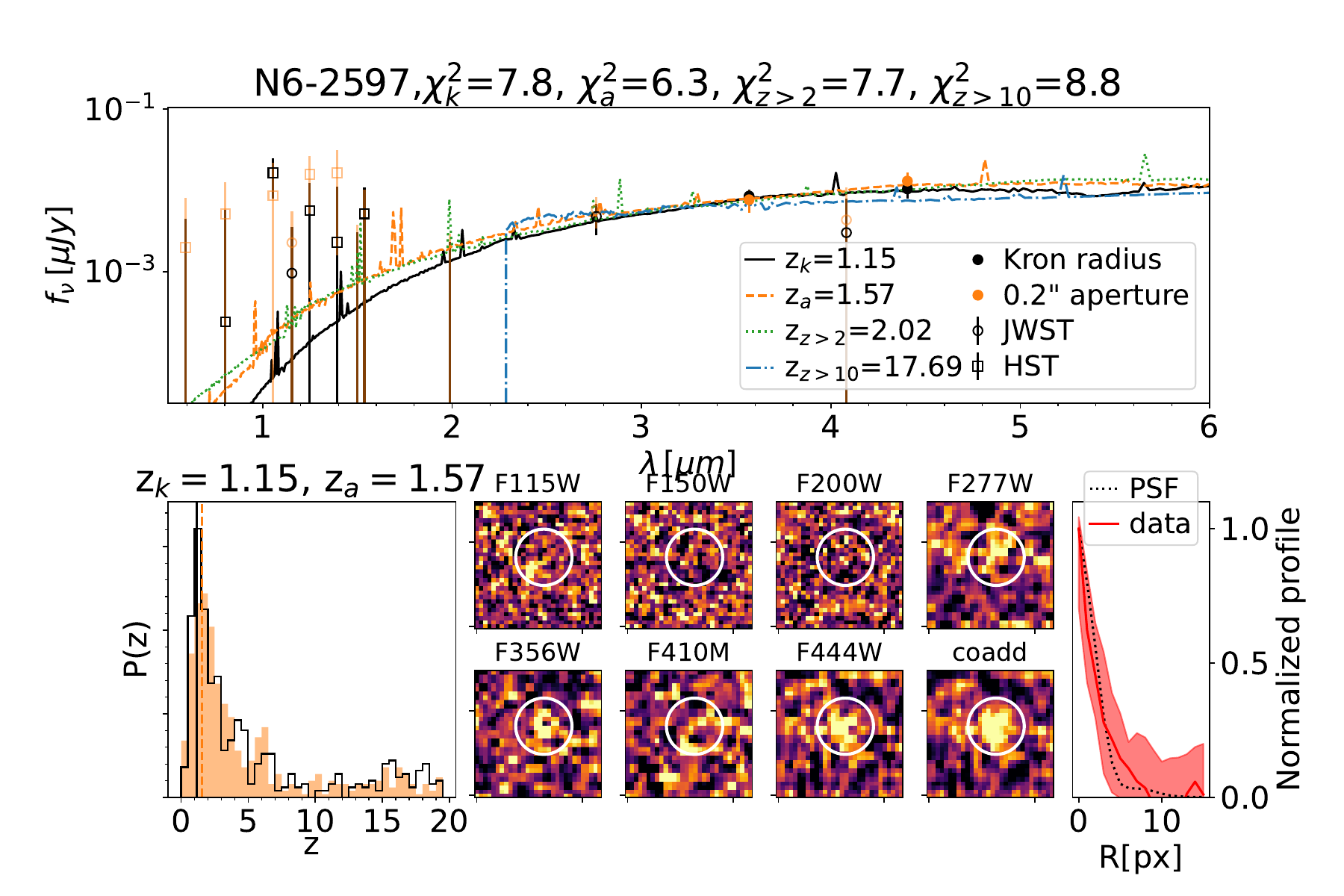}&
    \includegraphics[trim={20 10 50 40},clip,width=0.44\linewidth,keepaspectratio]{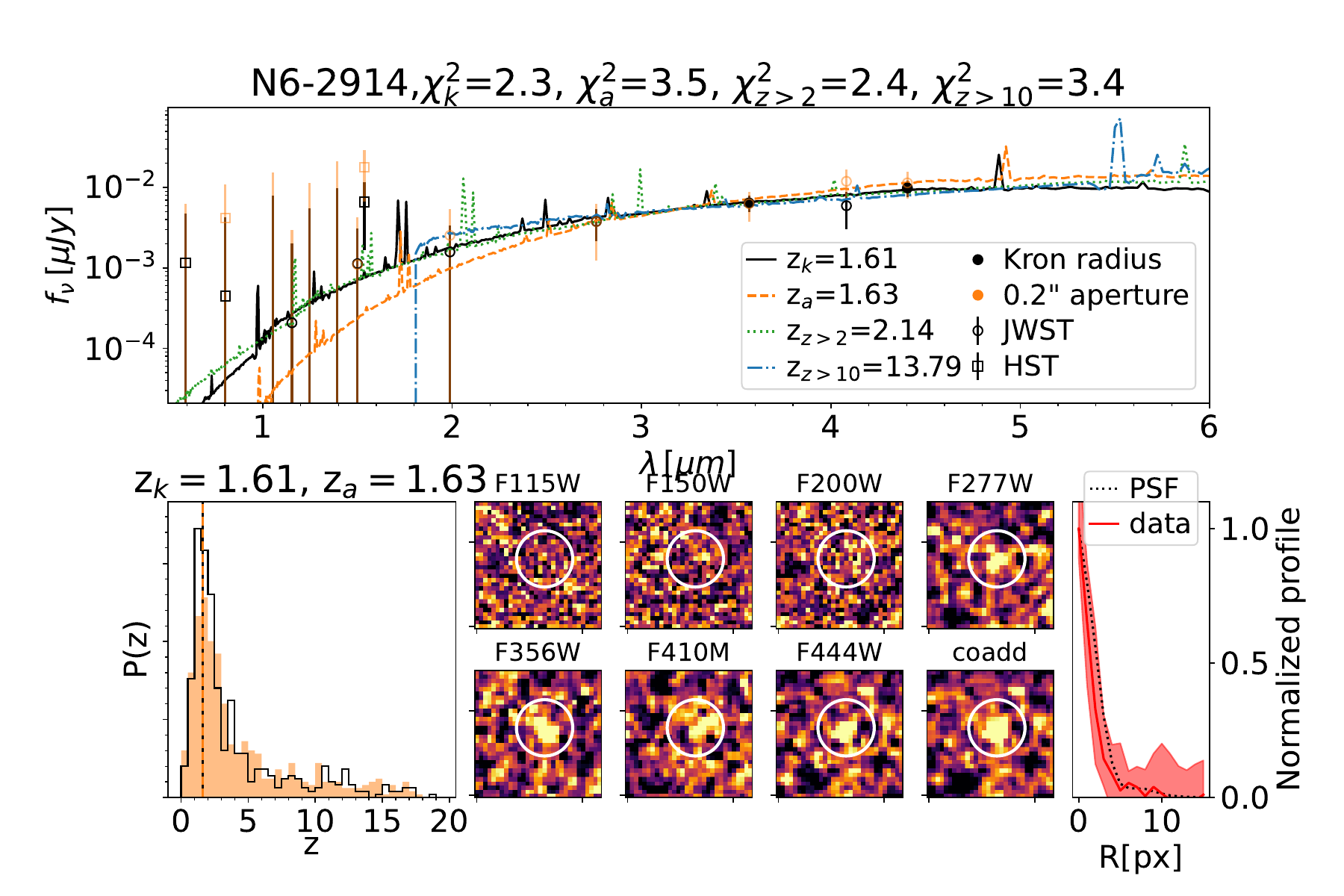}\\
      \caption{continued.}\\
    \includegraphics[trim={20 10 50 40},clip,width=0.44\linewidth,keepaspectratio]{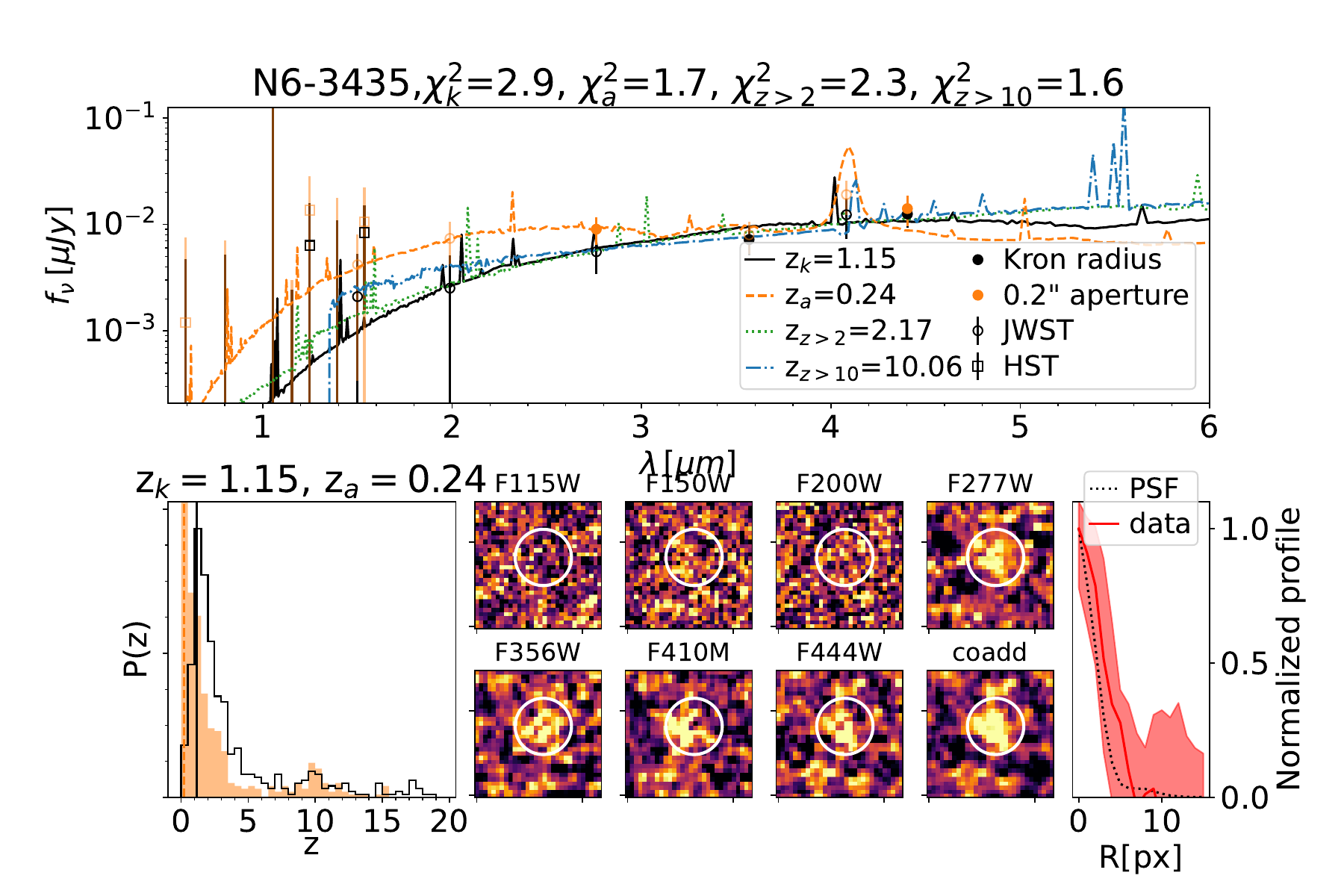}&
    \includegraphics[trim={20 10 50 40},clip,width=0.44\linewidth,keepaspectratio]{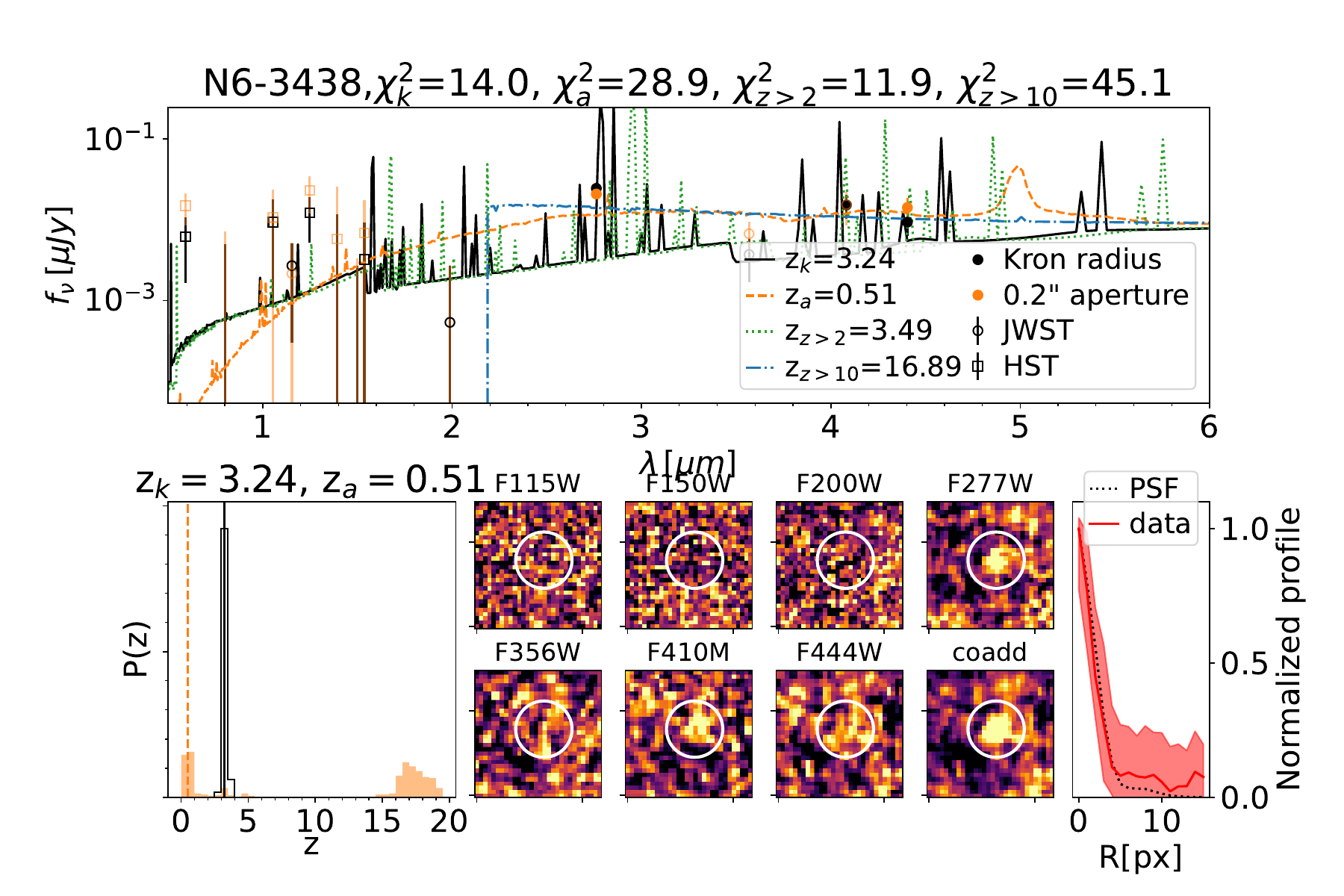}\\
    \includegraphics[trim={20 10 50 40},clip,width=0.44\linewidth,keepaspectratio]{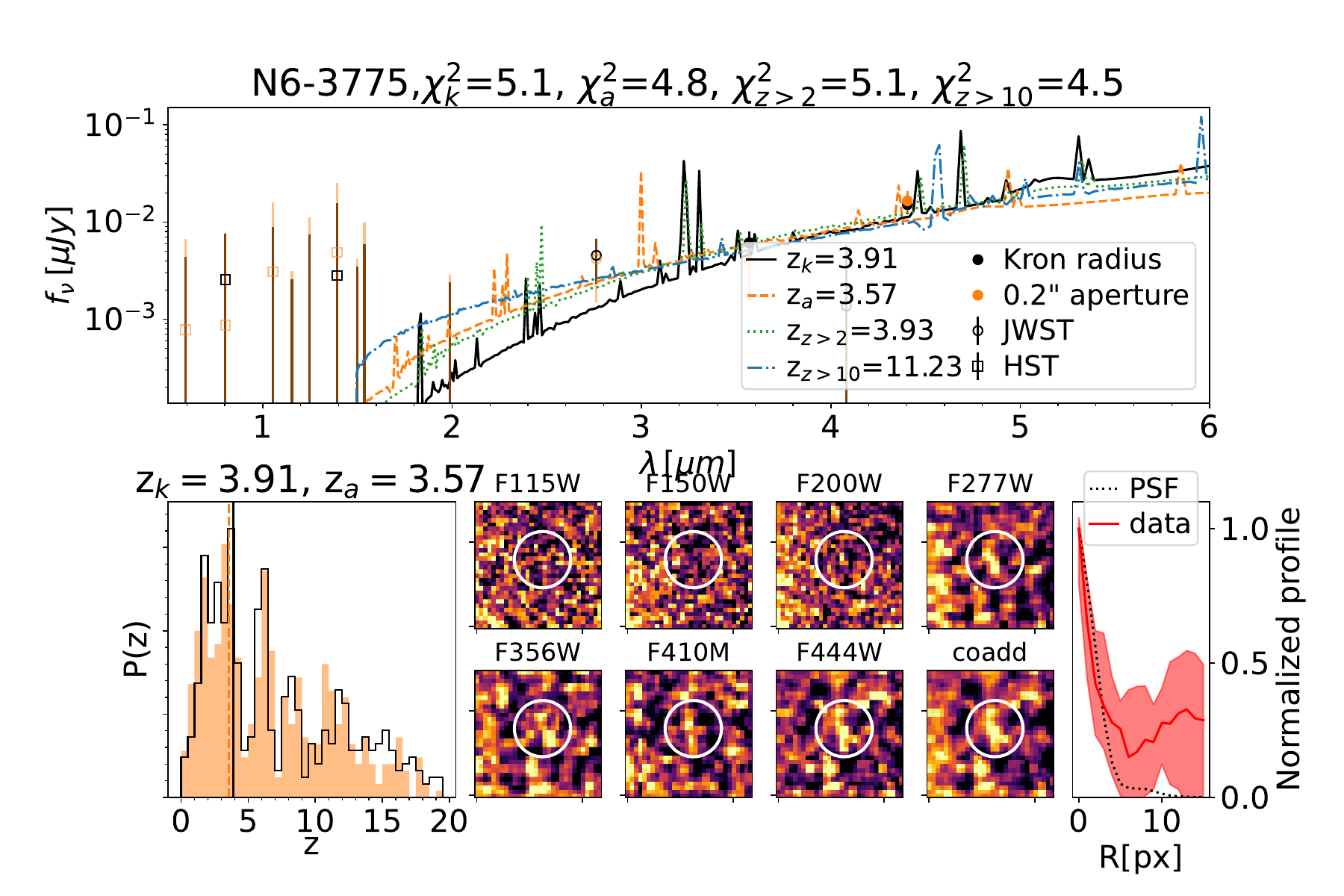}&
    \includegraphics[trim={20 10 50 40},clip,width=0.44\linewidth,keepaspectratio]{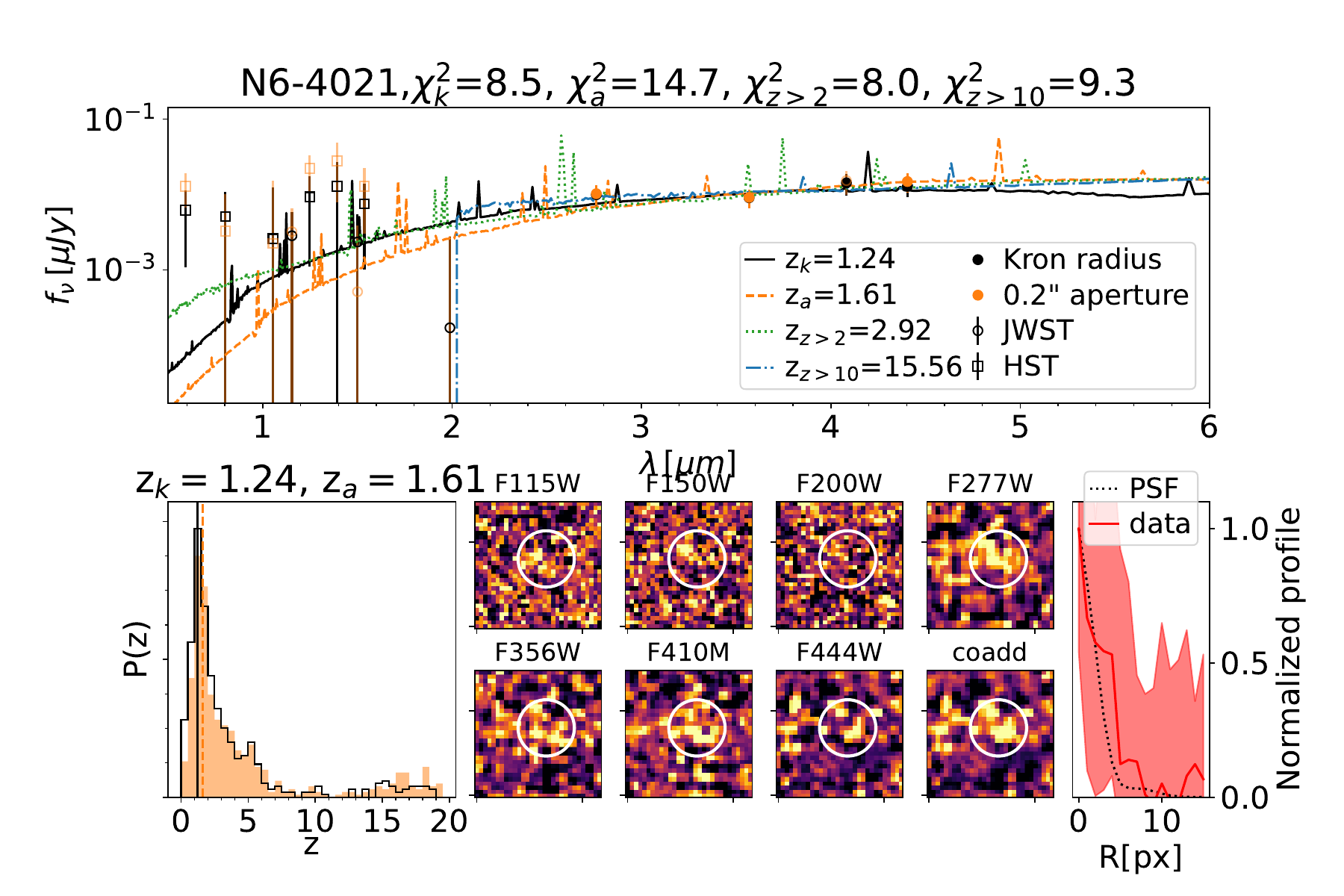}\\
    \includegraphics[trim={20 10 50 40},clip,width=0.44\linewidth,keepaspectratio]{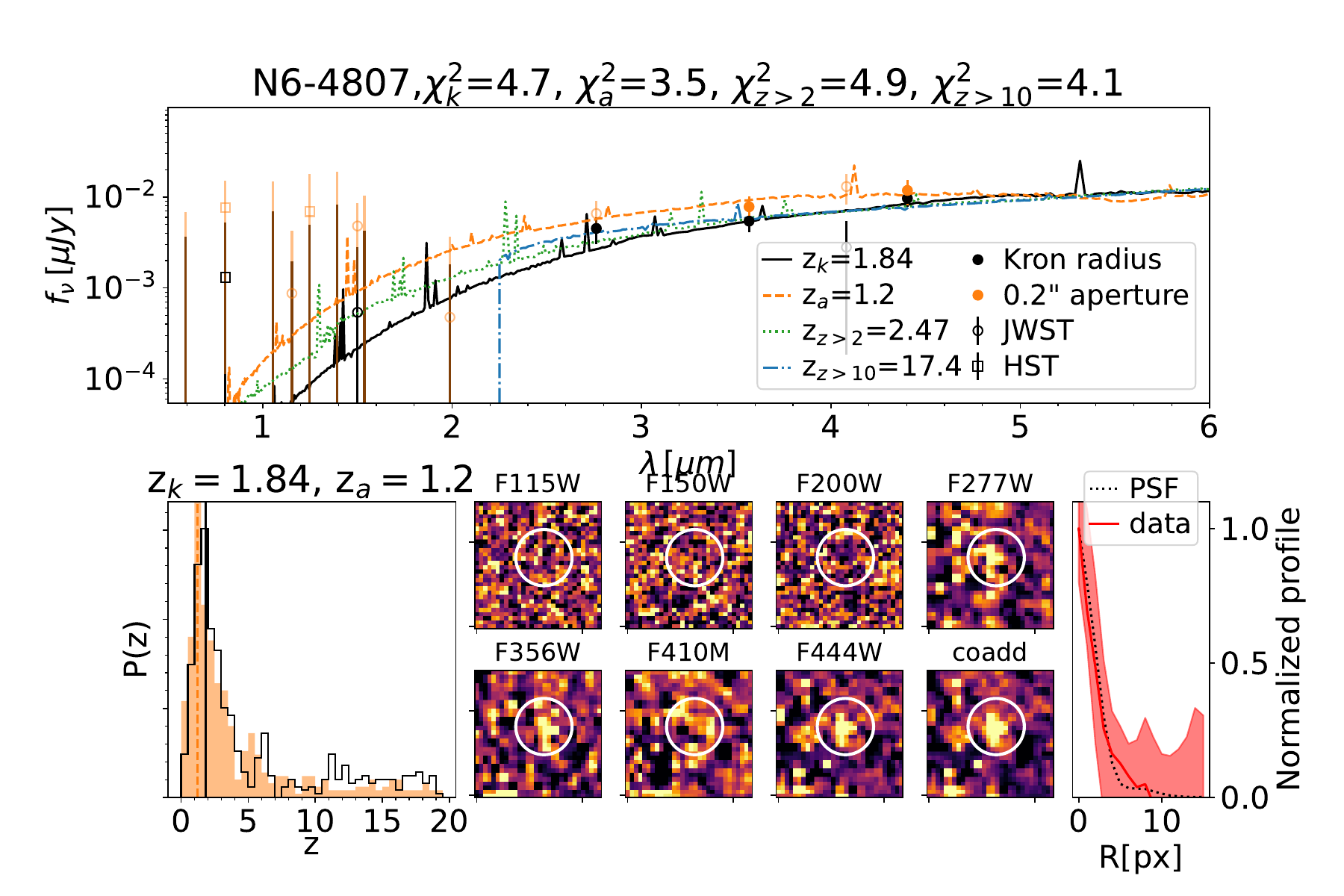}&
    \includegraphics[trim={20 10 50 40},clip,width=0.44\linewidth,keepaspectratio]{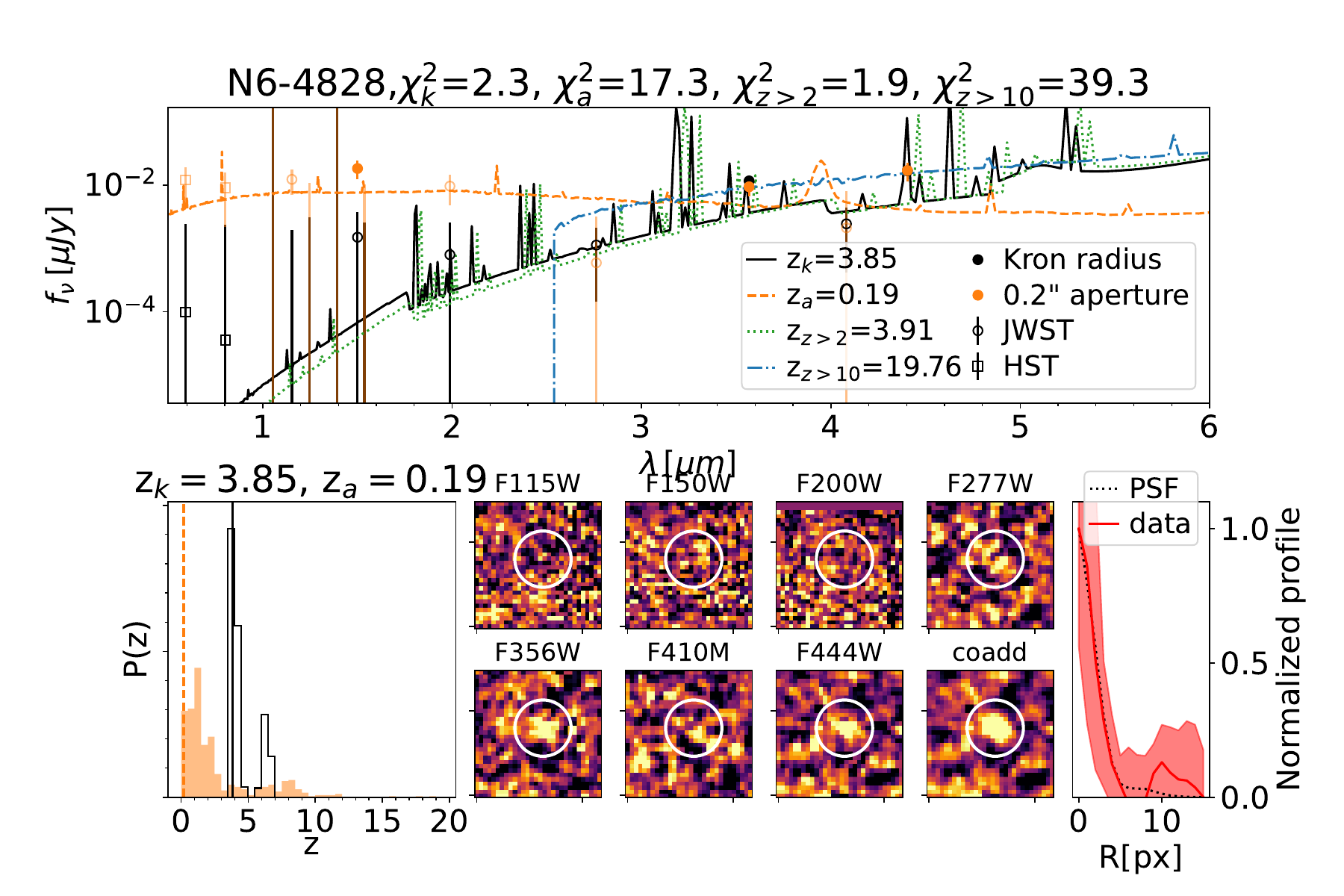}\\
    \includegraphics[trim={20 10 50 40},clip,width=0.44\linewidth,keepaspectratio]{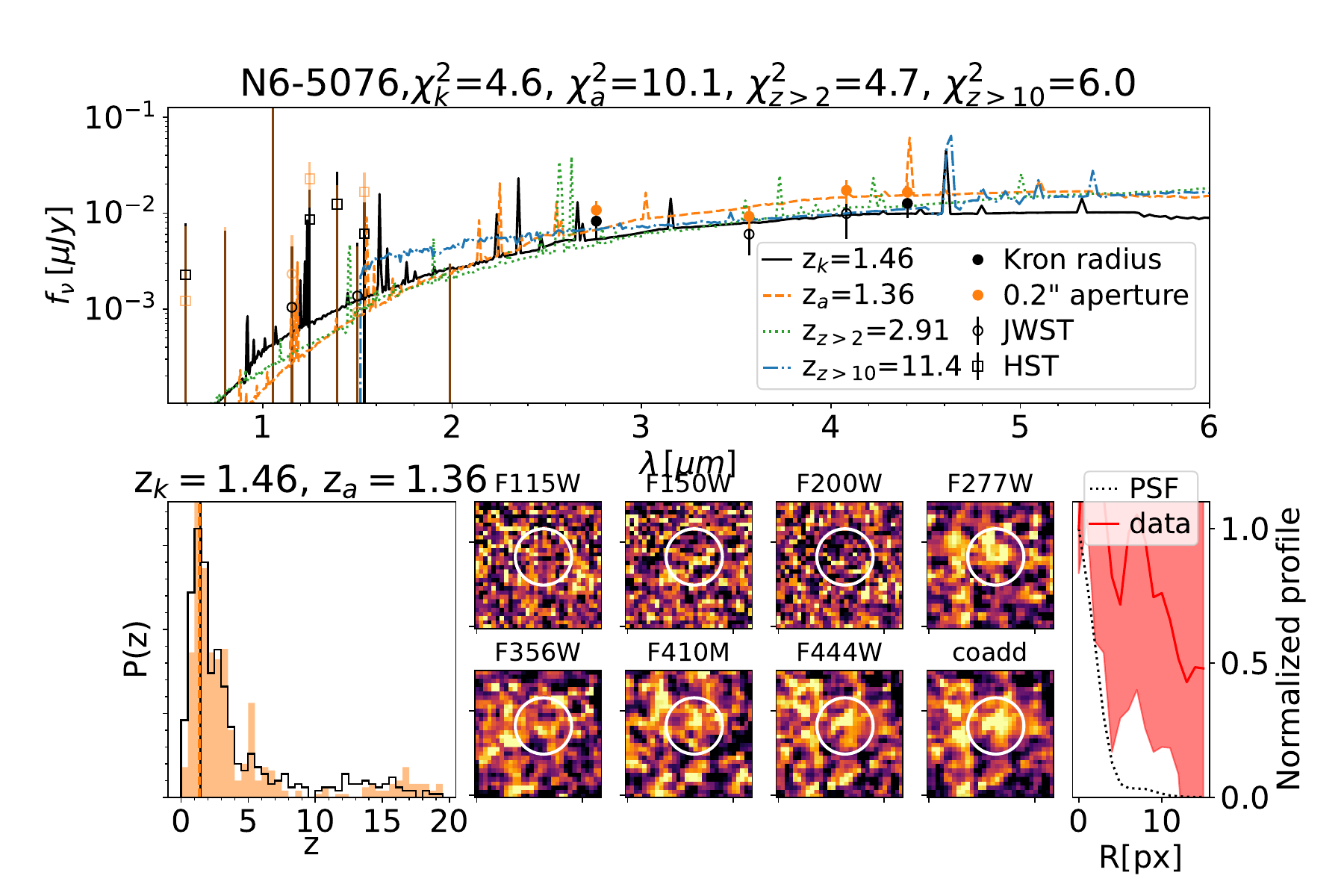}&
    \includegraphics[trim={20 10 50 40},clip,width=0.44\linewidth,keepaspectratio]{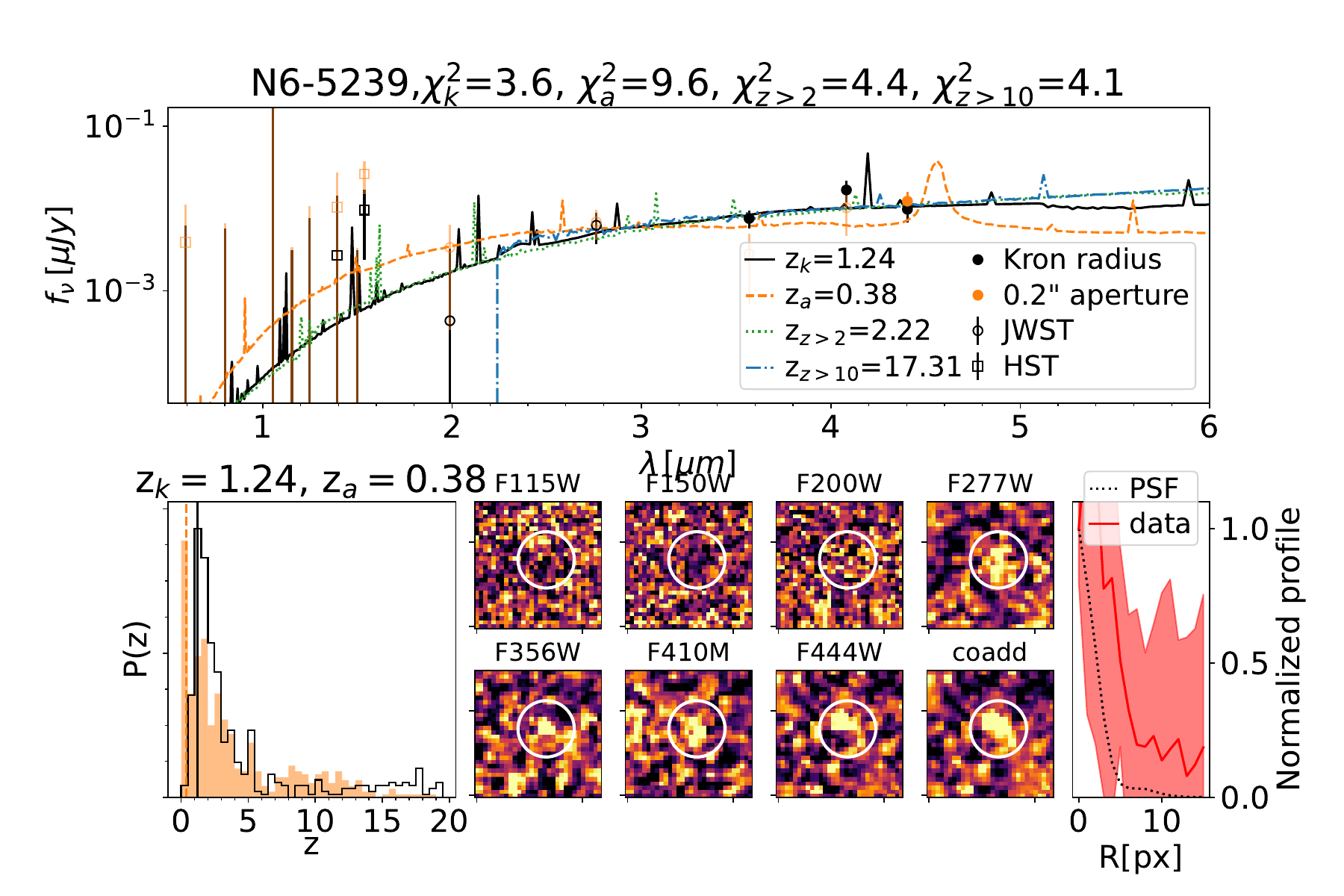}\\
      \caption{continued.}\\
    \includegraphics[trim={20 10 50 40},clip,width=0.44\linewidth,keepaspectratio]{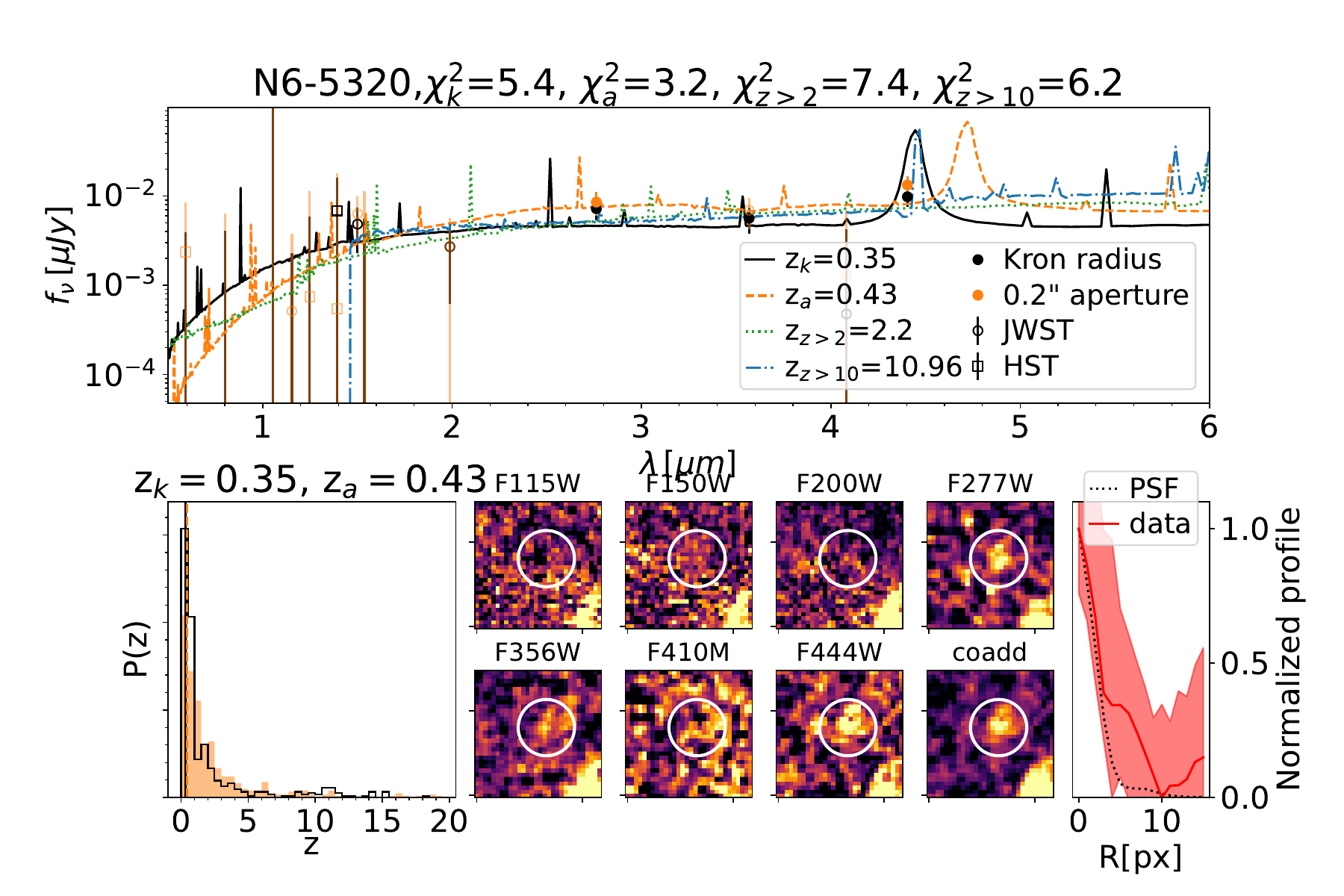}&
    \includegraphics[trim={20 10 50 40},clip,width=0.44\linewidth,keepaspectratio]{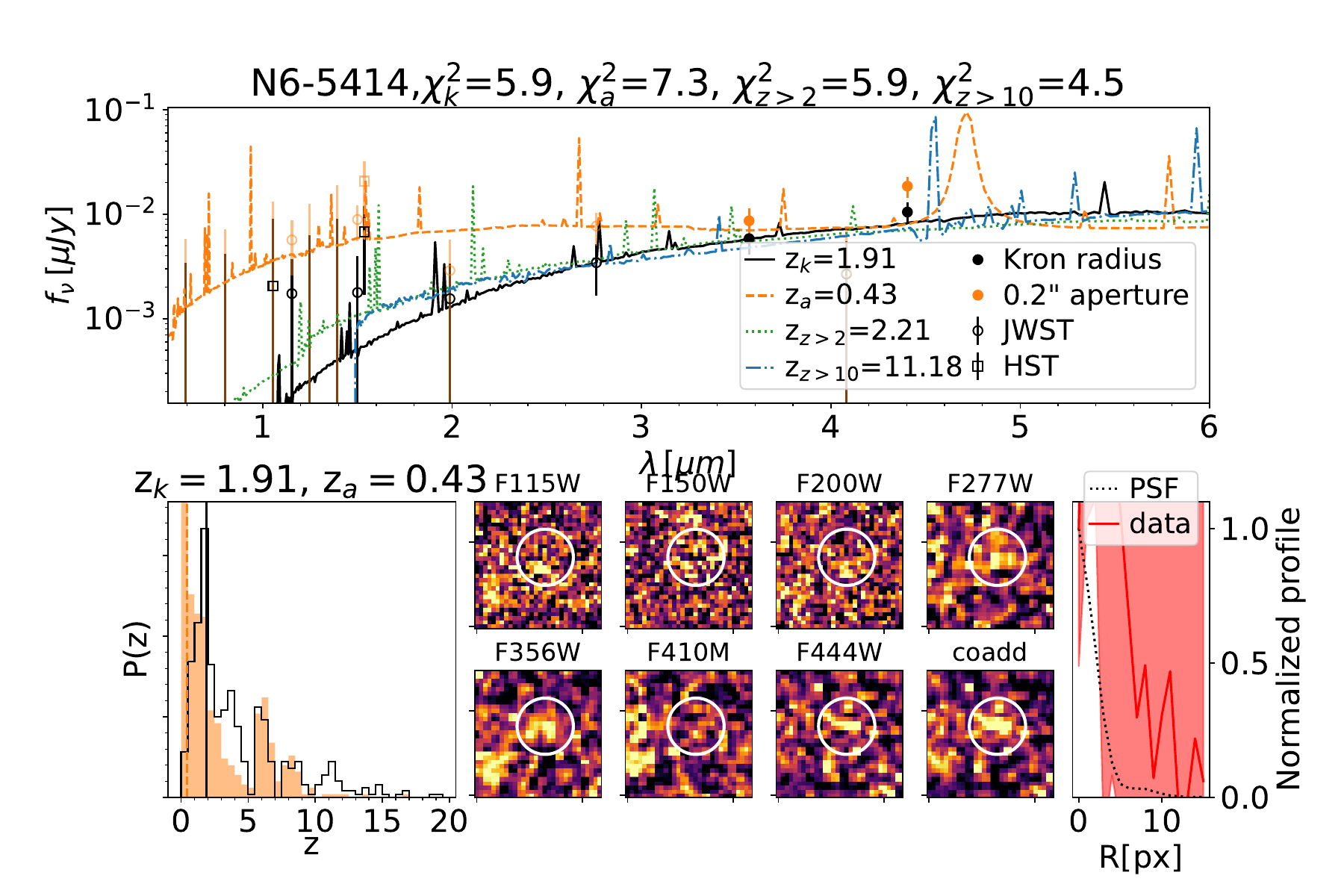}\\
    \includegraphics[trim={20 10 50 40},clip,width=0.44\linewidth,keepaspectratio]{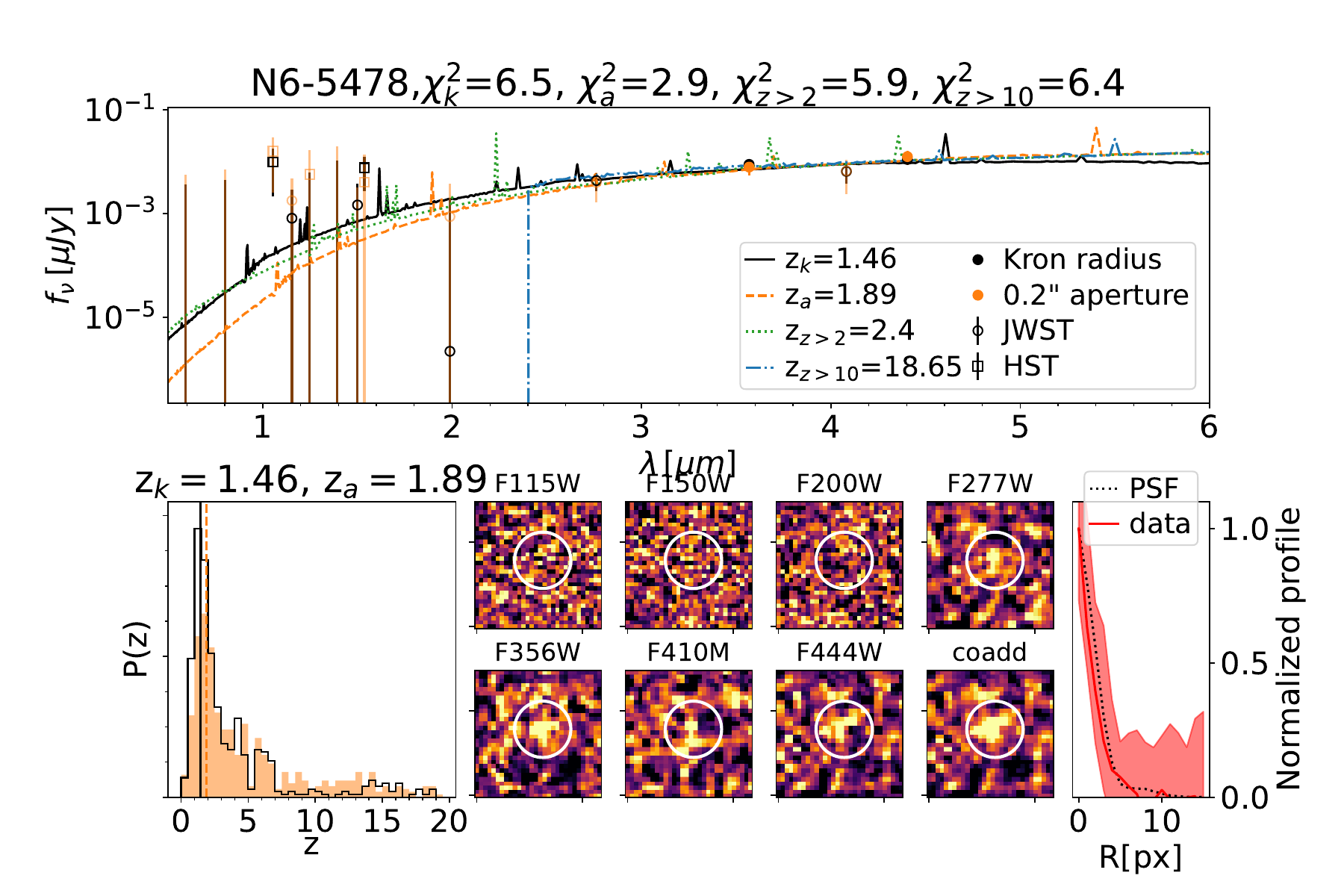}&
    \includegraphics[trim={20 10 50 40},clip,width=0.44\linewidth,keepaspectratio]{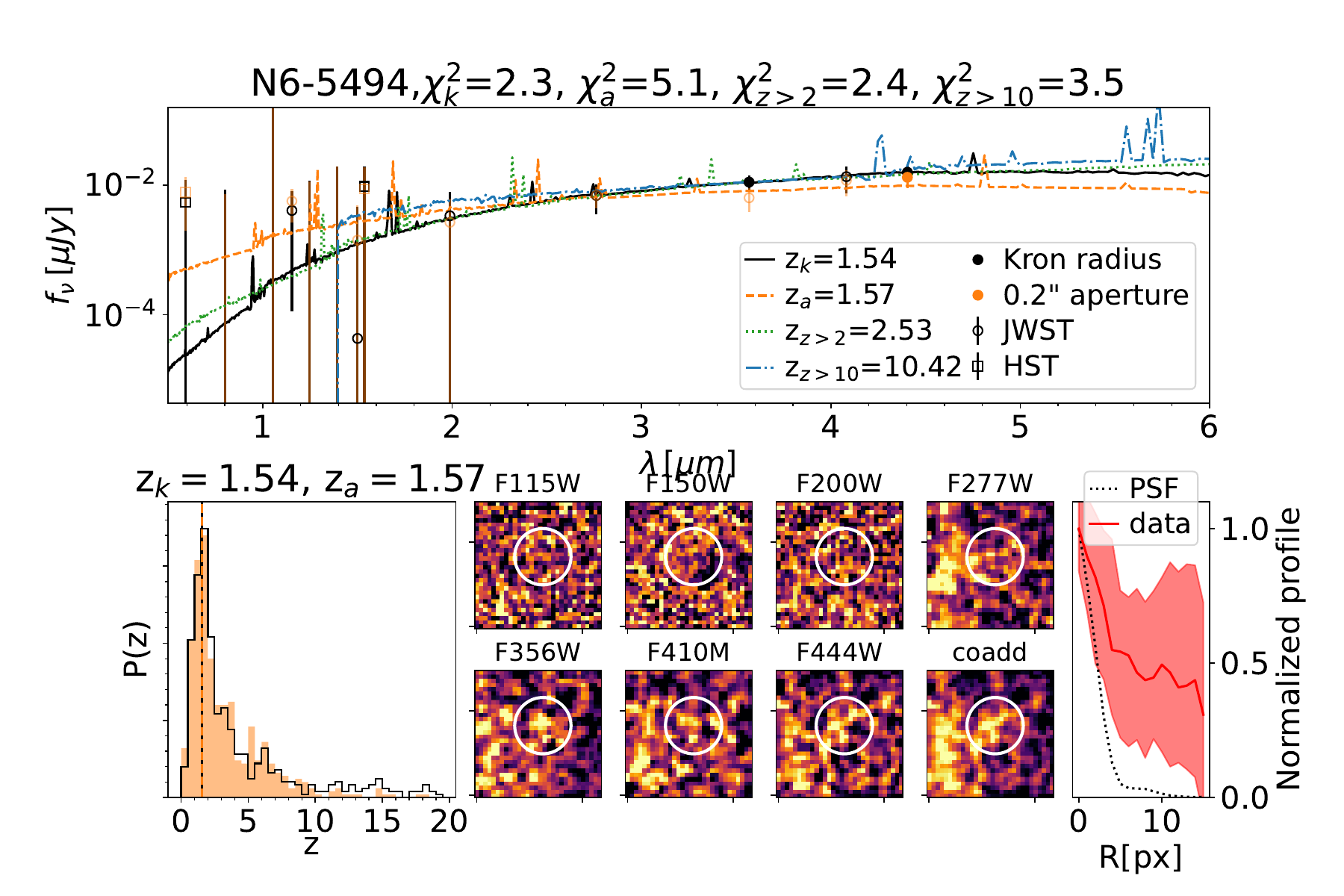}\\
    \includegraphics[trim={20 10 50 40},clip,width=0.44\linewidth,keepaspectratio]{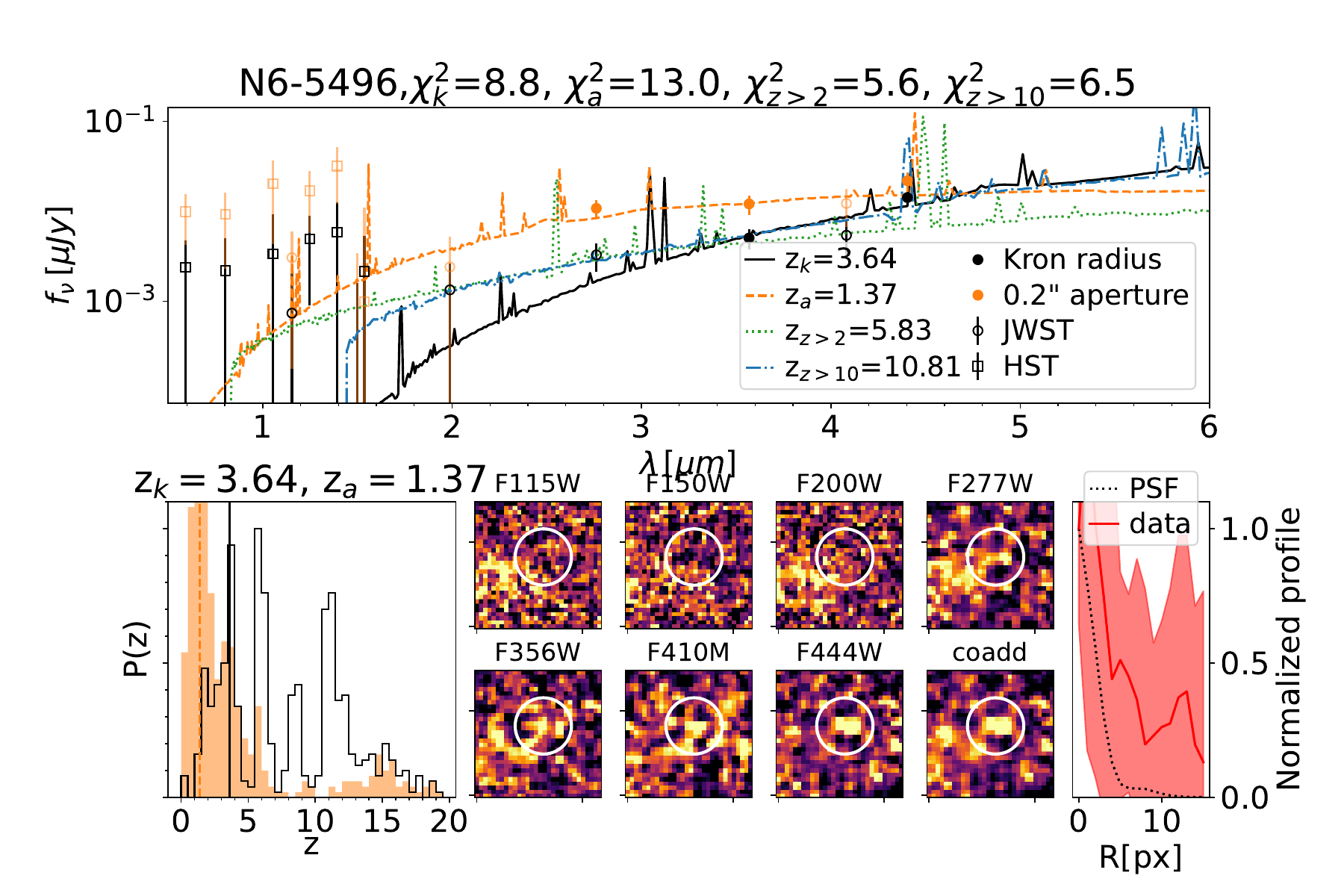}&
    \includegraphics[trim={20 10 50 40},clip,width=0.44\linewidth,keepaspectratio]{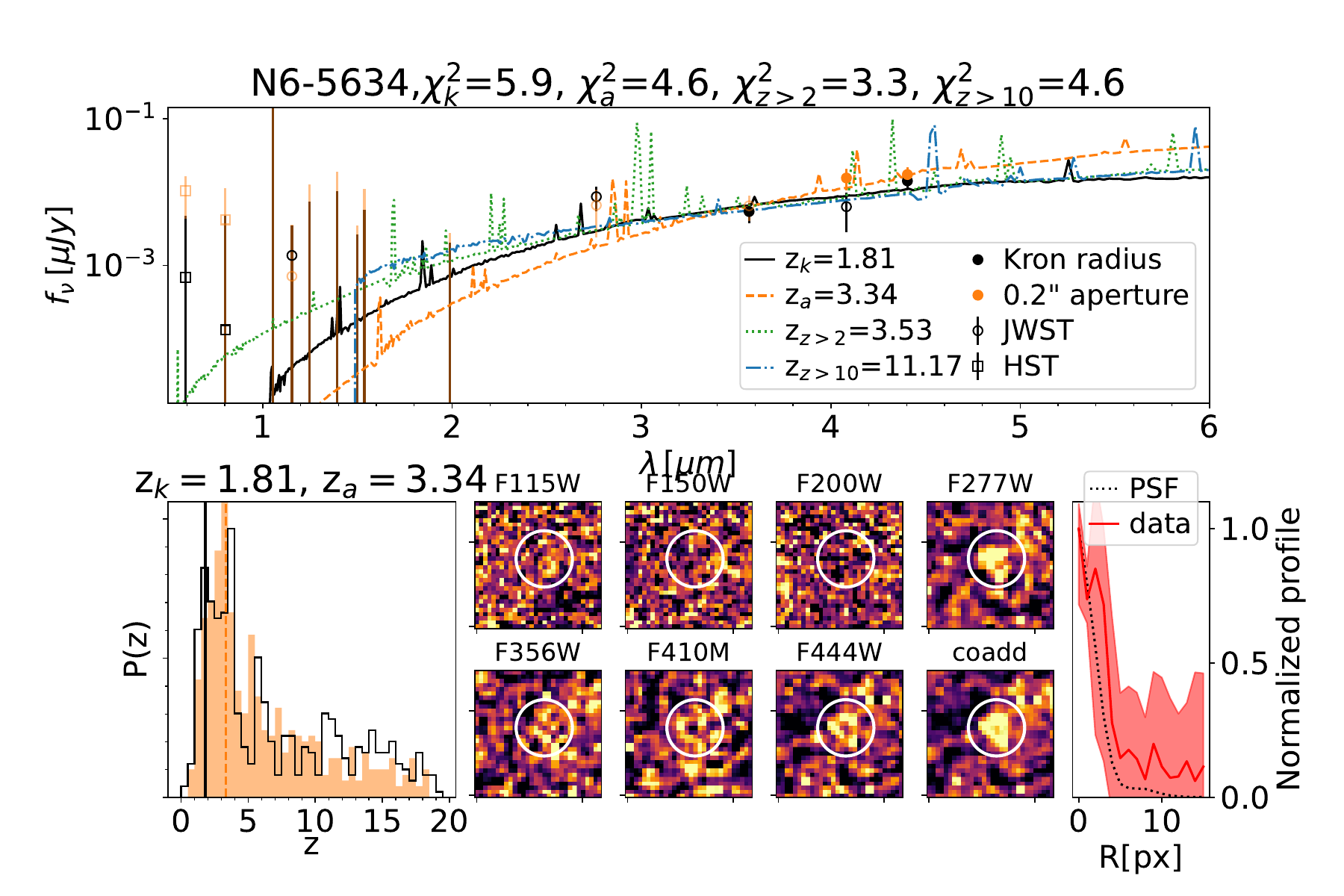}\\
    \includegraphics[trim={20 10 50 40},clip,width=0.44\linewidth,keepaspectratio]{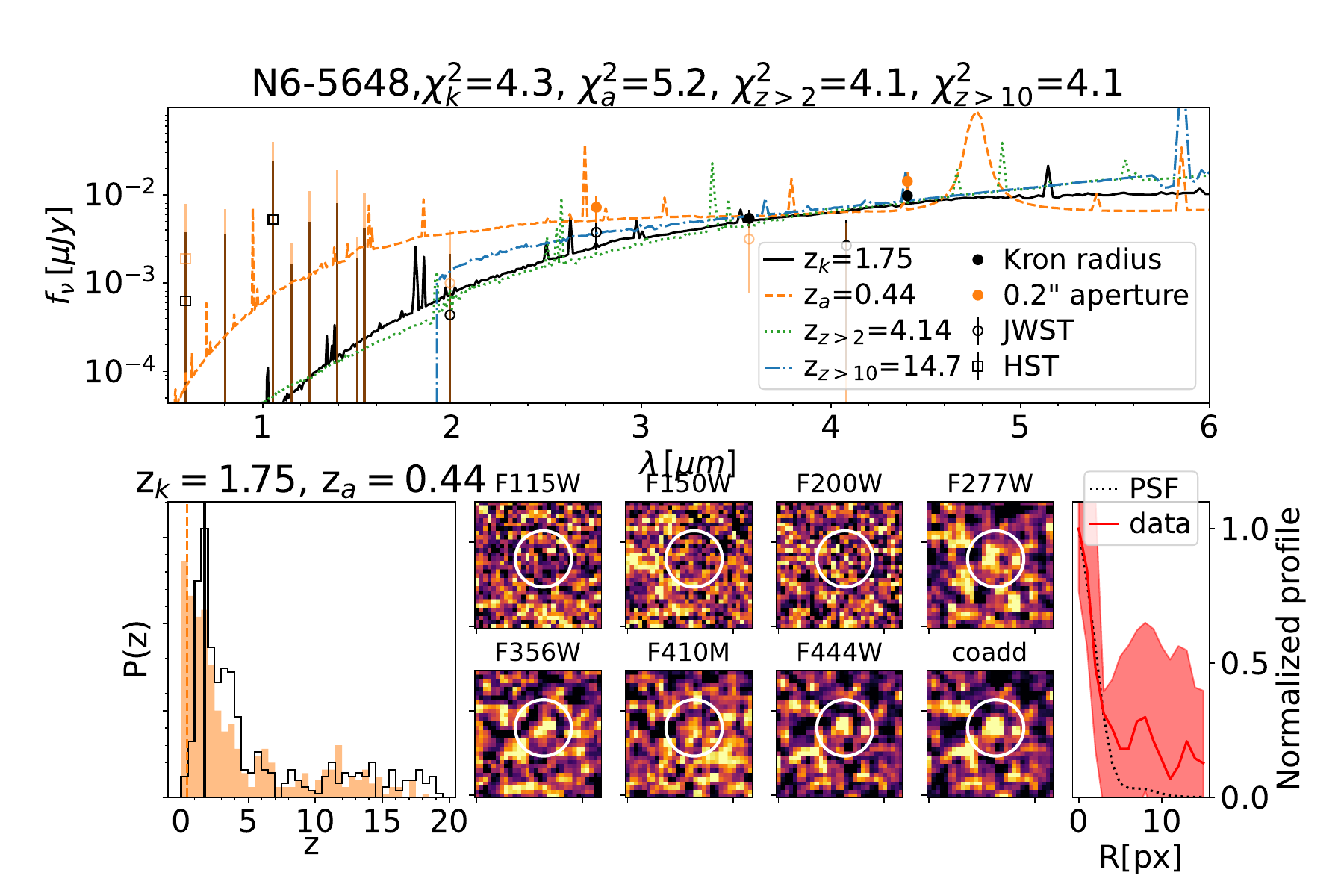}&
    \includegraphics[trim={20 10 50 40},clip,width=0.44\linewidth,keepaspectratio]{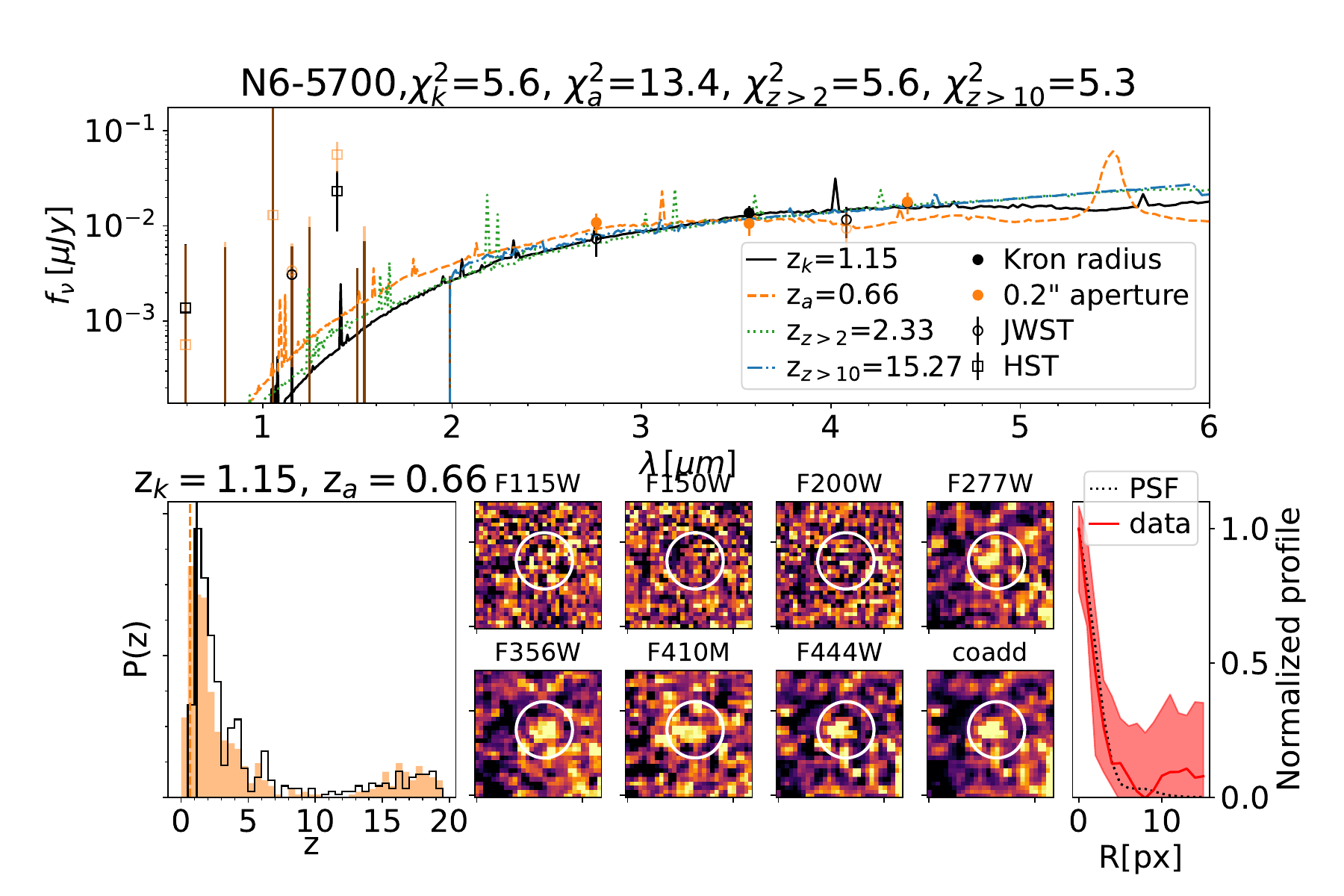}\\
      \caption{continued.}\\
    \includegraphics[trim={20 10 50 40},clip,width=0.44\linewidth,keepaspectratio]{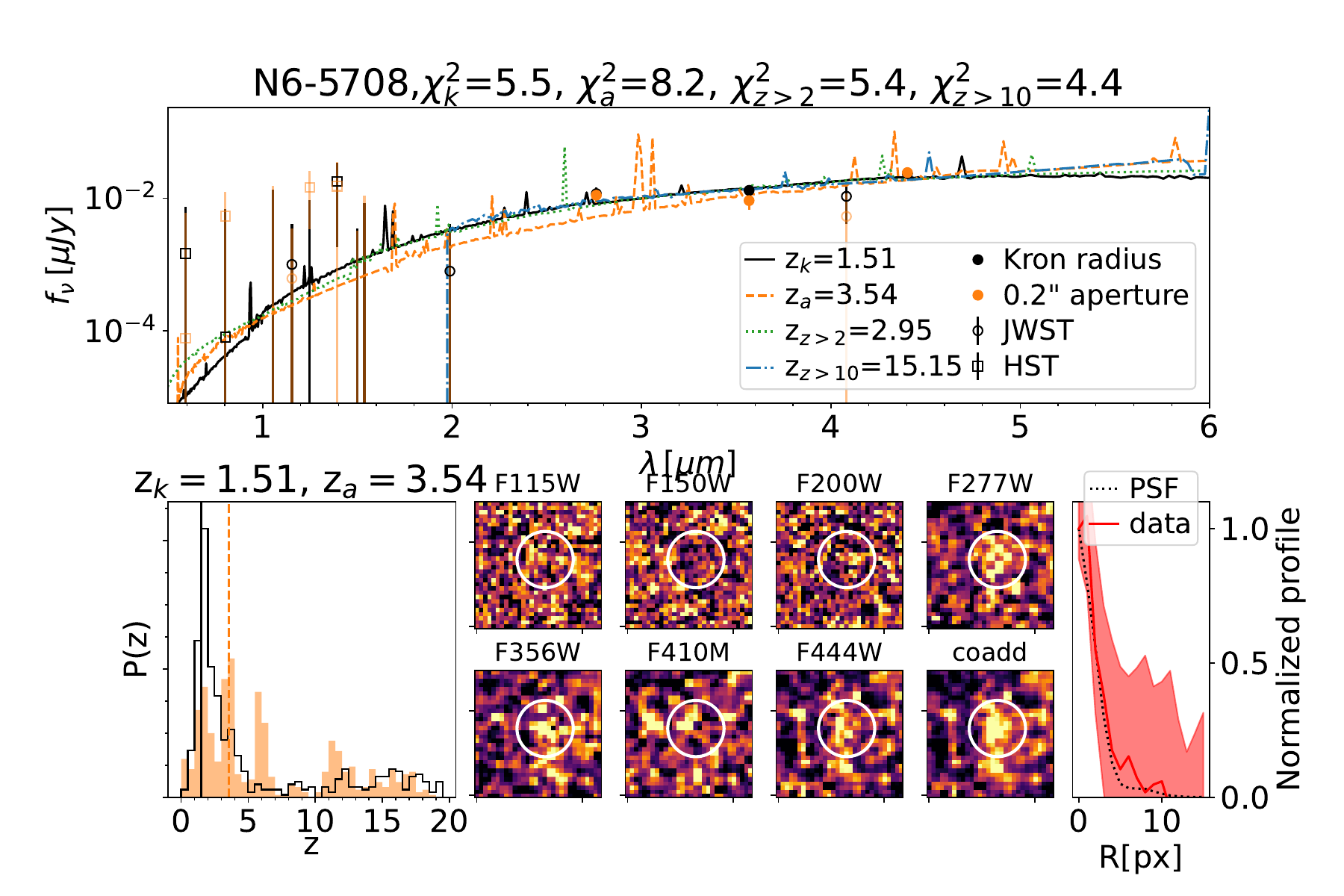}&
    \includegraphics[trim={20 10 50 40},clip,width=0.44\linewidth,keepaspectratio]{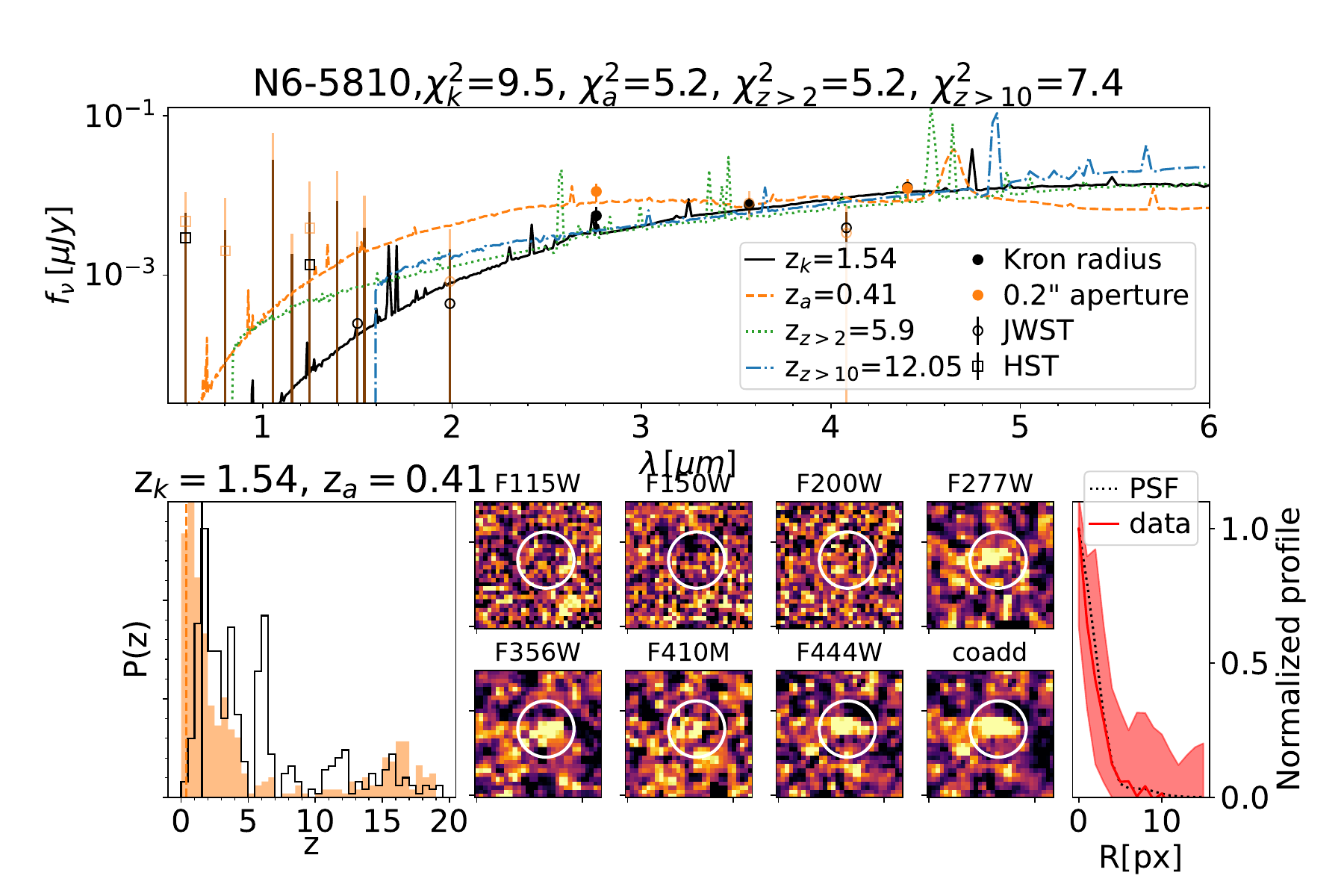}\\
    \includegraphics[trim={20 10 50 40},clip,width=0.44\linewidth,keepaspectratio]{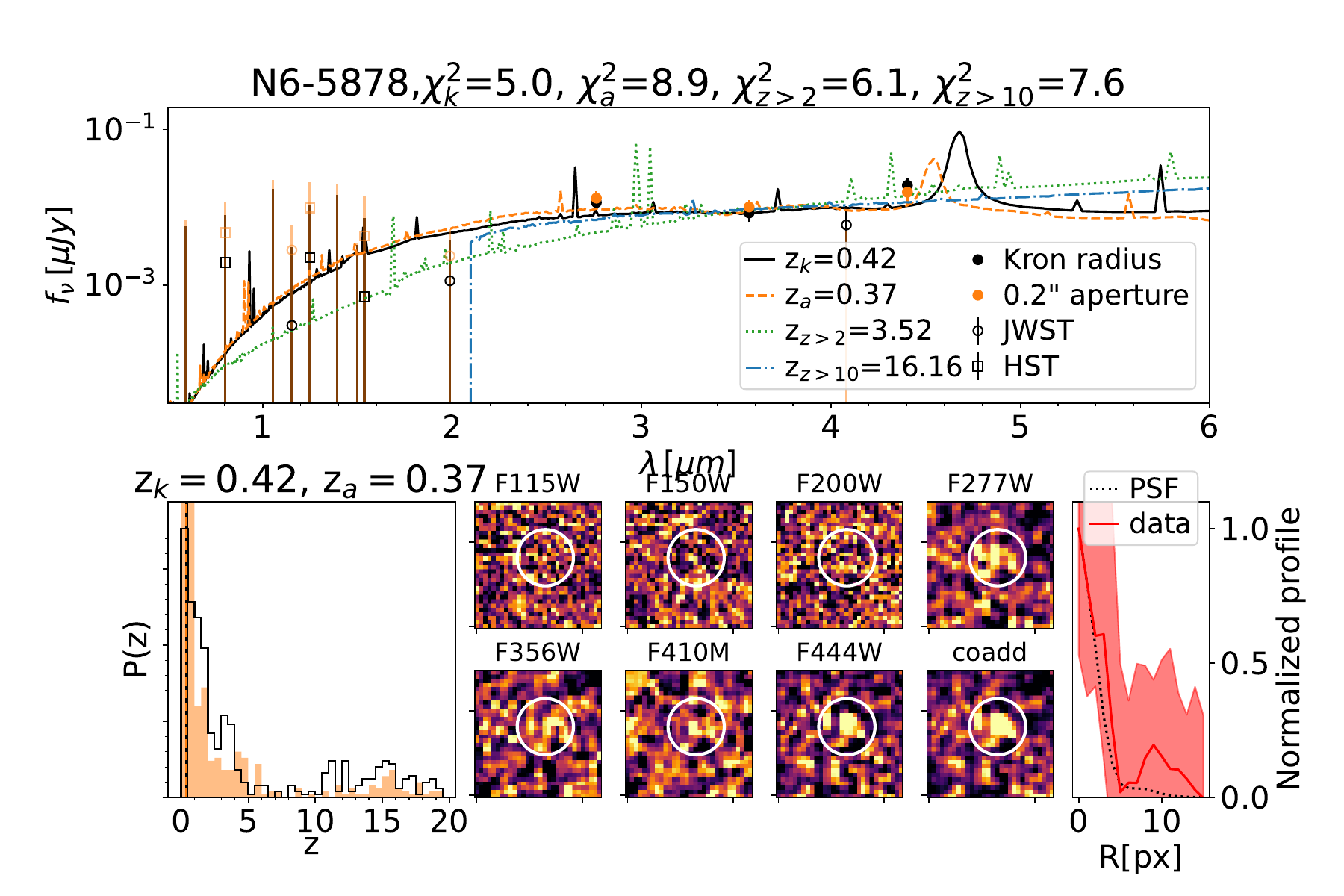}&
    \includegraphics[trim={20 10 50 40},clip,width=0.44\linewidth,keepaspectratio]{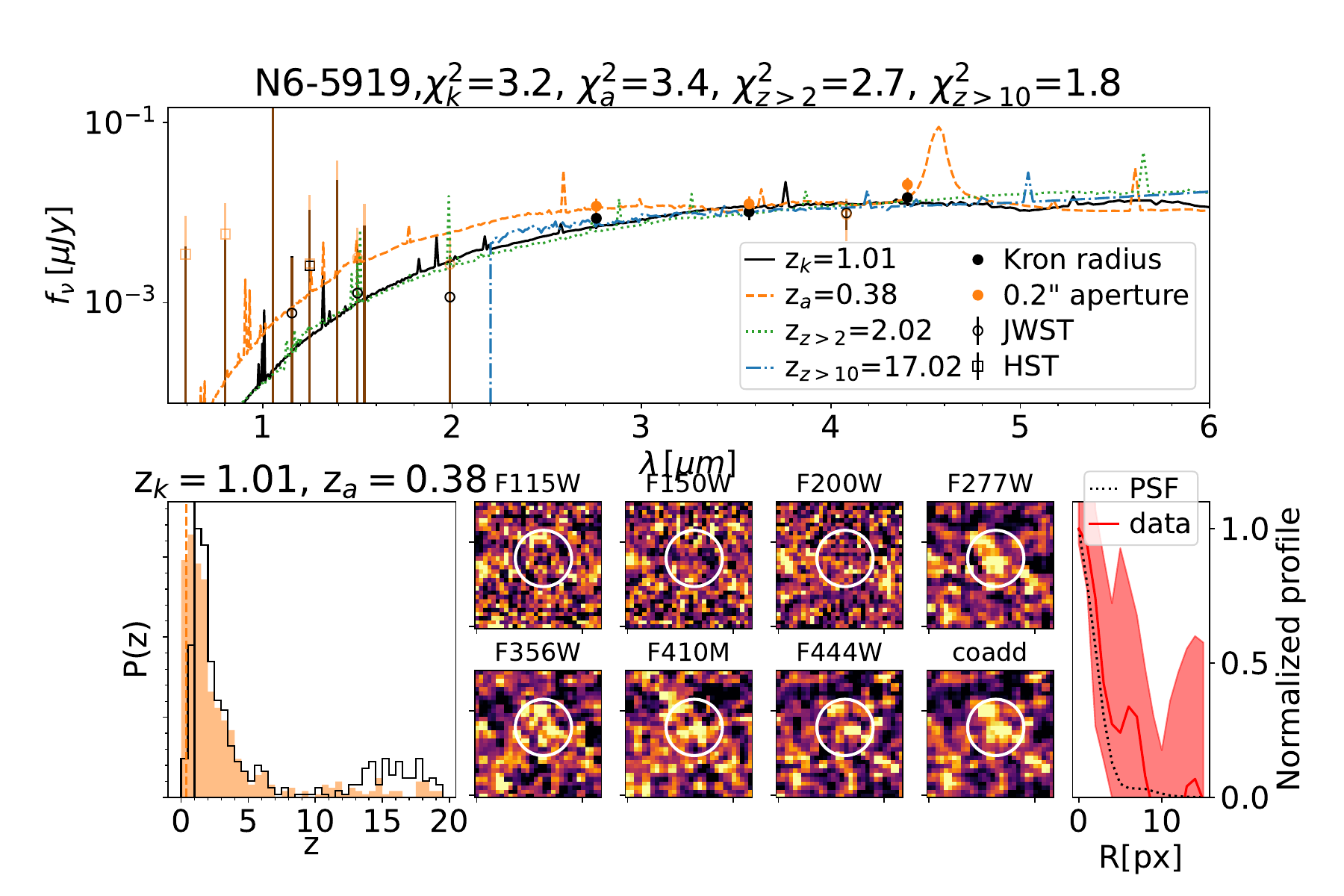}\\
    \includegraphics[trim={20 10 50 40},clip,width=0.44\linewidth,keepaspectratio]{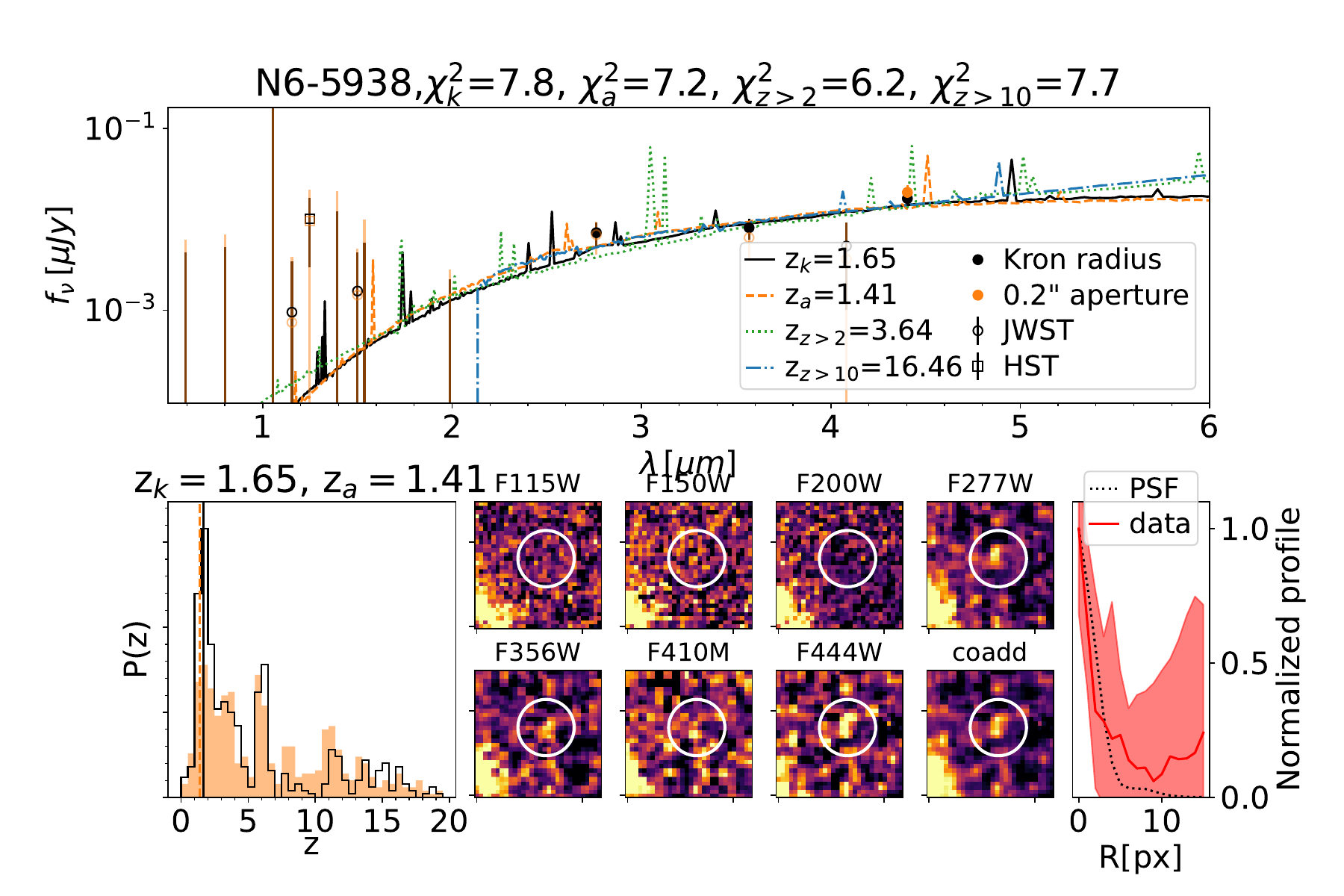}&
    \includegraphics[trim={20 10 50 40},clip,width=0.44\linewidth,keepaspectratio]{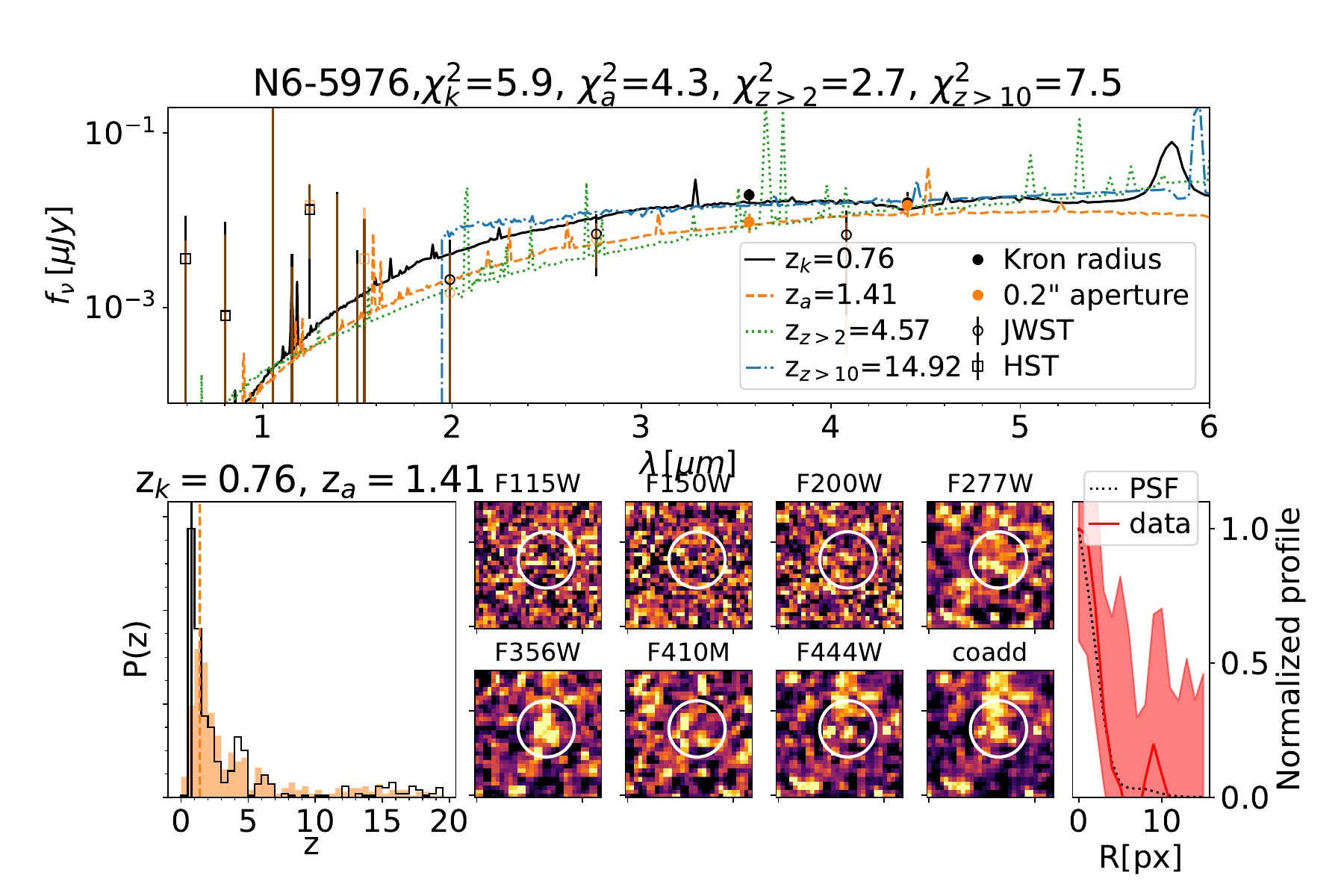}\\
    \includegraphics[trim={20 10 50 40},clip,width=0.44\linewidth,keepaspectratio]{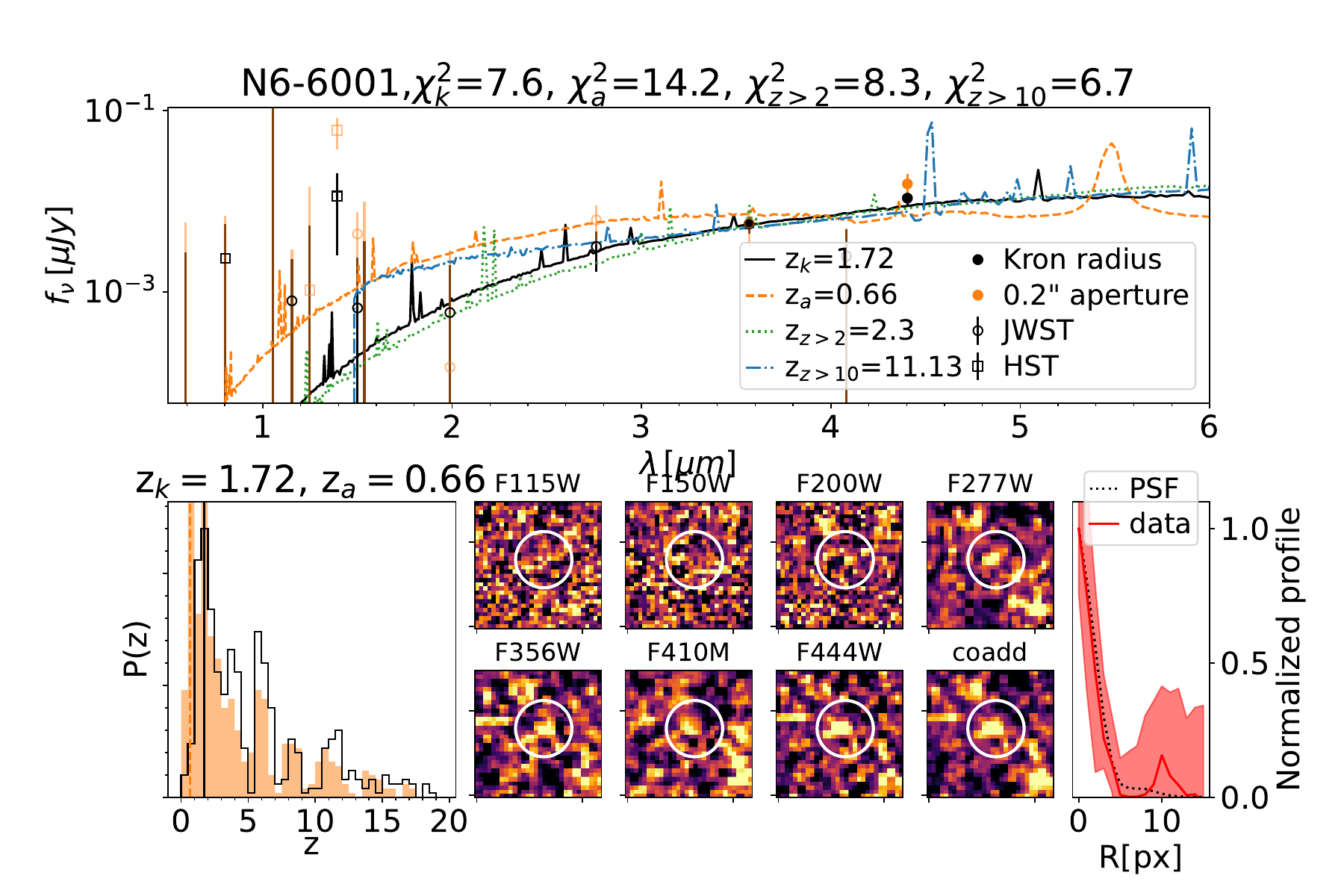}&
    \includegraphics[trim={20 10 50 40},clip,width=0.44\linewidth,keepaspectratio]{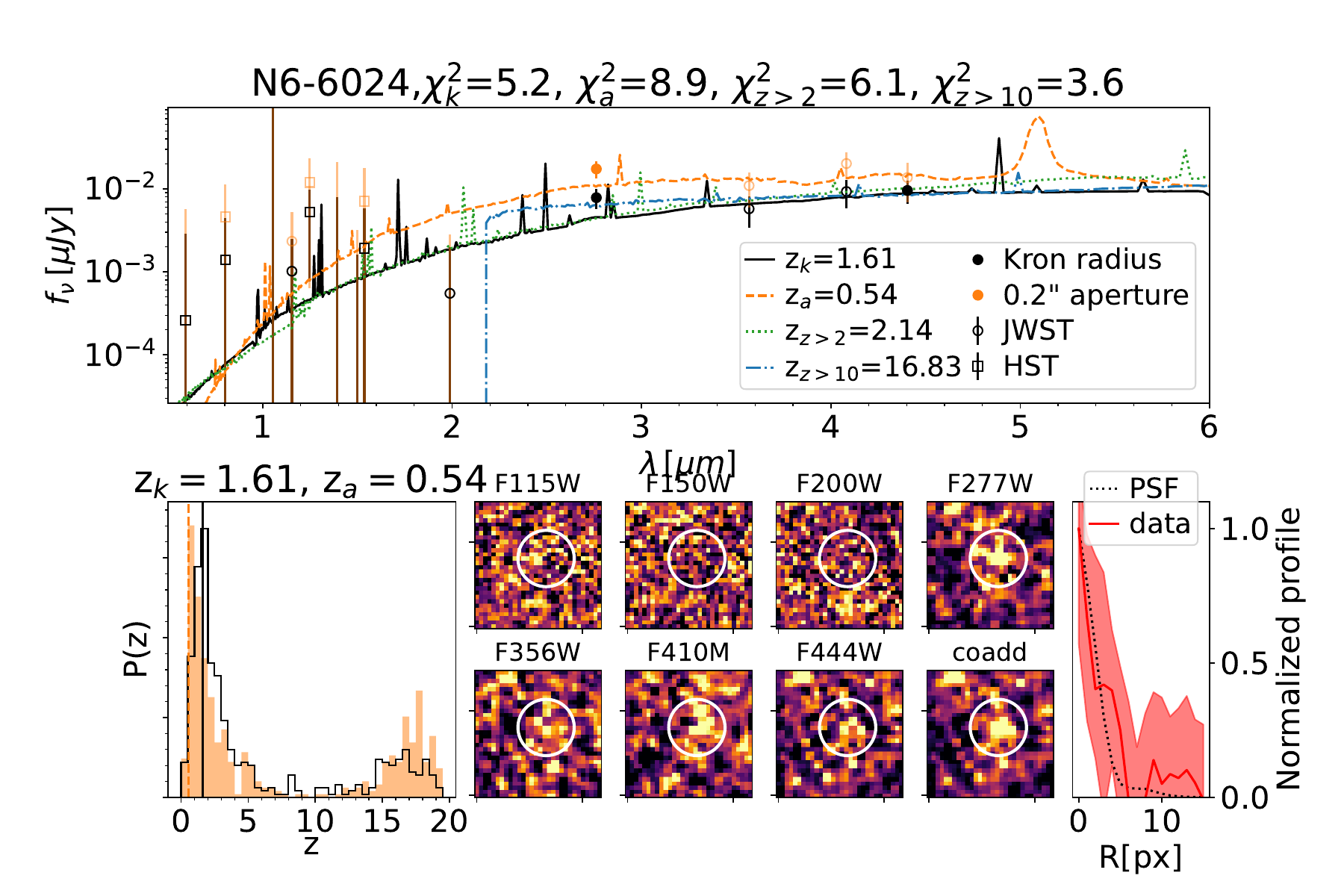}\\
      \caption{continued.}\\
    \includegraphics[trim={20 10 50 40},clip,width=0.44\linewidth,keepaspectratio]{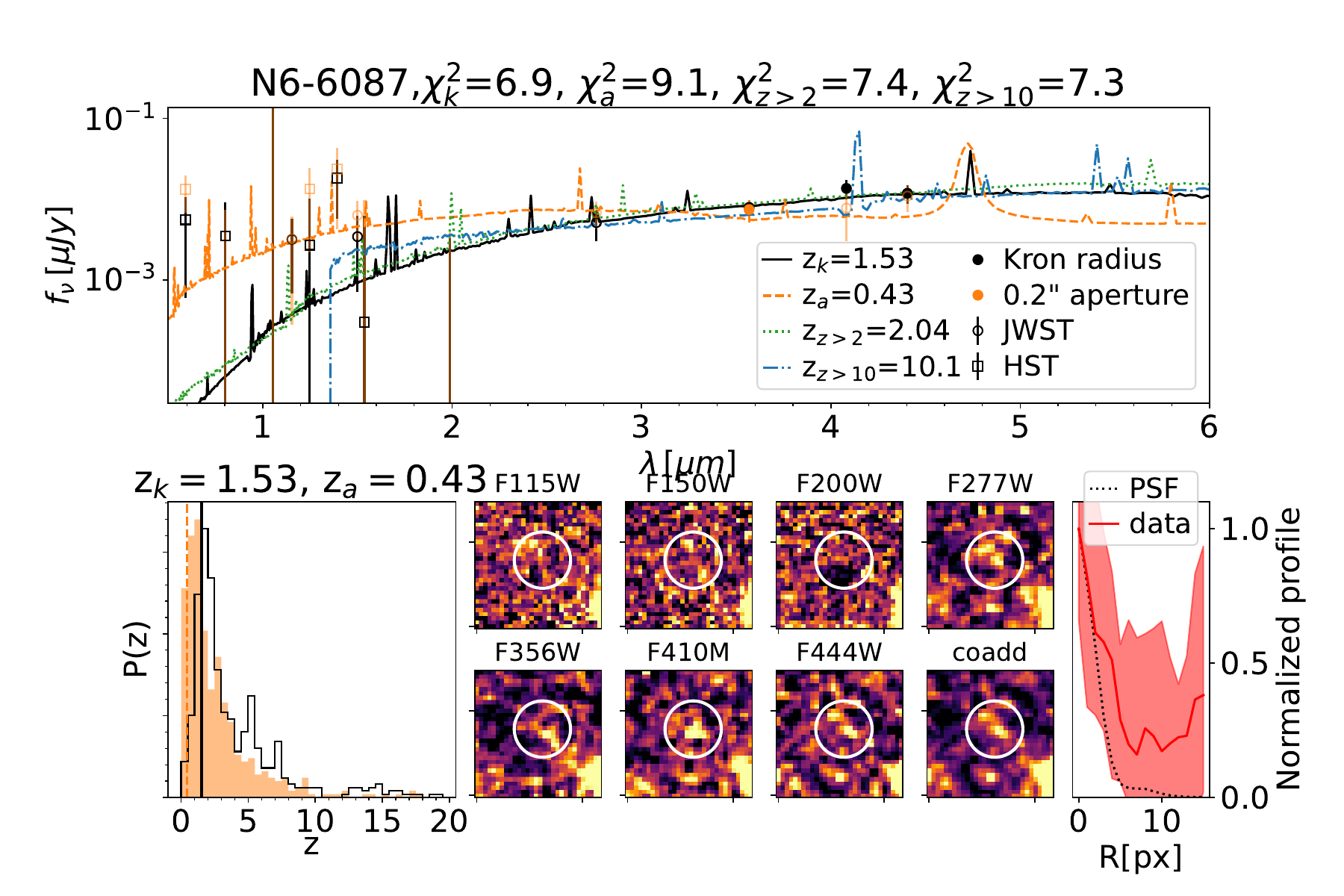}&
    \includegraphics[trim={20 10 50 40},clip,width=0.44\linewidth,keepaspectratio]{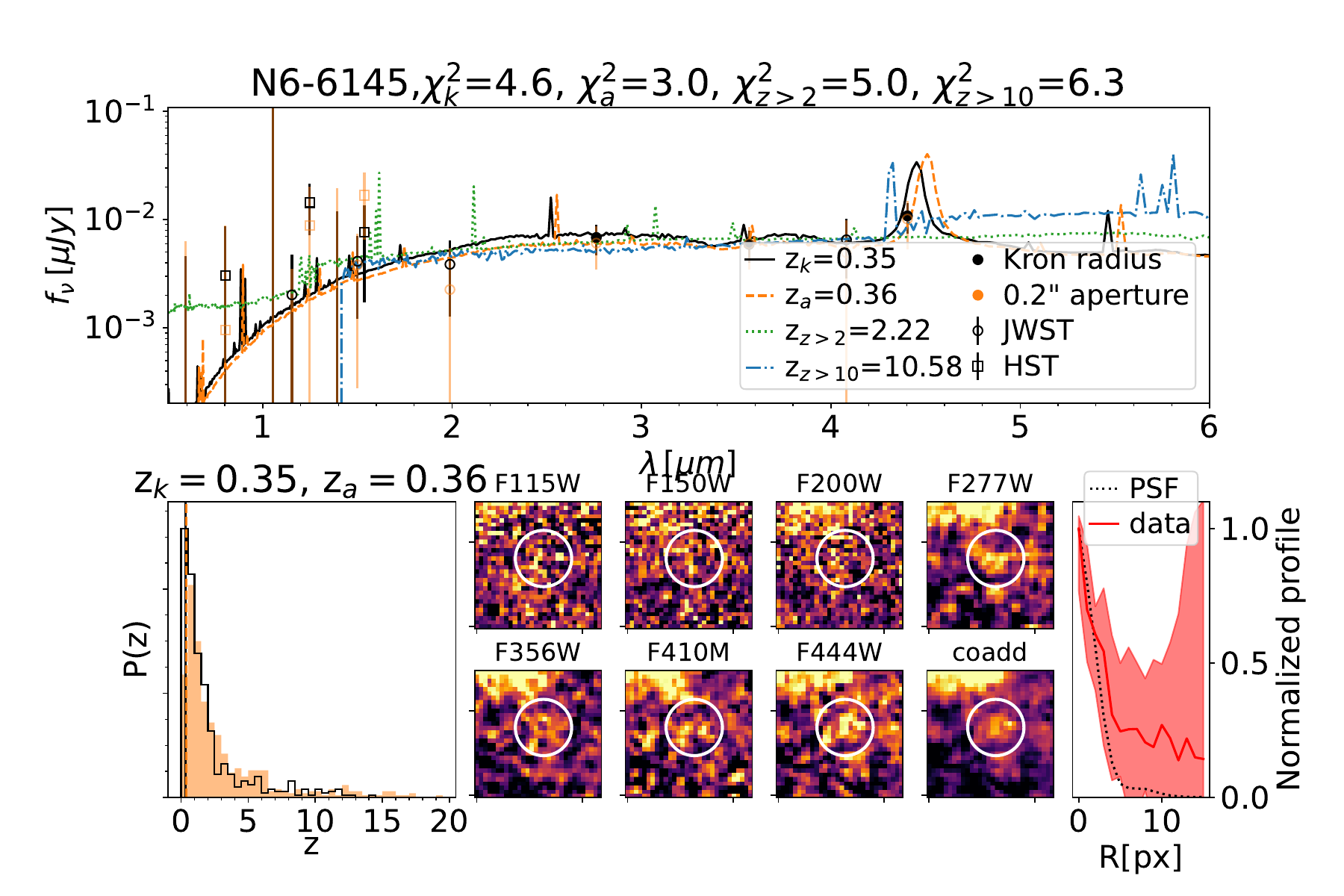}\\
    \includegraphics[trim={20 10 50 40},clip,width=0.44\linewidth,keepaspectratio]{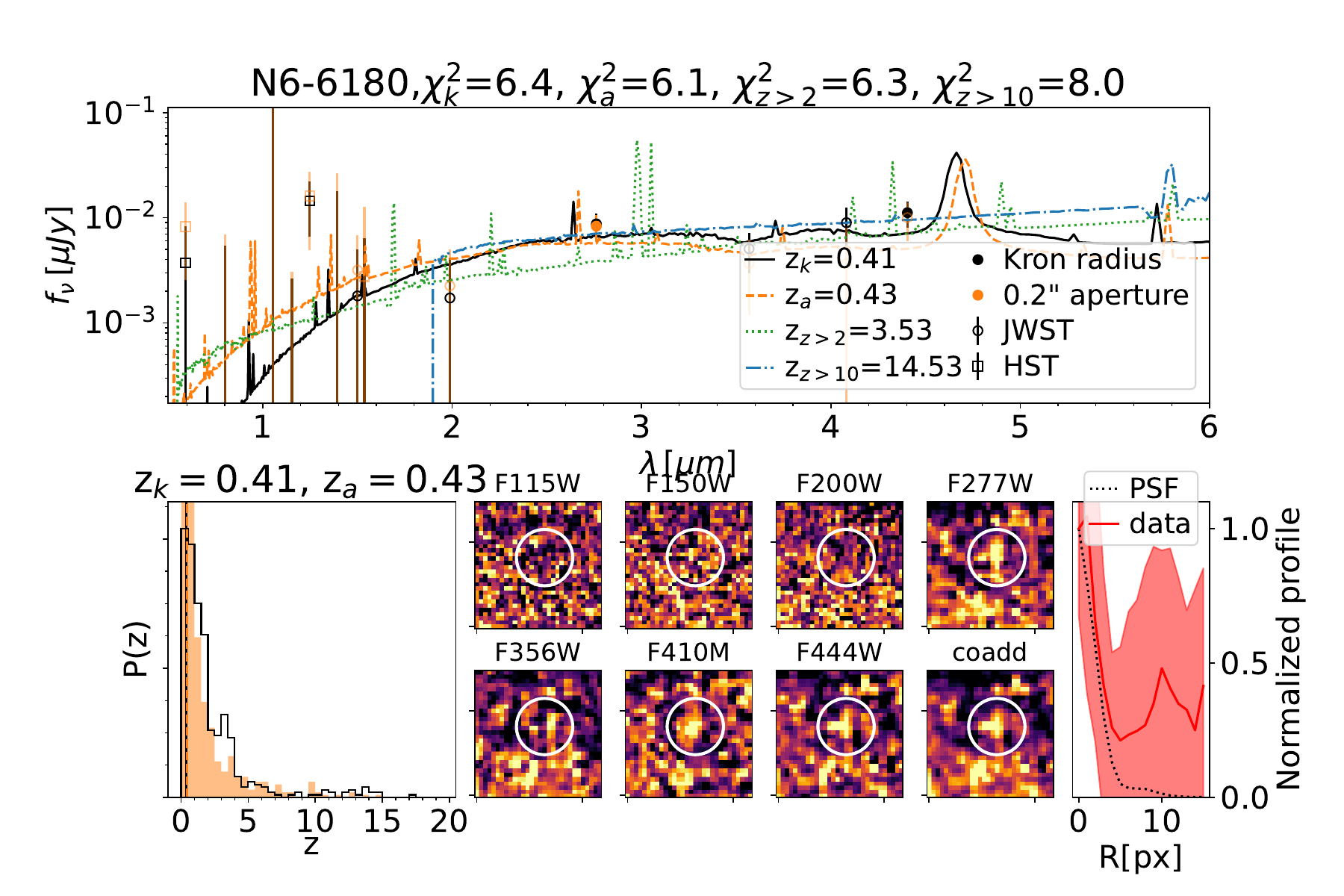}&
    \includegraphics[trim={20 10 50 40},clip,width=0.44\linewidth,keepaspectratio]{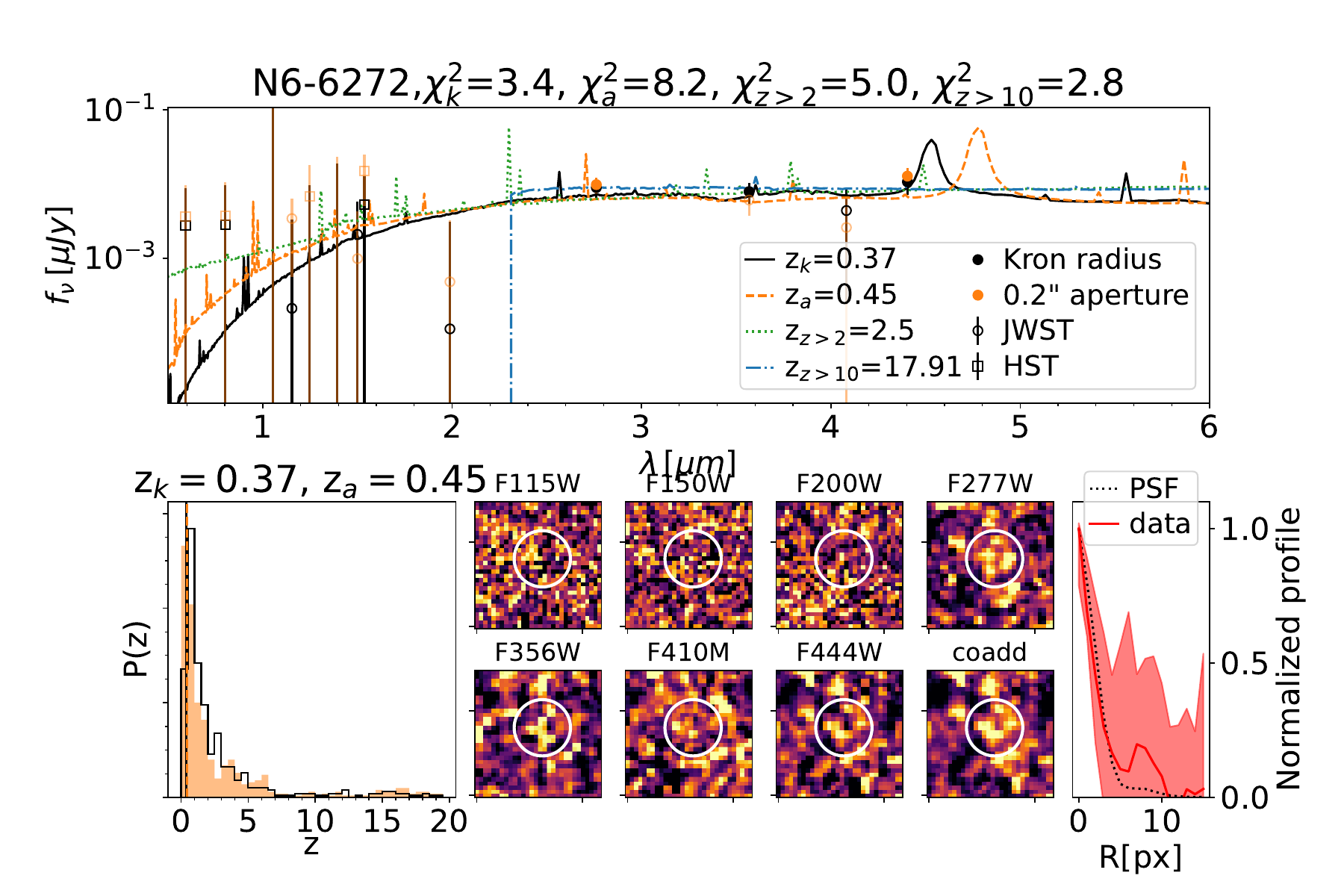}\\
    \includegraphics[trim={20 10 50 40},clip,width=0.44\linewidth,keepaspectratio]{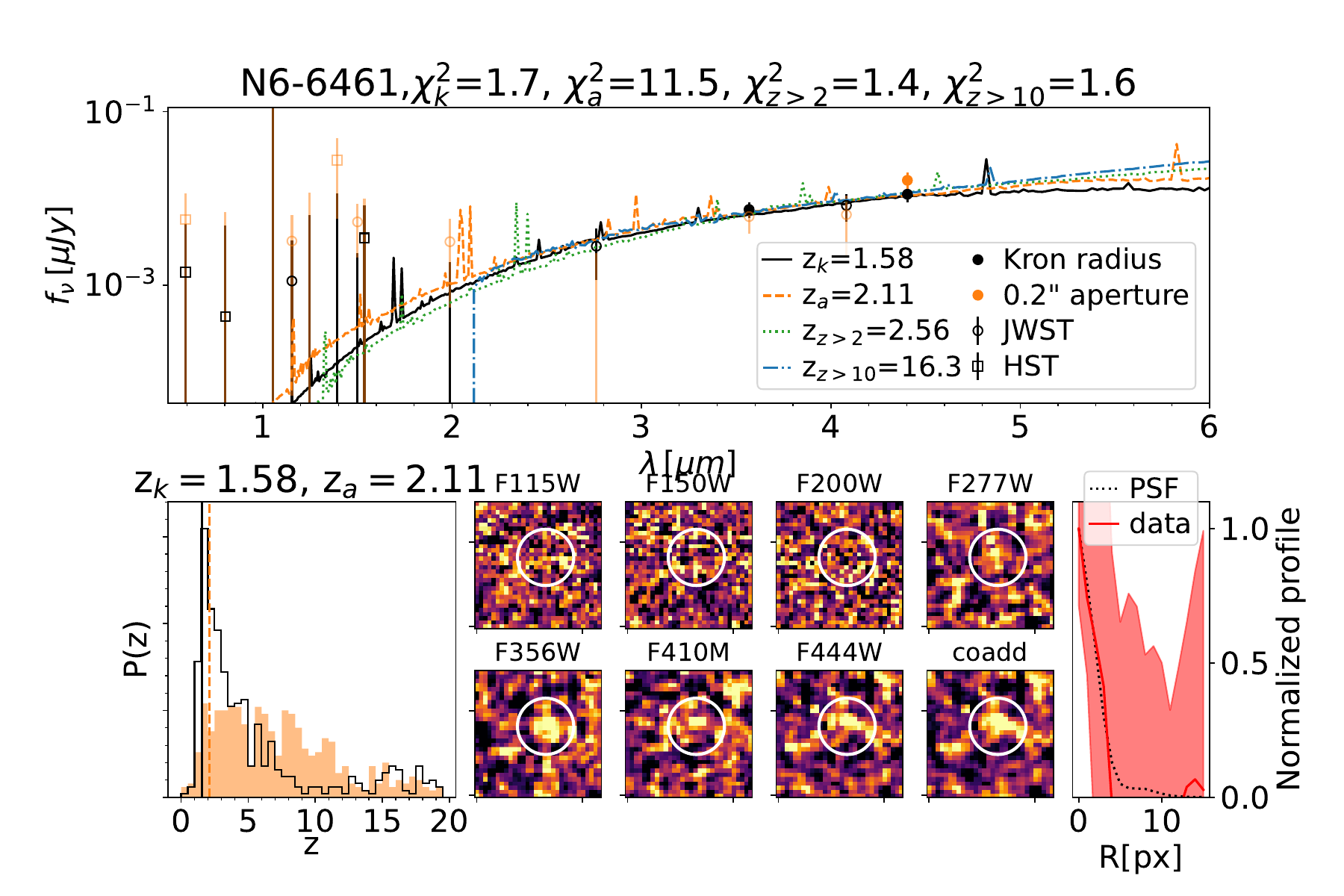}&
    \includegraphics[trim={20 10 50 40},clip,width=0.44\linewidth,keepaspectratio]{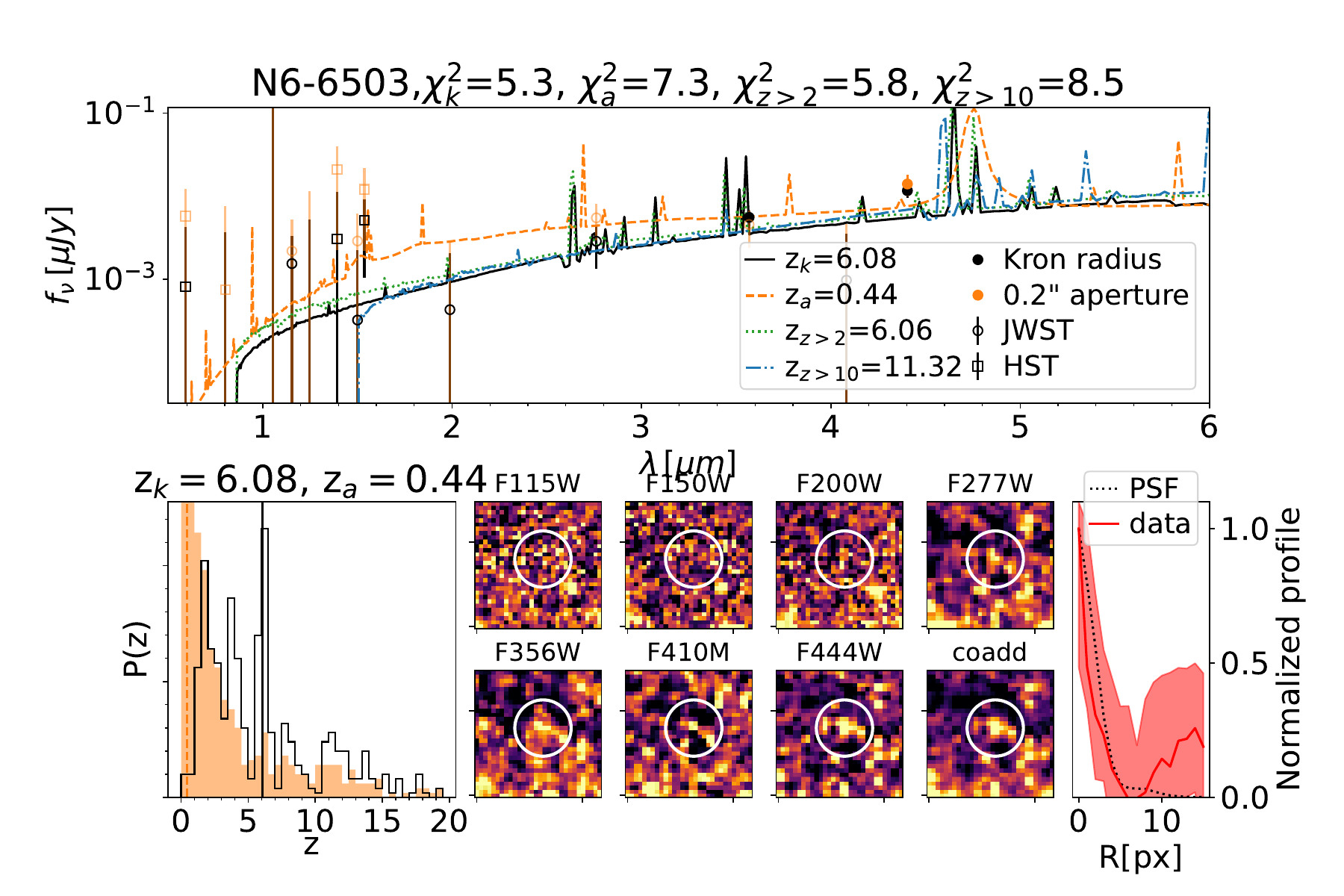}\\
    \includegraphics[trim={20 10 50 40},clip,width=0.44\linewidth,keepaspectratio]{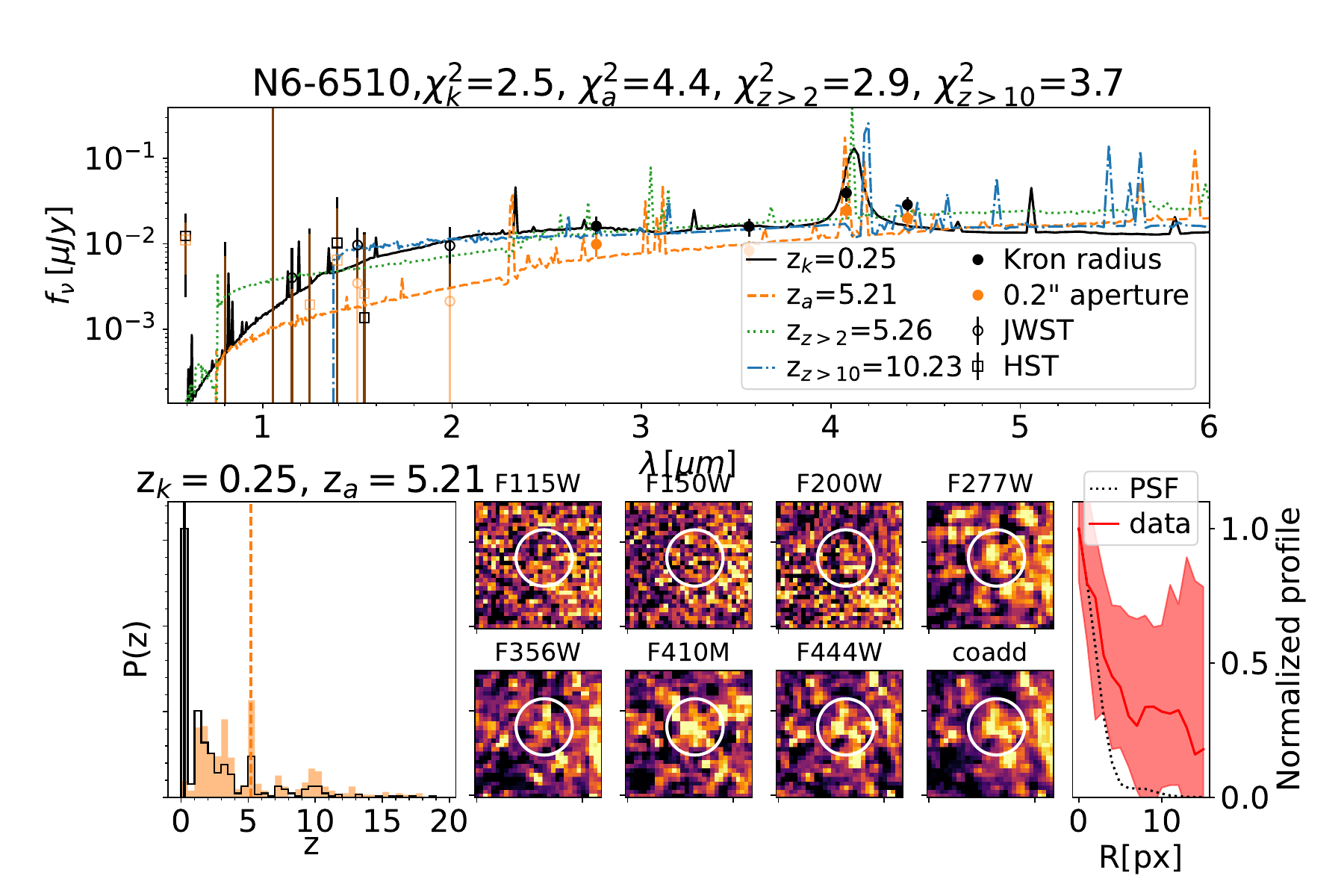}&
    \includegraphics[trim={20 10 50 40},clip,width=0.44\linewidth,keepaspectratio]{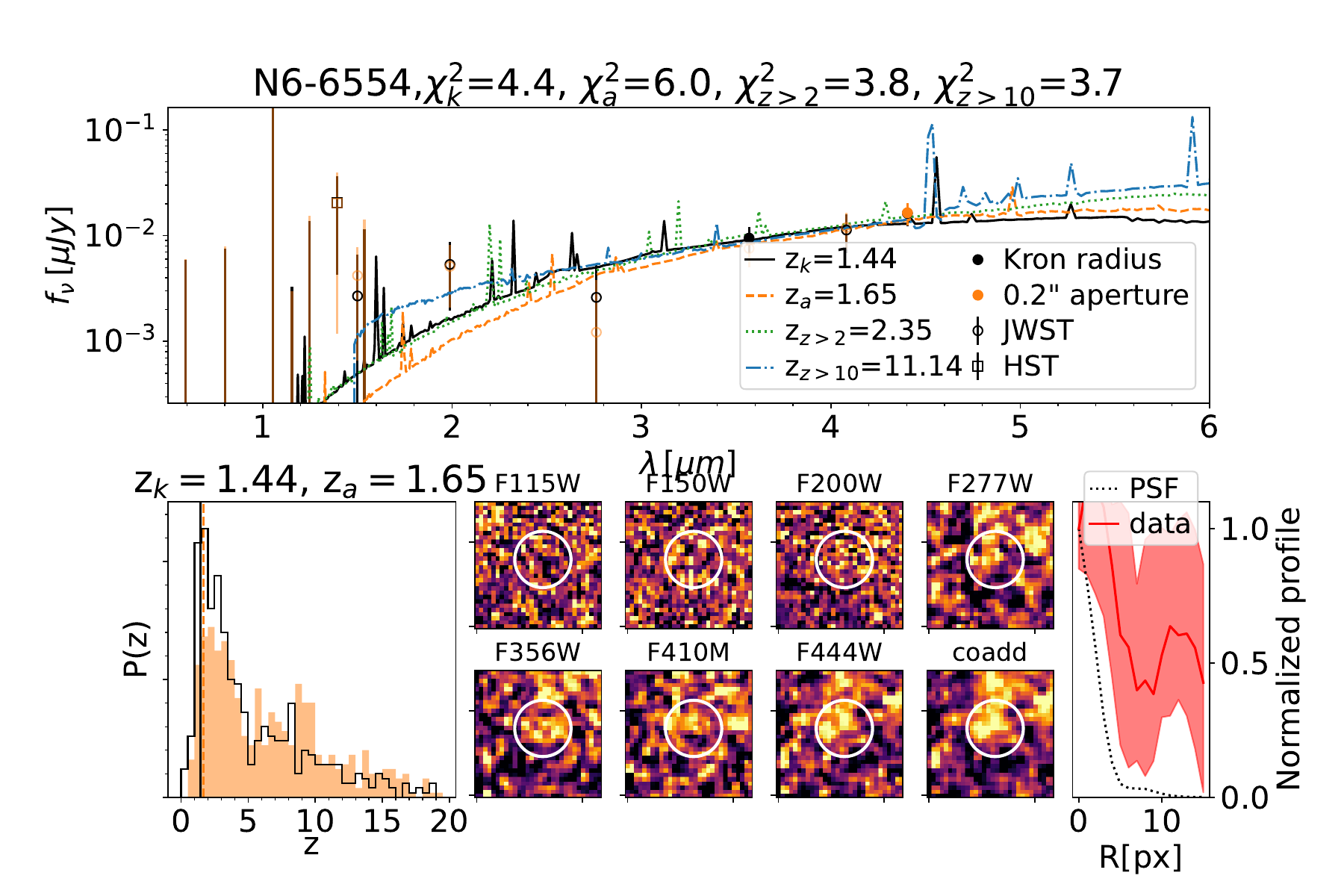}\\
      \caption{continued.}\\
    \includegraphics[trim={20 10 50 40},clip,width=0.44\linewidth,keepaspectratio]{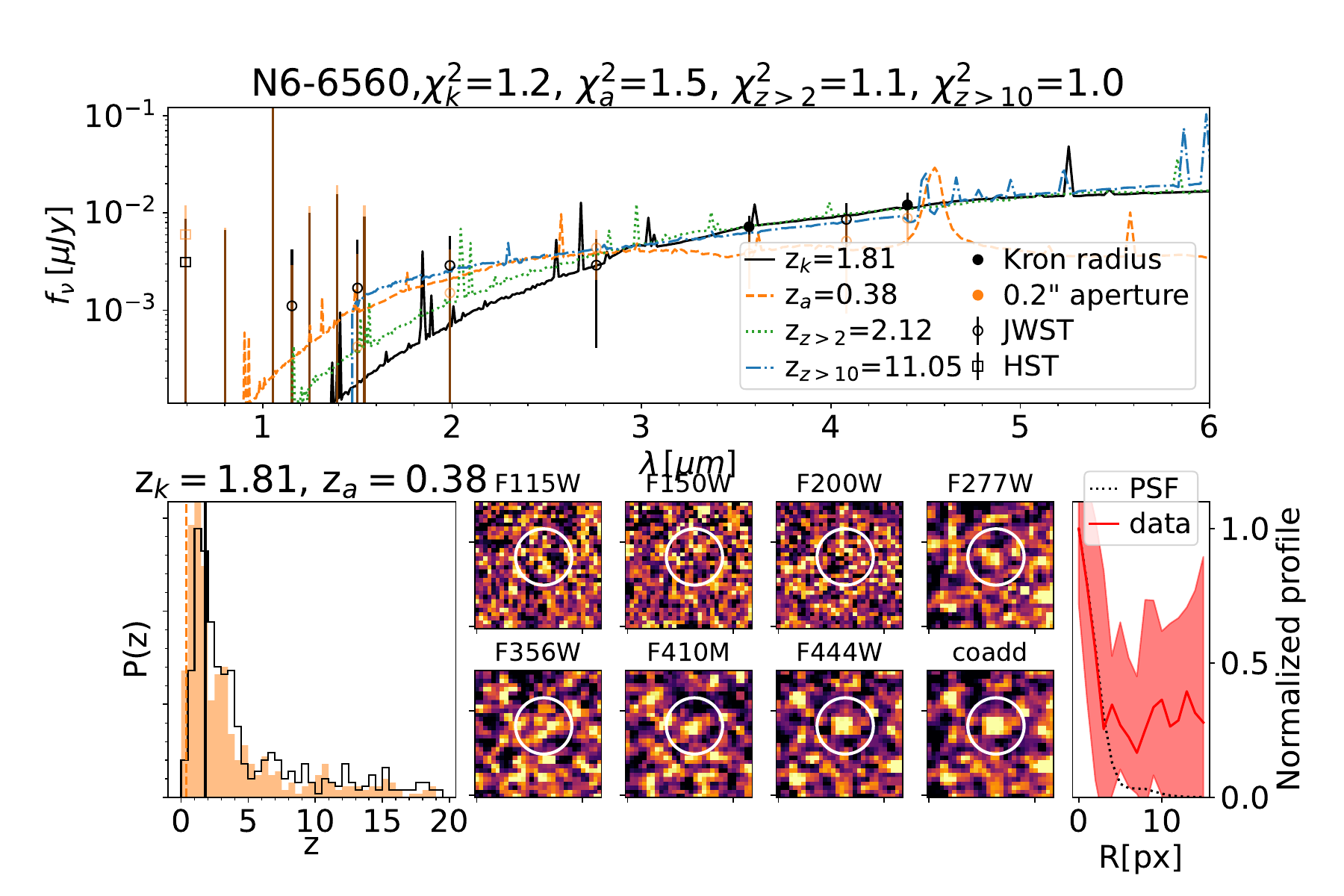}&
    \includegraphics[trim={20 10 50 40},clip,width=0.44\linewidth,keepaspectratio]{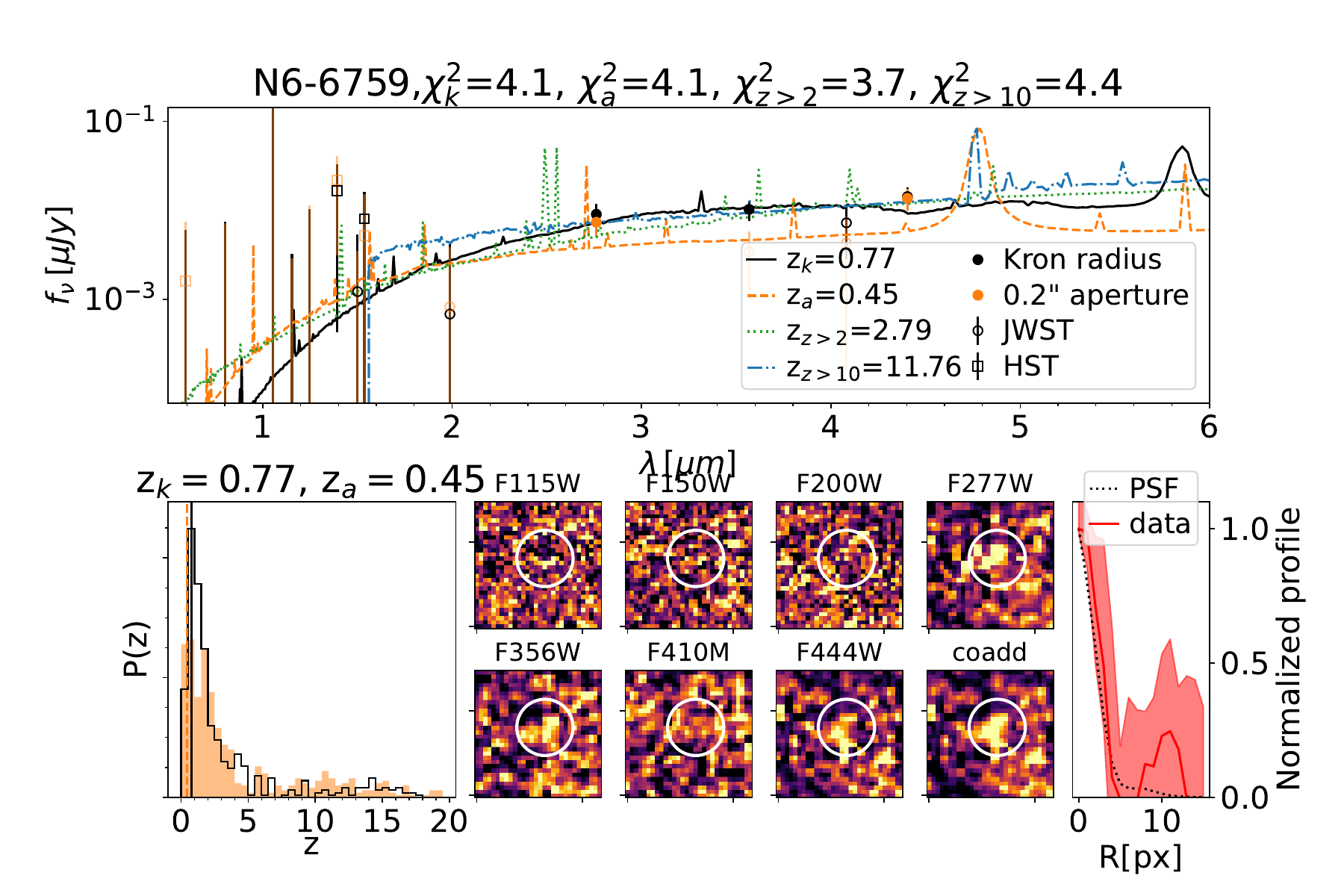}\\
    \includegraphics[trim={20 10 50 40},clip,width=0.44\linewidth,keepaspectratio]{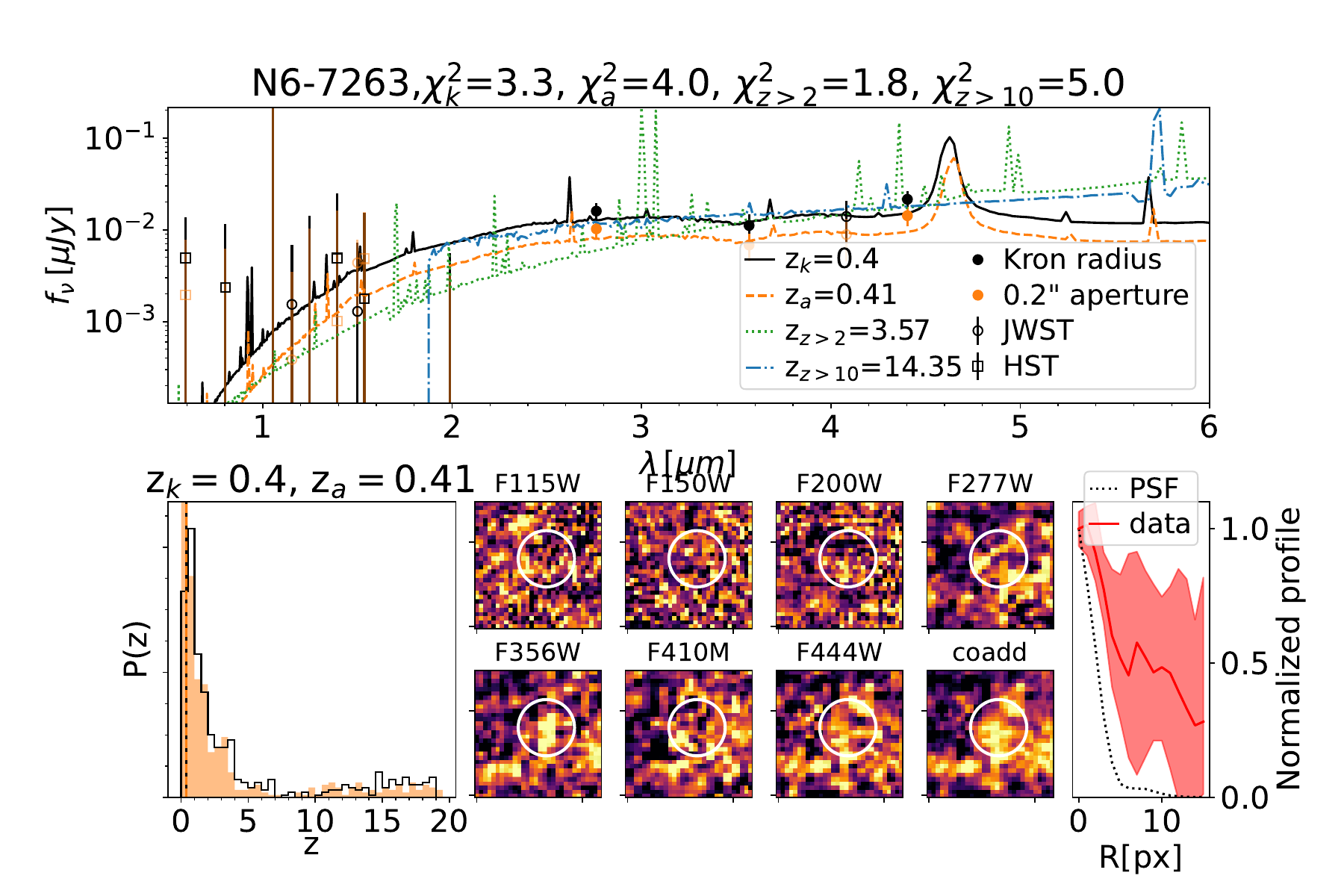}&
    \includegraphics[trim={20 10 50 40},clip,width=0.44\linewidth,keepaspectratio]{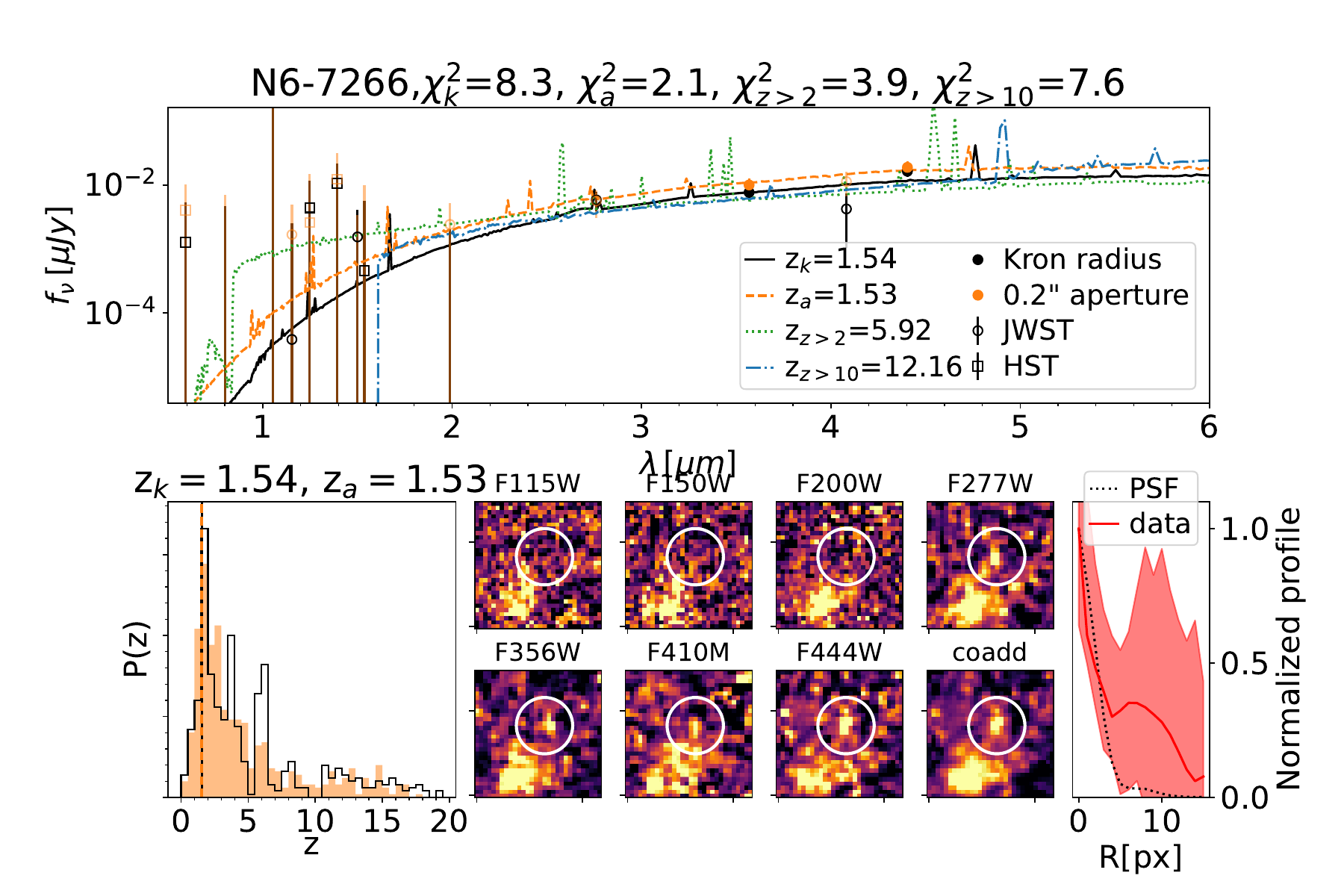}\\
    \includegraphics[trim={20 10 50 40},clip,width=0.44\linewidth,keepaspectratio]{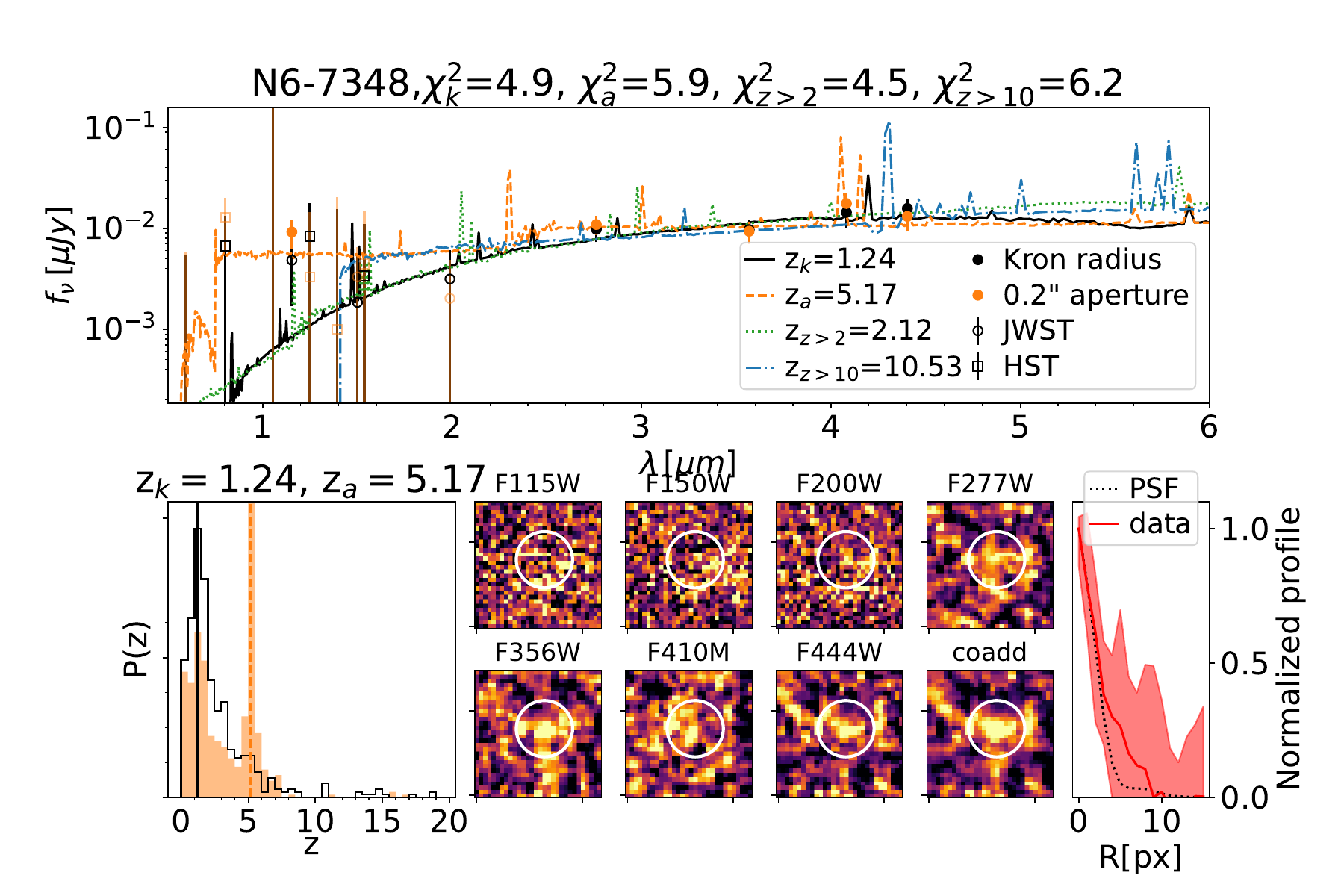}&
    \includegraphics[trim={20 10 50 40},clip,width=0.44\linewidth,keepaspectratio]{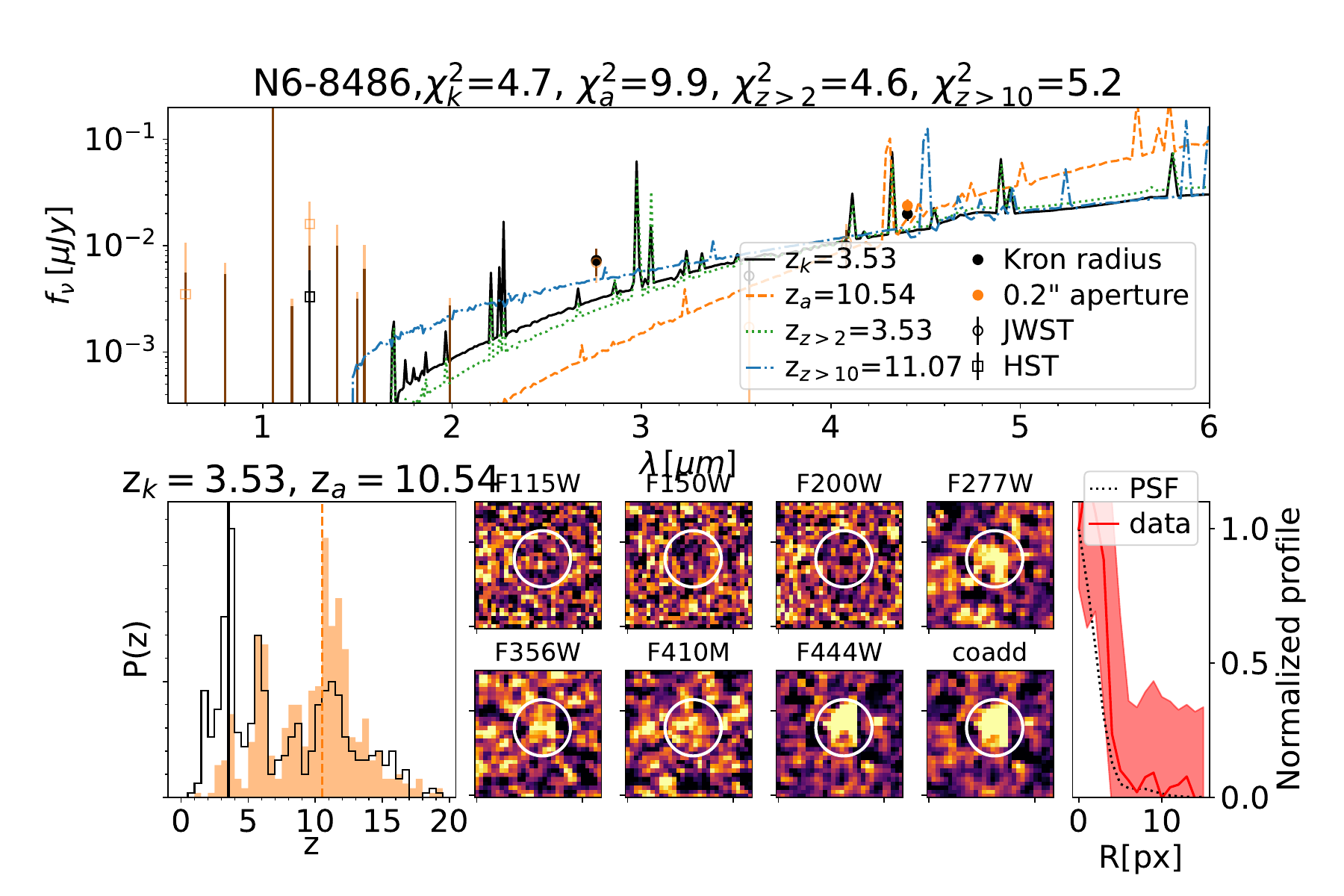}\\
    \includegraphics[trim={20 10 50 40},clip,width=0.44\linewidth,keepaspectratio]{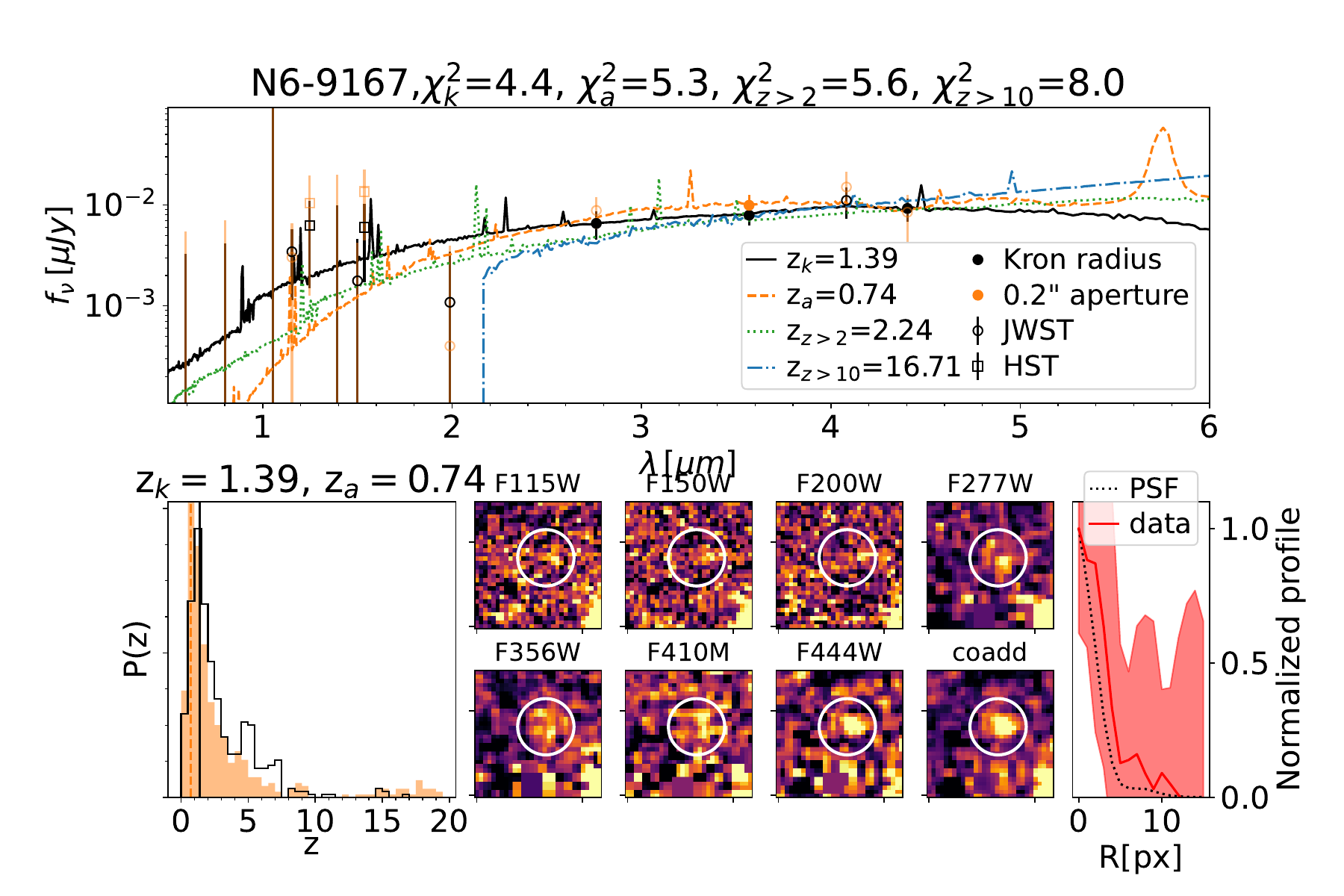} & \\
      \caption{continued.}\\
\end{longfigure}

\end{appendix} 
\end{document}